\documentclass[final,3p]{elsarticle}

\usepackage{amssymb}
\usepackage{amsthm}
\usepackage{natbib}
\usepackage[usenames,dvipsnames]{color}
\usepackage[colorlinks]{hyperref}

\usepackage{graphicx}  
\usepackage{epstopdf}
\usepackage{dcolumn}   
\usepackage{bm}        
\usepackage{amsmath}
\usepackage{amssymb}   
\usepackage{array}
\usepackage[caption=false]{subfig}
\usepackage{exscale,relsize}
\usepackage{lipsum}
\usepackage{pdfpages}
\usepackage{setspace}
\usepackage{bbm}
\usepackage{perpage} 
\usepackage[title,titletoc]{appendix}
\usepackage{bibentry}
\usepackage{soul}

\journal{Physic Reports}
\biboptions{sort&compress,numbers}

\newcommand{\Eq}[1]{Eq.~(\ref{#1})}

\def \ShiftFig{6 mm}

\def \g{\gamma}
\def \c{\chi}
\def \eps{{\varepsilon}}

\def \tr{{\rm Tr}}
\def \sgn{{\rm sgn}}
\def \Real{{\rm Re}}
\def \Imag{{\rm Im}}

\def \ua{{\uparrow}}
\def \da{{\downarrow}}
\def \half{\frac{1}{2}}

\def \bs#1{{\boldsymbol#1}}
\def \tx#1{{\rm#1}}
\def \mc#1{\mathcal{#1}}
\def \mbb#1{\mathbb{#1}}

\def \nd{{^{\vphantom{\dagger}}}}

\def \ket#1{{\,|\,#1\,\rangle\,}}

\def \eps{{\varepsilon}}

\def \w{{\rm w}}
\def \s{{\rm sc}}
\def \k{{k}}
\def \top{{\rm T}}
\def \mk{{-k}}
\def \tw{\textwidth}
\def \hs{\hskip}
\def \nabR{\nabla_{\bs{R}}}
\def\R{\bs{R}(t)}
\def\bmat{\begin{pmatrix}}
\def\emat{\end{pmatrix}}
\def\be{\begin{equation}}
\def\ee{\end{equation}}
\def \Rup{\psi_{\tx{R}\ua}}
\def \Lup{\psi_{\tx{L}\ua}}
\def \Rdn{\psi_{\tx{R}\da}}
\def \Ldn{\psi_{\tx{L}\da}}

\begin{document}
\hypersetup{linkcolor=[rgb]{0,0.47,0.69},citecolor=[rgb]{0,0.47,0.69},urlcolor=[rgb]{0,0.47,0.69}}
\begin{frontmatter}

\title{Time-reversal-invariant topological superconductivity}

\author[CALTECH]{Arbel Haim}
\author[WIS]{and Yuval Oreg}
\address[CALTECH]{Walter Burke Institute for Theoretical Physics and Institute for Quantum Information and Matter, California Institute of Technology, Pasadena, CA 91125, USA}
\address[WIS]{Department of Condensed Matter Physics, Weizmann Institute of Science, Rehovot 7610001, Israel}

\begin{abstract}
A topological superconductor is characterized by having a pairing gap in the bulk and gapless self-hermitian Majorana modes at its boundary. In one dimension, these are zero-energy modes bound to the ends, while in two dimensions these are chiral gapless modes  traveling along the edge. Majorana modes have attracted a lot of interest due to their exotic properties, which include non-abelian exchange statistics. Progress in realizing topological superconductivity has been made by combining spin-orbit coupling, conventional superconductivity, and magnetism. The existence of protected Majorana modes, however, does not inherently require the breaking of time-reversal symmetry by magnetic fields. Indeed, pairs of Majorana modes can reside at the boundary of a \emph{time-reversal-invariant} topological superconductor (TRITOPS). It is the time-reversal symmetry which then protects this so-called Majorana Kramers' pair from gapping out. This is analogous to the case of the two-dimensional topological insulator, with its pair of helical gapless boundary modes, protected by time-reversal symmetry. Realizing the TRITOPS phase will be a major step in the study of topological phases of matter. In this paper we describe the physical properties of the TRITOPS phase, and review recent proposals for engineering and detecting them in condensed matter systems, in one and two spatial dimensions. We mostly focus on extrinsic superconductors, where superconductivity is introduced through the proximity effect. We emphasize the role of interplay between attractive and repulsive electron-electron interaction as an underlying mechanism. When discussing the detection of the TRITOPS phase, we focus on the physical imprint of Majorana Kramers' pairs, and review proposals of transport measurement which can reveal their existence.
\end{abstract}


\end{frontmatter}

\setcounter{tocdepth}{3}
\tableofcontents

\section{Introduction}
\label{sec:Intro}

Topological phases in condensed matter are generally characterized by having unique surface properties which are dictated by the topological properties of the bulk. Probably the best known example of such a topological phase is the quantum Hall effect (QHE)~\cite{Klitzing1980new,Laughlin1981quantized,Thouless1982quantized}, in which gapless chiral edge modes, protected only by topology, reside on the edges of a two-dimensional system and give rise to a quantized Hall conductivity. Remarkably, these edge modes cannot be removed by perturbing the system locally. Their presence is guaranteed by the topology of the band structure characterizing the bulk. This is a manifestation of the so called bulk-edge correspondence~\cite{Wen04}.

Upon considering the presence of various symmetries, a rich variety of topological phases can 
emerge~\cite{schnyder2008classification,kitaev2009periodic,Ryu2010a,Qi2011topological} beyond the example of the QHE. These 
phases also contain gapless boundary\footnote{We refer to the boundary of a system in one, two, and three dimensions as 
an \emph{end}, \emph{edge}, and \emph{surface}, respectively, and use the word \emph{boundary} when not restricting to a 
certain dimensionality.} modes which are related to the topological nature of the bulk, however, they are only protected in 
the presence of some imposed symmetries, and could otherwise become gapped. Here, the paradigmatic example is the topological 
insulator 
(TI)~\cite{Kane2005quantum,Bernevig2006quantum,konig2007quantum,Fu2007topological,Moore2007topological,hsieh2008topological,Hasan2010}
 which in two dimensions can be thought of as two copies of the QHE, related by time-reversal. The gapless edge modes of the 
system are now helical, rather than chiral, and they are protected by the presence of time-reversal symmetry (TRS).

In the case of quadratic Hamiltonians of fermions\footnote{That is Hamiltonians of free fermions or of systems which are described within mean-field theory, such as Bardeen Cooper Schrieffer (BCS) superconductors.} a full topological classification exists~\cite{schnyder2008classification,kitaev2009periodic,Ryu2010a}. It is based on the presence or absence of time-reversal symmetry, particle-hole symmetry and their combination - the chiral symmetry. Time-reversal symmetry is defined as an anti-unitary operator, $\Theta$, which commutes with the Hamiltonian, $[\mc{H},\Theta]=0$. Particle-hole symmetry (PHS) is an anti-unitary symmetry which anticommutes with the Hamiltonian, $\{\mc{H},\Xi\}=0$. The multiplication of these two symmetries forms a unitary operator, $\Pi=\Theta\Xi$. Chiral symmetry is said to exists if $\{\mc{H},\Pi\}=0$. It can be shown~\cite{Ryu2010a} that in the absence of ordinary symmetries (i.e. unitary operators which commute with the Hamiltonian), there can be at most one TRS, and one PHS. It can further be shown, in this case, that acting with the same anti-unitary symmetry twice is equivalent to the identity operator up to a sign, $\Theta^2=\pm1$, $\Xi^2=\pm1$. One then obtains overall ten different symmetry classes, depending on whether each of the symmetries exists and whether it squares to $+1$ or $-1$~\cite{Altland1997}. The dimensionality, together with the symmetry class, determine how many topologically-distinct phases are possible~\cite{schnyder2008classification,kitaev2009periodic,Ryu2010a}.

The topological superconductor (TSC) phase of class D have attracted a lot of attention~\cite{Read2000paired,Kitaev2001unpaired,Alicea2012,Beenakker2013}. This phase, which exists in one and two dimensions, host exotic boundary modes which have received the name Majorana modes due to their self-hermitian nature. In 1d, the boundary mode is a zero-energy bound state (termed Majorana bound state or Majorana zero mode). In two dimensions, the boundary hosts propagating chiral modes (Majorana chiral modes), and Majorana bound states exist inside the core of quantum vortices~\cite{Kopnin1991mutual,Read2000paired,Ivanov2001non}. 
Part of the attention gained by this phase is owed to the potential application of Majorana bound states (MBSs) in topological quantum computation~\cite{Freedman2003topological,Kitaev2003,nayak2008non,Bonderson2008measurement,Karzig2017Scalable}.

Systems belonging to class D lack TRS, and have a PHS which squares to $1$. This PHS symmetry is special as it exists in all superconducting systems; it is an immediate consequence of the mean-field description of the Hamiltonian (see Sec.~\ref{subsec:Intro_top_prot}). Therefore, it cannot truly be broken. This makes its boundary modes - the Majorana bound state (in 1d) and the Majorana chiral mode (in 2d) - extremely robust. In that sense, the class-D topological superconductor can be viewed as the superconducting analog of the QHE.

\begin{figure}
\begin{center}
\begin{tabular}{lcr}
\rlap{\hskip -0.03\textwidth \parbox[c]{\textwidth}{\vspace{-0cm}(a)}}
\includegraphics[clip=true,trim =-2mm 0mm -2mm 0mm,width=0.26\textwidth]{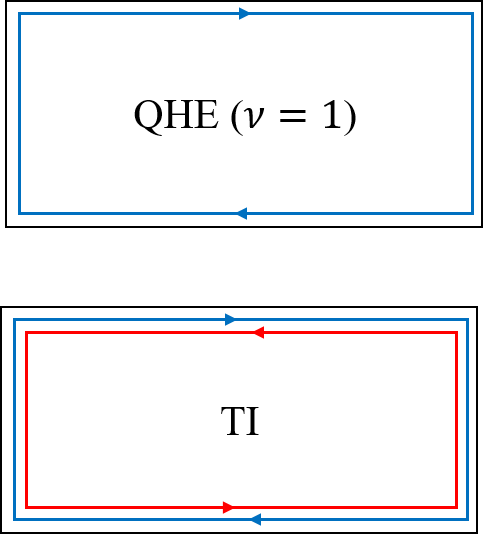}
\hs 2mm
&
\hs 2mm
\rlap{\hskip -0.03\textwidth \parbox[c]{\textwidth}{\vspace{-0cm}(b)}}
\includegraphics[clip=true,trim =-2mm 0mm -2mm 0mm,width=0.26\textwidth]{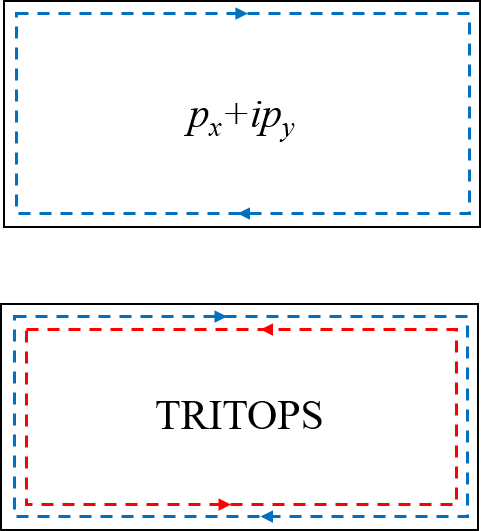}
\hs 2mm
&
\hs 2mm
\rlap{\hskip -0.015\textwidth \parbox[c]{\textwidth}{\vspace{-0cm}(c)}}
\includegraphics[clip=true,trim =0mm -12mm 0mm 0mm,width=0.26\textwidth]{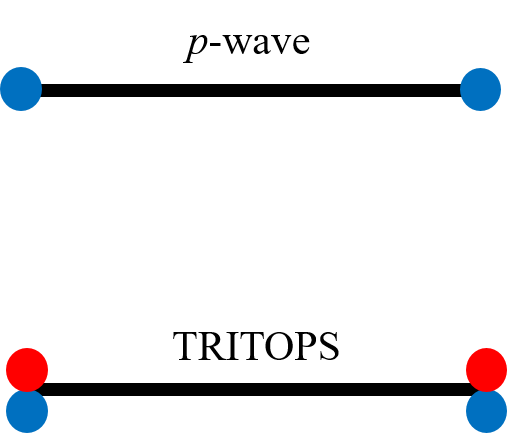}
\end{tabular}
\end{center}
\caption{(a) Schematic description of the integer quantum Hall effect and its time-reversal-symmetric version - the topological insulator (TI). For filling $\nu=1$, the integer quantum hall phase hosts a chiral edge mode. The topological insulator can be viewed as composed of two copies of the quantum Hall phase, related by time-reversal symmetry (TRS). Accordingly, the TI hosts \emph{counter-propagating} helical edge modes. (b) The (TRS-broken) topological $p_x+ip_y$ superconductor and its time-reversal symmetric version - the \emph{time-reversal-invariant topological superconductor} (TRITOPS), in two dimensions. The $p_x+ip_y$ superconductor is a superconducting analog of the quantum Hall effect. The chiral edge modes now become \emph{Majorana} modes (denoted by a dashed line). The TRITOPS phase is equivalent to two time-reversal-related copies of the $p_x+ip_y$ superconductor, with counter-propagating Majorana modes which are protected by TRS. (c) The TRS-broken topological superconductor and the TRITOPS in one dimension. The topologically-protected boundary modes are now zero-energy Majorana bound states. In the TRITOPS phase, they come in spatially overlapping pairs, known as Majorana Kramers pairs.}\label{fig:Intro_QHE_TI_analogy}
\end{figure}

A natural question to ask is then: what is the superconducting analog of the topological insulator? This would be the 
\emph{time-reversal-invariant topological superconductor} (TRITOPS)~\cite{schnyder2008classification,Qi2009time,Fu2010}. This 
phase belongs to symmetry class DIII, which on top of the above-mentioned PHS, also has a TRS, squaring to $-1$. In one or two 
dimensions, it can be described as two copies of a class D Topological superconductor, related by time-reversal 
transformation. Each edge (or end) of this phase hosts a \emph{pair} of time-reversal related Majorana modes, analogous to the 
pair of helical edge modes of the two-dimensional TI. This is depicted in Fig.~\ref{fig:Intro_QHE_TI_analogy}. The TRITOPS 
phase can also exist in 3d, although making an analogy with the TRS-broken phase is no longer possible in this case. 
Interestingly, the B phase of He-$3$ is an example of a 3d time-reversal-invariant topological 
\emph{superfluid}~\cite{volovik2003universe,vollhardt2013superfluid}.

Experimentally realizing the TRITOPS phase in condensed matter systems is a major outstanding challenge in the study of 
topological phases. To date, however, attempts have been focused on realizing the TSC of class D. An important breakthrough in 
this context was the understanding that one can \emph{engineer} this phase by combining relatively well-understood building 
block, such as magnetism, spin-orbit coupling and conventional $s$-wave superconductivity~
\cite{Fu2008superconducting,Fu2009josephson,Sau2010generic,Alicea2010Majorana,Lutchyn2010majorana,Oreg2010helical,Duckheim2010Andreev,Pientka2013Topological,Nadj-Perge2013proposal,Klinovaja2013topological,Braunecker2013interplay,Vazifeh2013self,Lutchyn2018majorana}. These predictions led to a series of experiments which have shown evidence of Majorana bound 
states~\cite{mourik2012signatures,deng2012anomalous,Das2012zero,churchill2013superconductor,Finck2013anomalous,Nadj-Perge2014observation,Pawlak2016probing,Ruby2015end,Albrecht2016exponential,Deng2016Majorana,Chen2017Experimental,gul2018ballistic,Suominen2017zero,Nichele2017scaling,zhang2017quantized}.

In this paper, we describe in detail the properties of the TRITOPS phase and review the various theoretical proposals for its realization and detection in one- and two-dimensional condensed matter systems. Borrowing from the experience of the class-D (TRS-broken) TSC, these proposals will follow the concept of engineering the TRITOPS phase. While magnetism breaks TRS, and should therefore be avoided\footnote{Nevertheless, in some cases magnetism can exist in the system while still having an \emph{emergent} time-reversal symmetry (squaring to $-1$) within the low-energy description of the system~\cite{Schrade2015proximity}.}, the proximity effect and spin-orbit coupling will still be main tools in achieving the goal.  For this reason, we focus on TRITOPS in 1d and 2d, where using the superconducting proximity effect is naturally most relevant. Progress towards realizing a 3d TRITOPS in condensed matter system has nevertheless been made in recent years. The newly discovered superconductor, Cu$_x$Bi$_2$Se$_3$~\cite{Hor2010}, has been suggested to realize several topological superconducting phases~\cite{Fu2010,Brydon2014Odd}, with recent experiments~\cite{Kriener2011bulk,Bay2012superconductivity,matano2016spin} possibly supporting a fully-gapped TRITOPS phase~\cite{Fu2014Odd,Venderbos2016Odd,fu2016superconductivity}.

The structure of this review is as follows. In the remaining part of this section we describe the general properties of the TRITOPS phase. 
In particular, we present simple models to describe the TRITOPS phase and use them to obtain and analyze the Majorana boundary modes. In Sec.~\ref{sec:TopInv}, we construct the general $\mbb{Z}_2$ topological invariant for 1d and 2d systems in class DIII, which determines whether a given system is in the topological phase. In Sec.~\ref{sec:realizations} we review various proposals for realizing the TRITOPS phase, by analyzing their microscopic models. We put emphasis on the role of repulsive electron-electron interactions in these proposals. In Sec.~\ref{sec:Signa} we describe possible experimental signatures of the TRITOPS phase. Specifically, we examine different ways in which probing the Majorana boundary modes can distinguish the system from a topologically-trivial one. We then go on to analyze the braiding properties of Kramers pairs of Majorana bound states (the topological boundary modes of 1d TRITOPS) in Sec.~\ref{sec:Braid}. 
While the exchange of MKPs generally affects the ground-state manifold in a nontrivial way, the resulting unitary operation is nonuniversal unless additional symmetries are present. 
Finally we conclude and discuss future prospects in Sec~\ref{sec:Summary}.

\subsection{Minimal low-energy model}
\label{subsec:Intro_min_model}

\begin{figure}
\begin{center}
\begin{tabular}{lr}
\rlap{\hskip -0.01\textwidth \parbox[c]{\textwidth}{\vspace{-2mm}(a)}}
\includegraphics[clip=true,trim =0mm -12mm 0mm 0mm,width=0.475\textwidth]{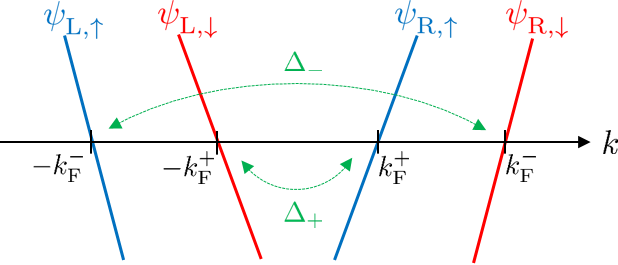}
\hs 3mm
&
\hs 3mm
\rlap{\hskip -0.01\textwidth \parbox[c]{\textwidth}{\vspace{-2mm}(b)}}
\includegraphics[clip=true,trim =0mm 0mm 0mm 0mm,width=0.35\textwidth]{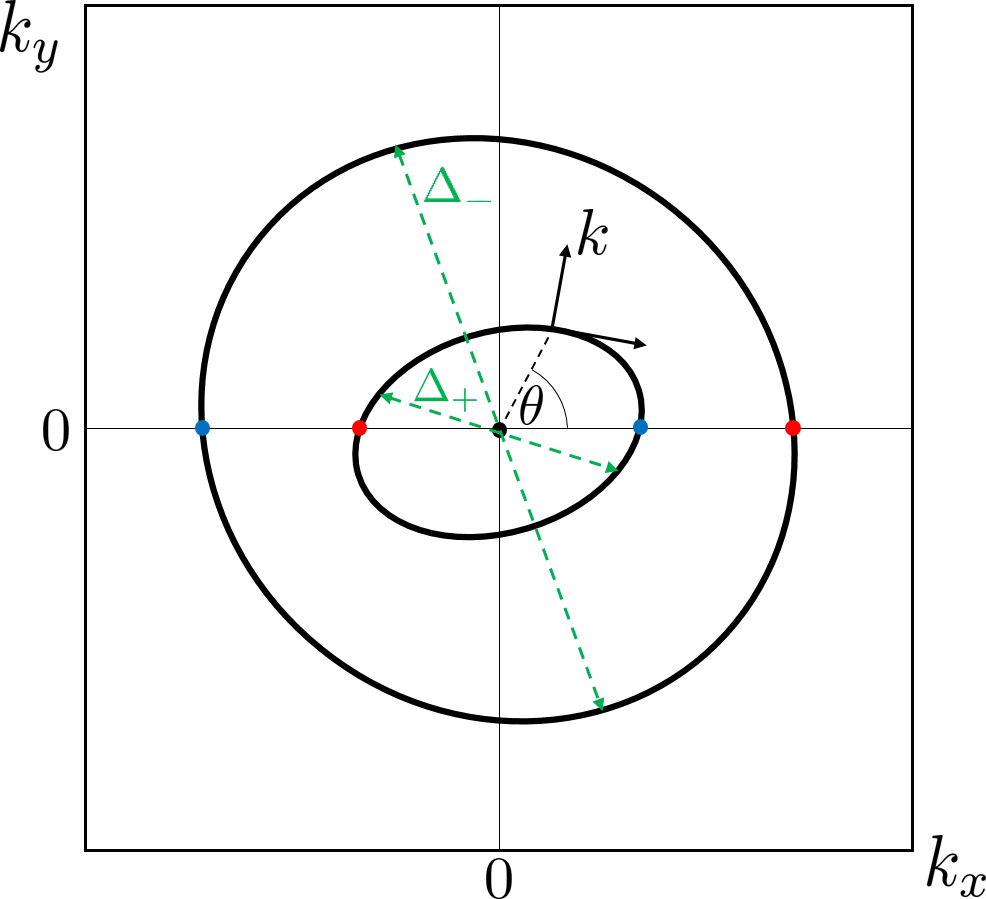}
\end{tabular}
\end{center}
\caption{(a) Dispersion of the one-dimensional low-energy Hamiltonian $H_0$, having two right-moving modes and two left-moving modes [see Eq.~\eqref{eq:minimal_H}]. The Hamiltonian $H_\Delta$ describes induced superconductivity. The pairing potential $\Delta_+$ couples the modes of positive helicity, while $\Delta_-$ couples the modes of negative helicity. The system is in its topologically nontrivial phase when $\sgn(\Delta_+)\sgn(\Delta_-)=-1$ [see Eq.~\eqref{eq:top_inv_1}]. (b) Generalization of the low-energy model to two dimensions. The black contours represents the Fermi surfaces (or Fermi contours) of the normal part of the Hamiltonian in Eq.~\eqref{eq:minimal_H_2d}. The states near each Fermi surface are parameterized using $\theta$ and $k$. The dashed green lines shows the electronic states connected by the pairing term of the Hamiltonian. None of the pairing potentials, $\Delta_+(\theta)$ and $\Delta_-(\theta)$, can change sign as a function of $\theta$ without closing the superconducting gap. The topological criterion then stays the same as in the 1d case, $\sgn(\Delta_+)\sgn(\Delta_-)=-1$. The blue and red dots represent the Fermi points in the 1d system obtained by taking only $\theta=0,\pi$.}\label{fig:TRITOPS_low_E_H_0_spectrum}
\end{figure}

In Sec.~\ref{sec:realizations}, we shall be concerned with various \emph{microscopic} models for systems in the TRITOPS phase. These models would attempt to capture correctly the microscopic properties of these system, such as spin-orbit coupling, electron-electron interactions, and proximity effect.  It is instructive, however, to start by introducing the most simple low-energy model which can describe the TRITOPS phase. First, such a model can serve as a convenient platform for examining the most generic properties of the phase; for example, its boundary modes. Second, as we will see, the low-energy degrees of freedom of all the above-mentioned microscopic models will be described by this much more simple model. This minimal low-energy model for the TRITOPS phase is given by
\begin{equation}\label{eq:minimal_H}
\begin{split}
&H = H_0+H_\Delta,\\
&H_0 = -i\int \tx{d}x
\left\{ v_+  \left[\psi_{\tx{R}\ua}^\dag(x)\partial_x \psi_{\tx{R}\ua}^\nd(x) -\psi_{\tx{L}\da}^\dag(x)\partial_x \psi_{\tx{L}\da}^\nd(x)\right]
+ v_- \left[\psi_{\tx{R}\da}^\dag(x)\partial_x \psi_{\tx{R}\da}^\nd(x) -\psi_{\tx{L}\ua}^\dag(x)\partial_x \psi_{\tx{L}\ua}^\nd(x)\right]\right\},\\
&H_\Delta = \int \tx{d}x \left[\Delta_+ \psi_{\tx{R}\ua}^\dag(x)\psi_{\tx{L}\da}^\dag(x) + \Delta_- \psi_{\tx{L}\ua}^\dag(x)\psi_{\tx{R}\da}^\dag(x) + \tx{h.c.} \right],
\end{split}
\end{equation}
where $\psi_{\tx{R},s}$ ($\psi_{\tx{L},s}$) is an annihilation operator of a right- (left-) moving fermionic mode of spin $s$. Here, $\Delta_+$ and $\Delta_-$ are two induced pairing potentials\footnote{We refer to these pairing potentials as \emph{induced} since 
to have a gapped superconducting phase in 1d, one has to rely on proximity to a higher-dimensional superconductor for inducing superconductivity. The possibility of a \emph{gapless} TRITOPS phase has also been proposed~\cite{Keselman2015gapless,Kainaris2015Emergent,kainaris2017interaction}.}.
$\Delta_+$ describes pairing between the modes of positive helicity, $\Rup^\nd$ and $\Ldn^\nd$, while $\Delta_-$ describes pairing between the modes of negative helicity, $\Lup^\nd$ and $\Rdn^\nd$\footnote{It should be noted that the identification of the index $s=\ua,\da$ as the spin is not crucial. One can instead consider a model with modes $\psi_{\tx{R},1}^\nd(x)$, $\psi_{\tx{R},2}^\nd(x)$ and their time-reversal partners $\psi_{\tx{L},2}^\nd(x)$, $\psi_{\tx{L},1}^\nd(x)$, respectively. While in this case the physical meaning of helicity is absent, one can still refer to $\psi_{\tx{R},1}$ and $\psi_{\tx{L},2}$ as having ``positive helicity", and to $\psi_{\tx{R},2}$ and $\psi_{\tx{L},1}$ as having ``negative helicity", or the other way around.}. Similarly, $v_\pm$ are the velocities of the modes with positive and negative helicity, respectively. The dispersion of $H_0$ is shown in Fig.~\hyperref[fig:TRITOPS_low_E_H_0_spectrum]{\ref{fig:TRITOPS_low_E_H_0_spectrum}(a)}.

We are interested in systems obeying time-reversal symmetry. We define this symmetry operation, $\mbb{T}$, by its form when acting on the annihilation operators and on $c$-numbers\footnote{We distinguish between the TRS operator, $\mbb{T}$, which acts on the second-quantized states, and the operator $\Theta$, which acts on first-quantized states (see Sec.~\ref{subsec:Intro_top_prot} below).},
\begin{equation}\label{eq:TR_operation}
\mathbb{T}\psi_{\tx{R},s}^\nd(x)\mathbb{T}^{-1} = i\sigma^y_{ss'} \psi_{\tx{L},s'}^\nd(x) \hs 4mm ; \hs 4mm
\mathbb{T}\psi_{\tx{L},s}^\nd(x)\mathbb{T}^{-1} = i\sigma^y_{ss'} \psi_{\tx{R},s'}^\nd(x) \hs 4mm ; \hs 4mm
\mathbb{T}i\mathbb{T}^{-1}= -i,
\end{equation}
where $\{\sigma^i\}_{i=x,y,z}$ is the set of Pauli matrices. Namely, TRS reverse the propagation of the particle as well as its spin. The last part of Eq.~\eqref{eq:TR_operation} signifies that $\mbb{T}$ is an \emph{anti-unitary} transformation, taking $c$-numbers to their complex conjugates. Requiring that $H$ obeys time-reversal symmetry, $\mathbb{T}H\mathbb{T}^{-1}=H$, imposes the constraints that both $\Delta_+$ and $\Delta_-$ are real\footnote{Alternatively, the pairing potentials $\Delta_+$ and $\Delta_-$ can be complex numbers having the same phase, $\Delta_\pm=|\Delta_\pm|e^{i\varphi}$, in which case $H$ would be symmetric under a slightly modified TRS, given by $\mbb{T}'\psi_{\tx{R},s}^\nd{\mbb{T}'}^{-1}=ie^{i\varphi}\sigma^y_{ss'} \psi_{\tx{L},s'}^\nd$, $\mbb{T}'\psi_{\tx{L},s}{\mbb{T}'}^{-1}=ie^{i\varphi}\sigma^y_{ss'} \psi_{\tx{R},s'}$.}. Since operating twice with $\mbb{T}$ on an annihilation operator, $\mbb{T}^2\psi_{\rho,s}\mbb{T}^{-2}=-\psi_{\rho,s}$ (for $\rho=\tx{R},\tx{L}$), results in a minus sign, one says that the symmetry $\mbb{T}$ squares to $-1$.

In the absence of inversion symmetry, $H$ is the most general low-energy quadratic Hamiltonian which describes a single-channel 1d system with TRS. If the system also had inversion symmetry, namely symmetry under $x\rightarrow-x$, the Fermi momenta would necessarily be equal, $k_\tx{F}^+=k_\tx{F}^-$, which would allow for additional terms. For example, the term 
$(V\Rup^\dag \Rdn^\nd + V^\ast \Lup^\dag \Ldn^\nd + \tx{h.c.})$ is allowed by TRS, but as long as $k_\tx{F}^+\neq k_\tx{F}^-$, this term is suppressed at low energies due to momentum mismatch.

Since the TRS obeyed by $H$ squares to $-1$, the system belongs to symmetry class DIII of the Altland-Zirnbauer classification~\cite{Altland1997}. The symmetry class determines the number of topologically-distinct phases in which the system can, in principle, be. One-dimensional Hamiltonians in symmetry class DIII are characterized by a $\mathbb{Z}_2$ topological invariant~\cite{schnyder2008classification,kitaev2009periodic,Ryu2010a}, which means that the system can be in one of two topologically-inequivalent phases. What physically distinguishes theses phases is the presence or absence of protected boundary modes - zero-energy Majorana Kramers pairs (MKPs).

\begin{figure}
\begin{center}
\includegraphics[clip=true,trim =0mm 0mm 0mm 0mm,width=0.7\tw]{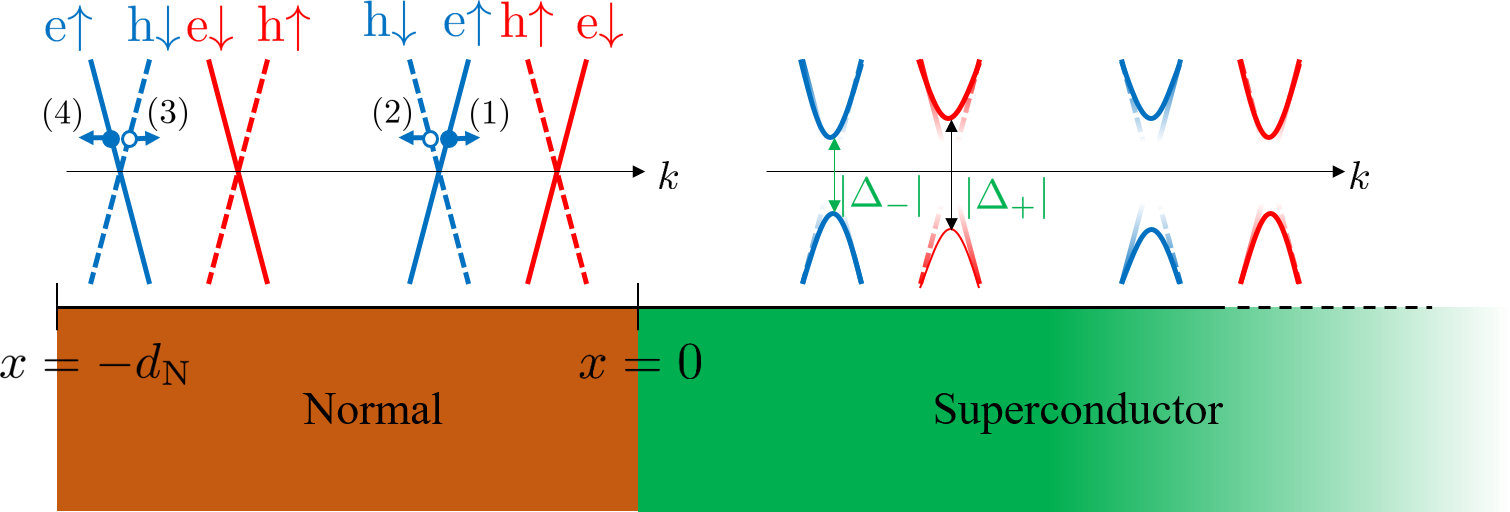}
\end{center}
\caption{Construction for obtaining the criterion for the low-energy Hamiltonian $H=H_0+H_\Delta$ to be in the topologically nontrivial phase [see Eq.~\eqref{eq:minimal_H}]. The semi-infinite region $x>0$ is described by the Hamiltonian $H=H_0+H_\Delta$, while the region $-d_\tx{N}<x\le0$ is described by $H_0$. A spin-$\ua$ electron moving to the right (1) in the normal region goes through a series of scattering events: Andreev reflection to a spin-$\da$ hole moving to the left (2), normal reflection to a spin-$\da$ hole moving to the right (3), Andreev reflection to a spin-$\ua$ electron moving to the left (4), and finally a normal reflection back to a spin-$\ua$ electron moving to the right, thereby returning to (1). For a bound state to exist, the overall phase acquired during this scattering process should be a multiple of $2\pi$. This results in the condition for a zero-energy bound state: $\sgn(\Delta_+)=-\sgn(\Delta_-)$. A second bound state is obtained by considering the time-reversed scattering process, starting with a spin-$\da$ electron moving to the left.}\label{fig:TRITOPS_top_crit}
\end{figure}

As we now show, the topologically trivial phases of $H$ corresponds to the cases $\sgn(\Delta_+)=\sgn(\Delta_-)$, while the topologically non-trivial phase corresponds to the case $\sgn(\Delta_+)=-\sgn(\Delta_-)$~\cite{Qi2010topological}. To see this, let us consider the system in a semi-infinite geometry with a boundary at $x=0$, and look for the condition for the system to have zero-energy modes at the boundary~\cite{Fulga2011scattering}. It is convenient to do so by attaching a normal-metal stub to the system, such that the overall system is described by $H=H_0+H_\Delta$ for $x>0$, and by $H_0$ for $-d_N<x<0$, as depicted in Fig.~\ref{fig:TRITOPS_top_crit}. At the end, the normal-metal stub can be removed by taking $d_N\rightarrow0$.

Let us concentrate on an electron in the normal stub which propagates to the right, towards the NS interface. For energies smaller than the induced pairing potentials (and in particular for zero energy), the electron goes through a series of scattering processes before returning to its original state: (i) Andreev reflection, $e\rightarrow h$, at the NS interface, (ii) normal reflection, $h\rightarrow h$, at the vacuum interface on the left, (iii) Andreev reflection, $h\rightarrow e$ at the NS interface, and finally (iv) normal reflection, $e\rightarrow e$, at the vacuum interface. This is depicted in Fig.~\ref{fig:TRITOPS_top_crit}. For a bound state to exist, the overall phase acquired by the electron during this process should be a multiple of $2\pi$.

To calculate the overall phase, we begin by considering a spin-$\ua$ electron moving to the right and being Andreev reflected at the NS interface into a spin-$\da$ left-moving hole. The Andreev-reflection amplitude for this process is given by $e^{i\varphi_\tx{I}}=\sgn(\Delta_+)e^{-i\cos^{-1}(\eps/|\Delta_+|)}$~\cite{andreev1964thermal}. Notice that since this process involves positive-helicity modes, the expression for the amplitude contains $\Delta_+$. Next, the spin-$\da$ hole propagates towards the $x=-d_\tx{N}$ boundary where it is normally reflected as a spin-$\da$ hole and then propagates back towards the NS interface. In this process it acquires a phase $e^{i\varphi_\tx{II}}=-e^{-i(k^+_\tx{F}-\eps/v_+)d_\tx{N}}e^{-i(k^-_\tx{F}-\eps/v_-)d_\tx{N}}$. The right-moving spin-$\da$ hole is then Andreev reflected into a left-moving spin-$\ua$ electron, this time with an amplitude $e^{i\varphi_\tx{III}}=\sgn(\Delta_-)e^{-i\cos^{-1}(\eps/|\Delta_+|)}$. Finally, it propagates to the left interface and back acquiring a phase $e^{i\varphi_\tx{IV}}=-e^{i(k^-_\tx{F}+\eps/v_-)d_\tx{N}}e^{i(k^+_\tx{F}+\eps/v_+)d_\tx{N}}$. At zero energy, the overall phase gained during the process is simply 
$e^{i(\varphi_\tx{I}+\varphi_\tx{II}+\varphi_\tx{III}+\varphi_\tx{IV})}=-\sgn(\Delta_+)\sgn(\Delta_-)$, which means that for a zero-energy bound state to exist the signs of the pairing potentials need to be opposite.

In reaching this criterion for a zero-energy bound state, we have chosen to track the path of a right-moving spin-$\ua$. Exactly the same criterion is obtained by considering the time-reversed process, starting with a spin-$\da$ electron moving to the left. Namely, In the topological phase there are actually \emph{two} zero energy MBSs at the system's boundary, in accordance with Kramers' degeneracy theorem. These are the so-called Majorana Kramers pair.

Notice that, since $\Delta_+$ and $\Delta_-$ are real numbers, their sign can only change if they go through zero, namely if the energy gap closes. We can thus define a topological invariant for the Hamiltonian at hand~\cite{Qi2010topological},
\begin{equation}\label{eq:top_inv_1}
\nu=\sgn(\Delta_+)\sgn(\Delta_-),
\end{equation}
which takes the value $1$ when the system is trivial, and $-1$ when the system is topological.

It might seemed that the obtained result depends on specific details in our construction. For example, we have implicitly assumed that the NS interface is smooth, and that the normal-metal stub is clean etc. As we demonstrate in Secs.~\ref{subsec:Intro_top_prot} and~\ref{subsec:Intro_top_prot_TRS_anom}, however, once we have established the existence of the zero-energy Majorana Kramers pair, it cannot be removed without closing the bulk gap or breaking the TRS. This means, in particular, that our conclusions do not depend on the specific microscopic details of the system's boundary.

\subsubsection{Triplet versus singlet pairing}
\label{subsubsec:Intro_min_model_Triplet_Singlet}

Some insight into the topological invariant, Eq.~\eqref{eq:top_inv_1}, can be gained by rewriting the superconducting part of the Hamiltonian in the following form
\begin{equation}\label{eq:H_delta_2}
\begin{split}
H_\Delta =& \int \tx{d}x \left\{\Delta_\tx{s}\left[\Rup^\dag(x)\Ldn^\dag(x) - \Rdn^\dag(x)\Lup^\dag(x)\right]
+\Delta_\tx{t}\left[\Rup^\dag(x)\Ldn^\dag(x) + \Rdn^\dag(x)\Lup^\dag(x)\right] +\tx{h.c.} \right\},
\end{split}
\end{equation}
where $\Delta_\tx{s,t}=(\Delta_+ \pm \Delta_-)/2$ are the singlet and triplet pairing potentials respectively. Inserting this in Eq.~\eqref{eq:top_inv_1} results in
\begin{equation}\label{eq:top_inv_2}
\nu=\sgn(\Delta_\tx{s}^2-\Delta_\tx{t}^2).
\end{equation}
Namely, the topological phase ($\nu=-1$) is obtained when the triplet pairing term exceeds in magnitude the singlet pairing term. This formulation will help us understand the role played by short-range electron-electron interactions, when we discuss realizations of TRITOPS in Sec.~\ref{sec:realizations}.

\subsubsection{Two dimensions}
\label{subsubsec:Intro_min_model_2d}

Before moving on, let us generalize the low-energy minimal model of Eq.~\eqref{eq:minimal_H}, and the criterion to be in the topological phase, to the case of two dimensions. This is written most easily in momentum space,
\begin{equation}\label{eq:minimal_H_2d}
H^\tx{2d}=\int_0^{2\pi} \tx{d}\theta\sum_{\nu=\pm}\sum_{k} \left\{ v_\nu(\theta)k \psi^\dag_{\theta,\nu,k} \psi^\nd_{\theta,\nu,k}
+\left[\Delta_\nu(\theta) e^{i\chi_\nu(\theta)}\psi^\dag_{\theta,\nu,k}\psi^\dag_{\theta+\pi,\nu,k} +\tx{h.c.}\right] \right\}.
\end{equation}
This Hamiltonian describes two Fermi surfaces (contours), denoted by $\nu=\pm$, which are gapped by superconductivity. The modes belonging to each Fermi surface are parameterized by the angle $\theta$, as depicted in Fig.~\hyperref[fig:TRITOPS_low_E_H_0_spectrum]{\ref{fig:TRITOPS_low_E_H_0_spectrum}(b)}. The momentum, $k$, is measured from the Fermi surface in the direction perpendicular to the surface. Notice that the one-dimensional model of Eq.~\eqref{eq:minimal_H} is obtained from $H^\tx{2d}$ by keeping only the angles $\theta=0,\pi$, and identifying $\psi_{\tx{R},k,\ua}=\psi_{0,+,k}$, $\psi_{\tx{R},k,\da}=\psi_{0,-,k}$, $\psi_{\tx{L},k,\ua}=\psi_{\pi,-,k}$ and $\psi_{\tx{L},k,\da}=\psi_{\pi,+,k}$.

Under time-reversal symmetry, the fermionic fields transform as $\mbb{T}\psi_{\theta,\nu,k}\mbb{T}^{-1}=e^{-i\chi(\theta)}\psi_{\theta+\pi,\nu,k}$. For the TRS to square to $-1$, we therefore require $\chi_\nu(\theta+\pi)-\chi_\nu(\theta)=(2m+1)\pi$, for integer $m$. Requiring time-reversal symmetry,  $\mbb{T}H^\tx{2d}\mbb{T}^{-1}=H^\tx{2d}$, then translates to the conditions: $v_\nu(\theta)=v_\nu(\theta+\pi)\in\mbb{R}$, $\Delta_\nu(\theta)\in\mbb{R}$. Notice that since $e^{i\chi_\nu(\theta)}\psi^\dag_{\theta,\nu,k}\psi^\dag_{\theta+\pi,\nu,k}=e^{i\chi_\nu(\theta+\pi)}\psi^\dag_{\theta+\pi,\nu,k}\psi^\dag_{\theta,\nu,k}$, it is implied in Eq.~\eqref{eq:minimal_H_2d} that $\Delta_\nu(\theta)=\Delta_\nu(\theta+\pi)$\footnote{Alternatively stated, if one splits $\Delta_\nu(\theta)$ to a part which is periodic in $\pi$ and a part which is antiperiodic, $\Delta_\nu(\theta)=[\Delta_\nu(\theta)+\Delta_\nu(\theta+\pi)]/2+[\Delta_\nu(\theta)-\Delta_\nu(\theta+\pi)]/2$, then the antiperiodic part cancels under the integration over $\theta$ in Eq.~\eqref{eq:minimal_H_2d}.}. We also note that since the Hamiltonian is by assumption completely gapped in the bulk, $\Delta_{\nu}(\theta)$ does not switch sign as a function of $\theta$.

To obtain the criterion for the system to be in the topological phase, we can use the result of the one-dimensional Hamiltonian. We consider the system with open boundary conditions in the $x$ direction and periodic boundary conditions in the $y$ direction, while keeping the system infinite. In the topological phase, there should be helical counter-propagating modes along the edge [see also Fig.~\hyperref[fig:exc_spec_triv_and_top_1d]{\ref{fig:exc_spec_triv_and_top_1d}(d)} below]. In particular, at $k_y=0$, there should be two modes at zero energy. The condition for these to exist can be obtained by setting $k_y=0$  (namely $\theta=0,\pi$) and considering the resulting 1d Hamiltonian. We can then use the 1d result, namely that the signs of $\Delta_+(0)=\Delta_+(\pi)$ and $\Delta_-(0)=\Delta_-(\pi)$ must be opposite. As noted above, $\Delta_\pm(\theta)$ cannot switches sign as a function of $\theta$ while keeping the bulk gap, so one can omit the $\theta$ argument and write~\cite{Qi2010topological}
\be\label{eq:top_inv_2d}
\nu_\tx{2d}=\sgn(\Delta_+)\sgn(\Delta_-),
\ee
similarly to the topological invariant in the 1d case.

\subsection{Lattice model}
\label{subsec:Lattice_model}

The low-energy continuum model of Eq.~\eqref{eq:minimal_H} will help us analyze and understand the microscopic systems which will be introduced in Sec.~\ref{sec:realizations}. Nevertheless, it is beneficial to have also a simple lattice model describing TRITOPS. Such a model will be useful, for example, when we want to numerically simulate the TRITOPS phase (see Appendix~\ref{sec:JJ_lattice}). As we see next, it can also help in understanding the source for the topologically-protected modes - the Majorana Kramers pair.

In constructing the lattice model, we are assisted by the conclusions drawn from the low-energy model of Eq.~\eqref{eq:minimal_H}. First, it tells us that to describe a system in the topological phase, the model has to break inversion symmetry (otherwise one necessarily has $\Delta_+=\Delta_-$). Furthermore, since the pairing potential has to be momentum dependent, one has to go beyond on-site pairing of electrons, and consider at least nearest-neighbor pairing. Inserting these ingredients, we arrive at
\be\label{eq:H_min-latt}
\begin{split}
H_\tx{Latt}= \sum_{n}\left\{-\mu \bs{c}^\dag_{n} \bs{c}_{n} + [\bs{c}^\dag_n(-t + iu\sigma^z)\bs{c}_{n+1}
+\half\Delta_0 \bs{c}^\dag_n i\sigma^y \bs{c}^{\dag\top}_{n} + \half \bs{c}^\dag_n (\Delta'_1 + \Delta''_1\sigma^z)i\sigma^y \bs{c}^{\dag\top}_{n+1} +\tx{h.c.}]\right\},
\end{split}
\ee
where $\bs{c}^\dag_n=(c^\dag_\ua,c^\dag_\da)$ is a vector of electron creation operators. Here, $\mu$ is the chemical potential, $t$ is the hopping parameter, $u$ is a spin-orbit coupling term, $\Delta_0$ is the on-site pairing potential, $\Delta'_1$ is the nearest neighbor 
singlet pairing potential, and $\Delta''_1$ is a nearest neighbor 
triplet-component pairing potential. Under time-reversal symmetry, the electron annihilation operators transform as  $\mbb{T}\bs{c}_n\mbb{T}^{-1}= i\sigma^y \bs{c}_n$. 
One can then check that this model is time-reversal symmetric, $\mbb{T}H_\tx{Latt}\mbb{T}^{-1}=H_\tx{Latt}$ , so long as the coefficients $t$, $u$, $\Delta_0$, $\Delta'_1$, and $\Delta''_1$ are real.

We note that, for the sake of simplicity, we chose to consider a model having a $U(1)$ spin-rotation symmetry. Indeed, the Hamiltonian is invariant under $\bs{c}_n\rightarrow \exp(i\theta\sigma^z) \bs{c}_n$, namely this model conserves $\sigma^z$. This, however, is not an essential property for a model describing TRITOPS. One can, for example, 
add a SOC term, $iu'\bs{c}^\dag_n\sigma^{x,y}\bs{c}_{n+1}+\tx{h.c.}$ (or alternatively an additional triplet-component pairing term, $\Delta'\bs{c}^\dag_n\sigma^{x,y}i\sigma^y \bs{c}^{\dag\top}_{n+1} +\tx{h.c.}$), thereby breaking spin-rotation symmetry completely, while still keeping TRS intact. As long as such a change does not close the superconducting gap, the topological properties of the model are not affected. While in most cases a $\sigma^z$-conserving model will suffice to describe the relevant physics, there are some cases where breaking this symmetry will introduces new features to the phenomenology. An example of this will be the Josephson junctions between two superconductors in the TRITOPS phase (see Sec.~\ref{subsec:Signa_JJ}).

We can make a connection with the low-energy model of Eq.~\eqref{eq:minimal_H} by going to momentum space and linearizing the lattice model near the Fermi points. Assuming periodic boundary conditions, we can write the lattice Hamiltonian as
\be\label{eq:H_min-latt_k}
\begin{split}
H_\tx{Latt}= \sum_{k}\left\{ \xi_\ua(k) c^\dag_{k\ua} c_{k\ua} + \xi_\da(k) c^\dag_{k\da} c_{k\da}
+ [\Delta(k)c^\dag_{k\ua} c^\dag_{-k\da} +\tx{h.c.}]\right\},
\end{split}
\ee
where $\xi_s(k)=-\mu -2t\cos(ka) + 2u\sin(ka)\sigma^z_{ss}$ and $\Delta(k)=\Delta_0 + \Delta'_1\cos(ka) + \Delta''_1\sin(ka)$. Here, $c_{ks}=L^{-\half}\sum_k e^{ina}c_{ns}$, with $L$ being the number of sites in the system, and $a$ the lattice constant.

The four Fermi points, defined by $\xi_\ua(\mp k^\pm_\tx{F})=0$ and $\xi_\da(\pm k^\mp_\tx{F})=0$ [see also Fig.~\hyperref[fig:TRITOPS_low_E_H_0_spectrum]{\ref{fig:TRITOPS_low_E_H_0_spectrum}(a)}], are given by $\pm k_\tx{F}^\pm$ with $k_\tx{F}^\pm a=\cos^{-1}[-\mu/(2\sqrt{t^2+u^2})]\mp \lambda$ , where $\tan\lambda=u/t$. Assuming the pairing potential is small compared with the bandwidth, $4\sqrt{t^2+u^2}$ (weak pairing limit), we can linearize the Hamiltonian by approximating $\psi_{\tx{R},\ua,k} \simeq c_{k^{+}_{\rm F}+k,\uparrow}$, $\psi_{\tx{L},\da,k} \simeq c_{-k^{+}_{\rm F}+k,\downarrow}$, $\psi_{\tx{R},\da,k} \simeq c_{k^{-}_{\rm F}+k,\downarrow}$, and $\psi_{\tx{L},\ua,k} \simeq c_{-k^{-}_{\rm F}+k,\uparrow}$, which results in
\begin{equation}\label{eq:H_lin}
\begin{split}
H^\tx{lin}_\tx{Latt} = \sum_{k} &\Big\{ vk\sum_{s=\ua,\da} (\psi^\dag_{\tx{R},s,k}\psi^\nd_{\tx{R},s,k}-\psi^\dag_{\tx{L},s,k}\psi^\nd_{\tx{L},s,k})
+ ( \Delta_+ \psi^\dag_{\tx{R},\ua,k}\psi^\dag_{\tx{L},\da,-k}
+\Delta_-\psi^\dag_{\tx{L},\ua,k}\psi^\dag_{\tx{R},\da,-k}+{\rm h.c.} ) \Big\}.
\end{split}
\end{equation}
where the pairing potentials are given by $\Delta_+=\Delta(k_{\rm F}^{+})$ and $\Delta_-=\Delta(-k_{\rm F}^{-})$, and the Fermi velocity is given by $v=|\partial_k\xi_s(k^\pm_\tx{F})|=2\sqrt{t^2+u^2-(\mu/2)^2}$. This Hamiltonian is indeed the momentum representation of the low-energy model, Eq.~\eqref{eq:minimal_H}, with $v_+=v_-=v$.

We can thus use the topological invariant, Eq.~\eqref{eq:top_inv_1}, to obtain a condition for $H_\tx{Latt}$ to be in the topological phase in the weak pairing limit. For $\mu=0$, for example, this results in the condition $|\Delta'_1\sin\lambda + \Delta''_1\cos\lambda|>|\Delta_0|$. Notice that to be in the topological phase one needs either a nonvanishing (and large enough) $\Delta''_1$~\cite{Dumitrescu2013topological}  or nonvanishing $u$ and $\Delta'_1$~\cite{Wong2012majorana,Zhang2013time}. This conclusion holds also for $\mu\neq0$. In Sec.~\ref{sec:TopInv} we derive a general expression for the topological invariant of a Hamiltonian which goes beyond the weak pairing limit, and can be applied, in particular, to the lattice model Eq.~\eqref{eq:H_min-latt}.

\subsubsection{TRITOPS as two copies of spinless $p$-wave superconductor}
\label{subsubsec:Intro_lattice_2_copy_p_wave}

It was mentioned earlier that the TRITOPS phase can be thought of as composed of two copies of the spinless $p$-wave superconductor. To see this connection, let us go back to the real-space representation, Eq.~\eqref{eq:H_min-latt}, and focus on the special case of $u=\Delta_0=\Delta'_1=0$. By making the transformation $\tilde{c}_n=\exp(i\pi\sigma_y/4)c_n$, the Hamiltonian can be written as~\cite{Dumitrescu2013topological}
\be
\begin{split}
H_\tx{Latt}= \sum_{s=\ua,\da}\sum_{n}\left\{-\mu \tilde{c}^\dag_{n,s} \tilde{c}^\nd_{n,s} +[-t\tilde{c}^\dag_{n,s} \tilde{c}^\nd_{n+1,s}
+\Delta''_1\tilde{c}^\dag_{n,s} \tilde{c}^\dag_{n+1,s} +\tx{h.c.}]\right\},
\end{split}
\ee
which is indeed two copies (denoted by $s=\ua,\da$) of the Kitaev chain model~\cite{Kitaev2001unpaired}, describing a spinless $p$-wave superconductor. The two copies are related by time-reversal symmetry, which takes $\tilde{c}_{n\ua}\rightarrow\tilde{c}_{n\da}$ and $\tilde{c}_{n\da}\rightarrow-\tilde{c}_{n\ua}$.

To access the zero-energy Majorana Kramers pair (which are present when the system is in the topological phase), we consider the system with open boundary conditions. Following Kitaev~\cite{Kitaev2001unpaired}, we focus on the case $\mu=0$, $\Delta'_1=-t$. In this case the Hamiltonian takes a very revealing form,
\be
H_\tx{Latt}^\tx{obc}=it\sum_{s=\ua,\da}\sum_{n=1}^{L-1}\tilde\beta_{n,s}\tilde\alpha_{n+1,s}
\ee
where we have defined the Majorana operators $\alpha_{n,s}=\tilde{c}^\nd_{n,s}+\tilde{c}^\dag_{n,s}$ and $\beta_{n,s}=(\tilde{c}^\nd_{n,s}-\tilde{c}^\dag_{n,s})/i$, which obey the commutation relations, $\{\alpha_{n,s},\beta_{m,s}\}=0$ and $\{\alpha_{n,s},\alpha_{m,s}\}=\{\beta_{n,s},\beta_{m,s}\}=2\delta_{mn}$. Importantly, the Majorana operators at left end of the chain, $\alpha_{1\ua}$, $\alpha_{1\da}$, and at the right end of the chain, $\beta_{L\ua}$, $\beta_{L\da}$, do not appear in the Hamiltonian. They therefore commute with the Hamiltonian and constitute zero-energy modes of the Hamiltonian. In terms of the original electron creation and annihilation operators these modes are given by
\be
\begin{split}
&\gamma_\tx{L}\equiv\alpha_{1\ua} = (c^\nd_{1\ua} - c^\nd_{1\da} + c^\dag_{1\ua} - c^\dag_{1\da})/\sqrt{2}
\hs 4mm ; \hs 4mm
\gamma_\tx{R}\equiv\beta_1\ua = (c^\nd_{L\ua} - c^\nd_{L\da} - c^\dag_{L\ua} + c^\dag_{L\da})/(\sqrt{2}i)\\
&\tilde\gamma_\tx{L}\equiv\alpha_{1\da} = (c^\nd_{1\ua} + c^\nd_{1\da} + c^\dag_{1\ua} + c^\dag_{1\da})/\sqrt{2}
\hs 4mm ; \hs 4mm
\tilde\gamma_\tx{R}\equiv\beta_{1\da} = (c^\nd_{L\ua} + c^\nd_{L\da} - c^\dag_{L\ua} - c^\dag_{L\da})/(\sqrt{2}i).
\end{split}
\ee
Notice that under time-reversal symmetry, $\gamma_\tx{L,R}\rightarrow\tilde\gamma_\tx{L,R}$, $\tilde\gamma_\tx{L,R}\rightarrow-\gamma_\tx{L,R}$, namely the zero modes $\gamma_\tx{L}$ and $\tilde\gamma_\tx{L}$ are in fact a Kramers pair (and similarly $\gamma_\tx{R}$ and $\tilde\gamma_\tx{R}$). 

\subsubsection{Lattice model in two dimensions}
\label{subsubsec:Intro_lattice_2d}

Finally, the lattice model of Eq.~\eqref{eq:H_min-latt} can be generalized to two dimensions in a straight forward manner,
\be\label{eq:H_min-latt_2d}
\begin{split}
H^\tx{2d}_\tx{Latt}= \sum_{n_x,n_y}&\left\{-\mu c^\dag_\bs{n} c_\bs{n} +\left[\half\Delta_0 c^\dag_\bs{n} i\sigma^y c^{\dag\top}_\bs{n} +\tx{h.c.}\right] \phantom{\sum_{\alpha,\beta\in \{x,y\}}}\right.\\
 &+ \left.\sum_{\alpha,\beta\in \{x,y\}}\left[c^\dag_\bs{n}(-t\delta_{\alpha\beta} + iu_{\alpha\beta}\sigma^\beta)c_{\bs{n}+\hat{e}_\alpha} + \half c^\dag_\bs{n} (\Delta'_1\delta_{\alpha\beta} + \Delta''_{1,\alpha\beta}\sigma^\beta)i\sigma^y c^{\dag\top}_{\bs{n}+\hat{e}_\alpha} +\tx{h.c.}\right]\right\},
\end{split}
\ee
where $n=(n_x,n_y)$ runs over the sites of a square lattice, and where $\hat{e}_{\alpha=x,y}$ are unit vectors in the Cartesian directions.  The spin-orbit coupling term, $u_{\alpha\beta}$, and the triplet pairing term, $\Delta''_{1,\alpha\beta}$, are now $2\times2$ matrices. For example, the case $u_{\alpha\beta}=u\eps_{\alpha\beta}$ corresponds to a Rashba spin-orbit coupling term, with $\eps_{\alpha\beta}$ the antisymmetric tensor, and the case $\Delta''_{1,\alpha\beta}=\Delta''_1\delta_{\alpha\beta}$ corresponds to a $p_x\pm ip_y$ superconducting term.

~\\

\subsection{Topological and symmetry protection}
\label{subsec:Intro_top_prot}

We have mentioned above that once the system is in its topological phase, local perturbations to the Hamiltonian do not affect its topological properties, and in particular, do not split the zero-energy Majorana Kramers pairs. For quadratic Hamiltonians [such as those considered in Eqs.~\eqref{eq:minimal_H} and \eqref{eq:H_min-latt}] this can be understood by examining the symmetries of the excitation spectrum.

Consider a general quadratic Hamiltonian of fermions, written in the following Bogoliubov-de Gennes (BdG) form,
\be\label{eq:H_quad_gnrl}
H = \sum_{i,j=1}^N h_{ij}a^\dag_i a^\nd_j + \half (\Delta_{ij}a^\dag_i a^\dag_j + \tx{h.c.}) = \frac{1}{2}\sum_{i,j=1}^N
(a_i^\dag,a_i)\mc{H}_{ij}
\begin{pmatrix}a_j \\ a_j^\dag\end{pmatrix} + \frac{1}{2}\tr(h)
\hs 3mm ; \hs 3mm
\mc{H}_{ij}=\begin{pmatrix} h_{ij} & \Delta_{ij} \\ \Delta^\ast_{ji} & -h_{ji} \end{pmatrix}
\ee
where $a_i$ is a fermionic annihilation operator for a single-particle states denoted by the index $i$ (which generally includes spin), $h$ is a hermitian matrix, and $\Delta$ is a matrix which can be chosen antisymmetric, without loss of generality, thanks to the anticommutation relations of the fermions. As a result, the so-called BdG Hamiltonian, $\mc{H}$, automatically obeys
\be\label{eq:PHS_BdG}
\Xi\mc{H}\Xi^{-1} = -\mc{H} \hs 4mm ; \hs 4mm \Xi=\tau_x\mc{K},
\ee
where $\tau_{\alpha=x,y,z}$ is the set of Pauli matrices operating in the space connecting particles and holes, and where $\mc{K}$ stands for the complex conjugation operator (namely $\mc{K}\mc{H}\mc{K}=\mc{H}^\ast$ and $\mc{K}\vec{w}=\vec{w}^\ast$). The operator $\Xi$ constitute a particle-hole symmetry of $\mc{H}$, that is an anti-unitary transformation which anticommutes with the Hamiltonian.

This particle-hole symmetry has consequences regarding the spectrum of $\mc{H}$. It dictates that the eigenstates of $\mc{H}$ come in pairs having opposite energies,
\be\label{eq:PHS_spect}
\mc{H}\vec{w} = \eps \vec{w}
\hs 4mm \Leftrightarrow \hs 4mm
\mc{H}(\tau_x\vec{w}^\ast) = -\eps (\tau_x\vec{w}^\ast),
\ee
as follows from Eq.~\eqref{eq:PHS_BdG}. Notice that the above PHS is quite artificial; it resulted from our construction of $\mc{H}$. Indeed, we did not have to assume anything about $H$, except for being quadratic. The relation, Eq.~\eqref{eq:PHS_spect}, is the first-quantized representation of the statement that if $\Gamma^\dag$ is an excitation of the system, $[H,\Gamma^\dag]=\eps\Gamma^\dag$, then $\Gamma$ can be viewed as an excitation with opposite energy, $[H,\Gamma]=-\eps\Gamma$, which follows immediately by hermitian conjugation. The connection between the representations is made by identifying $\Gamma^\dag=\sum_i u_i a_i^\dag + v_i a_i$, where $\vec{w}^\top=(\vec{u}^\top,\vec{v}^\top)$.

Now let us assume that, in addition to the above PHS, the system also obey a TRS. In terms of second-quantized operators, this means that the Hamiltonian is invariant under $\mbb{T}H\mbb{T}^{-1}=H$, where $\mbb{T}$ is an anti unitary symmetry defined by $\mbb{T}a_i\mbb{T}^{-1}= U_{ij}a_j$, $\mbb{T}i\mbb{T}^{-1}=-i$, with $U$ being an $N\times N$ unitary matrix. Let us further assume that the TRS squares to $-1$, namely that applying it twice takes $a_i\rightarrow -a_i$. This translates to $UU^\ast=-1$. Together with the unitarity of $U$, this means that $U$ is antisymmetric (In the case examined in Sec.~\ref{subsec:Lattice_model}, for example, one had $U=i\sigma_y$ ).

Demanding that $H$ is invariant under this TRS yields the following condition on the first-quantized BdG Hamiltonian,
\be\label{eq:TRS_BdG}
\Theta\mc{H}\Theta^{-1} = \mc{H} \hs 4mm ; \hs 4mm \Theta=\mc{U}\mc{K},
\ee
where $\mc{U}$ is a $2N\times2N$ unitary antisymmetric matrix given by 
$\mc{U}=\tx{diag}\{U,U^\ast\}$. From Eq.~\eqref{eq:TRS_BdG}, it follows that if $\vec{w}$ is an eigenstate of $\mc{H}$, then so is $\Theta\vec{w}$, namely
\be\label{eq:Kramers_degen}
\mc{H}\vec{w} = \eps \vec{w}
\hs 4mm \Leftrightarrow \hs 4mm
\mc{H}(\mc{U}\vec{w}^\ast) = \eps (\mc{U}\vec{w}^\ast).
\ee
To show that $\vec{w}$ and $\mc{U}\vec{w}^\ast$ are linearly independent, we use the fact that the TRS squares to $-1$ (and therefore $\mc{U}$ is antisymmetric) to arrive at\footnote{The indices $I$ and $J$ run over $2N$ states, which include both particle states and hole states, to be distinguished from $i$ and $j$, which run only over the $N$ particle states.}
\be\label{eq:Kram_pair_orth}
\vec{w}^\dag\cdot(\mc{U}\vec{w}^\ast) = \sum_{I,J=1}^{2N}u^\ast_I \mc{U}_{IJ} u^\ast_J = \sum_{I,J=1}^{2N}u^\ast_I (-\mc{U}_{JI}) u^\ast_J = -\vec{w}^\dag\cdot(U\vec{w}^\ast),
\ee
namely $\vec{w}^\dag(\mc{U}\vec{w}^\ast)=0$, which means that $\vec{w}$ and $\mc{U}\vec{w}^\ast$ are not only linearly independent, but in fact orthogonal. This two-fold degeneracy of the excitation spectrum is known as Kramers degeneracy, and we shall refer to such a degenerate pair of states as a Kramers pair.

\begin{figure}
\begin{center}
\begin{tabular}{cccc}
\rlap{\hskip -0.01\textwidth \parbox[c]{\textwidth}{\vspace{0\textwidth}(a)}}
\includegraphics[clip=true,trim =0mm -2mm 0mm 0mm,width=0.15\tw]{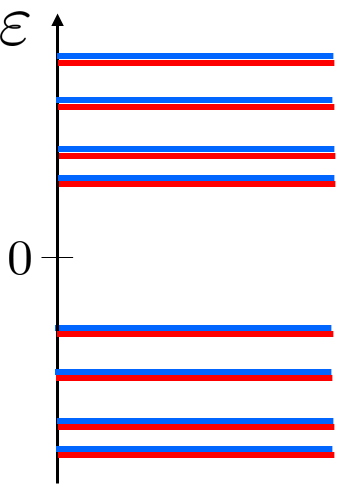}
\hs 0cm & \hs 0cm
\rlap{\hskip -0.01\textwidth \parbox[c]{\textwidth}{\vspace{0\textwidth}(b)}}
\includegraphics[clip=true,trim =0mm -2mm 0mm 0mm,width=0.15\tw]{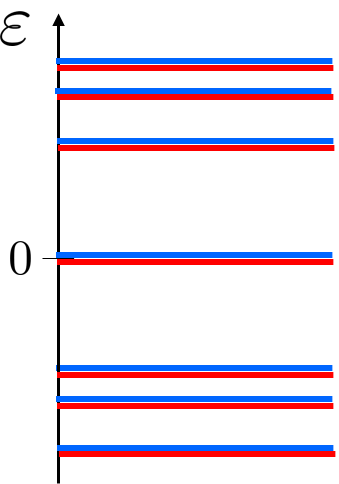}
\hs 0.5cm & \hs 0.5cm
\rlap{\hskip -0.01\textwidth \parbox[c]{\textwidth}{\vspace{0\textwidth}(c)}}
\includegraphics[clip=true,trim =0mm 0mm 0mm 0mm,width=0.245\tw]{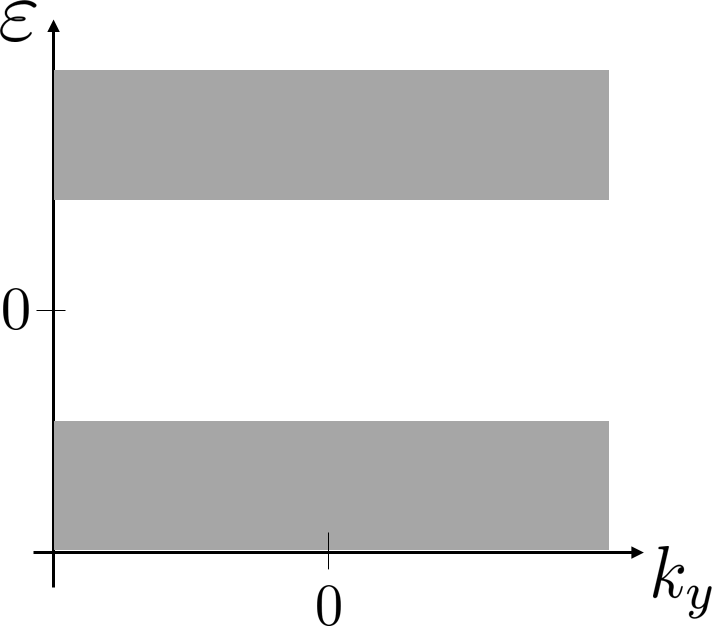}
&
\rlap{\hskip -0.01\textwidth \parbox[c]{\textwidth}{\vspace{0\textwidth}(d)}}
\includegraphics[clip=true,trim =0mm 0mm 0mm 0mm,width=0.245\tw]{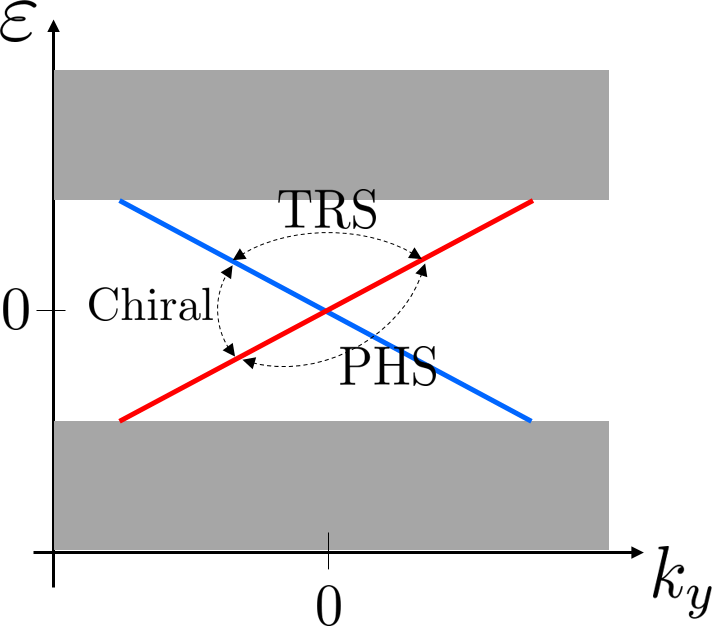}
\end{tabular}
\end{center}
\caption{(a-b) Two types of spectra for a semi-infinite (single boundary) 1d system with particle-hole symmetry (PHS) and time-reversal symmetry (TRS). Each eigenstate is simultaneously a part of a degenerate pair and of an opposite-energy pair. (a) a system in the trivial phase. The spectrum is completely gapped. (b) a system in the topological phase. While the system is gapped in its bulk, at the end of the system there is a (Kramers) pair of zero-energy Majorana states. The two spectra cannot be adiabatically connected to each other without breaking either PHS or TRS. (c,d) Two types of spectra for a semi-infinite 2d system with PHS and TRS. The system has open boundary conditions in the $x$ direction and periodic boundary conditions in the $y$ direction. The gray regions represent the bulk states. In the trivial phase (c) the spectrum is completely gap, with no states crossing zero, while in the topological phase there are two counter-propagating states crossing zero energy. TRS relates states with opposite momenta and equal energy, and PHS relates states with opposite momenta and opposite energy. The combination of these symmetries - the chiral symmetry - relates states at the same momenta and opposite energies. The crossing 
cannot be moved or gapped without breaking either TRS or PHS. \label{fig:exc_spec_triv_and_top_1d}}
\end{figure}

The combination of PHS and TRS dictates that a single isolated Kramers pair of zero-energy states cannot be gapped. To see this, consider a semi-infinite one-dimensional system with a gapped bulk, and having the above-mentioned symmetries. Two types of spectra are possible for such a system, shown in Fig.~\ref{fig:exc_spec_triv_and_top_1d}. The spectrum for a system in the trivial phase, without zero-energy end states, is shown in Fig.~\hyperref[fig:exc_spec_triv_and_top_1d]{\ref{fig:exc_spec_triv_and_top_1d}(a)}, while the spectrum for a system in the topological phase, having a Kramers pair of zero modes, is shown in Fig.~\hyperref[fig:exc_spec_triv_and_top_1d]{\ref{fig:exc_spec_triv_and_top_1d}(b)}. The crucial point to notice is that these two spectra cannot be smoothly connected without breaking either PHS or TRS. Indeed, since any eigenstate must be part of a degenerate pair (TRS), as well as a part of an opposite-energy pair (PHS), the Majorana Kramers pair at zero energy cannot be removed.

One might wonder what happens to the spectrum when the system nevertheless goes through a phase transition from the topological to the trivial phase. To answer that, we must consider the system with both ends having open boundary conditions (keeping the system infinite). In the topological phase, there necessarily exists another Majorana Kramers pair at the other end of the system. The spectrum then has overall \emph{two} pairs of zero-energy states. For an infinite system, the bulk energy gap which imposes a finite correlation length, prevents any local perturbation from connecting these pairs of states, which are located infinitely apart. When the system goes through a topological phase transition, on the other hand, the bulk gap closes, allowing for hybridization of the Majorana Kramers pairs at both ends, and consequently splitting of them to finite energy. When the bulk gap reopens in the trivial phase, the spectrum resembles that of Fig.~\hyperref[fig:exc_spec_triv_and_top_1d]{\ref{fig:exc_spec_triv_and_top_1d}(a)}.

\subsection{Time-reversal anomaly}
\label{subsec:Intro_top_prot_TRS_anom}

More insight into the Majorana Kramers pair and its topological protection can be gained from the so-called \emph{time-reversal anomaly}~\cite{Qi2009time,chung2013time}. This anomaly is the statement that, locally, time-reversal symmetry anticommutes with the fermion-number parity, As we will see, this anomaly 
assures that a \emph{single} Majorana Kramers pair cannot exist by itself; it must be accompanied by a second pair at the other end of the system. A consequence of this is that the MKP cannot be removed by any local perturbation, since it can only be removed together with its partner on the other (far away) end of the system.

Consider a system in the TRITOPS phase. In order to focus on a single MKP, let us consider the system in a semi-infinite geometry, such that it has open boundary conditions at its left end. The MKP is described by the Majorana operators $\gamma_\tx{L}$ and $\tilde\gamma_\tx{L}$. Under TRS, these transform as $\mbb{T}\gamma_\tx{L}\mbb{T}^{-1}=\tilde\gamma_\tx{L}$ and $\mbb{T}\tilde\gamma_\tx{L}\mbb{T}^{-1}=-\gamma_\tx{L}$. We can construct the following regular fermion, $f^\nd_\tx{L}=(\gamma_\tx{L}+i\tilde\gamma_\tx{L})/2$, $f_\tx{L}^\dag=(\gamma_\tx{L}-i\tilde\gamma_\tx{L})/2$. Under TRS, this fermion transforms in a very special way,
\be
\mbb{T}f^\nd_\tx{L}\mbb{T}^{-1} = if^\dag_\tx{L}
\hs 5mm ; \hs 5mm
\mbb{T}f^\dag_\tx{L}\mbb{T}^{-1} = -if^\nd_\tx{L},
\ee
namely, TRS changes the occupation of the fermion, 
$f^\dag_\tx{L}f^\nd_\tx{L}\rightarrow1-f^\dag_\tx{L}f^\nd_\tx{L}$.

Since both $\gamma_\tx{L}$ and $\tilde\gamma_\tx{L}$ are zero modes of the Hamiltonian, the ground state is doubly degenerate with the two states related by $f^\dag_\tx{L}$,
\be\label{eq:GS_deg_left}
|G_1\rangle
\hs 5mm ; \hs 5mm
|G_2\rangle = f^\dag_\tx{L}|G_1\rangle.
\ee
While in our description of the system the total fermion number $N_\tx{F}$  is not conserved, the fermion-number parity, $(-1)^{N_\tx{F}}$, is conserved. We can therefore choose the (many-body) eigenstates of the Hamiltonian to be states of definite fermion-number parity. Let $|G_1\rangle$ have, without loss of generality, even fermion parity, then $|G_2\rangle$ necessarily has odd fermion parity. Therefore, at low energies the fermion-number parity of the system is determined by the occupation of the $f_\tx{L}$ fermion\footnote{For a (semi-infinite) system described by a quadratic Hamiltonian this is true not only at low energies. Indeed, in this case all many-body eigenstates come in degenerate opposite-parity pairs which are distinguished by whether the fermion $f_\tx{L}$ is occupied or empty.}, namely $(-1)^{N_\tx{F}}=2f^\dag_\tx{L}f^\nd_\tx{L}-1=i\gamma_\tx{L}\tilde\gamma_\tx{L}$. From the transformation of $f_\tx{L}$ under time-reversal, it then follows that the fermion parity and the TRS anticommute\cite{Qi2009time}
\be\label{eq:TR_anomaly}
\mbb{T}(-1)^{N_\tx{F}}=-(-1)^{N_\tx{F}}\mbb{T},
\ee
which is refereed to as the time-reversal anomaly~\cite{Qi2009time}.

Apparently, Eq.~\eqref{eq:TR_anomaly} seems to contradict the fact that TRS should clearly commute with the total number of fermions in the system, $N_\tx{F}$. To resolve this we must consider a closed system, with open boundary conditions both on the left and on the right. The existence of a second MKP at the right end of the system, $\gamma_\tx{R}, \tilde\gamma_\tx{R}$, then saves us from having a contradiction. The total fermion-number parity is now determined by the occupations of both $f_\tx{L}$ and $f_\tx{R}=\gamma_\tx{R}+i\tilde\gamma_\tx{R}$,
\be
(-1)^{N_\tx{F}}=(i\gamma_\tx{L}\tilde\gamma_\tx{L})(i\gamma_\tx{R}\tilde\gamma_\tx{R}),
\ee
where $P_\tx{L}\equiv i\gamma_\tx{L}\tilde\gamma_\tx{L}$ and $P_\tx{R}\equiv i\gamma_\tx{R}\tilde\gamma_\tx{R}$ determine the fermion parity at the left and right ends of the system, respectively. Notice that since the system is assumed to be gapped to fermionic excitations in the bulk, this distinction is well defined. While TRS locally anticommutes with fermion parity on the left and on the right, $\{\mbb{T},P_\tx{L}\}=\{\mbb{T},P_\tx{R}\}=0$, it commutes with the total fermion parity.

Finally, for systems which conserve one component of the total spin (e.g. $S^z$), an additional interesting phenomenon exist, which is related to the time-reversal anomaly. In the TRITOPS phase of such systems, the ground states exhibit a non-zero spin expectation value near the two ends of the systems, such that each end accumulates a spin $\pm1/4$~\cite{Nakosai2013majorana,Keselman2013inducing,Nakosai2014theoretical,Keselman2015gapless}. This is very different than the case of a time-reversal-invariant system in the trivial phase. There, since the ground state (which is unique) is time-reversal symmetric, the total spin is zero.

To understand this phenomenon, let us again consider a semi-infinite system with a single boundary. The zero-energy fermion, $f^\dag_\tx{L}$, either creates or destroys a spin $1/2$\footnote{For a quadratic system conserving $S^z=\half\sum_{i}a^\dag_{i\ua}a^\nd_{i\ua}-a^\dag_{i\da}a^\nd_{i\da}$, the operator $f^\dag_\tx{L}$ must have one of the following forms: $f^\dag_\tx{L}=\sum_i u_{i\ua} a_{i\ua}^\dag + v_{i\da} a^\nd_{i\da}$ or $f^\dag_\tx{L}=\half\sum_i u_{i\da} a_{i\da}^\dag + v_{i\ua} a^\nd_{i\ua}$, which correspond to creating and destroying a spin $1/2$, respectively. For non-quadratic systems, $f^\dag_\tx{L}$ will be dressed by particle-hole and Cooper-pair excitations, however these would not change the property, $[S^z,f^\dag_\tx{L}]=\pm\half f^\dag_\tx{L}$, so long as $f^\dag_\tx{L}$ has a nonzero single-particle weight. Notice the difference in notation with respect to Sec.~\ref{subsec:Intro_top_prot_TRS_anom}; there the index $i$ included the spin, while here we have separated the spin from the rest of the degrees of freedom.}, namely the ground states $|G_1\rangle$ and $|G_2\rangle=f^\dag_\tx{L}|G_1\rangle$ differ by a spin $\pm1/2$. On the other hand, these states are related by TRS (which flips the occupation of $f_\tx{L}$), and therefore they must have opposite spin. The expectation value of the total spin, $S^z$, in the two ground states is thus $\pm1/4$. Since $f^\dag_\tx{L}$ has support only near the boundary, this spin is localized at the system's boundary. For a system with two boundaries, there are altogether four ground states, corresponding to the different possibilities of having spin $\pm1/4$ at each end.

\section{Topological invariants}
\label{sec:TopInv}

Above, in Sec.~\ref{subsec:Intro_min_model}, 
we introduced a minimal low-energy model, Eq.~\eqref{eq:minimal_H}, and obtained the condition for this model to be in the TRITOPS phase. This condition is expressed as a $\mathbb{Z}_2$ topological invariant whose value can only change during a topological phase transition, accompanied by a closing of the gap. In this section, we go beyond the low-energy model, and derive such an expression for a more general quadratic Hamiltonian. In Sec.~\ref{subsec:TopInv_wind_num} we obtain the topological invariant for a 1d system. We follow the derivation presented in Refs.\cite{Keselman2013inducing,Haim2016NoGo}; alternative approaches can be found in Refs.~\cite{Qi2010topological,Fulga2011scattering,budich2013topological,Gaidamauskas2014majorana,Mandal2015counting}. We then make use of the 1d result in order to construct the topological invariant for a 2d system in Sec.~\ref{subsec:TopInv_1dto2d}. Finally, in Sec~\ref{subsec:TopInv_WeakPair} we present a simplified version of the topological invariant, correct in the limit where superconductivity is weak~\cite{Qi2010topological}.

\subsection{One dimension topological invariant}
\label{subsec:TopInv_wind_num}

Consider a general translationally-invariant quadratic Hamiltonian in 1d. Written in momentum space, this is given by
\begin{equation}\label{eq:H_1d}
H = \sum_\k \left[ \psi_\k^\dag h_\k \psi_\k + \frac{1}{2}(\psi_\k^\dag \Delta_\k \psi_\mk^{\dag\top} + \tx{h. c.})\right],
\end{equation}
where for every $\k$, $\psi^\dag_\k$ is a $2M$-dimensional vector of fermionic creation operators which includes all degrees of freedom within a unit cell, including spin, transverse modes, sublattice sites, atomic orbitals etc. (the dimension of the vector $\psi^\dag_\k$ has to be even due to the spin degree of freedom). Here, $h_\k$ and $\Delta_\k$ are $2M\times 2M$ matrices operate on these degrees of freedom, describing the normal and pairing parts of the Hamiltonian, respectively\footnote{Namely, the expression $\psi_\k^\dag h_\k \psi_\k$ is shorthand writing for $\sum_{\alpha,\beta=1}^M\psi_{\k;\alpha}^\dag h^{\phantom{\dag}}_{\k;\alpha\beta} \psi_{\k;\beta}^{\phantom{\dag}}$.}. Due to the anticommutativity of the fermionic operators, $\psi^\dag_\k$, one can choose the pairing matrix to obey $\Delta^\top_\mk=-\Delta_\k$\footnote{To be more specific, if we write $\Delta_\k$ as composed of two parts, $\Delta_\k=(\Delta_\k+\Delta_\mk^\top)/2+(\Delta_\k-\Delta_\mk^\top)/2$, then the first term cancels upon summing over $k$ in Eq.~\eqref{eq:H_1d}. We are therefore allowed to consider only the second part (which obeys $\Delta^\top_\mk=-\Delta_\k$) to begin with.}, where the superscript stands for the transpose of a matrix.

We are interested in Hamiltonians belonging to class DIII, that is obeying a PHS that squares to $+1$, and a TRS that squares to $-1$. Let us start by constructing the TRS. As in Sec.~\ref{sec:Intro}, we define it by its operation on the second-quantized annihilation operators, and on $c$ numbers,
\begin{equation}\label{eq:wind_num_TR}
\mathbb{T}\psi_\k \mathbb{T}^{-1}=\mc{T} \psi_\mk  \hskip 5mm ; \hskip 5mm
\mathbb{T} i \mathbb{T}^{-1}=-i,
\end{equation}
where $\mc{T}$ is a unitary $2M\times 2M$ matrix. Namely, this transformation reverse the momentum, as well as operating on the degrees of freedom of the unit cell (such as on the spin). The requirement that $\mathbb{T}$ squares to $-1$ means that operating with it twice should take $\psi_\k\to-\psi_\k$, which results in the condition $\mc{T}\mc{T}^\ast=-1$.  Enforcing this TRS on the Hamiltonian, $\mbb{T}H\mbb{T}^{-1}=H$, translates to conditions on $h_\k$ and $\Delta_\k$,
\begin{subequations}\label{eq:wind_num_TR_on_matrices}
\begin{align}
&\mc{T}^\dag h^\ast_\mk \mc{T} = h_\k \label{eq:wind_num_TR_on_h}, \\ 
&\mc{T}^\dag \Delta^\ast_\mk \mc{T}^\ast = \Delta_\k.\label{eq:wind_num_TR_on_delta}
\end{align}
\end{subequations}

To obtain the PHS, we first need to construct the BdG form of the Hamiltonian. Defining the Nambu spinor as $\Psi^\dag_k=(\psi^\dag_k , \psi^\top_\mk\mc{T})$, we can write\footnote{Notice the construction of the BdG Hamiltonian here is somewhat different than in Eq.~\eqref{eq:H_quad_gnrl} of Sec.~\ref{subsec:Intro_top_prot}, as here we incorporate the matrix $\mc{T}$ (whose role is played by $U$ in Sec.~\ref{subsec:Intro_top_prot}) into the definition of the basis.}
\begin{equation}\label{eq:wind_num_H_BdG}
H=\frac{1}{2}\sum_\k \Psi_\k^\dag \mc{H}_\k \Psi_\k
\hskip 4mm ; \hskip 4mm
\mc{H}_\k= \tau^z \otimes h_\k + \tau^x \otimes \Delta_\k \mc{T},
\end{equation}
where we have made use of the relations given in Eq.~\eqref{eq:wind_num_TR_on_matrices}. Notice in particular that Eq.~\eqref{eq:wind_num_TR_on_delta} together with $\Delta^\top_\mk=-\Delta_\k$ imply that $\Delta_\k \mc{T}$ is a hermitian matrix (and therefore also $\mc{H}_k$). In terms of the BdG Hamiltonian, $\mc{H}_k$, the PHS is expressed as $\Xi\mc{H}_k\Xi^{-1}=-\mc{H}_{-k}$, with $\Xi=\tau^y\otimes\mc{T}\mc{K}$, and where as before $\mc{K}$ stands for complex conjugation. TRS is expressed in these terms as $\Theta\mc{H}_k\Theta^{-1}=\mc{H}_{-k}$, with $\Theta=\mathbbm{1}_{2\times2}\otimes\mc{T}\mc{K}$. Notice that our choice of the Nambu spinor makes the chiral symmetry of $\mc{H}_k$ particularly very apparent,  
$\tau^y\mc{H}_k\tau^y=-\mc{H}_k$.

In the next step, we write the Hamiltonian in the basis which diagonalizes the chiral symmetry, where it has the following block off-diagonal form:
\begin{equation}\label{eq:H_rot}
e^{i\frac{\pi}{4}\tau^x} \mc{H}_\k e^{-i\frac{\pi}{4}\tau^x} = \begin{pmatrix} 0 & Q_\k \\ Q_\k^\dag & 0 \end{pmatrix}.
\end{equation}
To make progress in analyzing the matrix $Q_k$, we consider its singular value decomposition, $Q_k=U^\dag_\k D_\k V_\k$, where $U_\k$, $V_\k$ are unitary matrices and $D_\k$ is a square diagonal matrix with non-negative elements on its diagonal. By squaring both sides of Eq.~\eqref{eq:H_rot}, it becomes apparent that the elements of $D_\k$ are the positive eigenvalues of $\mc{H}_\k$. As long as $\mc{H}_\k$ is gapped, we can therefore further conclude that the diagonal elements of $D_\k$ are all nonzero.

We are allowed to make a smooth deformation to $\mc{H}_k$, as long as this does not cause its gap to close and keep its symmetries intact, since such a deformation cannot affect the topological invariant. We do that by smoothly deforming $D_\k$ to the identity matrix without closing the gap (which is possible due to all its diagonal elements being positive). This in turn deforms the Hamiltonian, $\mc{H}_\k \rightarrow \tilde{\mc{H}}_\k$ , such that $\tilde{\mc{H}}_\k$ has two flat bands at energies $\pm 1$ (in the appropriate units), but the same eigenstates as $\mc{H}_\k$ (and therefore the same symmetries). The deformed Hamiltonian, $\tilde{\mc{H}}_\k$, is given by Eq.~\eqref{eq:H_rot} with $Q_\k$ replaced by $\tilde{Q}_\k = U_\k^\dag V_\k$. The advantage of making the above deformation is that $\tilde{Q}_k$, is a unitary matrix, therefore obeying the spectral theorem. We now use the spectrum of $\tilde{Q}_k$ to construct he topological invariant. 

The spectrum of $\tilde{Q}_k$ is constraint by the TRS of the Hamiltonian, $\mc{T}^\dag\tilde{\mc{H}}^\ast_\mk\mc{T}=\tilde{\mc{H}}_\k$, which implies 
\be\label{eq:Q_k_TRS}
\mc{T}^\dag \tilde{Q}^\ast_\mk \mc{T} = \tilde{Q}^\dag_\k.
\ee
Together with the unitarity of $\tilde{Q}_\k$, Eq.~\eqref{eq:Q_k_TRS} dictates that its eigenstates and eigenvalues come in pairs, related by TRS. 
Namely, if $| \alpha \rangle$ is an eigenstate of $\tilde{Q}_k$ with eigenvalue $e^{i\theta}$, then  $\mc{T}^\dag | \alpha \rangle^\ast$ is an eigenstate of $\tilde{Q}_\mk$ with the same eigenvalue. We can therefore divide spectrum of $\tilde{Q}_\k$ into two sectors, $\{ \exp(i\theta_{n,\k}^\tx{I}) \}_{n=1}^M$ and $\{ \exp(i\theta_{n,\k}^\tx{II}) \}_{n=1}^M$, related by time reversal, $\theta_{n,\k}^\tx{II}=\theta^\tx{I}_{n,\mk}$. This means, in particular, that at the time-reversal-invariant momenta, $k=0,\pi$, the eigenvalues of $\tilde{Q}_k$ come in Kramers' degenerate pairs.

\begin{figure}
\begin{center}
\begin{tabular}{lr}
\rlap{\hskip 0\textwidth \parbox[c]{\textwidth}{\vspace{-0.35\textwidth}(a)}}
\includegraphics[clip=true,trim =0mm 0mm 0mm 0mm,width=0.29\textwidth]{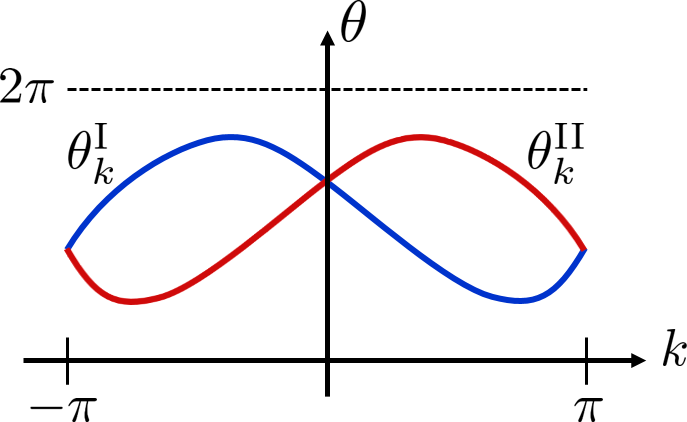}
\hskip 0.1\textwidth
&
\hskip 0.1\textwidth
\rlap{\hskip 0\textwidth \parbox[c]{\textwidth}{\vspace{-0.35\textwidth}(b)}}
\includegraphics[clip=true,trim =0mm 28.3mm 0mm 0mm,width=0.29\textwidth]{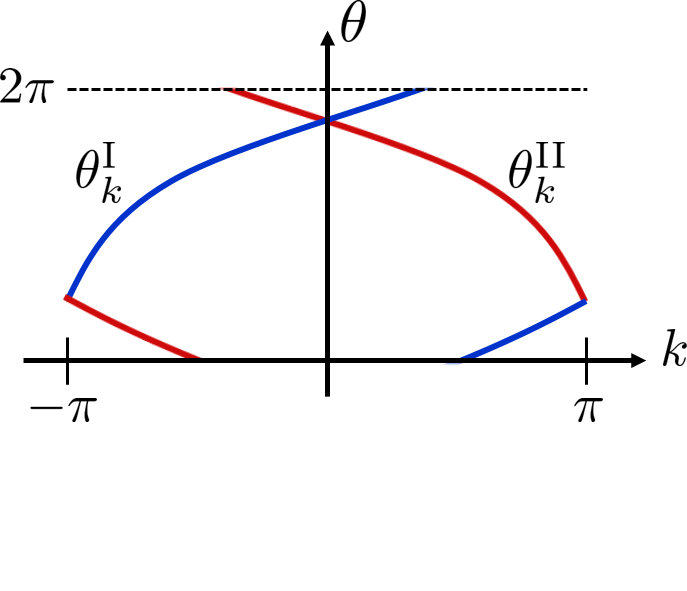}
\end{tabular}
\end{center}
\caption{Examples of spectra of the unitary matrix $\tilde{Q}_\k$ [see Eq.~\eqref{eq:H_rot} and below], corresponding to (a) a topologically-trivial case, and (b) a topologically-nontrivial case. Due to time-reversal symmetry, the eigenvalues of $\tilde{Q}_\k$ come in pairs, $\{\exp(i\theta^{\rm I}_{n,\k})\}_{n=1}^M$ and $\{\exp(i\theta^{\rm II}_{n,\k})\}_{n=1}^M$ (blue and red lines, respectively), where $\theta^{\rm II}_{n,k}=\theta^{\rm I}_{n,-k}$. The parity of the winding number of the blue (or red) line gives the class-DIII topological invariant in 1d. In the examples shown here $M=1$.
}\label{fig:TRITOPS_necess_int_theta_spectra}
\end{figure}

Consider now the spectrum of $\tilde{Q}_\k$ as a function of $k\in[-\pi,\pi]$. From the TRS of the spectrum it follows that the number of pairs of degenerate states at a given value $\theta$ cannot change by an odd number during an adiabatic change which leaves the gap of $\mc{H}_k$ open. The parity of the number of degenerate pairs is therefore a topological invariant. Alternatively stated, the topological invariant is given by
\begin{equation}\label{eq:Winding_parity}
\nu = (-1)^W \hskip 5mm ;\hskip 5mm
W = \sum_{n=1}^M \frac{1}{2\pi} \int_{k=-\pi}^{k=\pi} \tx{d}\theta_{n,\k}^\tx{I},
\end{equation}
namely, the parity of the sum of windings of $\{\theta_{n,\k}^\tx{I}\}_{n=1}^M$ (A similar results is guaranteed if one considers $\{\theta_{n,\k}^\tx{II}\}_{n=1}^M$ instead). Figure~\ref{fig:TRITOPS_necess_int_theta_spectra} presents examples of trivial and topological spectra of $\tilde{Q}_\k$, for the case $M=1$.

\subsection{From one to two dimensions}
\label{subsec:TopInv_1dto2d}

Having obtained an expression for the class DIII topological invariant in 1d, we now show how it can be used for obtaining the invariant for the 2d case as well. Consider the two-dimensional BdG Hamiltonian, $\mc{H}_{k_x,k_y}$, defined by Eqs.~\eqref{eq:H_1d} and \eqref{eq:wind_num_H_BdG} with the replacement $k_x\to(k_x,k_y)$. We argue that the $\mathbb{Z}_2$ topological invariant in 2d is given by\footnote{Note that most generally $k_x$ and $k_y$ should be considered as coordinates along the reciprocal primitive vectors, $\bs{G}_{1,2}$, in units of $|\bs{G}_{1,2}|/2\pi$, respectively.}
\begin{equation}\label{eq:2d_top_inv}
\nu_\tx{2d} = \nu[\mc{H}_{k_x,k_y=0}]\cdot\nu[\mc{H}_{k_x,k_y=\pi}],
\end{equation}
where $\nu[\mc{H}_{k_x,k_y=0}]$ is the topological invariant of the 1d Hamiltonian obtained by setting $k_y=0$  in $\mc{H}_{k_x,k_y=0}$ (and similarly for $\mc{H}_{k_x,k_y=\pi}$). 
Notice that, since we are only concerned with the time-reversal-invariant momenta, $k_y=0,\pi$, the Hamiltonians $\mc{H}_{k_x,k_y=0,\pi}$ obey the TRS and PHS described below Eq.~\eqref{eq:wind_num_H_BdG}, making $\nu[\mc{H}_{k_x,k_y=0,\pi}]$ well defined.

To derive Eq.~\eqref{eq:top_inv_2d}, let us consider a semi-infinite system with periodic boundary conditions in the $y$ direction, and an edge along the line $x=0$ [see also Fig.~\hyperref[fig:exc_spec_triv_and_top_1d]{\ref{fig:exc_spec_triv_and_top_1d} (c,d)}]. 
The non-trivial phase is characterized by having an odd number of pairs of helical edge modes\footnote{While an even number of pairs of helical edge modes can generally be completely gapped by a symmetry-allowed perturbation, an \emph{odd} number of pairs dictates the survival of at least one pair.}. At the edge of such a system, at every energy inside the bulk gap, there must therefore be an odd number of Kramers' pairs of states 
[degenerate states at momenta $k_y$ and $-k_y$ related by TRS as depicted in Fig.~\hyperref[fig:exc_spec_triv_and_top_1d]{\ref{fig:exc_spec_triv_and_top_1d} (d)}].
Let us further focus on the middle of the gap, which is necessarily at $E=0$ due to particle-hole symmetry. We can infer the number of pairs of helical modes crossing the gap from the number of degenerate Kramers' pairs of states at $E=0$.

At $k_y=0$, the number of Kramers' pairs is equal to the $\mathbb{Z}_2$ invariant of the corresponding DIII one-dimensional Hamiltonian $\mathcal{H}_{k_x, k_y=0}$.  The same is true at the other time-reversal-invariant momentum, $k_y=\pi$. At momenta away from $k_y=0,\pi$, the number of zero-energy Kramers' pairs must be even due to time-reversal and chiral-symmetries\footnote{The chiral symmetry dictates that for each state with energy $E=0$ and momentum $k_y$, there is another state with the same energy and momentum. TRS dictates that for each such state there is another state with energy $E=0$ and momentum $-k_y$.}.  Therefore, the parity of the total number of Kramers' pairs at $E=0$ (which equals the number of pairs of gapless helical modes) is given by $\nu[\mc{H}_{k_x,k_y=0}]\cdot\nu[\mc{H}_{k_x,k_y=\pi}]$, which is indeed the right hand side of Eq.~\eqref{eq:2d_top_inv}. Finally, we note that using the same arguments, one can show that the topological invariant can equivalently be written as $\nu_\tx{2d} = \nu[\mc{H}_{k_x=0,k_y}]\cdot\nu[\mc{H}_{k_x=\pi,k_y}]$.

Examining Eq.~\eqref{eq:2d_top_inv}, we see that there are two ways in which one can arrive at a topologically-trivial index, $\nu_\tx{2d}=1$: either $\nu[\mc{H}_{k_x,k_y=0}]=\nu[\mc{H}_{k_x,k_y=\pi}]=1$, or $\nu[\mc{H}_{k_x,k_y=0}]=\nu[\mc{H}_{k_x=,k_y\pi}]=-1$. The latter scenario is rather interesting, since it implies that the edge spectrum (still assuming periodic boundary conditions in the $y$ direction) of the system contains two pairs of helical gapless modes crossing the gap, one pair at $k_y=0$ and one pair at $k_y=\pi$. In the absence of perturbations connecting states at $k_y=0$ with states at $k_y=\pi$, these helical modes are protected by the topological invariants $\nu_{y,0}\equiv\nu[\mc{H}_{k_x,k_y=0}]$ and $\nu_{y,\pi}\equiv\nu[\mc{H}_{k_x=,k_y\pi}]$. Accordingly, $\nu_{y,0}$ and $\nu_{y,\pi}$  are called ``weak'' topological indices, as they only predict the existence of gapless edge modes in the presence of translation invariance\footnote{It has been shown, however, that weak phases are protected even in the presence of disorder, as long as translation invariance is maintained \emph{on average}~\cite{Ringel2012strong,Fu2012Topology,Fulga2014statistical}}. Similarly, given periodic boundary conditions in the $x$ direction and open boundary conditions in the $y$ direction, the ``weak'' topological indices, $\nu_{x,0}\equiv\nu[\mc{H}_{k_x=0,k_y}]$ and $\nu_{x,\pi}\equiv\nu[\mc{H}_{k_x=\pi,k_y}]$, predict the existence of helical edge modes for a system having translation invariance in the $x$ direction.

\subsection{Weak-pairing limit}
\label{subsec:TopInv_WeakPair}

The procedure for obtaining the topological invariant can be greatly simplified in the so-called weak-pairing limit, where $\Delta_k$ is small. Consider again the 1d Hamiltonian $\mathcal{H}_k$ of Eq.~\eqref{eq:wind_num_H_BdG}, composed of a normal block, $h_k$, and a superconducting block, $\Delta_k\mathcal{T}$. We begin by diagonalizing the normal block,
\begin{equation}
h_{k}=\sum_{n}\sum_{\sigma\in\{\rm I,II\}}\varepsilon_{n,k,\sigma}|n,k,\sigma\rangle\langle n,k,\sigma|,
\end{equation}
where we have used the label $\sigma={\rm I, II}$ to divide all states into two sectors related by TRS, namely $|n,k,{\rm II}\rangle=\mathcal{T}|n,-k,{\rm I}\rangle^\ast$, and accordingly $\varepsilon_{n,k,{\rm I}}=\varepsilon_{n,-k,\rm {II}}$ [similar to the way we divided the spectrum of $\tilde{Q}_{k}$, in Fig.~\ref{fig:TRITOPS_necess_int_theta_spectra}]. Next we write the superconducting block, $\Delta_k\mathcal{T}$, in the same basis,
\begin{equation}\label{eq:Delta_in_h_basis}
\Delta_{k}\mathcal{T}=\sum_{mn}\sum_{\sigma\sigma'\in\{\rm I,II\}}|n,k,\sigma\rangle\langle n,k,\sigma|\Delta_{k}\mathcal{T}|m,k,\sigma'\rangle\langle m,k,\sigma'|.
\end{equation}

In the weak-pairing limit, we keep only the diagonal elements in Eq.~\eqref{eq:Delta_in_h_basis}. This is justified since only pairing between $|n,k,\sigma\rangle$ and $\mathcal{T}|n,k,\sigma\rangle$ opens a gap at the Fermi energy\footnote{When making this statement we assume the spectrum of $h_k$ is nondegenerate at the Fermi level. If it is, one has to first diagonalize $\Delta_k\mathcal{T}$ in that degenerate subspace.}, assuming the off-diagonal elements of $\Delta_k$ are small enough. Within this approximation, the  matrix $Q_k$ of Eq.~\eqref{eq:H_rot} is given by
\begin{equation}
Q_{k}=\sum_{n}\sum_{\sigma\in\{\rm I,II\}}|n,k,\sigma\rangle(\delta_{n,k,\sigma}+i\varepsilon_{n,k,\sigma})\langle n,k,\sigma|
\end{equation}
where we define $\delta_{n,k,\sigma}\equiv\langle n,k,\sigma|\Delta_{k}\mathcal{T}|n,k,\sigma\rangle$.

According to Eq.~\eqref{eq:Winding_parity}, the topological invariant is then given by the parity of the sum (over $n$) of winding numbers
of $z_{n,k}\equiv\delta_{n,k,{\rm I}}+i\varepsilon_{n,k,{\rm I}}$, upon sweeping $k$ from $-\pi$ to $\pi$. The parity of each such winding number can be obtained by examining
the sign of $\delta_{n,k,{\rm I}}$ at the momenta for which
$\varepsilon_{n,k,{\rm I}}$ vanishes: if the product of $\delta_{n,k,{\rm I}}$
at these momenta is negative, then $z_{n,k,{\rm I}}$ winds an
odd number of times, while if the product is positive it winds an
even number of times.

Finally, since TRS dictates that $\delta_{n,k,{\rm I}}=\delta_{n,-k,{\rm II}}$, we can include both sectors I and II, and instead restrict ourselves only to the Fermi momenta between $k=0$ and $k=\pi$. We then obtain~\cite{Qi2010topological}
\begin{equation}\label{eq:top_inv_weak_pair_1d}
\nu=\prod_{s}{\rm sgn}(\delta_s),
\end{equation}
where $s$ labels all the Fermi points in $k\in[0,\pi]$, and it includes all bands $\{n,\sigma\}$. The expression for the topological invariant of the minimal model, Eq.~\eqref{eq:top_inv_1} is a special case of Eq.~\eqref{eq:top_inv_weak_pair_1d}.

This result can also be extended to the two-dimensional case, with the help of Eq.~\eqref{eq:2d_top_inv}, which states that $\nu_{\rm 2d}=\nu[\mc{H}_{k_x,k_y=0}]\cdot\nu[\mc{H}_{k_x,k_y=\pi}]$. In the 2d case, we take $s$ to labels the Fermi contours, as opposed to Fermi points. Notice also that ${\rm sgn}(\delta_s)$ cannot change along a given Fermi contour as long as the system is gapped, and it is therefore well defined. We next make the observation that ${\rm sgn}(\delta_s)$ belonging to a Fermi contour encircling an even number of TRIM [$\boldsymbol{k}=(0,0)$, $(0,\pi)$,  $(\pi,0)$ and $(\pi,\pi)$] necessarily appears an even number of times in the expression for $\nu_{\rm 2d}$, and therefore does not contribute. We can therefore write~\cite{Qi2010topological}
\begin{equation}\label{eq:top_inv_weak_pair_2d}
\nu_{\rm 2d}=\prod_{s}[{\rm sgn}(\delta_s)]^{m_s},
\end{equation}
where $m_s$ is the number of TRIM enclosed by the $s$'th Fermi contour.

\section{Realizations of Time-Reversal-Invariant Topological Superconductors (TRITOPS)}
\label{sec:realizations}

In Sec.~\ref{subsec:Intro_min_model} we introduced a minimal model for the TRITOPS phase. In this section we study specific microscopic models which can be realized in currently-available experimental setups, and which are described by the minimal model at low energies. As we saw, the TRITOPS phase is obtained when the pairing potential of the positive-helicity modes, $\Delta_+$, is opposite to that of the negative-helicity modes, $\Delta_-$ [see Eqs.~\eqref{eq:minimal_H} and \eqref{eq:top_inv_1}]. In searching for microscopic realizations of TRITOPS, the challenge would therefore be to find a mechanism creating this sign change in the superconducting pair potentials.

We focus here on three general mechanisms where this occurs. In Sec.~\ref{subsec:realize_pi_junc} we show how coupling a normal system to two SCs having a $\pi$ phase difference \cite{Fu2008superconducting,Fu2009josephson,Dahlhaus2010,Keselman2013inducing,Seradjeh2012majorana} essentially translates to a sign change in momentum space, between $\Delta_+$ and $\Delta_-$. This, however, generally requires fine-tuning of the superconducting phase, where deviations from a $\pi$ phase difference breaks TRS and therefore lift the protection of the MKPs.  
Then, in Sec.~\ref{subsec:realize_inter_syst}, we show how repulsive electron-electron interactions can stabilize the required sign change without any fine tuning, even when coupling the system to a single conventional $s$-wave superconductor~\cite{Gaidamauskas2014majorana,Haim2014time,Klinovaja2014Kramers,Klinovaja2014time,Danon2015interaction,Schrade2015proximity,Thakurathi2018majorana}. Finally, in Sec.~\ref{subsec:realize_unconv_sc} we consider proximity coupling to unconventional SCs~\cite{Nakosai2012topological,Wong2012majorana,Nakosai2013majorana,Zhang2013time,Nakosai2014theoretical,Chen2018helical}, which while being topologically trivial, contain a sign change of the pairing potential inside the Brillouin zone. This sign change can then be induced in the normal system through proximity, thereby realizing the TRITOPS phase.

\subsection{Proximity to superconducting $\pi$ junctions}
\label{subsec:realize_pi_junc}

In this section we examine systems in 1d and 2d, coupled to two SCs with a $\pi$ phase difference. As we will see, the combination of the superconducting $\pi$ phase difference, together with the spin-orbit interaction in the system, can cause a sign difference inside the Brillouin zone between the pairing potentials $\Delta_+$ and $\Delta_-$, of the low-energy model, Eq.~\eqref{eq:minimal_H}. We concentrate on two examples of such systems. (i) a finite-width two-dimensional topological insulator (in which the gapless edges serve as an effective 1d system)~\cite{Fu2009josephson,Keselman2013inducing} and (ii) a Rashba spin-orbit-coupled semiconductor wire~\cite{Keselman2013inducing,Kotetes2015topological}. In both cases, coupling the system to a superconducting $\pi$ junction can realize the 1d TRITOPS phase. The same effect can occur also in 2d, either in a finite-thickness three-dimensional TI~\cite{Fu2008superconducting,Dahlhaus2010,Wang2016Electrically,Parhizgar2017highly,Schrade2015proximity,Parhizgar2014unconventional}, or in a Rashba 2DEG~\cite{Deng2012majorana,Deng2013Multiband,Parhizgar2014unconventional,Kotetes2015topological}.

\subsubsection{Finite-width 2d topological insulator}
\label{subsec:realize_pi-n2dTI}

\begin{figure}
\begin{center}
\begin{tabular}{lcr}
\rlap{\hskip 0\tw \parbox[c]{\tw}{\vspace{-0.37\tw}(a)}}
\includegraphics[clip=true,trim =0cm 0cm 0cm 0.0cm,height=0.2\tw]{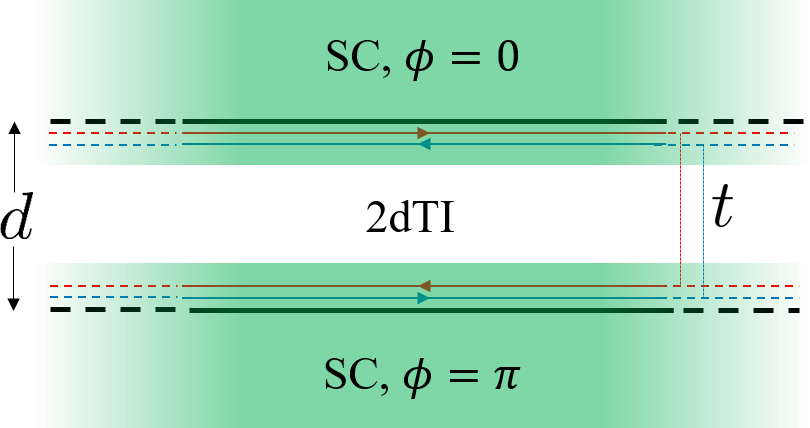}
&
\rlap{\hskip -0.02\tw \parbox[c]{\tw}{\vspace{-0.37\tw}(b)}}
\includegraphics[clip=true,trim =0cm 0cm 0cm 0.0cm,height=0.2\tw]{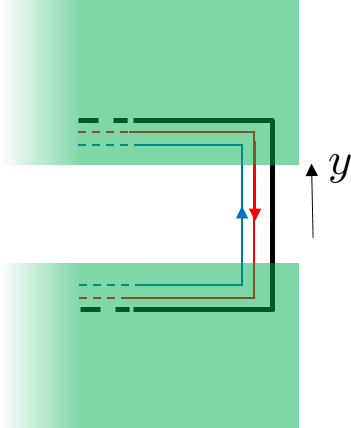}
&
\rlap{\hskip 0\tw \parbox[c]{\tw}{\vspace{-0.37\tw}(c)}}
\includegraphics[clip=true,trim =0cm 0cm 0cm 0cm,width=0.22\tw]{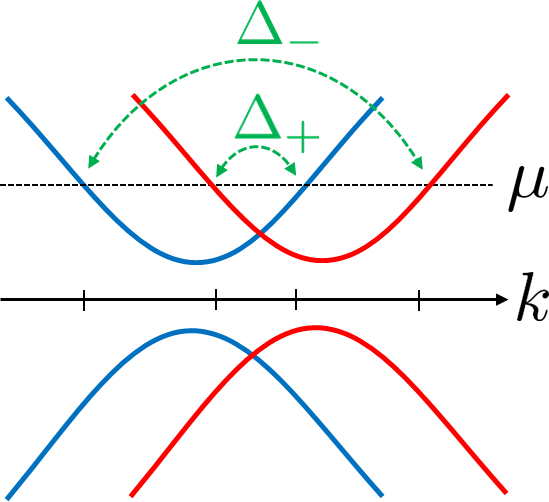}
\end{tabular}
\end{center}
\caption{(a,b) A finite-width two dimensional topological insulator (2dTI) in proximity to two $s$-wave superconductors, with a $\pi$ phase difference. The system's bulk is shown in (a), while (b) shows the system's end. (c) The low-energy spectrum of the 2dTI in the absence of induced pairing. We allow for a different chemical potential on each edge, as well as coupling between the edges (which creates a gap of size $2|t|)$. If the chemical potential does not lie inside the gap, then the system is described at low energies by the minimal model of Eq.~\eqref{eq:minimal_H}. Since the lower edge of the 2dTI host modes with positive helicity, while the upper edge host modes with negative helicity, the respective induced pairing potential, $\Delta_+$ and $\Delta_-$ have opposite signs.}\label{fig:TRITOPS_micro_Narrow_2DTI}
\end{figure}

We are looking for a system that will be described at low energies by the model of Eq.~\eqref{eq:minimal_H} with opposite pairing potentials, $\sgn(\Delta_+\Delta_-)=-1$. That is, it should contain two right-moving modes, and two left-moving modes, where the pairing between positive-helicity modes, $\Delta_+$, is opposite to that of the negative-helicity modes, $\Delta_-$. Perhaps the most natural setup meeting these requirements is a finite-width strip of a 2d topological insulator (2DTI)~\cite{Kane2005quantum,Bernevig2006quantum,konig2007quantum}, placed between two superconductors whose phase difference is fine-tuned to $\pi$, as depicted in Fig.~\hyperref[fig:TRITOPS_micro_Narrow_2DTI]{\ref{fig:TRITOPS_micro_Narrow_2DTI}(a)}.

The 2DTI phase is characterized by a pair of counter-propagating helical modes on each edge. Importantly, modes on the lower edge have positive helicity (a right-moving spin-$\ua$ mode and a left-moving spin-$\da$ mode), while modes on the upper edge have negative helicity (a right-moving spin-$\da$ mode and a left-moving spin-$\ua$ mode). If we couple each edge to a SC, with a $\pi$ phase difference between them, one immediately has $\sgn(\Delta_+)=-\sgn(\Delta_-)$, thereby realizing the TRITOPS phase.

The same conclusion can be reached by focusing on the boundary of this quasi-1d system~\cite{Fu2009josephson}. In the TRITOPS phase, the boundary must host a Kramers pair of zero-energy Majorana bound states. As depicted in Fig.~\hyperref[fig:TRITOPS_micro_Narrow_2DTI]{\ref{fig:TRITOPS_micro_Narrow_2DTI}(b)}, the boundary is described by a \emph{single pair} of helical modes, connecting the two superconductors. One can easily obtained the excitation spectrum of such a junction. In the limit of a short junction it is given by~\cite{Kwon2004fractional,Fu2009josephson}
\begin{equation}\label{eq:pi_junc_spect}
\varepsilon = \pm|\Delta_0| \cos(\phi/2),
\end{equation}
where $\Delta_0$ is the pair potential of the superconductors, and $\phi$ is their phase difference. Indeed, when the phase difference between the superconductors is $\phi=\pi$, there is a pair of zero energy states in the junction (the Majorana Kramers pair), indicating that the system is in the TRITOPS phase. 
Notice that once a MKP is present in the junction, it cannot be removed by any local perturbation, as long as TRS is preserved\footnote{In contrast, in a trivial SNS junction (where the normal part is \emph{not} described by a pair of helical modes), a similarly-looking excitation spectrum is obtained, however, importantly the number of states there is doubled compared to Eq.~\eqref{eq:pi_junc_spect}. At $\phi=\pi$, there are therefore \emph{two} pairs of Majorana bound states, which are not protected and can split due to perturbation (for example weak disorder).}. Namely our conclusions still holds, even if the Hamiltonian describing the system's boundary is modified, for example by going away from the short junction limit.

\subsubsection{semiconductor nanowires}
\label{subsec:realize_pi-J_wires}

A similar effect to that described above can take place when a \emph{semiconductor nanowire} is coupled to a superconducting $\pi$ junction~\cite{Keselman2013inducing,Kotetes2015topological}. As in before, the real-space $\pi$ phase difference translates into a sign difference between the induced pairing potentials of the positive- and negative-helicity modes, $\Delta_\pm$. This happens due to the difference in the spatial profile of the wave functions of these modes, which is a result of spin-orbit interaction.

\begin{figure}
\begin{center}
\includegraphics[clip=true,trim =0cm 0cm 0cm 0cm,width=0.4\tw]{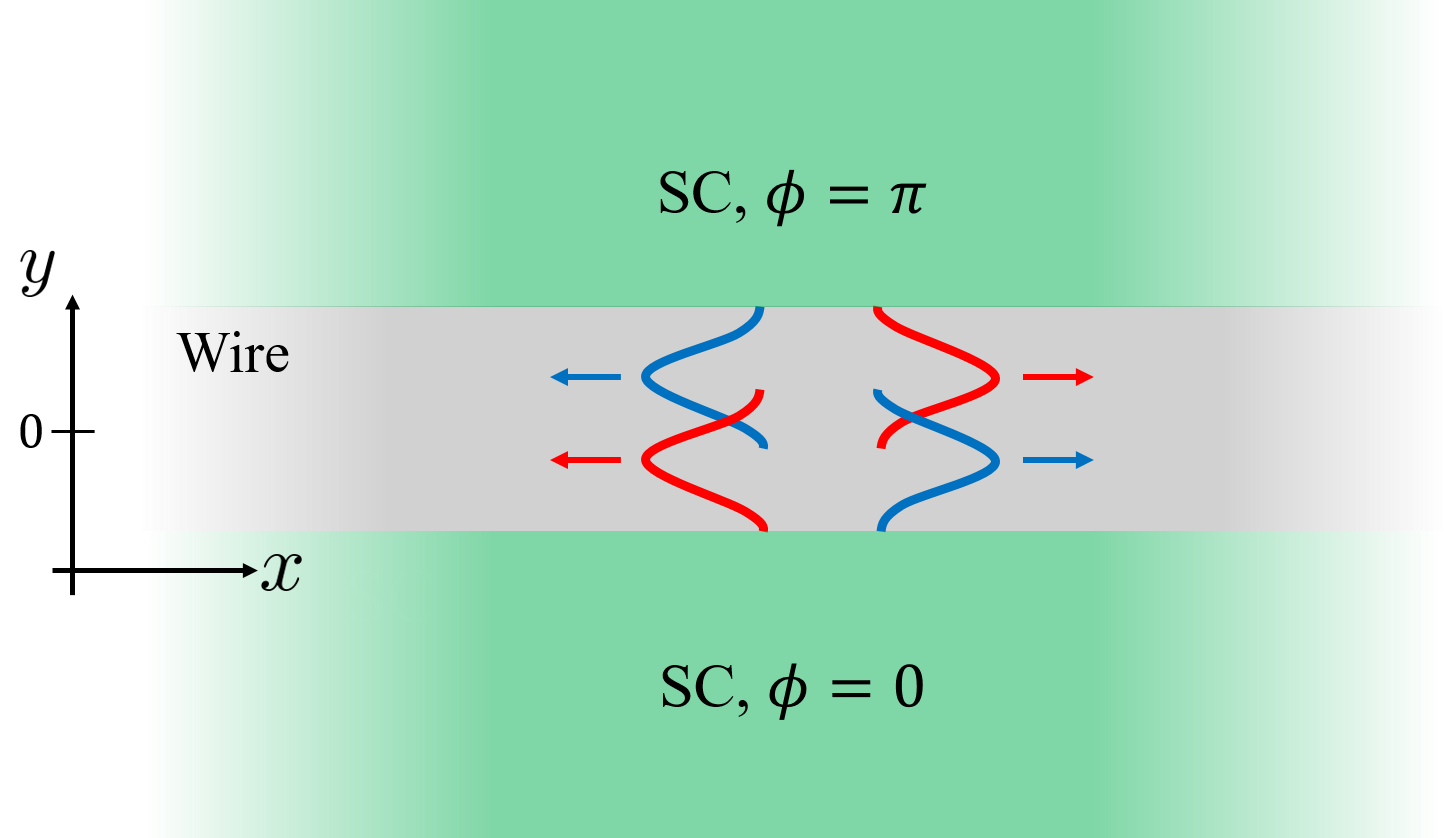}
\end{center}
\caption{A semiconductor quasi one-dimensional nanowire coupled to two bulk $s$-wave superconductors, having a $\pi$ phase difference. As a result of spin-orbit coupling, the spatial profile of the electronic wave functions depends on the factor $ks$, with $s=1$ for spin $\ua$, and $s=-1$ for spin $\da$, and with $k$ being the momentum in the $x$ direction. Wave functions with positive helicity ($ks>0$) are pushed towards the lower superconductor, while wave functions with negative helicity ($ks<0$) are pushed towards the upper SC.}\label{fig:TRITOPS_micro_wire_dispersion}
\end{figure}

To understand this effect better, let us consider the following simplified model for describing the nanowire \cite{Haim2016interaction}. We assume the confining potential of electrons in the wire is described by an harmonic potential, $V_\tx{c}(y)=m^\ast\omega_\tx{c}^2y^2/2$, where $y=0$ is at the center of the wire, as depicted in Fig.~\ref{fig:TRITOPS_micro_wire_dispersion}. Here, $m^\ast$ is the effective mass of the electron, and $\omega_\tx{c}$ is related to the width of the wire through $w_y\sim1/\sqrt{m^\ast\omega_\tx{c}}$. The spin-orbit coupling in the wire contributes a term of the form $\mc{H}_\tx{so}=u\partial_y V_\tx{c}(y) \hat{p}_x\sigma^z$. Ignoring the $z$ direction (justified when $w_z\ll w_y$), the electrons in the wire are governed by the first-quantized Hamiltonian
\begin{equation}\label{eq:H_wire_simplified}
\mc{H}_\tx{wire}=-\frac{\nabla^2}{2m^\ast} + \frac{1}{2}m^\ast\omega_c^2(y-iu\sigma^z\partial_x)^2.
\end{equation}
The eigenfunctions of the lowest-energy transverse band are given by 
\begin{equation}
\phi_{k,s}(x,y)=e^{ikx} \cdot e^{-\frac{m^\ast\omega_\tx{c}}{2}(y+uks)^2}, 
\end{equation}
up to normalization, where $s=\pm 1$ corresponds here to spin $\ua$ and spin $\da$, respectively. 
It is now apparent that states with $ks>0$ are shifted towards $y<0$, while states with negative $ks<0$ are shifted towards $y>0$~\cite{Moroz1999Effect}. This is illustrated in Fig.~\ref{fig:TRITOPS_micro_wire_dispersion}
, where the blue and red curves qualitatively describe the spin-$\ua$ and spin-$\da$ wave functions, respectively.

Upon coupling the two SCs to the wire, modes with $ks>0$ are therefore better coupled to the lower SC, while modes with $ks<0$ are better coupled to the upper SC. Since the two SCs have a $\pi$ phase difference, modes with positive helicity ($ks>0$) experience an induce pairing potential, $\Delta_+$, with opposite sign to the pairing potential of the negative-helicity ($ks>0$) mode, $\Delta_-$. If the chemical potential is inside the lowest-energy transverse band, one therefore expect the system to be in the TRITOPS phase.

\subsection{Interacting proximity-coupled systems}
\label{subsec:realize_inter_syst}

The $\pi$ junctions considered above provide a very intuitive platform for realizing the TRITOPS phase. They nevertheless require fine-tuning of superconducting phases. Indeed, any deviation from a phase difference of $\phi=\pi$ would break time-reversal symmetry, and thereby lift the protection of the topological boundary modes - the Majorana Kramers pair. Moreover, creating a superconducting phase difference experimentally usually involves applying a magnetic field, or forcing a current in the system, both of which break TRS.   

Below we show how repulsive e-e interactions can stabilize the TRITOPS phase in a normal system coupled to a single conventional $s$-wave SC~\cite{Gaidamauskas2014majorana,Haim2014time,Klinovaja2014time,Klinovaja2014Kramers,Danon2015interaction,Schrade2015proximity,Thakurathi2018majorana}. We begin in secs.~\ref{subsubsec:real_int-induced_pi_junc} and \ref{subsubsec:real_dir_vs_cross_AR} by motivating the inclusion of repulsive interactions on a qualitative level. 
In sec.~\ref{subsubsec:realize_inter_syst_eff_of_int} we then adopt a more formal approach, studying the low-energy model of Eq.~\eqref{eq:minimal_H} in the presence of all the interaction terms allowed by symmetry. Finally, we present numerical evidence showing that repulsive interactions can drive a proximity-coupled Rashba wire into the TRITOPS phase. One might raise the question of whether including interactions is necessary for ending up in the TRITOPS phase. In appendix~\ref{subsec:realize_inter_syst_necess_int} we show that indeed, a non-interacting system coupled to a single conventional $s$-wave SC is always in the topologically-trivial phase~\cite{Zhang2013time,Gaidamauskas2014majorana,Haim2016NoGo}.

\subsubsection{Interaction-induced $\pi$ junction}
\label{subsubsec:real_int-induced_pi_junc}

\begin{figure}
	\begin{center}
		\begin{tabular}{lcr}
			\rlap{\hskip -4mm \parbox[c]{0mm}{\vspace{0cm}(a)}}
			\includegraphics[clip=true,trim =0mm -13mm 0mm 0mm,width=0.32\textwidth]{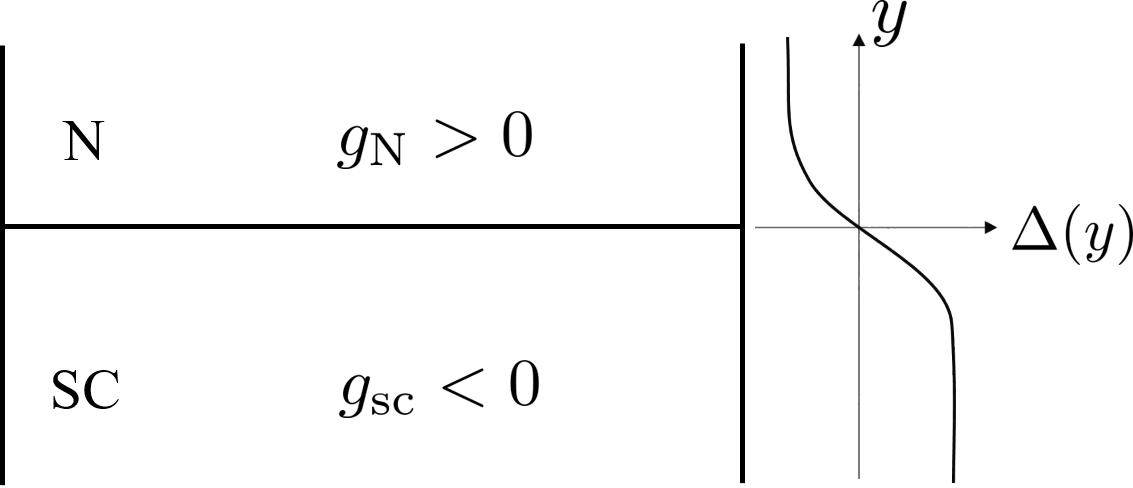}
			&
			\rlap{\hskip -3mm \parbox[c]{0mm}{\vspace{0cm}(b)}}
			\includegraphics[clip=true,trim =0mm -16mm 0mm 0mm,width=0.30\textwidth]{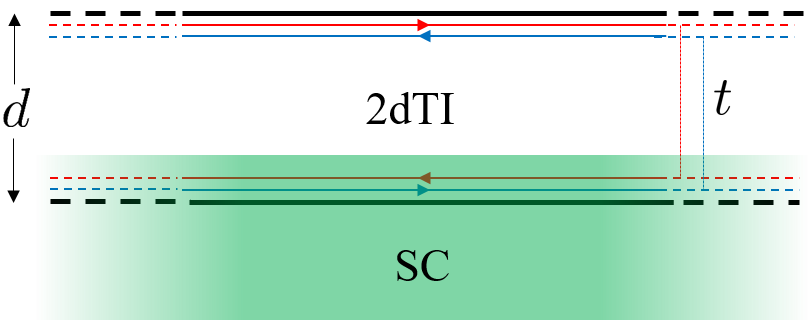}
			&
			\rlap{\hskip -2mm \parbox[c]{0mm}{\vspace{0cm}(c)}}
			\includegraphics[clip=true,trim =0mm -20mm 0mm 0mm,width=0.30\textwidth]{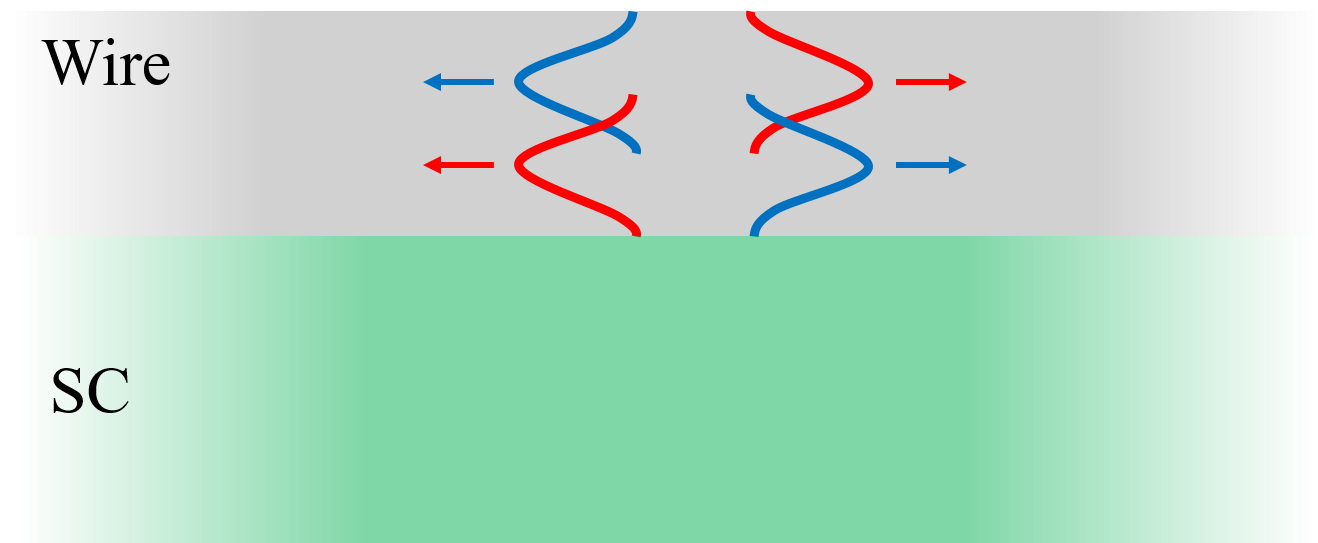}
		\end{tabular}
	\end{center}
	\caption{(a) Interface between a superconductor (SC), where e-e interactions are effectively attractive ($g_{\rm sc}<0$), and a Normal metal, where e-e interactions are repulsive ($g_{\rm N}>0$). The combination of the proximity effect and the repulsive interactions in the normal metal causes the pairing potential to change sigh across the interface. (b) a finite-width two-dimensional topological insulator (2dTI), coupled to a single SC. The lower-edge modes play the role of the SC, while the upper-edge modes play the role of the Normal metal. The proximity effect is achieved due to tunneling, $t$ between the edges. The emergent $\pi$ junction then stabilizes the TRITOPS phase (see also Fig.~\ref{fig:real_inter_ind_pi_junc}). (c) A similar effect can take place in a Rashba wire, where the positive-helicity modes (which are pushed downwards) play the role of the SC, and the negative-helicity modes (which are pushed upwards) play the role of the normal metal.\label{fig:real_inter_ind_pi_junc}}
\end{figure}

Consider an interface between a conventional superconductor and a normal metal, as depicted in Fig.~\hyperref[fig:real_inter_ind_pi_junc]{\ref{fig:real_inter_ind_pi_junc}(a)}. Inside the SC, the phonon-mediated e-e interactions are attractive, represented by $g_{\rm sc}<0$, while inside the normal metal the e-e interactions are repulsive, $g_{\rm N}>0$. This problem was addressed by de Gennes~\cite{DeGennes1964Boundary} who showed that the pairing potential in such a setup changes sign across the interface, as shown qualitatively in the right panel of Fig.~\hyperref[fig:real_inter_ind_pi_junc]{\ref{fig:real_inter_ind_pi_junc}(a)}.

One can easily understand the origin of this result. First, the attractive interactions in the SC generate a non-zero mean-field pairing potential, $\Delta_\tx{sc}=g_\tx{sc}F_\tx{sc}$, where $F_\tx{sc}=\langle \psi_\da(\bs{r})\psi_\ua(\bs{r}) \rangle|_{y<0}$ is the pair correlation function. Then, the superconductor induces by proximity a non-zero pairing \emph{correlation}, $F_\tx{N}=\langle \psi_\da(\bs{r})\psi_\ua(\bs{r}) \rangle|_{y>0}$, of the same phase as  $F_\tx{sc}$. Finally, this gives rise to a pairing \emph{potential}, $\Delta_\tx{N}=g_\tx{N} F_\tx{N}$. Since $g_\tx{N}$ and $g_\tx{sc}$ have opposite signs, $\Delta_\tx{sc}$ and $\Delta_\tx{N}$ also differ by a $\pi$ phase\footnote{For more details on the behavior of the pair correlations vs. the pair potential see the supplemental material of Ref.~\cite{Haim2014time}.}.   

We can now reconsider the systems discussed in Sec.~\ref{subsec:realize_pi_junc} - the finite-width 2dTI and the Rashba wire - but now coupled only to a single SC, as shown in Fig.~\hyperref[fig:real_inter_ind_pi_junc]{\ref{fig:real_inter_ind_pi_junc}(b,c)}. As explained above, in both these cases the positive-helicity modes are somewhat separated (spatially) from the negative-helicity modes. By coupling the SC as depicted in Fig.~\hyperref[fig:real_inter_ind_pi_junc]{\ref{fig:real_inter_ind_pi_junc}(b,c)}, one obtain a situation resembling the SC-N interface of Fig.~\hyperref[fig:real_inter_ind_pi_junc]{\ref{fig:real_inter_ind_pi_junc}(a)}, where the positive-helicity modes play the role of the SC, and the negative-helicity modes play the role of the Normal metal. The combination of the proximity between the modes and the repulsive e-e interaction in the 2dTI (or Rashba wire) then effectively create the sought $\pi$ junction. In sec.~\ref{subsubsec:realize_inter_syst_eff_of_int_R_wire} we present numerical results corroborating these conclusions for the proximity-coupled Rashba wire.

\subsubsection{Local versus Crossed Andreev reflection}
\label{subsubsec:real_dir_vs_cross_AR}

\begin{figure}
	\begin{center}
		\begin{tabular}{c}
			\includegraphics[clip=true,trim =0mm 0mm 0mm 0mm,width=0.35\textwidth]{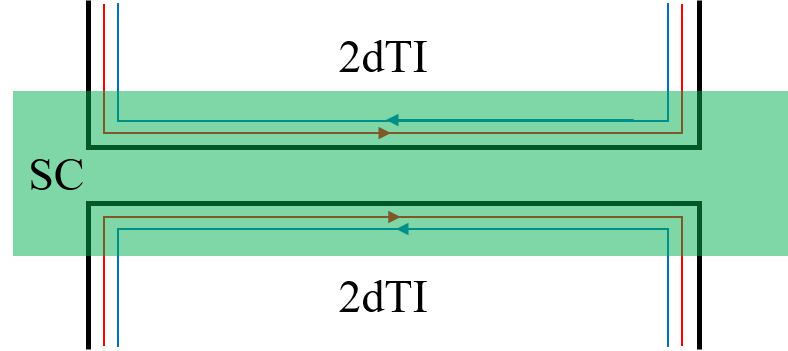}
		\end{tabular}
	\end{center}
	\caption{Two topological insulators coupled through a single conventional superconductor. In the absence of normal tunneling between the edges, the system is in the TRITOPS phase when the crossed-Andreev-reflection term is stronger than the direct-Andreev reflection term. The latter is expected to be suppressed by repulsive electron-electron interactions.}\label{fig:real_loc_vs_cross_AR}
\end{figure}

Another way in which interactions can drive a system into the TRITOPS phase is by affecting the competition between two superconducting proximity mechanisms: \emph{local} Andreev reflection and \emph{crossed} Andreev reflection~\cite{Gaidamauskas2014majorana,Klinovaja2014Kramers,Klinovaja2014time,Danon2015interaction,Parhizgar2017highly,Reeg2017destructive}. To understand this better, we focus again on a system composed of two edges of a 2dTI. This time, however, we consider the case where the edges belong to different 2dTI samples with opposite topological indices, as depicted in Fig.~\ref{fig:real_loc_vs_cross_AR}. The two right-moving modes then have spin $\ua$, while the two left-moving modes have spin $\da$.

We label the samples by $j=1,2$, such that $\psi_{\tx{R},j}$ ($\psi_{\tx{L},j}$) denote the right- (left)-moving mode on the edge belonging to sample $j$. We couple the two edges through a single conventional $s$-wave SC (see Fig.~\ref{fig:real_loc_vs_cross_AR}). Assuming there is no normal tunneling between the edges, the Hamiltonian describing the system at low energies is $H=H_0 +H_\Delta$, with
\begin{equation}\label{eq:H_delta_cross}
\begin{split}
H_0 &= -iv\int \tx{d}x
 \sum_{j=1,2} \left[\psi_{\tx{R},j}^\dag(x)\partial_x \psi_{\tx{R},j}^\nd(x) -\psi_{\tx{L},j}^\dag(x)\partial_x \psi_{\tx{L},j}^\nd(x)\right]\\
H_\Delta &= \int \tx{d}x \left[\Delta_1 \psi_{\tx{R},1}^\dag(x)\psi_{\tx{L},1}^\dag(x)
+ \Delta_2 \psi_{\tx{R},2}^\dag(x)\psi_{\tx{L},2}^\dag(x)
+ \Delta_\tx{c} \psi_{\tx{R},1}^\dag(x)\psi_{\tx{L},2}^\dag(x) +\Delta^\ast_\tx{c}\psi_{\tx{R},2}^\dag(x)\psi_{\tx{L},1}^\dag(x)
+ \tx{h.c.} \right],
\end{split}
\end{equation}
where $\Delta_{1,2}$ are the induced pairing potentials on the edge of sample '1' and '2', respectively, and $\Delta_\tx{c}$ is the induced crossed pairing potential term involving both edges simultaneously. While $\Delta_1$ ($\Delta_2$) arise as a result of a \emph{local} AR process, where a Cooper pair tunnels to the edge of sample '1' ('2'), the term $\Delta_\tx{c}$ arise as a result of a \emph{crossed} AR process, where a Cooper pair is split between sample '1' and sample '2'\footnote{Notice that the form of the crossed pairing term is dictated by TRS, which takes $\psi_{\rm R,j}\to\psi_{\rm L,j}$ and $\psi_{\rm L,j}\to-\psi_{\rm R,j}$. TRS also dictates that $\Delta_j=\Delta_j^\ast$ for $j=1,2$.}.

To analyze the system, let us rewrite the pairing term in a matrix form,
\begin{equation}
H_\Delta=\int {\rm d}x \left\{[\psi_{\rm R,1}^{\dagger}(x),\psi_{\rm R,2}^{\dagger}(x)]\overset{\leftrightarrow}{\Delta}\begin{bmatrix}\psi_{\rm L,1}^{\dagger}(x)\\
\psi_{\rm L,2}^{\dagger}(x)
\end{bmatrix} + {\rm h.c.} \right\}
\hskip 5mm ; \hskip 5mm
\overset{\leftrightarrow}{\Delta}=\begin{pmatrix}\Delta_{1} & \Delta_{\rm c}\\
\Delta_{\rm c}^{\ast} & \Delta_{2}.
\end{pmatrix}
\end{equation}
When we now compare the Hamiltonian of Eq.~\eqref{eq:H_delta_cross} with that of the minimal model, Eq.~\eqref{eq:minimal_H}, we recognize that the two are related by a unitary transformation that diagonalize $\overset{\leftrightarrow}{\Delta}$ . In matrix form, the topological criterion, Eq.~\eqref{eq:top_inv_1}, can be expressed as the determinant of the pairing matrix. Since the latter is invariant under unitary transformations, one infers~\cite{Gaidamauskas2014majorana,Klinovaja2014Kramers}
\begin{equation}\label{eq:top_inv_3}
\nu=\det(\overset{\leftrightarrow}\Delta)=\Delta_1\Delta_2-|\Delta_{\rm c}|^2.
\end{equation}

For simplicity, we can make the reasonable assumption that $\Delta_1=\Delta_2\equiv\Delta_{\rm loc}$, in which case we realize that the system is topological whenever $|\Delta_{\rm c}|>|\Delta_{\rm loc}|$, namely when the \emph{crossed} Andreev reflection dominates over \emph{local} Andreev reflection. In the absence of e-e interactions in the proximitized system (here the 2dTIs), this will never be the case, as shown in Appendix~\ref{subsec:realize_inter_syst_necess_int}. In contrast, if short-range repulsive e-e interactions exist, they are expected to suppress the local Andreev reflection process responsible for $\Delta_{\rm loc}$, as it requires paring of electrons on the same edge. If the later are strong enough [such that $\nu=-1$ in Eq.~\eqref{eq:top_inv_3}], they can thereby drive the system into the TRITOPS phase. While this effect was discussed here in the context of edges of 2dTIs~\cite{Klinovaja2014Kramers}, the same is true for semiconductor nanowires~\cite{Gaidamauskas2014majorana,Danon2015interaction,Klinovaja2014time,Ebisu2016theory}.

In 2d, a related effect can occur in coupled semiconducting layers (or one layer with two orbitals), even without proximity to an external superconductor. In these cases, superconductivity is intrinsically generated as a result of \emph{interactions}, and the competition is now between the inter-layer and intra-layer interaction terms~\cite{Nakosai2012topological,Wang2014two,Nakosai2014theoretical,Yang2015time}.

\subsubsection{Low-energy interacting model}
\label{subsubsec:realize_inter_syst_eff_of_int}

\begin{figure}[t]
\begin{center}
\begin{tabular}{lr}
\rlap{\hskip -3mm \parbox[c]{\textwidth}{\vspace{0cm}(a)}}
\includegraphics[clip=true,trim = 5.7mm 10mm 16.5mm 15mm,width=0.5\tw]{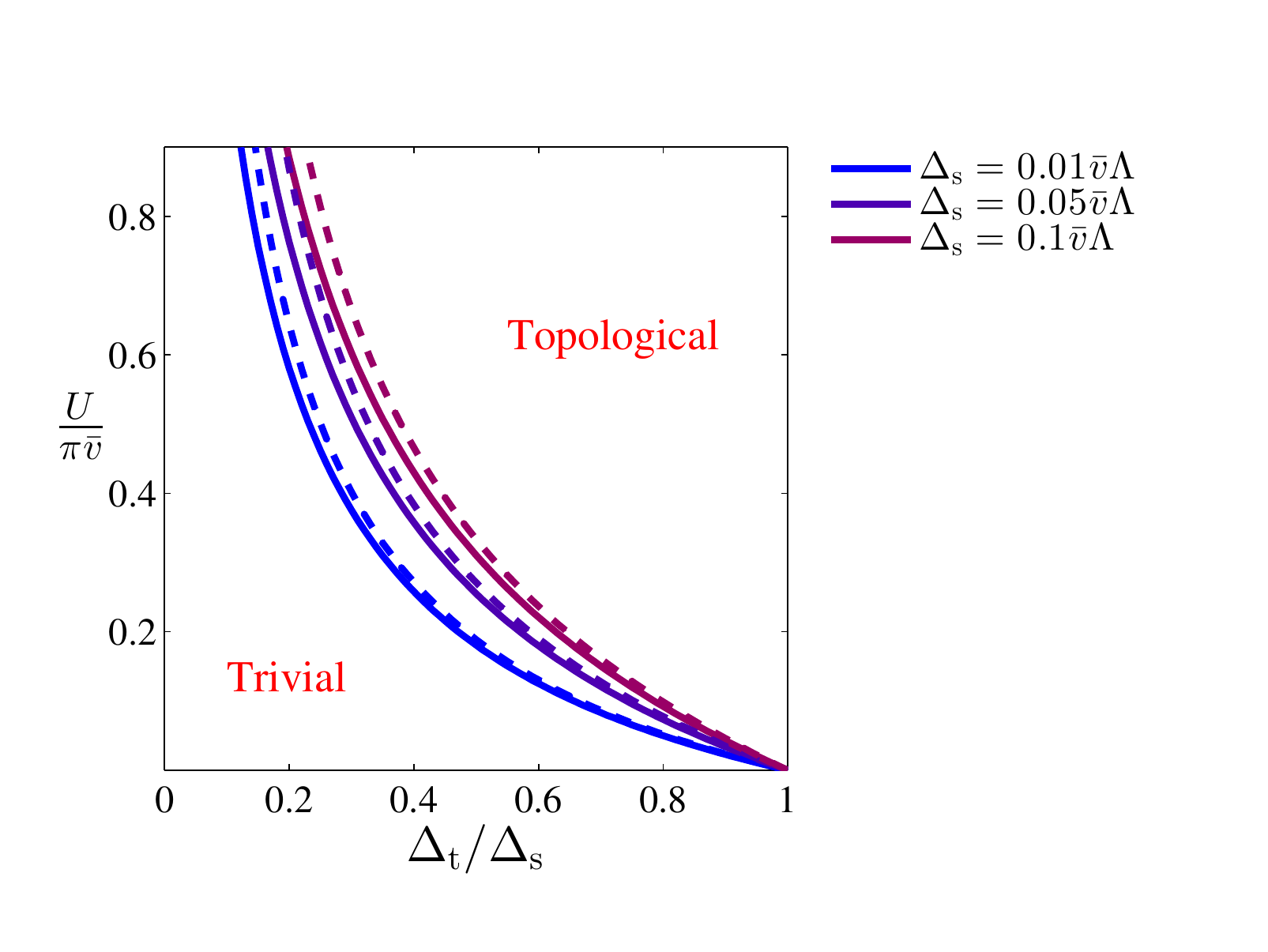}
\hs 2mm
&
\hs 0mm
\rlap{\hskip -3.2mm \parbox[c]{\textwidth}{\vspace{0cm}(b)}}
\includegraphics[clip=true,trim = 45mm -1mm 50mm 5mm,width=0.35\textwidth]{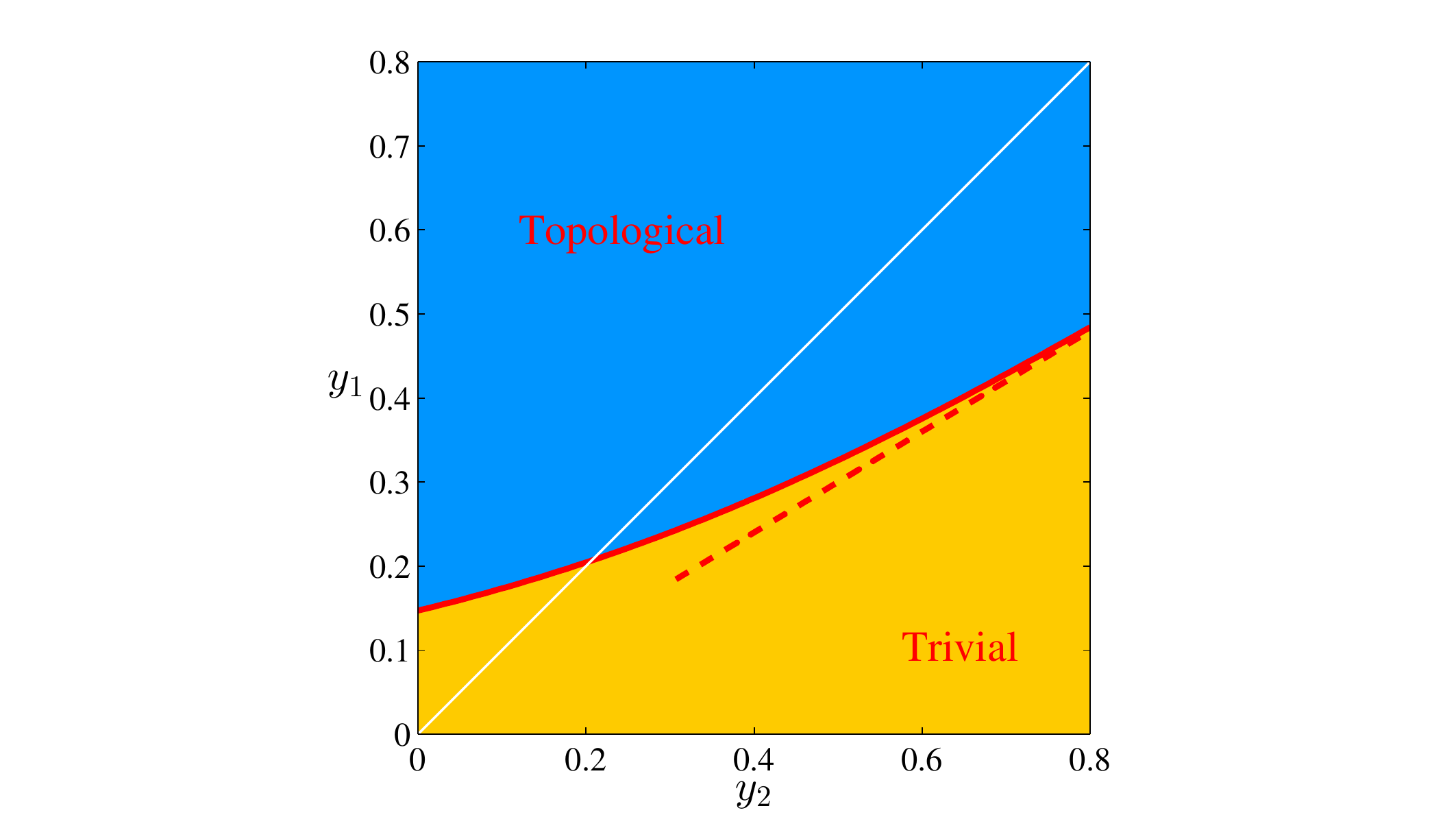}
\end{tabular}
\end{center}
\caption{(a) Phase diagram of the interacting model described in Eq.~\eqref{eq:H_int}. The phase diagram is analyzed as a function of the interaction strength $U=g_1=g_2^+=g_2^-$, and the ratio $\Delta_\tx{t}/\Delta_\tx{s}$, for different fixed values of $\Delta_\tx{s}$. $\Delta_\tx{s}$ and $\Delta_\tx{t}$ are the singlet and triplet induced pairing potentials, respectively (referred to $\Delta^0_\tx{s,t}$ in the context of the RG analysis), and are related to the pairing potentials $\Delta_\pm$ through $\Delta_\tx{s,t}=(\Delta_+\pm\Delta_-)/2$. The solid lines are the phase boundaries calculated using weak-coupling RG, while the dashed lines are those calculated from Eq.~\eqref{eq:MF_ph_bound_U}, obtained from a mean-field treatment.
Notice that for $\Delta_\tx{t}=0$ the system cannot be driven into the topological phase for any interaction strength, i.e., some initial triplet pairing term is required. For a nonzero $\Delta_\tx{t}$, the system goes through a topological phase transition at a finite value of $U$ which increases with $\Delta_\tx{s}$. For $\Delta_\tx{t}=\Delta_\tx{s}$ the system is on the verge of becoming topological, and any finite interaction will drive it to the topological phase. (b) Phase diagram as a function of forward ($y_2$) and backward ($y_1$) scattering interaction terms. The solid red line shows the phase boundary, calculated using the RG flow equations, Eq.~\eqref{eq:flow_eqs}, and the topological invariant, Eq.~\eqref{eq:top_inv_RG}. The dashed red line indicates the long RG-time approximation for the phase boundary, obtained by setting $\Delta_{\rm t}(\ell)=\Delta_{\rm s}(\ell)$ in Eq.~\eqref{eq:r_t_s_ell}. It agrees with the numerical result when $\Delta_\pm^0 \ll v_\pm\Lambda \exp(-1/A)$, where $A^2=y_2^2-y_1^2$. The white solid line corresponds to the separatrix of the Kosterlitz-Thouless flow, above which $y_1$ and $y_2$ flow to strong coupling. In obtaining this phase diagram, we have used $v_-=v_+=\bar{v}$, $y_2^+=y_2^-=y_2$, and the initial singlet and triplet pairing potentials were taken to be $\Delta^0_{\mathrm{t}}=0.01\bar{v}\Lambda$ and $\Delta^0_{\mathrm{s}}=0.02\bar{v}\Lambda$, respectively.}
\label{fig:TRITOPS_low_E_phase_diagram}
\end{figure}

We now move on to study the effect of interactions more formally. This is done by considering the minimal model of Eq.~\eqref{eq:minimal_H}, and adding to it possible interaction terms which are allowed by symmetry. We then analyze the model using a mean-field approach and using the renormalization-group. As we will see, the presence of short-range repulsive interactions can drive the system from the trivial to the topological phase [see Fig.~\hyperref[fig:TRITOPS_low_E_phase_diagram]{\ref{fig:TRITOPS_low_E_phase_diagram}(a)}].   

This will be understood in terms of the competition between singlet and triplet pairing. Due to spin-orbit coupling, proximity-induced superconductivity is generally described by both a singlet and a triplet pairing potential, $\Delta_\tx{s}$ and $\Delta_\tx{t}$, respectively. For a noninteracting system in proximity to a conventional $s$-wave SC the system will always be in the topologically trivial phase, with $|\Delta_\tx{s}|\ge |\Delta_\tx{t}|$ (see Appendix~\ref{subsec:realize_inter_syst_necess_int}). However, short-range repulsive interactions effectively suppress the singlet pairing term in comparison with the triplet term, and can therefore drive the system into the TRITOPS phase.

The Hamiltonian we consider is given by $H_0+H_\Delta+H_\tx{int}$, with
\begin{equation}\label{eq:H_int}
\begin{split}
H_0 =& -i\int \tx{d}x
\left\{ v_+  \left[\psi_{\tx{R}\ua}^\dag(x)\partial_x \psi_{\tx{R}\ua}^\nd(x) -\psi_{\tx{L}\da}^\dag(x)\partial_x \psi_{\tx{L}\da}^\nd(x)\right]
+ v_- \left[\psi_{\tx{R}\da}^\dag(x)\partial_x \psi_{\tx{R}\da}^\nd(x) -\psi_{\tx{L}\ua}^\dag(x)\partial_x \psi_{\tx{L}\ua}^\nd(x)\right]\right\},\\
H_\Delta =& \int \tx{d}x \left[\Delta_+ \psi_{\tx{R}\ua}^\dag(x)\psi_{\tx{L}\da}^\dag(x) + \Delta_- \psi_{\tx{L}\ua}^\dag(x)\psi_{\tx{R}\da}^\dag(x) + \tx{h.c.} \right],\\
H_\tx{int} =& \int\tx{d}x \left\{ g_1^\perp \left[ \psi^\dag_{\tx{R},\ua}(x)\psi^\dag_{\tx{L},\da}(x)\psi_{\da,R}(x)\psi_{L,\ua}(x) + \tx{h.c.}\right]
+ g_2^+\rho_{\tx{R}\ua}(x)\rho_{\tx{L}\da}(x)+ g_2^-\rho_{\tx{R}\da}(x)\rho_{\tx{L}\ua}(x) \right.\\
& \hs 7mm + \left. g_2^\parallel\left[\rho_{\tx{R}\ua}(x)\rho_{\tx{L}\ua}(x) + \rho_{\tx{L}\da}(x)\rho_{\tx{R}\da}(x)\right] \right\},
\end{split}
\end{equation}
where $\rho_{\tx{R}s}(x)=\psi^\dag_{\tx{R},s}(x)\psi_{\tx{R},s}(x)$ and $\rho_{\tx{L}s}(x)=\psi^\dag_{\tx{L},s}(x)\psi_{\tx{L},s}(x)$. Here $H_\Delta$ describes induced pairing in the normal system due to proximity to a superconductor, and $H_{\rm int}$ describes e-e interactions inside the normal system, where $g_1^\perp$ is a backscattering interaction term, and $g_2^+$, $g_2^-$, and $g_2^\parallel$ are forward scattering interaction terms. In the absence of symmetry under inversion ($x\to-x$), $H_\tx{int}$ is the most general low-energy time-reversal symmetric Hamiltonian describing interaction between modes of opposite chirality. Interaction terms between modes of the same chirality can exist, however, they would not affect the RG flow, nor would they contribute to our mean-field solution, and therefore we do not include them here~\cite{Haim2016interaction}.

\paragraph{Mean-field theory}
\label{parag:realize_inter_syst_eff_of_int_MF}

In this analysis we replace the low-energy interacting Hamiltonian by the quadratic part of the Hamiltonian, but with new \emph{effective} pairing potentials, $\bar\Delta_+$ and $\bar\Delta_-$. Upon determining $\bar\Delta_\pm$ from self-consistent equations, one can then extract the topological invariant using Eq.~\eqref{eq:top_inv_1}. In the mean-field approximation one assumes that the system has a superconducting order, and accordingly the averages of the pairing terms, $\langle \Ldn^\nd(x) \Rup^\nd(x) \rangle$ and $\langle \Rdn^\nd(x) \Lup^\nd(x) \rangle$, are large compared to their respective fluctuations, $\delta_+\equiv\psi_{\tx{L},\da}(x)\psi_{\tx{R},\ua}(x)-\langle \Ldn^\nd(x) \Rup^\nd(x) \rangle$ and $\delta_-\equiv \psi_{\tx{R},\da}(x)\psi_{\tx{L},\ua}(x)-\langle \Rdn^\nd(x) \Lup^\nd(x) \rangle$. We therefore expand the Hamiltonian in Eq.~\eqref{eq:H_int}, to first order in $\delta_\pm(x)$, resulting in the mean-field Hamiltonian, $H^\tx{MF}=H_0+H^\tx{MF}_\Delta$, with\footnote{The $g_2^\parallel$ term in Eq.~\eqref{eq:H_int} involves interaction between electrons of the same spin species. It therefore does not affect $\bar{\Delta}_\pm$, and its sole effect would be to change the effective chemical potential. Hence, we ignore it in the present mean-field treatment.}
\begin{equation}\label{eq:H_MF}
H^\tx{MF}_\Delta = \int \tx{d}x \left[\bar\Delta_+ \Rup^\dag(x)\Ldn^\dag(x) + \bar\Delta_- \Lup^\dag(x)\Rdn^\dag(x) + \tx{h.c.} \right],
\end{equation}
where
\begin{equation}\label{eq:Deltas_bar}
\begin{split}
&\bar\Delta_+=\Delta_+ + g_1^\perp\langle \Rdn(x) \Lup(x) \rangle + g_2^+\langle \Ldn(x) \Rup(x) \rangle\\
&\bar\Delta_-=\Delta_- + g_1^\perp\langle \Ldn(x) \Rup(x) \rangle + g_2^-\langle \Rdn(x) \Lup(x) \rangle.
\end{split}
\end{equation}

Since $H^\tx{MF}$ is a quadratic Hamiltonian, one can easily calculate the above pair correlation functions and arrive at the following self-consistent equations for $\bar\Delta_+$ and $\bar\Delta_-$~\cite{Haim2016interaction},
\begin{equation}\label{eq:self_consist}
\bar{\Delta}_\pm = \Delta_\pm
- \frac{g_1^\perp}{2\pi v_\mp} \bar{\Delta}_\mp \sinh^{-1}\left(v_\mp\Lambda/|\bar{\Delta}_\mp|\right)
- \frac{g_2^\pm}{2\pi v_\pm} \bar{\Delta}_\pm \sinh^{-1}\left(v_\pm\Lambda/|\bar{\Delta}_\pm|\right),
\end{equation}
These coupled equations can be solved numerically for $\bar \Delta_\pm$, after which the topological invariant of $H^\tx{MF}$ is obtained by $\nu=\sgn(\bar{\Delta}_+)\sgn(\bar{\Delta}_-)$.

One can, however, make further analytical progress by searching for the phase boundary between $\nu=1$ and $\nu=-1$. This occurs when either $\bar{\Delta}_-=0$, or $\bar{\Delta}_+=0$. By plugging $\bar\Delta_\pm=0$ in Eq.~\eqref{eq:self_consist}, one obtains the conditions on the parameters of the original Hamiltonian, Eq.~\eqref{eq:H_int}, to be on the phase boundary. One obtains
\begin{equation}\label{eq:phase_bound}
\frac{v_\mp\Lambda g_1^\perp}{|g_1^\perp \Delta_\mp - g_2^\mp \Delta_\pm |} =
\sinh\left( \frac{2\pi v_\mp\Delta_\pm}{g_1^\perp \Delta_\mp - g_2^\mp \Delta_\pm} \right),
\end{equation}
where the two options correspond to the phase boundary occurring at $\bar\Delta_{\pm}=0$, respectively.

As a relevant example we can consider a Hubbard-type interaction, $g_1=g_2^+=g_2^-=U$, and furthermore $v_+=v_-=\bar{v}$. Let us assume without loss of generality that $|\Delta_+|>|\Delta_-|$. This means that the phase boundary will occur when $\bar{\Delta}_-=0$, namely when~\cite{Haim2016interaction}
\begin{equation}\label{eq:MF_ph_bound_U}
\frac{U}{\pi \bar{v}} = \frac{\Delta_\tx{s}/\Delta_\tx{t} - 1}{\sinh^{-1}\left( \bar{v}\Lambda/2|\Delta_\tx{t}| \right)}.
\end{equation}
Figure~\hyperref[fig:TRITOPS_low_E_phase_diagram]{\ref{fig:TRITOPS_low_E_phase_diagram}(a)} presents the topological phase diagram, obtained using Eq.~\eqref{eq:MF_ph_bound_U} (see dashed line), as a function of $U$ and the ratio $\Delta_\tx{t}/\Delta_\tx{s}$, for different values of $\Delta_\tx{s}$. For $\Delta_\tx{t}/\Delta_\tx{s}\to0$ no finite amount of interactions can bring the system to the topological phase. In contrast, when $\Delta_\tx{t}=\Delta_\tx{s}$, the system is already at a phase transition, and any nonzero $U$ suffices to drive the system to the topological phase. In the intermediate regime, the system will become topological for some finite interaction strength which increases with $|\Delta_\tx{s}/\Delta_\tx{t}|$.

\paragraph{Renormalization-group analysis}
\label{parag:realize_inter_syst_eff_of_int_RG}

We now move on to study the interacting Hamiltonian of Eq.~\eqref{eq:H_int} using the renormalization group (RG). We are interested in the RG flow close to the noninteracting fixed point of free electrons, described by $H_0$. Both the singlet and triplet induced pairing potentials, $\Delta_\tx{s,t}=(\Delta_+\pm\Delta_-$)/2, are relevant perturbations to $H_0$, namely this is an unstable fixed point. Below we show that the introduction of $H_\tx{int}$ causes the instability to be more towards triplet pairing with compare to singlet pairing.

The flow equations of the various terms in $H_\Delta$ and $H_\tx{int}$ can be obtained, for example, using perturbative momentum shell Wilsonian RG for Fermions~\cite{Shankar1994Renormalization}. This procedure results in the following flow equations~\cite{Haim2016interaction}
\begin{subequations}\label{eq:flow_eqs}
\begin{align}
&\dot{y}_1^\perp = -y_2 y_1^\perp ,\label{eq:KT_1}
\\
&\dot{y}_2 = -\frac{1}{2}\left(\frac{\bar{v}^2}{v_+v_-}+1\right) {y_1^\perp}^2 ,\label{eq:KT_2}
\\
&\dot{y}_2^+ = -\frac{1}{2}\frac{\bar{v}^2}{v_+v_-} {y_1^\perp}^2 ,\label{eq:y2_p_flow}
\\
&\dot{y}_2^- = -\frac{1}{2}\frac{\bar{v}^2}{v_+v_-} {y_1^\perp}^2 ,\label{eq:y2_pp_flow}
\\
&\dot{\Delta}_+ = \left( 1 - \frac{1}{2} y_2^+ \right)\Delta_+ - \frac{1}{2}\frac{\bar{v}}{v_-} y_1^\perp \Delta_- ,
\label{eq:Delta_p_flow}\\
&\dot{\Delta}_- = \left( 1 - \frac{1}{2} y_2^- \right)\Delta_- - \frac{1}{2}\frac{\bar{v}}{v_+} y_1^\perp \Delta_+ ,\label{eq:Delta_m_flow}
\end{align}
\end{subequations}
where we have defined $\bar{v}=(v_++v_-)/2$, and the dimensionless couplings $y_1^\perp=g_1^\perp/\pi\bar{v}$, $y_2^+=g_2^+/\pi v_+$, $y_2^-=g_2^-/\pi v_-$, and $y_2=g_2^+/2\pi v_+ + g_2^-/2\pi v_- - g_2^\parallel/\pi\bar{v}$. The above equations have been derived using a perturbative treatment and they are valid when $y_1$, $y_2^\parallel$, $y_2^\pm$ and $\Delta_\pm/v_\pm\Lambda$ are all smaller than 1.

Equations~(\ref{eq:KT_1},\ref{eq:KT_2}) give rise to a Kosterlitz-Thouless (KT) type of flow for $y_1^\perp$ and $y_2$. It is described by the constant of motion $A^2=y_2^2 - y_1^2$, where $y_1\equiv y_1^\perp\sqrt{(\bar{v}^2/v_+v_- + 1)/2}$. Of greatest interest for us is the region $y_2 > y_1\ge0$; this corresponds to an interaction which is repulsive on all length scales. In this case, the flow of $y_1$ and $y_2$ is given by
\begin{equation}\label{eq:KT_flow}
y_1(\ell) =  A \operatorname{csch} \left[ A \ell + \operatorname{arcoth} \frac {y_2(0)} {A} \right]
\hs 3mm ; \hs 3mm
y_2(\ell) = A  \coth \left[ A \ell + \operatorname{arcoth} \frac {y_2(0)} {A} \right].
\end{equation}
Both $y_1$ and $y_2$ flow down, saturating after an RG time $\ell_\tx{sat}\sim A^{-1}$, at $0$ and $A$, respectively. One can insert these solutions into Eqs.~\eqref{eq:y2_p_flow} and \eqref{eq:y2_pp_flow}, and integrate to obtain $y_2^+$ and $y_2^-$, respectively. The interaction couplings $y_1^\perp$, $y_2^+$, and $y_2^-$ can then be inserted into Eqs.~(\ref{eq:Delta_m_flow},\ref{eq:Delta_p_flow}) which generally require a numerical solution for $\Delta_\pm$.

We wish to determine the topological phase diagram of the system as a function of its initial couplings. We solve the above flow equations up to an RG time $\ell^\ast$, at which one of the pairing potential flows to strong coupling, namely $|\Delta_\pm(\ell^\ast)|/v_\pm\Lambda=1$. Beyond this point the perturbative RG treatment is not valid anymore. Let us assume, without loss of generality, that $\Delta_+$ flows to strong coupling first. This in particular means that the interaction couplings (which have flown down) are small in comparison to it, namely  $y_1^\perp, y_2^\parallel,y_2^\pm \ll |\Delta_+(\ell^\ast)|/v_+\Lambda=1$. If at this point $\Delta_-(\ell^\ast)/v_-\Lambda$ happens also to be large in comparison to $y_1^\perp, y_2^\parallel,y_2^\pm$, then we can neglect the interaction couplings. One can then use the topological invariant of a noninteracting system [see Eq.~\eqref{eq:top_inv_1}], $\nu = \sgn[\Delta_+(\ell^\ast)]\sgn[\Delta_-(\ell^\ast)]$. Generally, however, $\Delta_-(\ell^\ast)/v_-\Lambda$ can be small, and one has to modify the expression for $\nu$ to account for the non-negligible interaction terms.

To this end we note that since $\Delta_+(\ell^\ast)$ is large, the positive-helicity degrees of freedom [$\Rup^\nd(x)$ and $\Ldn^\nd(x)$] are gapped, and one can safely integrate them out. Upon doing so, one is left with an action containing only the negative-helicity fields [$R_\da(x)$ and $L_\ua(x)$], with a pairing potential $\Delta'_-=\Delta_-(\ell^\ast)+\delta\Delta_-$. To leading order in the interaction couplings, the correction is then given by~\cite{Haim2016interaction}
\begin{equation}\label{eq:correc_to_Delta_minus}
\begin{split}
\delta\Delta_- &= -\frac{\bar{v}}{2v_+}y_1^\perp(\ell^\ast)\Delta_+(\ell^\ast)\sinh^{-1}\left[\frac{v_+\Lambda}{|\Delta_+(\ell^\ast)|}\right]=
 -\frac{1}{2}y_1^\perp(\ell^\ast)\sgn[\Delta_+(\ell^\ast)]\sinh^{-1}(1)\bar{v}\Lambda.
\end{split}
\end{equation}
At this point we can continue the RG procedure, applied only to the negative-helicity degrees of freedom,
\begin{subequations}
\begin{align}\label{eq:flow_eqs_ssecnd_stp}
\dot{y}_2^- &= 0,\\
\dot\Delta'_- &= \left( 1 - \frac{1}{2} y_2^- \right)\Delta'_-,
\end{align}
\end{subequations}
namely $\Delta'_-$ flows to strong coupling (without changing sign), while $y_2^-$ remains perturbative. We can therefore use the topological invariant of noninteracting systems, only with $\Delta_-(\ell^\ast)$ substituted by $\Delta'_-$, $\nu=\sgn[\Delta_+(\ell^\ast)]\sgn[\Delta'_-]$. Finally, accounting also for the possibility that $\Delta_-$ flows to strong coupling before $\Delta_+$, one can write~\cite{Haim2016interaction}
\begin{equation}\label{eq:top_inv_RG}
\begin{split}
\nu = &\sgn\left\{\frac{\Delta_+(\ell^\ast)}{\bar{v}\Lambda} - \frac{\sinh^{-1}(1)}{2}y_1^\perp(\ell^\ast)\sgn[\Delta_-(\ell^\ast)]\right\}\times \\
&\sgn\left\{\frac{\Delta_-(\ell^\ast)}{\bar{v}\Lambda} - \frac{\sinh^{-1}(1)}{2}y_1^\perp(\ell^\ast)\sgn[\Delta_+(\ell^\ast)]\right\},
\end{split}
\end{equation}
where $\ell^\ast$ is the RG time when the first of $\Delta_+$ and $\Delta_-$ reaches strong coupling.

Let us consider again the case of a Hubbard-type interaction, $g_1^\perp=g_2^+=g_2^-=U$, and $g_2^\parallel=0$. Note that for $v_+=v_-$ this mean $y_2=y_1$, while for $v_+\neq v_-$, this means $y_2\ge y_1$ [see the definitions below Eq.~\eqref{eq:flow_eqs}]. Importantly, in both cases the KT flow equations dictates that the interaction couplings flow down. Figure~\hyperref[fig:TRITOPS_low_E_phase_diagram]{\ref{fig:TRITOPS_low_E_phase_diagram}(a)} shows the phase diagram for this Hubbard-type interaction, for $v_+=v_-$, calculated from Eq.~\eqref{eq:top_inv_RG}. The critical interaction strength $U$ which defines the phase boundary is obtained as a function of the initial ratio $\Delta^0_\tx{t}/\Delta^0_\tx{s}$, for different fixed values of $\Delta^0_\tx{s}$. Notice this phase boundary (solid lines) agrees well with that obtained from the mean field analysis (dashed lines), given in Eq.~\eqref{eq:MF_ph_bound_U}. It was estimated in Ref.~\cite{Haim2016interaction} that for typical proximity-coupled semiconducting systems, the dimensionless interaction strength, $U/(\pi\bar{v})$, should be of the order of $\sim0.1-1$. Figure~\hyperref[fig:TRITOPS_low_E_phase_diagram]{\ref{fig:TRITOPS_low_E_phase_diagram}(a)} suggests that such a system will be in the topological phase for a large range of the ratio $\Delta^0_\tx{t}/\Delta^0_\tx{s}$.

To better understand how repulsive interactions drive the system into the TRITOPS phase, let us reexamine the flow equations for the special case, $v_+=v_-$, $y_2^-=y_2^+$, for which Eqs.~(\ref{eq:Delta_m_flow},\ref{eq:Delta_p_flow}) reduce to
\begin{subequations}\label{eq:Delta_s_t_flow_eqs}
\begin{align}
&\dot{\Delta}_\tx{s} = \left( 1 - \frac{1}{2} y_2^+ - \frac{1}{2} y_1 \right)\Delta_\tx{s},
\label{eq:Delta_s_flow}\\
&\dot{\Delta}_\tx{t} = \left( 1 - \frac{1}{2} y_2^+ + \frac{1}{2} y_1 \right)\Delta_\tx{t}.
\label{eq:Delta_t_flow}
\end{align}
\end{subequations}
The effect of forward scattering and of backscattering on the pairing potentials is now apparent. The forward scattering term $y_2^+$ equally suppresses the singlet and triplet pairing terms. The backscattering term $y_1$, on the other hand, suppresses $\Delta_\tx{s}$, while strengthening $\Delta_\tx{t}$, causing the latter to flow faster to strong coupling. From Eq.~\eqref{eq:Delta_s_t_flow_eqs} one can extract the ratio between the triplet and singlet pairing terms as a function of RG time,
\begin{equation}\label{eq:r_t_s_ell}
    \frac {\Delta_{\mathrm{t}}(\ell)} {\Delta_{\mathrm{s}}(\ell)} = \frac{\Delta_{\mathrm{t}}^0} {\Delta_{\mathrm{s}}^0}
    \exp\left[\int_{0}^{\ell}\!\mathrm{d}\ell'y_1(\ell')\right].
\end{equation}

If the time it takes $y_1$ to flow to zero, $\ell_\tx{sat}$, is much shorter than $\ell^\ast$, we can approximate the ratio $\Delta_\tx{t}(\ell^\ast)/\Delta_\tx{s}(\ell^\ast)$ by taking the upper limit of the above integral to infinity. Using Eq.~\eqref{eq:KT_flow}, one obtains in this case
\begin{equation}\label{eq:Long_RG_time_approx}
    \frac {\Delta_\mathrm{t}(\ell^\ast)} {\Delta_{\mathrm{s}}(\ell^\ast)} \simeq\frac
    {\Delta_{\mathrm{t}}^0} {\Delta_{\mathrm{s}}^0}\sqrt{\frac {y_2^0+y_1^0}{y_2^0-y_1^0} }.
\end{equation}
Furthermore, since by our assumption $y_1(\ell^\ast)\simeq 0$ (follows from $\ell_\tx{sat}\ll\ell^\ast$), Eq.~\eqref{eq:top_inv_RG} tells us that the condition for the system to be topological is simply $|\Delta_\tx{t}(\ell^\ast)|>|\Delta_\tx{s}(\ell^\ast)|$. To understand when this approximation is valid, we can estimate the time it would take for one of the pairing potentials to reach strong coupling, $\ell^\ast\sim \ln(v_\pm\Lambda/\Delta_\pm^0)$\footnote{This estimation is obtained upon neglecting the second order terms in Eqs.~(\ref{eq:Delta_p_flow},\ref{eq:Delta_m_flow}) and integrating them up to $\Delta_\pm(\ell^\ast)=v_\pm\Lambda$.}. Namely, the above long RG-time approximation will be valid if the initial pairing potentials are small enough such that $\Delta_\pm^0 \ll v_\pm\Lambda \exp(-1/A)$. Note that the above approximation will necessarily be violated close to the separatrix of the KT flow, since there $A\to 0$.

In Fig.~\hyperref[fig:TRITOPS_low_E_phase_diagram]{\ref{fig:TRITOPS_low_E_phase_diagram}(b)} we present the topological phase diagram in the $y_2y_1$-plane for fixed initial values $\Delta_\tx{s}$ and $\Delta_\tx{t}$. The phase boundary is obtained by numerically solving Eq.~\eqref{eq:flow_eqs} up to a time $\ell^\ast$, and then invoking Eq.~\eqref{eq:top_inv_RG}, with $\ell^\ast$ being the RG time when the first (dimensionless) coupling reaches 1. The dashed red line shows the phase boundary in the long-RG-time approximation, obtained from Eq.~\eqref{eq:Long_RG_time_approx} and the condition $|\Delta_\tx{t}(\ell^\ast)|=|\Delta_\tx{s}(\ell^\ast)|$. As anticipated, it becomes more accurate as $A$ increases. We note that above the separatrix of the KT flow, $y_1$ and $y_2$ flow to strong coupling and the system is driven into an intrinsically topological phase~\cite{Keselman2015gapless,Kainaris2015Emergent}, irrespective of the initial induced potentials $\Delta_\pm$. Some nonvanishing induced pairing is however necessary to keep the system fully gapped.

\subsubsection{Numerical Analysis}
\label{subsubsec:realize_inter_syst_eff_of_int_R_wire}

In this section we concentrate on a given microscopic model - a proximity-coupled interacting nanowire [see Fig.~\hyperref[fig:real_inter_ind_pi_junc]{\ref{fig:real_inter_ind_pi_junc}(c)}], and numerically study its phase diagram using both a Hartree-Fock approximation and a DMRG analysis. We consider a semiconductor wire with strong spin-orbit coupling and in proximity to a conventional $s$-wave SC. We verify that upon including a sufficiently strong repulsive e-e interactions, the system realizes the TRITOPS phase. A similar effect has been shown to take place in semiconducting quantum wells coupled to an $s$-wave superconductor~\cite{Yu2016Gapped}, realizing a TRITOPS in 2d.

\begin{figure}
\begin{center}
\begin{tabular}{lr}
\rlap{\hskip 3mm \parbox[c]{\textwidth}{\vspace{-0cm}(a)}}
\includegraphics[clip=true,trim =0cm -2.5cm 0cm 0cm,width=0.45\tw]{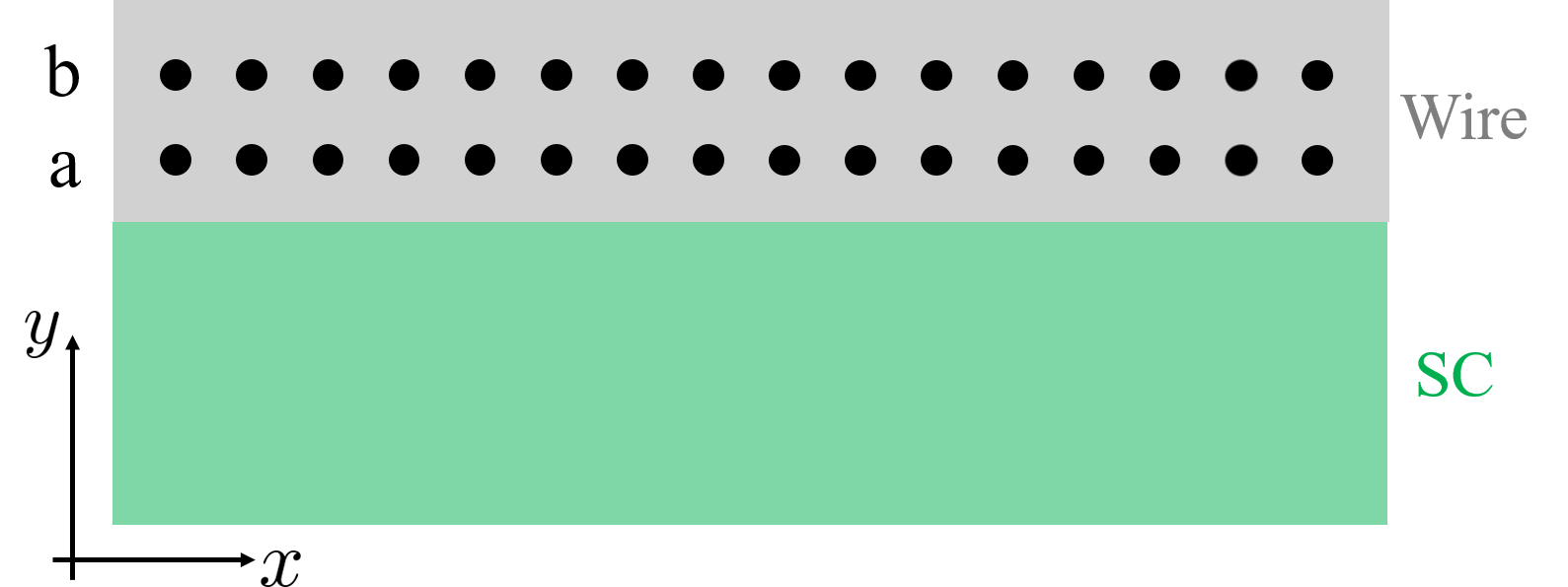}
\hs 2mm
&
\hs 2mm
\includegraphics[clip=true,trim =0.5cm 0cm 0.6cm 0.95cm,width=0.45\textwidth]{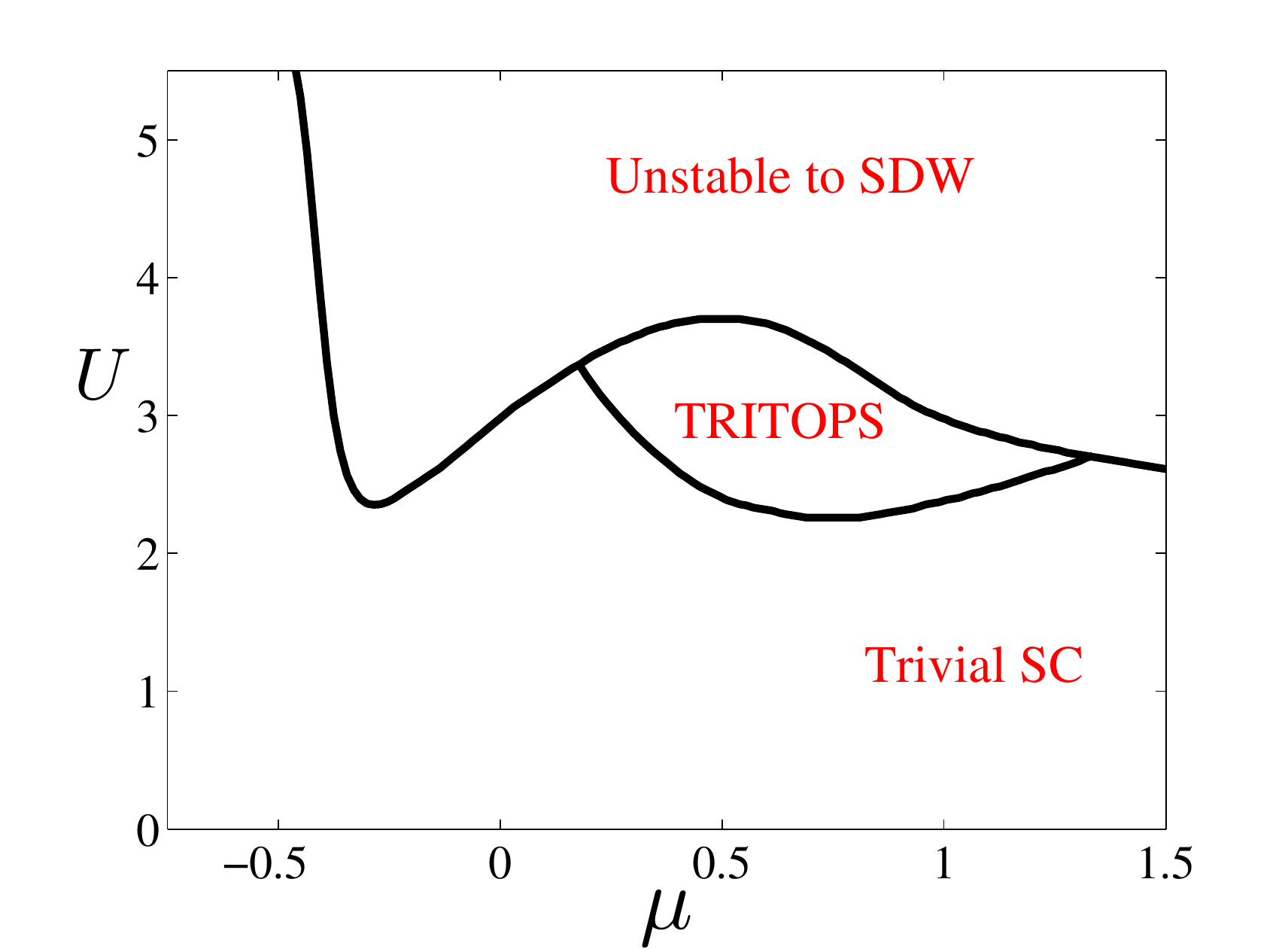}
\llap{\hskip 0mm \parbox[c]{0.45\textwidth}{\vspace{-0cm}(b)}}
\end{tabular}
\end{center}
\caption{(a) The system consists of a single quasi-1D wire (modelled by two chains) with SOC, coupled to a conventional {\it s}-wave superconductor. Integrating out the degrees of freedom of the superconductor generates a pairing potential $\Delta_{\rm{ind}}$ on the chain adjacent to the superconductor. (b) Hartree-Fock phase diagram as a function of chemical potential $\mu_a=\mu_b=\mu$, and interaction strength $U$, for $t_{\rm a}=t_{\rm b}=1, t_{\rm ab}=0.4, \alpha{\rm a}=0, \alpha_{\rm b}=0.6$, and $\Delta_{\rm ind}=1$. The diagram includes a time-reversal invariant-topological superconductor phase (TRITOPS), a trivial superconductor phase, and a region in which the Hartree-Fock solution is locally unstable to the formation of spin-density waves (see Ref.~\cite{Haim2014time} for details).}\label{fig:setup_and_num_phase_diagram_HF}
\end{figure}

We construct a lattice model for the nanowire, which is composed of two chains, as depicted in Fig.~\hyperref[fig:setup_and_num_phase_diagram_HF]{\ref{fig:setup_and_num_phase_diagram_HF}(a)}. The reason for using two chains is in order to simulate the effect described in Sec.~\ref{subsec:realize_pi-J_wires}, where modes of opposite helicity are separated in the $y$ direction. In a lattice model this can be most-simply captured by assuming two parallel chains, with two different strength of SOC, along each of them.

The Hamiltonian for the proximity-coupled nanowire in the presence of short-range interactions is then given by 
\begin{equation}
\begin{split}
&H=\frac{1}{2}\displaystyle\sum\limits_k \Psi_k^\dag\mathcal{H}^0_{k}\Psi_k + \sum_{i,\nu} U_{\nu}\hat{n}_{i\nu\uparrow}\hat{n}_{i\nu\downarrow}\\
&\mathcal{H}^0_{k}=\left[\bar{\xi}_k+\delta\xi_k\lambda_z-\left(\bar{\alpha}+\delta\alpha\lambda_z\right)\sin{k}\sigma_z +t_{ab}\lambda_x\right]\tau_z +\Delta_{\rm ind}/2~\cdot\left(1+\lambda_z\right)\tau_x ,
\end{split}\label{eq:H_NW_TB_Two_chain}
\end{equation}
where $\Psi_k^\dag=(\begin{matrix}\psi^\dag_k,&-i\sigma^y\psi_{-k}\end{matrix})$. The two spatially distinct chains are labeled, a and b, such that $\psi_k^\dag=(\begin{matrix}c_{{\rm a},k\uparrow}^\dag&c_{{\rm b},k\uparrow}^\dag&c_{{\rm a},k\downarrow}^\dag&c_{{\rm b},k\downarrow}^\dag\end{matrix})$. As before, $\tau_{x,y,z}$ and $\sigma_{x,y,z}$ are Pauli matrices in spin and PH basis, respectively. The Pauli matrices, $\lambda_{x,y,z}$, operate on the chain degree of freedom, $\nu={\rm a,b}$. Here, $\bar\xi_k, \delta\xi_k, \bar\alpha$ and $\delta\alpha$ are defined as  $(\xi_{k,{\rm a}}\pm\xi_{k,{\rm b}})/2$ and $\alpha_{\rm a}\pm\alpha_{\rm b}$, respectively, and $\xi_{k,\nu}=2t_\nu\left(1-\cos{k}\right)-\mu_\nu$. The parameters $t_\nu, \alpha_\nu, \mu_\nu$ and $U_\nu$ represent the hopping, SOC, chemical potential and on-site repulsion on chain $\nu={\rm a,b}$, while $t_{\rm ab}$ is the hopping between the chains. The operator $\hat{n}_{i,\nu,s}$ describes the number of electrons with spin $s$ on site $i$ of chain $\nu$.

\paragraph{Hartree-Fock}\label{parag:realize_inter_syst_eff_of_int_Num_HF}
In the Hartree-Fock analysis we consider a set of trial wave-functions which are ground states of the following quadratic Hamiltonian:
\begin{equation}
H_{\rm HF}=\frac{1}{2}\sum_k \Psi_k^\dag\mathcal{H}_k^{\rm HF}\Psi_k,
\hs 4mm ; \hs 4mm
\mathcal{H}_k^{\rm HF}=\tilde{\mathcal{H}}^0_{k}+\tilde{\Delta}_{\rm b}/2\cdot\left(1-\lambda^z\right)\tau^x,
\label{eq:H_HF}
\end{equation}
where $\tilde{\mathcal{H}}^0_{k}$ has the same form as $\mathcal{H}^0_{k}$, with effective parameters $\tilde\mu_{\rm a}, \tilde\mu_{\rm b}$ and $\tilde\Delta_{\rm a}$, where $\tilde\Delta_{\rm b}$ are effective pairing potentials on chains a and b, respectively. Upon determining the four effective parameters, the value of the topological invariant can be obtained by applying the results of Sec.~\ref{sec:TopInv} to Eq.~\eqref{eq:H_HF}.

We determine the effective parameters by numerically minimizing the expectation value of the full Hamiltonian in the ground state of $H_{\rm HF}$~\cite{Haim2014time},
\begin{equation}
\langle H\rangle_{\rm HF}=E_0+\frac{1}{L}\sum_{\nu={\rm a,b}} U_\nu\left(N_{\nu,\uparrow}N_{\nu,\downarrow}+\left|P_\nu\right|^2\right),
\end{equation}
with
\begin{equation}
\begin{split}
&N_{\nu,s}=\sum_k \langle c_{\nu,k,s}^\dag c_{\nu,k,s}\rangle_{\rm HF},\\
&P_\nu=\sum_k \langle c_{\nu,k,\uparrow}^\dag c_{\nu,-k\downarrow}^\dag\rangle_{\rm HF},\\
&E_0=\frac{1}{2}\sum_{k,m,n}\mathcal{H}^0_{k;mn}\langle \Psi_{k,m}^\dag\Psi_{k,n}\rangle_{\rm HF},
\end{split}\label{eq:H_avg}
\end{equation}
where $L$ is the number of sites in each chain, and we have used Wick's theorem, noting the exchange term vanishes due to the $\sigma^z$ conservation of $\mathcal{H}_k^{\rm HF}$.

Given the effective parameters, we are interested in the conditions under which $\mathcal{H}_k^{\rm HF}$ is in the topological phase. This Hamiltonian obeys the TRS $\Theta=i\sigma^yK$ and PH symmetry $\Xi=\tau^y\sigma^yK$, confirming it is in symmetry class DIII~\cite{Altland1997}. To obtain the topological invariant, we apply the procedure described in Sec.~\ref{subsec:TopInv_wind_num} for the Hamiltonian $\mathcal{H}_k^{\rm HF}$. The matrix $Q_k$, defined in Eq.~\eqref{eq:H_rot} is now given by
\begin{equation}
Q_k =
\frac{1}{2}(\tilde\Delta_{\rm a}+\tilde\Delta_{\rm b}) + \frac{1}{2}(\tilde\Delta_{\rm a}-\tilde\Delta_{\rm b})\lambda_z
+i\left[\bar{\xi}_k+\delta\xi_k\lambda_z-\left(\bar{\alpha}+\delta\alpha\lambda_z\right)\sin{k}\sigma_z +t_{ab}\lambda_x\right].
\end{equation}
The fact that $\sigma_z$ is a good quantum number allows us to easily obtain the topological invariant of, Eq.~\eqref{eq:Winding_parity}, as the parity of the winding number of~\cite{Haim2014time}
\begin{equation}\label{eq:det_Q_Z2}
\det[Q_k(\sigma_z=1)] = t_{ab}^2+\tilde\Delta_{\rm ind}\tilde\Delta_b -\tilde\varepsilon_{a,k}\tilde\varepsilon_{b,k}-i(\tilde\Delta_{\rm a}\tilde\varepsilon_{b,k}+\tilde\Delta_b\tilde\varepsilon_{a,k})
\end{equation}
where $\tilde\varepsilon_{\nu,k}=2t_\nu(1-\cos{k})-2\alpha_\nu\sin{k}-\tilde\mu_\nu$.

In Fig.~\hyperref[fig:setup_and_num_phase_diagram_HF]{\ref{fig:setup_and_num_phase_diagram_HF}(b)} we present the phase diagram obtained from Eq.~\eqref{eq:det_Q_Z2}, as a function of chemical potential and interaction strength for a specific set of wire parameters. The phase diagram includes a region in which the Hartree-Fock solution is locally unstable to formation of a \emph{spin-density wave} phase (see Ref.~\cite{Haim2014time} for more details).

\paragraph{Density matrix renormalization group}
\label{parag:realize_inter_syst_eff_of_int_Num_DMRG}
One can further verify the appearance of the topological phase using DMRG. We do this by studying the many-body spectrum of the system as a function of the system's length. As was explained in Sec.~\ref{sec:Intro}, the TRITOPS phase is characterized by a four-fold degenerated ground states, separated by an energy gap from the rest of the spectrum. Two of the states are of even fermion parity and two are of odd fermion parity. In a finite-size system, this degeneracy becomes an approximate one, with a splitting of the ground states which is exponentially small with the system size. However, the two odd-fermion-parity states will remain exactly degenerate for any system size due to Kramers' theorem. In contrast to the TRITOPS phase, in the trivial phase the spectrum is gapped with a \emph{single} ground state.

\begin{figure}
\begin{center}
\begin{tabular}{cc}
\hskip 0.07\textwidth
\includegraphics[clip=true, trim =-1cm 0cm 0cm 0cm,scale=0.5]{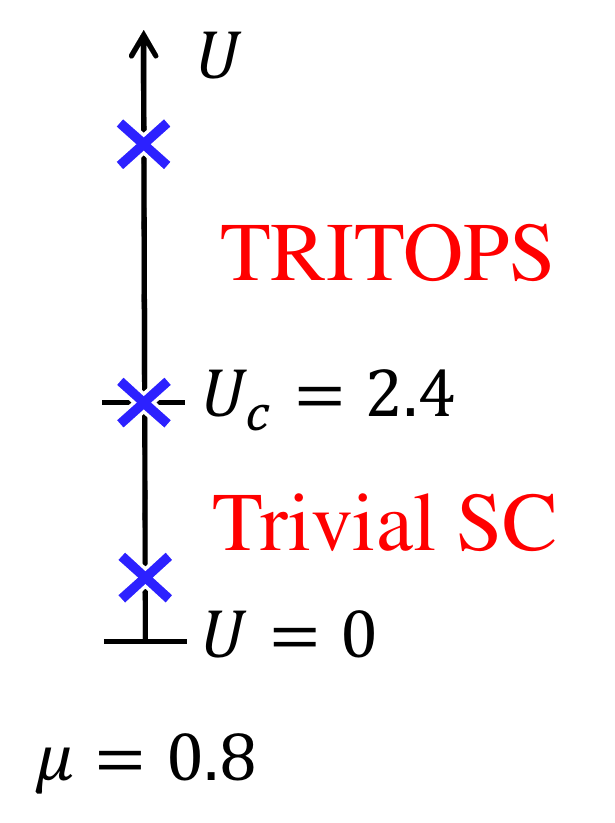}
\hs 0.05\textwidth
&
\hs 0.055\textwidth
\includegraphics[clip=true, trim =0cm 0cm 0.3cm 0.4cm,width=0.35\textwidth]{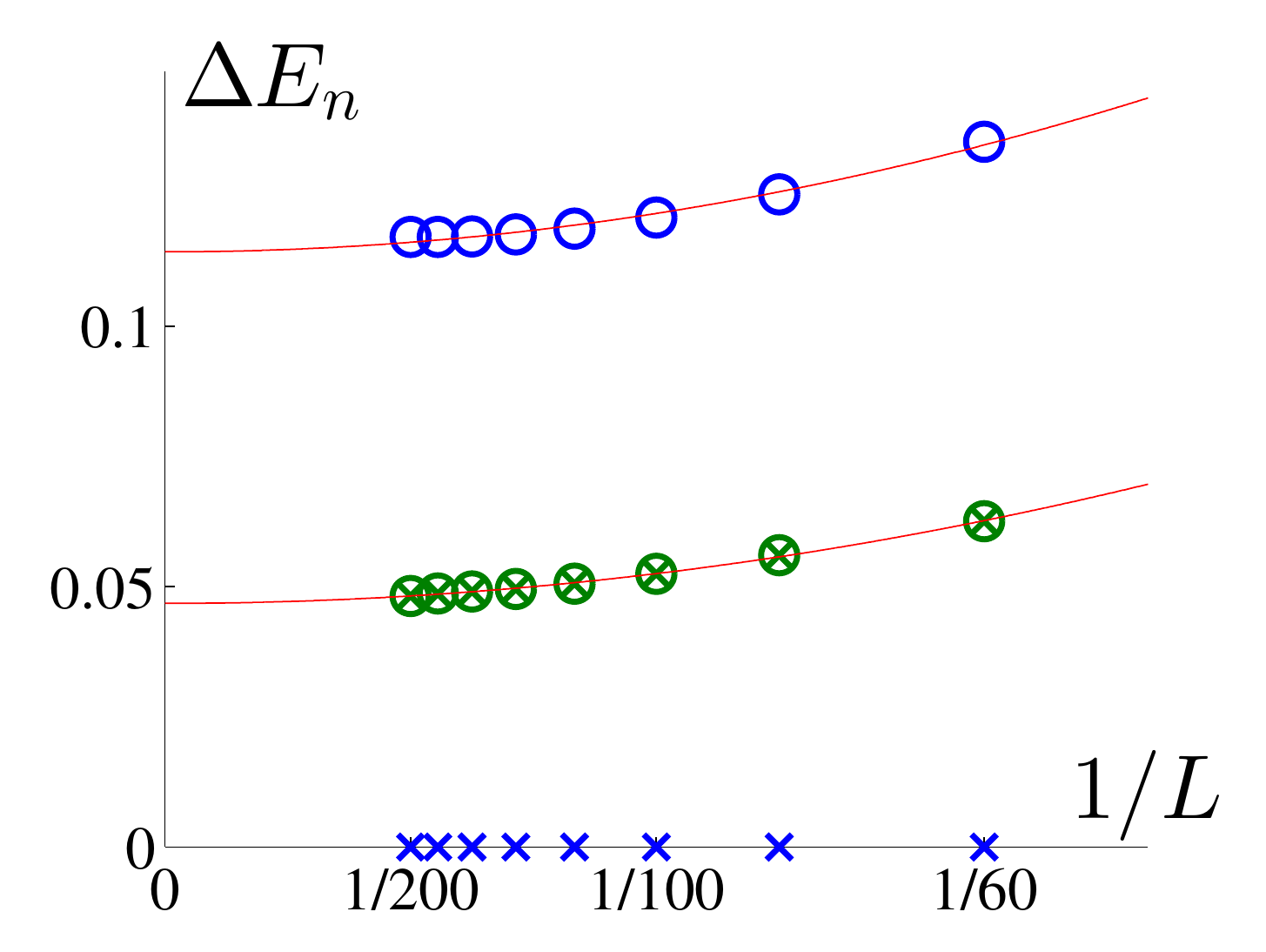} \\
\includegraphics[clip=true, trim =0cm 0.4cm 0cm 0cm,width=0.35\textwidth]{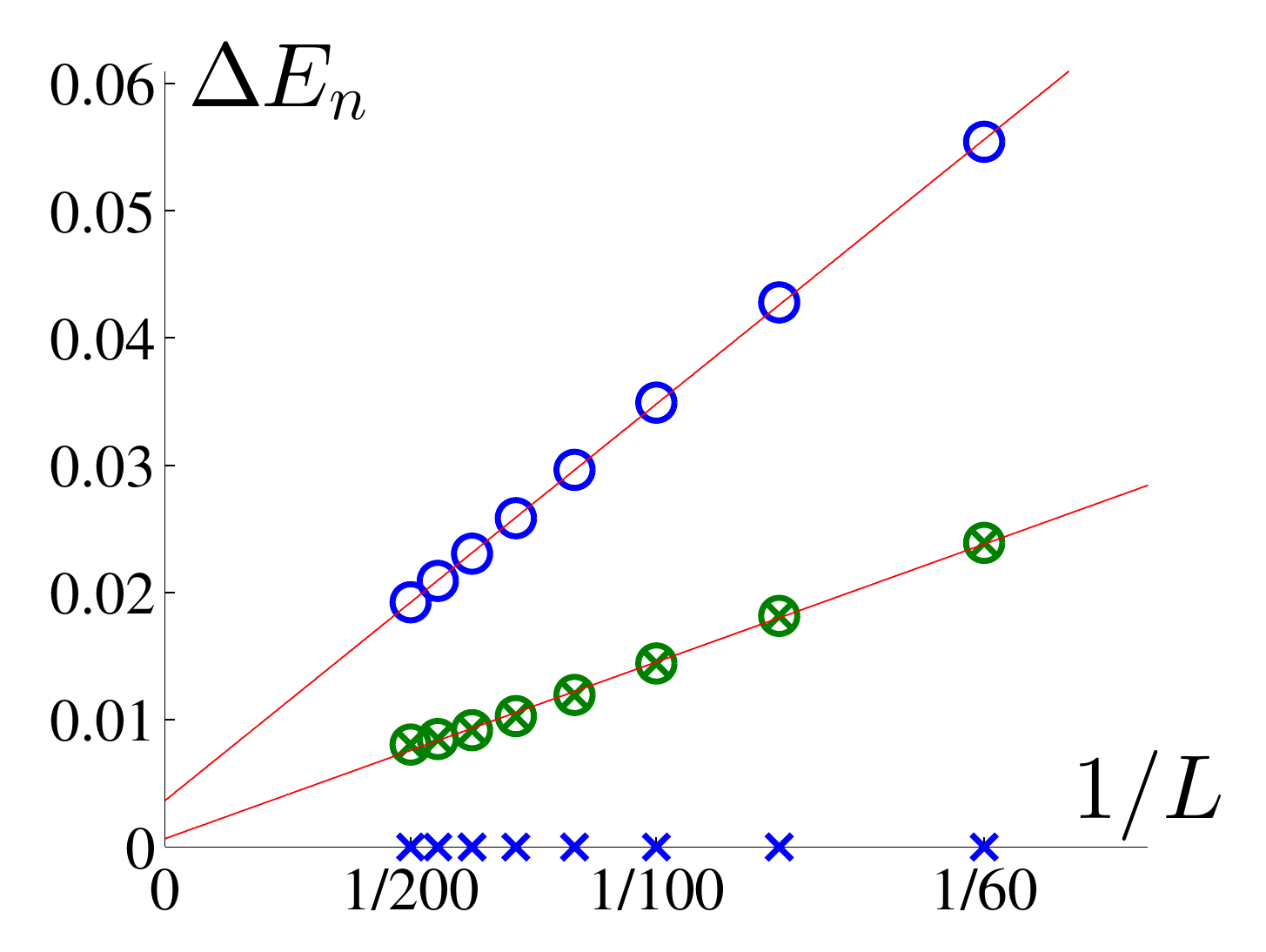}
\hskip 0.1\textwidth
&
\hskip 0.1\textwidth
\includegraphics[width=0.38\textwidth]{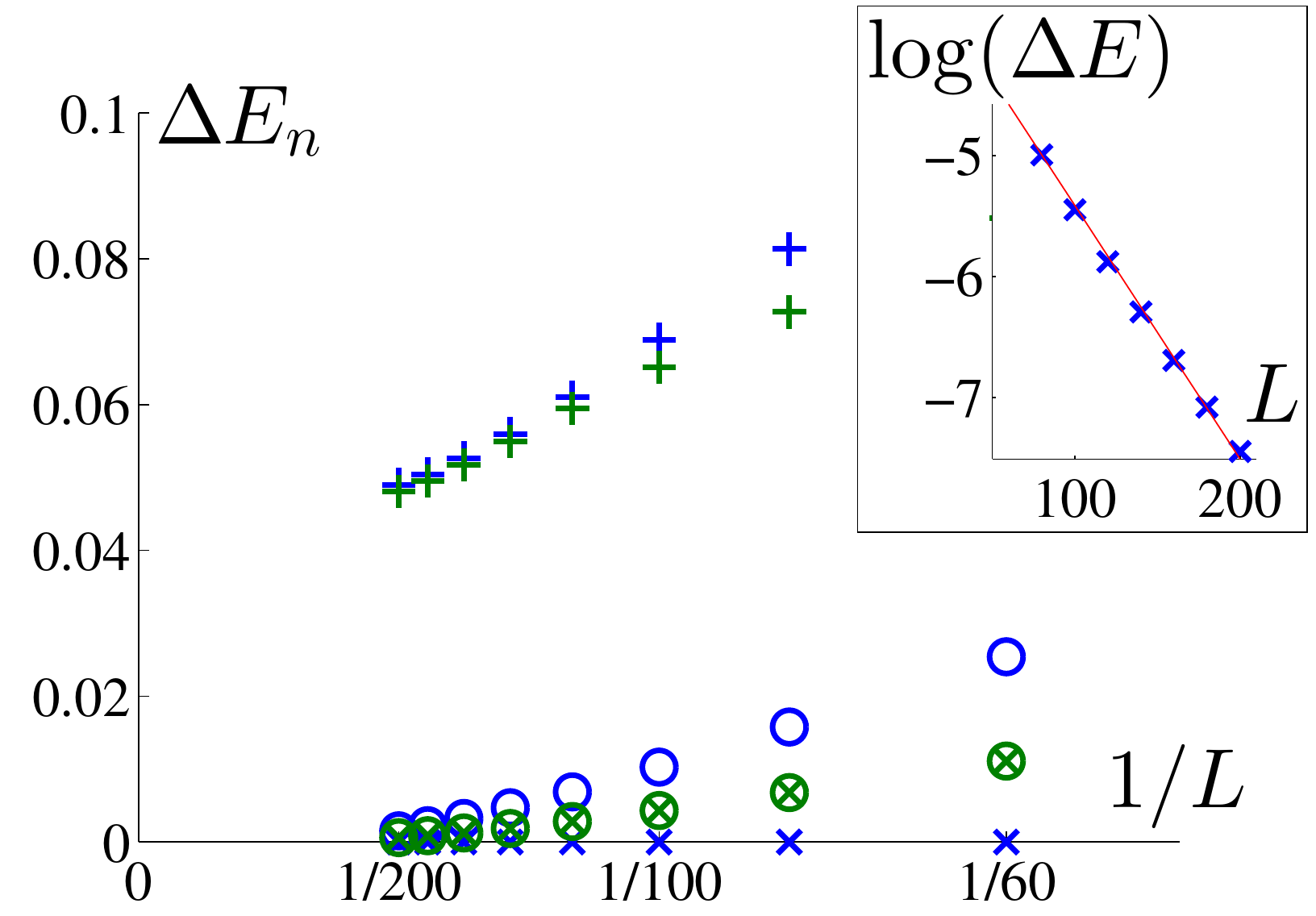}
\end{tabular}
\end{center}
\caption{(a) Phase diagram obtained using DMRG. The system is in the trivial superconducting phase for $U<U_\tx{c}$, and in the time-reversal-invariant topological superconductor (TRITOPS) phase for $U>U_\tx{c}$.
The system's parameters are $t_{\rm a}=t_{\rm b}=1,\ t_{\rm ab}=0.4,\ \alpha_{\rm a}=0,\ \alpha_{\rm b}=0.6,\ \Delta_{\rm{ind}}=1$, and $\mu_{\rm a}={\rm b}\equiv\mu=0.8$.
The low-energy many-body spectrum of the system vs. $1/L$, where $L$ is the length of the wire, for the three marked points is plotted in (b-d).
Energies plotted in blue (green) correspond to energy states in the even (odd) fermion parity sectors. All energies are plotted with respect to the energy of the ground state, which is in all cases the lowest energy state in the even fermion parity sector, $\Delta E_n=E_n-E^\tx{even}_0$.
(b) $U=0.5<U_\tx{c}$. The system is in the trivial superconducting phase. The first two states in the even and odd fermion parity sectors are shown. The ground state is unique and the gap tends to a constant as $L\rightarrow\infty$, with a quadratic correction in $1/L$ as expected (the red lines are quadratic fits). The first excited state which lies in the odd parity sector is doubly degenerate as expected from Kramers' theorem.
(c) $U=U_\tx{c}=2.4$. Phase transition point. Once again, the first two states in each fermion parity sector are shown. All gaps scale linearly with $1/L$ in agreement with the system being gapless in the infinite size limit (the red lines are linear fits).
(d) $U=5.5>U_\tx{c}$. The system is in the TRITOPS phase. Here, three lowest states in each fermion parity sector are shown. 
The result is consistent with a four-fold degenerate ground state in the thermodynamic limit, separated by a finite gap from the rest of the spectrum.
The inset shows the energy difference $\Delta E$ between the lowest states in the even and odd fermion parity sectors on a semi-log scale as a function of $L$. The result is consistent with an exponential dependence of $\Delta E$ on the system size.}\label{fig:TRITOPS_DMRG}
\end{figure}

A phase diagram obtained using DMRG is shown in Fig.~\hyperref[fig:TRITOPS_DMRG]{\ref{fig:TRITOPS_DMRG}(a)}~\cite{Haim2014time}.
Keeping the chemical potential $\mu=\mu_{\rm a}=\mu_{\rm b}$ constant we vary the on-site repulsive interaction strength $U$.
At $U=0$ the system is in a trivial superconducting phase with a finite gap for single particle excitations.
At a critical interaction strength, $U_\tx{c}$, a phase transition occurs and the gap closes.
For $U>U_\tx{c}$ the gap re-opens with the system now being in the TRITOPS phase.

Figures.~\hyperref[fig:TRITOPS_DMRG]{\ref{fig:TRITOPS_DMRG}(b-d)} present the scaling of the low-energy spectrum with the length of the wire at three different points in the phase diagram: one in the trivial superconducting phase, one in the TRITOPS phase and one at the critical point where the gap closes.
In the trivial superconducting phase [Fig.~\hyperref[fig:TRITOPS_DMRG]{\ref{fig:TRITOPS_DMRG}(b)}], the ground state is unique. The gap to the first excited state extrapolates to a finite value in the limit of an infinite system. Note that this state is doubly degenerate due to Kramers' theorem, as it is in the odd fermion parity sector. The gap to the first excited state in the even fermion parity sector is nearly twice as large, as expected. At the phase transition, the gap closes. For a finite 1D system this means that the gaps should be inversely proportional to the size of the system, as can be clearly seen in Fig.~\hyperref[fig:TRITOPS_DMRG]{\ref{fig:TRITOPS_DMRG}(c)}.
In the TRITOPS phase [Fig.~\hyperref[fig:TRITOPS_DMRG]{\ref{fig:TRITOPS_DMRG}(d)}] the ground state is four-fold degenerate up to finite size splitting.
The exponential dependence of the energy splitting on the length of the wire can be clearly seen in the inset.
The two lowest energy states in the odd fermion parity sector indeed remain degenerate for any system size.
Excited levels are separated from the ground state manifold by a finite gap. One thus concludes that the DMRG calculation supports the Hartree-Fock analysis of the system, confirming the appearance of the TRITOPS phase due to repulsive interactions.

\subsection{Proximity to unconventional superconductors}
\label{subsec:realize_unconv_sc}

Above we considered two mechanisms for realizing the TRITOPS phase, which included proximity to conventional superconductors. In the first, the $\pi$ phase difference between two SCs resulted in a $\pi$ phase difference in momentum space, between the positive-helicity and the negative-helicity modes, $\sgn(\Delta_+\Delta_-)=-1$. In the second mechanism, it were repulsive interactions which stabilized such a sign difference. In this section we explore the possibility of achieving the same effect by coupling the system to a single \emph{unconventional} SC~\cite{Nakosai2012topological,Wong2012majorana,Nakosai2013majorana,Zhang2013time,Chen2018helical}, such as for example an $s_\pm$-wave SC~\cite{Zhang2013time} or a $d_{x^2-y^2}$-wave superconductor~\cite{Wong2012majorana}.

\begin{figure}
	\begin{center}
		\begin{tabular}{cc}
			\rlap{\hskip 3mm \parbox[c]{\textwidth}{\vspace{-8cm}(a)}}
			\includegraphics[clip=true,trim =0mm -25mm 0mm 0mm,width=0.33\textwidth]{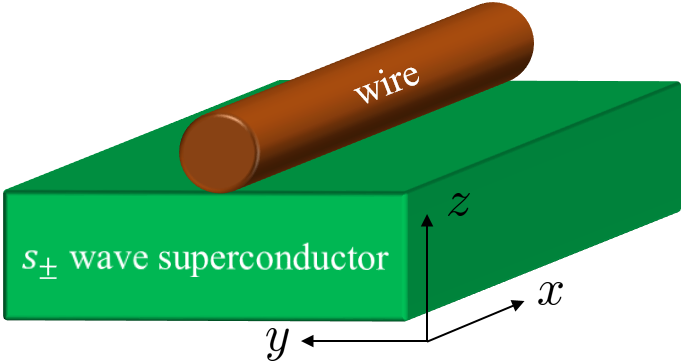}
			\hskip 0.3\textwidth
			&
			\hskip 0.0\textwidth
			\rlap{\hskip -1mm \parbox[c]{\textwidth}{\vspace{-10cm}(b)}}
			\includegraphics[clip=true,trim =0mm 0mm 0mm 0mm,width=0.375\textwidth]{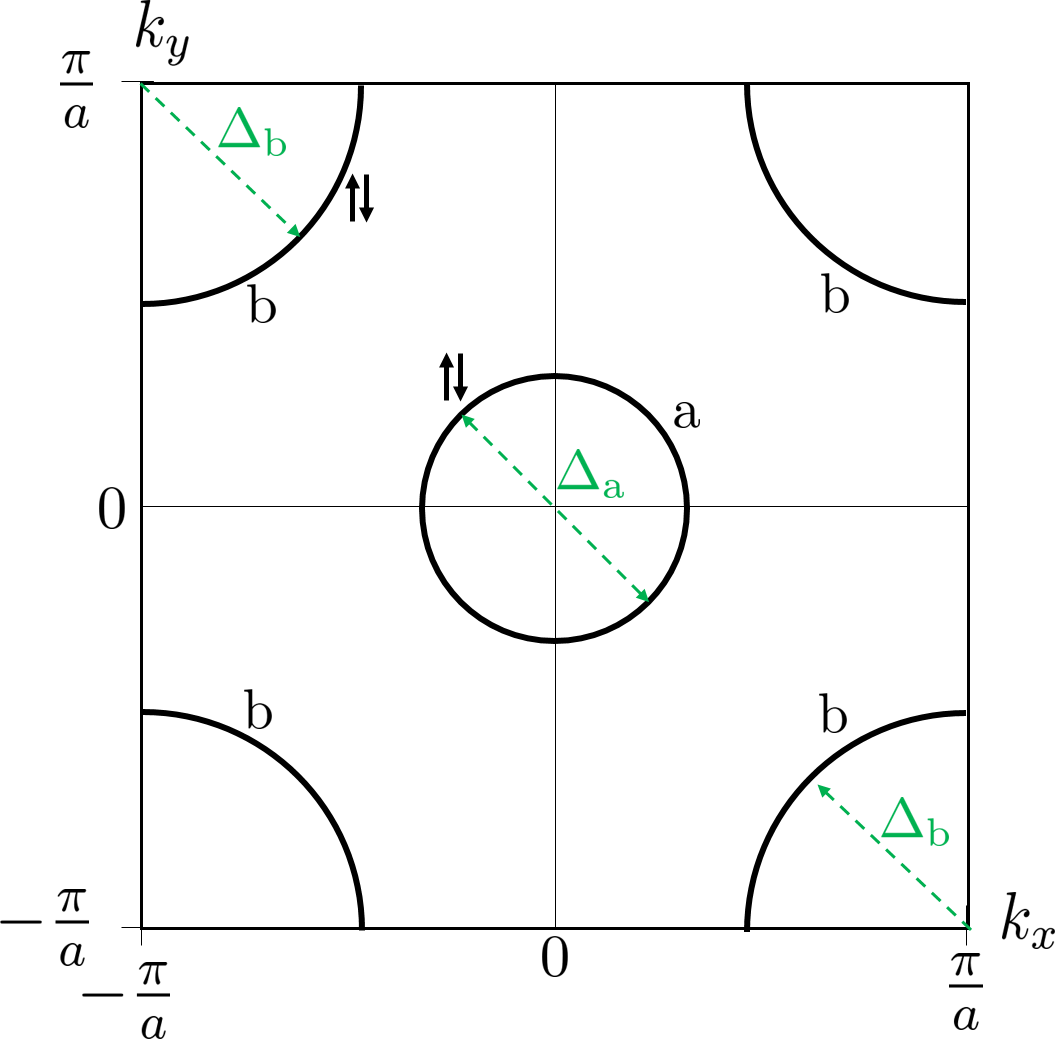}\\
			\hskip 0.0\textwidth
			&
			\hskip 0.0\textwidth
			\rlap{\hskip 3mm \parbox[c]{\textwidth}{\vspace{-4.5cm}(c)}}
			\hs 8mm
			\includegraphics[clip=true,trim =0mm 0mm 0mm 0mm,width=0.375\textwidth]{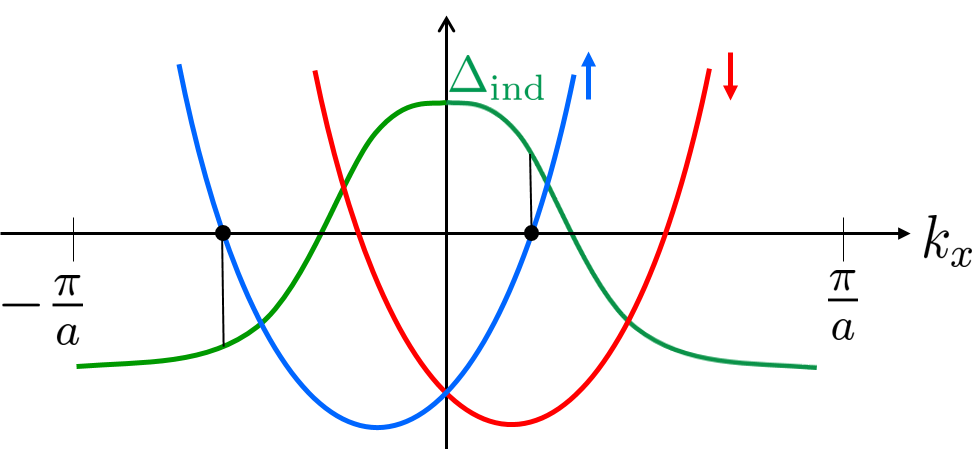}
		\end{tabular}
	\end{center}
	\caption{(a) A one-dimensional normal wire in proximity to a bulk three dimensional $s_\pm$-wave superconductor. (b) A two dimensional cut (constant $k_z$) of the First Brillouin Zone (FBZ) of the superconductor in its normal state.  The electronic dispersion is assumed to change only slightly in the $k_z$ direction. In the normal state, the SC has two spinful bands, labeled 'a' and 'b', whose Fermi surfaces are separated in momentum space. Importantly, the  pairing potentials in the two bands, $\Delta_\tx{a}$ and $\Delta_\tx{b}$, have opposite signs. (c) Since the contribution of each band to the induced superconductivity comes mostly from near the Fermi surface, the induced pairing potential has the sign of $\Delta_\tx{a}$ (assumed positive) at small momenta, and the sign of $\Delta_\tx{b}$ (assumed negative) at large momenta. If the wire in proximity to the SC has spin-orbit coupling, one can have $\Delta_\tx{ind}(k_x)$ with opposite signs at the two Fermi points (see the vertical black lines).}\label{fig:proximity_to_s_pm_SC}
\end{figure}

Consider a bulk three-dimensional $s_\pm$-wave SC in proximity to a normal 1d wire, as depicted in Fig.~\hyperref[fig:proximity_to_s_pm_SC]{\ref{fig:proximity_to_s_pm_SC}(a)}]. In the normal state of the SC, it has two spinful bands, labeled by 'a' and 'b', whose Fermi surfaces are separated in momentum space [see Fig.~\hyperref[fig:proximity_to_s_pm_SC]{\ref{fig:proximity_to_s_pm_SC}(b)}]. In the superconducting phase, the Fermi surfaces are gapped by a pairing potential of opposite signs. This can be described by the following Hamiltonian
\be
H_\tx{sc} = \sum_\bs{k} \left\{ \sum_{s=\ua,\da} \left[\xi_{\tx{a}}(\bs{k}) a^\dag_{\bs{k},s}a^\nd_{\bs{k},s} + \xi_\tx{b}(\bs{k}) b^\dag_{\bs{k},s}b^\nd_{\bs{k},s}\right]
+ [\Delta_\tx{a} a^\dag_{\bs{k},\ua}a^\dag_{-\bs{k},\da} + \Delta_\tx{b} b^\dag_{\bs{k},\ua}b^\dag_{-\bs{k},\da} +\tx{h.c.}] \right\},
\ee
where $a^\dag_{\bs{k},s}$ ($b^\dag_{\bs{k},s}$) creates an electron in band 'a' ('b') with momentum $\bs{k}=(k_x,k_y,k_z)$ and spin $s$. The dispersion relations for the two bands are $\xi_\tx{a}(\bs{k})$ and $\xi_\tx{b}(\bs{k})$, and their respective pairing potentials, $\Delta_\tx{a}$ and $\Delta_\tx{b}$, are assumed to have opposite signs, $\Delta_\tx{a}>0$, $\Delta_\tx{b}<0$.

The Hamiltonian for the combined system of the SC and the normal wire is $H=H_\tx{sc}+H_\tx{w}+H_\tx{t}$, with
\be
\begin{split}
	&H_\tx{w} = \sum_{k_x} c^\dag_{k_x,s}h_{ss'}(k_x) c^\nd_{k_x,s'},
	\\
	&H_\tx{t} = \sum_{\bs{k}} \sum_{s=\ua,\da} \left[t_\tx{a} c^\dag_{k_x,s}a_{\bs{k},s} + t_\tx{b} c^\dag_{k_x,s}b_{\bs{k},s} + \tx{h.c.}\right],
\end{split}
\ee
where $c^\dag_{k_x,s}$ creates an electron in the wire with momentum $k_x$ and spin $s$, $h(k_x)$ is the single-particle Hamiltonian (which we do not specify at the moment) describing the 1d system, and $t_\tx{a}$ ($t_\tx{b}$) is the coupling between the 1d system and the first (second) band of the superconductor.

Due to the proximity effect, superconductivity is induced in the wire. This is captured by the self-energy, $\Sigma(\omega,k_x)$, which results upon integrating out the SC's degrees of freedom (see for example Refs.~\cite{Alicea2012,Sau2010robustness}). At energies small compared with the bare SC gaps ($\omega\ll|\Delta_\tx{a}|,|\Delta_\tx{b}|$), the wire is described by the following effective BdG Hamiltonian
\begin{equation}
\begin{split}
&H^\tx{eff}_\tx{w} = \half\sum_{k_x}\psi^\dag_{k_x}\mc{H}^\tx{eff}_\tx{w}(k_x)\psi^\nd_{k_x}
\hs 4mm ; \hs 4mm
\psi^\dag_{k_x} = (c^\dag_{k_x,\ua},c^\dag_{k_x,\da},c^\nd_{-k_x,\da},-c^\nd_{-k_x,\ua}),\\
&\mc{H}^\tx{eff}_\tx{w}(k_x)=h(k_x)\tau^z + \Sigma(0,k_x),
\end{split}
\label{eq:integ_out}
\end{equation}
where the self-energy is given by
\begin{equation}
\begin{split}
&\Sigma(\omega,k_x)=\sum_{k_y,k_z} |t_\tx{a}|^2 G^{\rm sc}_\tx{a}(\omega,\boldsymbol{k}) + |t_\tx{b}|^2 G^{\rm sc}_\tx{b}(\omega,\boldsymbol{k}),\\
&G^{\rm sc}_i(\omega,\boldsymbol{k}) = \frac{\omega+\xi_i(\boldsymbol{k})\tau^z-\Delta_i\tau^x}{\omega^2-\xi^2_i(\bs{k})-\Delta_i^2}
\hs 4mm ; \hs 4mm
i=\tx{a}, \tx{b},
\end{split}
\end{equation}
where $G^{\rm sc}_\tx{a}(\omega,\boldsymbol{k})$ and $G^{\rm sc}_\tx{b}(\omega,\boldsymbol{k})$ are the Green functions for the first and second bands of the bare SC. One therefore has $\Sigma(0,k_x)=\Delta_\tx{ind}\tau^x$, with the induced pairing potential given by
\be
\Delta_\tx{ind}(k_x) =  \Delta_\tx{a}\sum_{k_y,k_z}\frac{|t_\tx{a}|^2}{\xi^2_\tx{a}(\bs{k})+\Delta_\tx{a}^2} + \Delta_\tx{b}\sum_{k_y,k_z}\frac{|t_\tx{b}|^2}{\xi^2_\tx{b}(\bs{k})+\Delta_\tx{b}^2},
\ee
where the two terms correspond to the contribution of the two bands.

Notice that the contribution of $\Delta_\tx{a}$ mainly comes from the momenta near the Fermi surface, $\xi^2_\tx{a}(\bs{k})=0$, and similarly the contribution of $\Delta_\tx{b}$ mainly comes from the momenta near the Fermi surface, $\xi^2_\tx{b}(\bs{k})$. The exact form of $\Delta_\tx{ind}(k_x)$ obviously depends on the exact values of $\Delta_\tx{a,b}$, and the exact form of $\xi_\tx{a,b}(\bs{k})$. Nevertheless, since, as depicted in Fig.~\hyperref[fig:proximity_to_s_pm_SC]{\ref{fig:proximity_to_s_pm_SC}(b)}, the Fermi surface of the a band occurs at small $k_x$, while the Fermi surface of the b band occurs at larger $k_x$, one generally expects $\Delta_\tx{ind}(k_x)$ to have the sign of $\Delta_\tx{a}$ at small $k_x$, and the sign of $\Delta_\tx{b}$ at large $k_x$. This is depicted in Fig.~\hyperref[fig:proximity_to_s_pm_SC]{\ref{fig:proximity_to_s_pm_SC}(c)}, with the green line marking $\Delta_\tx{ind}(k_x)$.

Having $\Delta_\tx{ind}(k_x)$ switching sign as a function of $k_x$ is by itself not a sufficient condition for being in the TRITOPS phase. In particular, notice that $\Delta_\tx{ind}(k_x)$ does not break inversion symmetry, which is a necessary condition (see Secs.~\ref{subsec:Intro_min_model} and~\ref{subsec:Lattice_model}). In order to be in the TRITOPS phase, the wire must therefore break inversion symmetry, for example due to an appreciable Rashba spin-orbit coupling. In Fig.~\hyperref[fig:proximity_to_s_pm_SC]{\ref{fig:proximity_to_s_pm_SC}(c)} the electronic dispersion of such a wire is depicted, with the blue and red line denoting the spin-$\ua$ and spin-$\da$ modes, respectively.

Zhang \emph{et al.}~\cite{Zhang2013time} have suggested describing the induced superconductivity from a $s_\pm$-wave superconductor using the lattice model of Eq.~\eqref{eq:H_min-latt_k}, (with $\Delta_1''=0$), which we rewrite here for convenience,
\be
\begin{split}
	H = \sum_{k}\left\{ \sum_{s,s'}(-\mu -2t\cos(k_xa) + 2u\sin(k_xa)\sigma^z_{ss}) c^\dag_{k_xs} c_{k_xs'} + [(\Delta_0 + \Delta'_1\cos(k_xa))c^\dag_{k_x\ua} c^\dag_{-k_x\da} +\tx{h.c.}]\right\}.
\end{split}
\ee
For $|\Delta_1'|>|\Delta_0|$, the induced pairing potential, $\Delta_\tx{ind} (k_x)=\Delta_0 + \Delta'_1\cos(k_xa))$, has the desired property of changing sign, when going from small $|k_x|$ to large and large. As noted above, however, this is not a sufficient condition. The pairing potential has to have different sign for the different Fermi momenta. This happens when~\cite{Zhang2013time}
\be
2|u|\sqrt{1-(\Delta_0/\Delta_1')^2}>|\mu - 2t\Delta_0/\Delta_1'|.
\ee
Namely, it is beneficial to have as large spin-orbit coupling and a large $\Delta_1'$.

The challenge in realizing a TRITOPS in this manner is in matching the Fermi momenta in the wire (typically a semiconductor), with the momentum at which $\Delta_\tx{ind}(k_x)$ changes sign, which is of the order of the Fermi momentum of the SC in its normal state. The latter is typically much larger than the former.

Another type of unconventional superconductor which can be used in proximity to the wire is a $d_{x^2-y^2}$-wave superconductor. In this kind of SC, the pairing potential changes sign (four times) when going along the Fermi contour in the $(k_x,k_y)$ plane. When placing the wire along the $x$ direction, small and large momenta then experience an induced superconductivity with opposite signs. An important thing to note is that, since the $d_{x^2-y^2}$-wave SC has gapless nodes in its spectrum, the zero-energy MKPs can hybridize with the gapless modes in the bulk SC. Namely they are no longer completely localized. Nevertheless, numerical simulations (in a clean system) show that this effect does not cause a strong coupling between the MKPs at the two ends of the wire~\cite{Wong2012majorana}.

\section{Signatures of Majorana Kramers pairs}
\label{sec:Signa}

Experimentally realizing the TRITOPS phase would of course be meaningless in the absence of a way to detect its physical properties. Since an obvious property distinguishing the TRITOPS phase from its topologically-trivial counterparts is the existence of protected boundary modes, one is encouraged to probe these modes when looking for distinct signatures of TRITOPS. In 1d systems these would be the (zero-dimensional) Kramers pairs of MBSs, while in 2d these would be the (one-dimensional) counter-propagating helical Majorana modes.

We focus in this review on three types of signatures. The first involves electronic transport from a metallic lead to the topological superconductor through the MKP which it hosts. The second kind of signature has to do with the unique way in which the Majorana boundary modes (either in 1d or 2d) behave under a Magnetic field. Finally, we consider Josephson junctions, either between two topological superconductors or between a trivial and a topological superconductor. As we will see, the existence of MKPs in the junction modifies the spectrum and current-phase relation compared with a Josephson junction of trivial superconductors.

\subsection{Conductance through a Majorana Kramers pair}
\label{subsec:Signa_Cond_MKP}

The Majorana Kramers pair (MKP), as discussed, is a zero-energy mode inside a superconductor. As such, one can expect to see a resonance in the differential conductance spectrum when tunneling from a metallic lead into the SC through the MKP. So much is true for any zero-energy mode which is a superposition of an electron and a hole. However, as we shall see below, the resonance due to a MKP is both robust and has a quantized amplitude, in a similar way to the case of a single MBS in a TRS-broken phase~\cite{Bolech2007Observing,Law2009majorana,Fidkowski2012universal,flensberg2010tunneling}.

We will begin with the case of a normal spinful lead, in which both spin species have right-moving and left-moving modes [see Fig.~\hyperref[fig:Signa_MKP_and_Lead]{\ref{fig:Signa_MKP_and_Lead}(a)}]. In this case, the model for the system can be decomposed into two copies, each describing a \emph{spinless} lead coupled to a \emph{single} MBS. It then follows that, due to perfect Andreev reflection, one obtains a zero-bias conductance peak quantized to $4e^2/h$~\cite{Wong2012majorana,Haim2014time}, each copy contributing $2e^2/h$ to the differential conductance.

We then move on to examine a helical lead, in which there is a right-moving spin-$\ua$ mode and a left-moving spin-$\da$ mode (or vice versa). This situation correspond corresponds to a MKP coupled to the edge of a 2d topological insulator (2dTI)~\cite{Pikulin2016Luttinger,Li2016detection}, as depicted in Fig.~\hyperref[fig:Signa_MKP_and_Lead]{\ref{fig:Signa_MKP_and_Lead}(b)}. In this case one can measure conductance in a three-terminal setup (one terminal being the SC hosting the MKP), opening the door for probing the scattering processes in more detail. As we will see, while electrons are still perfectly converted into holes, one can have either Andreev reflection or Andreev transmission.

\subsubsection{Normal spinful lead}
\label{subsubsec:Signa_norm_lead}

\begin{figure}
\begin{center}
\begin{tabular}{cc}
\hs -10mm
\rlap{\hskip -0.025\tw \parbox[c]{\tw}{\vspace{-0.1\tw}(a)}}
\includegraphics[clip=true,trim =0cm -3cm 0cm 0cm,width=0.43\tw]{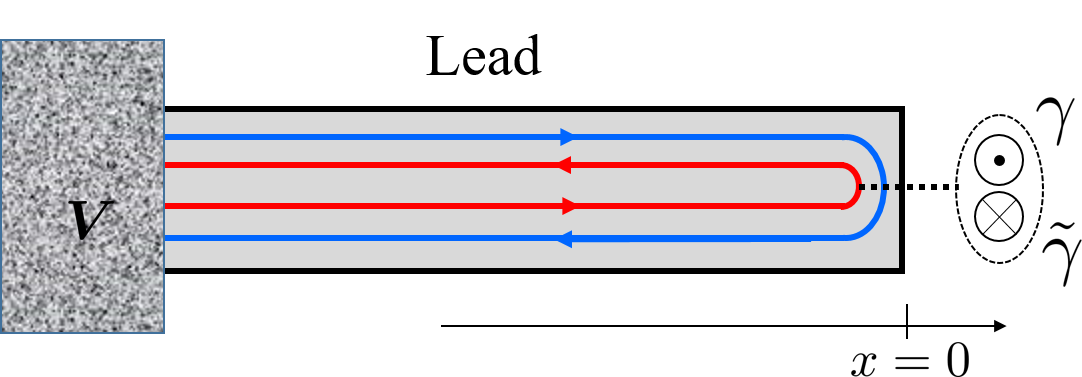}

&
\hs 10mm
\rlap{\hskip -0.025\tw \parbox[c]{\tw}{\vspace{-0.1\tw}(b)}}
\includegraphics[clip=true,trim =0cm 0cm 0cm 0cm,width=0.33\tw]{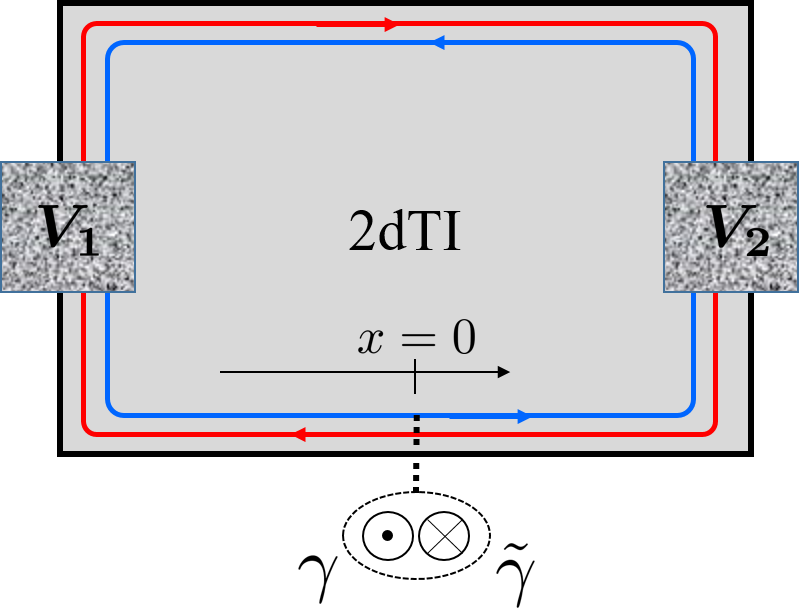}
\end{tabular}
\end{center}
\caption{Metallic leads coupled to a Kramers pair of Majorana bound states (MKP). (a) Normal spinful lead in which each spin has both a right-moving and a left-moving mode. Electrons incident with zero energy are perfectly Andreev reflected, giving rise to zero-bias conductance peak which is quantized to $4e^2/h$. (b) Helical lead in which modes of opposite spins move in opposite directions. Electrons with zero energy experience either Andreev reflection or Andreev transmission.\label{fig:Signa_MKP_and_Lead}}
\end{figure}

We consider a normal-metal spinful lead coupled to a Kramers pair of Majorana bound states. The Hamiltonian describing the lead in the wide-band limit is given by
\begin{equation}\label{eq:H_lead}
H_\tx{Lead} = -iv\sum_{s=\ua,\da}\int_{-\infty}^0\tx{d}x\left[ \psi_{\tx{R},s}^\dag(x)\partial_x\psi_{\tx{R},s}(x) - \psi_{\tx{L},s}^\dag(x)\partial_x\psi_{\tx{L},s}(x)\right],
\end{equation}
where $\psi_{\tx{R},s}^\dag$ and $\psi_{\tx{L},s}^\dag$ are creation operators in the lead for an electron with spin $s=\ua,\da$, moving in the right and left directions, respectively, as depicted in Fig.~\hyperref[fig:Signa_MKP_and_Lead]{\ref{fig:Signa_MKP_and_Lead}(a)}. Time-reversal symmetry relates the modes in the lead through
\be
\psi_{\tx{R},s}(x) \longrightarrow i\sigma^y_{ss'}\psi_{\tx{L},s'}(x)
\hs 5mm ; \hs 5mm
\psi_{\tx{L},s}(x) \longrightarrow i\sigma^y_{ss'}\psi_{\tx{R},s'}(x)
\ee

Before writing the Hamiltonian describing the coupling to the MKP, let us rewrite $H_\tx{Lead}$ in a more convenient form. We can ``unfold'' the modes in the lead by defining two chiral fields,
\begin{subequations}\label{eq:unfold_lead}
\be
\psi_\ua(x) = \left\{\begin{matrix} \psi_{\tx{R},\ua}(x), & x<0\\\psi_{\tx{L},\ua}(-x),& x>0 \end{matrix} \right. ,
\ee
and
\be
\psi_\da(x) = \left\{\begin{matrix} \psi_{\tx{L},\da}(x), & x<0\\{\psi_\tx{R},\da}(-x),& x>0 \end{matrix} \right. ,
\ee
\end{subequations}
which extend from $x\rightarrow-\infty$ to $x\rightarrow\infty$, such that the Hamiltonian for the lead reads
\be\label{eq:H_Lead}
H_\tx{Lead} = \sum_{s=\ua,\da}\int_{-\infty}^\infty \tx{d}x \psi^\dag_s(x)\partial_x\psi_s(x).
\ee
Time-reversal symmetry relates the two chiral fields to each other through
\be
\psi_\ua(x) \longrightarrow \psi_\da(x)
\hs 5mm ; \hs 5mm
\psi_\da(x) \longrightarrow -\psi_\ua(x).
\ee

We now turn to write the Hamiltonian describing coupling of the lead to the MKP. Using the new fields, the most general low-energy time-reversal-symmetric coupling Hamiltonian one can write is
\be\label{eq:H_coupling}
H_\tx{Coupling} = i\gamma [t_1\psi_\ua(0) + t_2 \psi_\da(0)] -i\tilde\gamma[t_1^\ast\psi_\da(0) - t_2^\ast \psi_\ua(0)] + \tx{h.c.}
\ee
where $\gamma$ and $\tilde\gamma$ are the Majorana operators creating the two MBSs of the Kramers pair. Notice that the form of $H_\tx{Coupling}$ (namely the relation between the coupling to $\gamma$ and the coupling to $\tilde\gamma$) is constrained by TRS, whose operation on the Kramers pair is given by $\gamma\longrightarrow\tilde\gamma$ and $\tilde\gamma\longrightarrow -\gamma$.

The form of the overall Hamiltonian, $H=H_\tx{Lead}+H_\tx{Coupling}$, allows one to write it as two copies, related by TRS, each describing a spinless lead coupled to a single MBS. To see this, one defines two new chiral fields,
\be\label{eq:chiral_fields_trans}
\bmat \psi(x) \\ \tilde\psi(x) \emat =U
\bmat \psi_\ua(x) \\ \psi_\da(x) \emat
\hs 5mm ; \hs 5mm
U=
\frac{1}{\sqrt{|t_1|^2+|t_2|^2}}\bmat t_1 & t_2 \\ -t_2^\ast & t_1^\ast \emat.
\ee
Notice that $U$ is a unitary matrix, making $\psi(x)$ and $\tilde\psi(x)$ obey fermionic commutation relations, and that under TRS, $\psi(x)\rightarrow\tilde\psi(x)$ ; $\tilde\psi(x)\rightarrow-\psi(x)$. Setting the transformation, \Eq{eq:chiral_fields_trans}, in Eqs.~\eqref{eq:H_Lead} and~\eqref{eq:H_coupling}, the system's Hamiltonian reduces to
\be\label{eq:two_spinless_MBS_copies}
\begin{split}
H =& -iv\int_{-\infty}^\infty \tx{d}x \psi^\dag(x)\partial_x\psi(x) + i\lambda\gamma[\psi(0) + \psi^\dag(0)]\\
& -iv\int_{-\infty}^\infty \tx{d}x \tilde\psi^\dag(x)\partial_x\tilde\psi(x) - i\lambda\tilde\gamma[\tilde\psi(0) + \tilde\psi^\dag(0)],
\end{split}
\ee
where $\lambda\equiv\sqrt{|t_1|^2+|t_2|^2}$. Indeed, Eq.~\eqref{eq:two_spinless_MBS_copies} describes two TR-related copies of a spinless lead coupled to a single MBS.

With the help of the decompositions of the Hamiltonian, Eq.~\eqref{eq:two_spinless_MBS_copies}, one can immediately predict several signatures of MKP which are generalizations of those for a single MBSs. For example, the zero-temperature differential conductance from a normal lead to SC through a single MBS exhibit a $2e^2/h$ zero-bias peak (ZBP), and therefore the differential conductance through a MKP has a ZBP which is quantized to $4e^2/h$. In the same way, signatures related to currant noise and current cross correlations can also be generalized in a straight-forward way.

More explicitly, we can obtain all transport quantities from the scattering matrix, relating the incoming and outgoing modes in the lead,
\be
\bmat \psi_\eps(0^+) \\ \tilde\psi_\eps(0^+) \\ \psi^\dag_\eps(0^+) \\ \tilde\psi^\dag_\eps(0^+)\emat
= S(\eps)
\bmat \psi_\eps(0^-) \\ \tilde\psi_\eps(0^-) \\ \psi^\dag_\eps(0^-) \\ \tilde\psi^\dag_\eps(0^-)\emat
\hs 5mm ; \hs 5mm
S(\eps) = \bmat S^\tx{ee}(\eps)&S^\tx{eh}(\eps)\\S^\tx{he}(\eps)&S^\tx{hh}(\eps)\emat
\ee
where $\psi_\eps(x)\equiv\int\tx{d}t\psi(x,t)\exp(-i\eps t)$, with a similar expression for $\tilde\psi_\eps(x)$, and where time evolution is according to $H$. The scattering matrix can be calculated  in a straight-forward way from Eq.~\eqref{eq:two_spinless_MBS_copies}, yielding
\be\label{eq:S_lead_MKP}
S(\eps) = \frac{1}{1+i\eps/\Gamma}
\bmat i\eps/\Gamma & 1 \\ 1 & i\eps/\Gamma \emat
\otimes\mathbbm{1}_{2\times2},
\ee
where $\Gamma\equiv2\lambda^2/v=2(|t_1|^2+|t_2|^2)/v$. Notice that for $\eps=0$ there's perfect Andreev reflection in both channels, $|S^\tx{he}_{11}(0)|^2=|S^\tx{he}_{22}(0)|^2 = 1$.

From the scattering matrix, one can obtain the differential conductance with the help of the generalized Landauer-B\"uttiker formalism~\cite{shelankov1980resistance,blonder1982transition,lesovik2011scattering}, which in this case yield
\be\label{eq:diff_conduc}
\frac{\tx{d}I}{\tx{d}V} = \frac{2e^2}{h} \sum_{ij}\int_{-\infty}^\infty \tx{d}\eps \left[\delta_{ij} - |S_{ij}^\tx{ee}(\eps)|^2 +|S_{ij}^\tx{he}(\eps)|^2\right]\left[-\frac{\partial f(\eps-eV)}{\partial\eps}\right],
\ee
where $f(\eps)=1/[1+\exp(\eps/k_\tx{B}T)]$ is the Fermi-Dirac distribution, and $T$ is the temperature. Inserting the expression for $S^\tx{he}(\eps)$ from Eq.~\eqref{eq:S_lead_MKP}, one obtains for $T=0$
\be
\frac{\tx{d}I}{\tx{d}V}=\frac{2e^2}{h} \left[|S^\tx{he}_{11}(eV)|^2+|S^\tx{he}_{22}(eV)|^2\right] = \frac{4e^2}{h} \frac{\Gamma^2}{(eV)^2+\Gamma^2},
\ee
which exhibits a resonance at $V=0$, with a peak quantized to $4e^2/h$~\cite{Wong2012majorana,Haim2014time}. Notice that while the width of the resonance is determined by $\Gamma$, the height of the resonance is independent of it. This is no longer the case at finite temperature, where for $T$ sufficiently larger than $\Gamma$ the height of the peak is decreased by a factor $\sim\Gamma/(k_\tx{B}T)$.

This behavior is in perfect analogy to the case of a lead coupled to a single MBS. From the fact that $S(\eps)$ is composed of two parts with each being the scattering matrix for a single MBS, one can also infer other transport quantities which are borrowed from the single MBS case. The zero-frequency shot noise at, for example, yields~\cite{Golub2011shot}
\be
P=\int_{-\infty}^\infty\tx{d}t \langle \delta\hat{I}(0)\delta\hat{I}(t) \rangle \xrightarrow[eV\ll\Gamma]{}
\frac{4e^2}{h} \frac{2(eV)^3}{3\Gamma^2}
\ee
at small bias voltage. Notice it decays to zero slower than linearly at small $V$, which is generally different than the case of an accidental low-energy (topologically trivial) Andreev bound state. This has to do with the perfect Andreev reflection at zero energy; shot noise is the result of the probabilistic nature of the scattering process.

Related to this effect is the behavior of cross correlation of currents in \emph{two} leads [each described by the Hamiltonian of Eq.~\eqref{eq:H_lead}] coupled to a MKP~\cite{Haim2015current},
\be
P_{12}=\int_{-\infty}^\infty\tx{d}t \langle \delta\hat{I}_1(0)\delta\hat{I}_2(t) \rangle = -\frac{4e^2}{h}\Gamma_1\Gamma_2\frac{eV}{(eV)^2+\Gamma_\tx{T}^2},
\ee
where here $\Gamma_1, \Gamma_2$, stand for the coupling of each lead to the MKP, and $\Gamma_\tx{T}=\Gamma_1+\Gamma_2$. The negative cross correlation which goes to zero at $eV\ll\Gamma_\tx{T}$ is again different than a trivial Andreev bound state, in which the cross correlation generally goes to a positive constant~\cite{Haim2015signatures,Haim2015current}. Importantly, this behavior survives even at finite temperatures ($k_\tx{B}T\gtrsim\Gamma$).

\subsubsection{Helical lead}
\label{subsubsec:Signa_helical_lead}

The above results were a straight-forward generalization of the case of a single MBS in a TRS-broken TSC. This was possible because, as we saw, the setup of Fig.~\hyperref[fig:Signa_MKP_and_Lead]{\ref{fig:Signa_MKP_and_Lead}(a)} is equivalent to two copies of a spinless lead coupled to a MBS. This analogy does not apply anymore when one considers a lead with counter-propagating helical modes, where each mode is leading to a different ohmic contact, as depicted in Fig.~\hyperref[fig:Signa_MKP_and_Lead]{\ref{fig:Signa_MKP_and_Lead}(b)}. This can be achieved when the edge of a two-dimensional topological insulator is coupled to a MKP~\cite{Pikulin2016Luttinger,Li2016detection}. One can then study the currents, $I_1$ and $I_2$, reaching the contacts as a function of their voltages, $V_1$ and $V_2$, respectively.

To obtain the scattering matrix for this setup, we first notice that the number of modes entering and exiting the junction ($x=0$) is the same as in the setup depicted in Fig.~\hyperref[fig:Signa_MKP_and_Lead]{\ref{fig:Signa_MKP_and_Lead}(a)} and analyzed above. the main difference is that here we cannot mix between the modes as we did in \Eq{eq:chiral_fields_trans}, since the modes are leading to different contacts. We can nevertheless use the results of the previous section to obtain the scattering matrix relevant for this setup. The sought scattering matrix relates the outgoing modes, $\psi_{\tx{R},\ua,\eps}(0^+)$ and $\psi_{\tx{L},\da,\eps}(0^-)$, to the incoming modes, $\psi_{\tx{R},\ua,\eps}(0^-)$ and $\psi_{\tx{L},\da,\eps}(0^+)$,
\be
\bmat \psi_{\tx{R},\ua,\eps}(0^+) \\ \psi_{\tx{L},\da,\eps}(0^-) \\ \psi^\dag_{\tx{R},\ua,\eps}(0^+) \\ \psi^\dag_{\tx{L},\da,\eps}(0^-) \emat =
\bar{S}(\eps)
\bmat \psi_{\tx{R},\ua,\eps}(0^-) \\ \psi_{\tx{L},\da,\eps}(0^+) \\ \psi^\dag_{\tx{R},\ua,\eps}(0^-)\\ \psi^\dag_{\tx{L},\da,\eps}(0^+)\emat.
\ee
Using the relation between $\psi_{\tx{R},\ua,\eps}(x), \psi_{\tx{L},\da,\eps}(x)$ and $\psi(x), \tilde\psi(x)$, as given in Eqs.~\eqref{eq:unfold_lead} and \eqref{eq:chiral_fields_trans}, one obtains
\be
\bar{S}(\eps) =
\bmat U^\dag &0 \\ 0 & U^\top \emat
S(\eps)
\bmat U &0 \\ 0 & U^\ast \emat
=\frac{1}{1+i\eps/\Gamma}
\bmat
\frac{i\eps}{\Gamma}\cdot\mathbbm{1}_{2\times2} & U^\dag U^\ast \\ U^\top U &  \frac{i\eps}{\Gamma}\cdot\mathbbm{1}_{2\times2}
\emat
\ee
The perfect Andreev scattering at zero energy is manifested by $|\bar{S}_{11}^\tx{he}(0)|^2+|\bar{S}_{21}^\tx{he}(0)|^2=1$, as can be inferred immediately by noting that $\bar{S}^\tx{he}(0)=-U^\dag U^\ast$ is a unitary matrix. Namely, a spin-$\ua$ electron approaching the MKP from the left can either be reflected as a spin-$\da$ hole (Andreev reflection) or transmitted as a spin-$\ua$ hole (Andreev transmission), but overall the probability of electron-to-all conversion is $1$. The two processes correspond to a Cooper pair being formed, either by two electrons which tunnel into the SC from the same contact, or from different contacts.

The probabilities for Andreev reflection and transmission depend, in general, on the microscopic details of the system (which in turn determine the couplings, $t_1$ and $t_2$, appearing in $U$.)\footnote{Notice, however, that if the system has a $U(1)$ spin-rotation symmetry, such that spin in the $z$ direction is conserved, then the scattering from a spin-$\ua$ electron to a spin-$\ua$ hole is forbidden and one necessarily has Andreev reflection rather than Andreev transmission.}. To probe these different processes one can study the elements of the differential conductance matrix
\be
\frac{\tx{d}I_i}{\tx{d}V_j} = \frac{e^2}{h} \int_{-\infty}^\infty \tx{d}\eps \left[\delta_{ij} - |\bar{S}^\tx{ee}_{ij}(\eps)|^2 + |\bar{S}^\tx{he}_{ij}(\eps)|^2\right]\left[-\frac{\partial f(\eps-eV)}{\partial\eps}\right].
\ee
Since at zero energy we have only Andreev processes, $\bar{S}^\tx{ee}(0)=0$, one obtains that (at zero temperature) the diagonal conductance,
\be
\left.\frac{\tx{d}I_1}{\tx{d}V_1}\right|_{V_1=0} = \frac{e^2}{h}[1 + |\bar{S}^\tx{he}_{11}(0)|^2],
\ee
is determined by the Andreev reflection amplitude, and the off-diagonal conductance,
\be\label{eq:diff_cross_cond}
\left.\frac{\tx{d}I_2}{\tx{d}V_1}\right|_{V_1=0} = \frac{e^2}{h}|\bar{S}^\tx{he}_{21}(0)|^2,
\ee
is determined by the Andreev transmission amplitude. Interestingly, the total current going into the SC, $I_\tx{S}=I_1+I_2$, still carries the information regarding the perfect Andreev scattering; the differential conductance of the current through SC (with respect to the voltage in either lead) is in fact quantized~\cite{Pikulin2016Luttinger},
\be
\left.\frac{\tx{d}I_\tx{S}}{\tx{d}V_1}\right|_{V_1=0} =\frac{\tx{d}I_\tx{S}}{\tx{d}V_2}= \frac{e^2}{h}[1+|\bar{S}^\tx{he}_{11}(0)|^2 + |\bar{S}^\tx{he}_{12}(0)|^2]=\frac{2e^2}{h}.
\ee
This is essentially because both processes (Andreev reflection and Andreev transmission) inject two electrons into the SC.

One can wonder about the effect of electron-electron interaction on the scattering amplitudes~\cite{Pikulin2016Luttinger,Aasen2016Interaction}. 
It was shown~\cite{Pikulin2016Luttinger} that the process of Andreev reflection is suppressed, while the process of Andreev transmission is strengthen (keeping the sum of their probability equal to unity). This can be understood intuitively, since in the presence of repulsive interactions, there is an energy cost for two electrons to be injected from the same lead, as compared to the case where each electron in the Copper pair comes from a different lead.

One of the advantages of using the edge of a 2dTI as a lead is that it can quite naturally embed the MKP in it~\cite{Pikulin2016Luttinger,Li2016detection}. Indeed, by proximitizing the edge of the 2dTI to two SC in a $\pi$ junction, one realizes a 1d TRITOPS with a MKP at its end~\cite{Fu2008superconducting,Keselman2013inducing}. Accordingly, Li \emph{et al.}~\cite{Li2016detection} suggested to combine a quantum point contact and a Josephson junction on the edge of a 2dTI [in a way which would essentially implement the setup of  Fig.~\hyperref[fig:Signa_MKP_and_Lead]{\ref{fig:Signa_MKP_and_Lead}(b)}], and monitor the zero-bias cross conductance as a function of the phase difference. When the phase difference is $\phi=\pi$, the MKP induces perfect Andreev scattering which yields a positive $\tx{d}I_2/\tx{d}V_1\propto|\bar{S}_{21}^\tx{he}(0)|^2$, as given in Eq.~\eqref{eq:diff_cross_cond}. When the phase difference is $\phi=0$, on the other hand, the MKP splits in energy and the cross conductance is dominated by normal scattering, giving rise to a negative cross conductance, $\tx{d}I_2/\tx{d}V_1\propto(|\bar{S}_{21}^\tx{he}(0)|^2 - |\bar{S}_{21}^\tx{ee}(0)|^2)$. The change of sign of $\tx{d}I_2/\tx{d}V_1$ as a function of $\phi$ can serve as a signature of the MKP.

\subsection{Anomalous Zeeman splitting}
\label{subsec:Signa_AnomZeemSplit}

The zero-bias conductance peak discussed above probes the existence the MKP, which is protected against splitting from zero energy as long as TRS in maintained. In this subsection we examine the behavior of the MKP under breaking of TRS by a magnetic field~\cite{Keselman2013inducing,Zhang2013time,Gaidamauskas2014majorana,Haim2014time,Dumitrescu2014magnetic,Aligia2018entangled}. The natural expectation is that, due to the Zeeman effect, the two MBSs composing the MKP will split in energy with a gap proportional to the magnetic field, $|\bs{B}|$. This would generally be the case for a low-energy Andreev bound states. It turns out, however, that for a MKP, the energy splitting depends on magnetic field in a unique way. While for a magnetic field in a general direction, the splitting is indeed linear, there is always a plane in which applying a magnetic field results in a splitting which goes like $\bs{B}^3$ or higher powers~\cite{Keselman2013inducing,Haim2014time,Dumitrescu2014magnetic}.

This anomalous Zeeman splitting can be used to distinguish between a trivial low-energy resonance and the topologically nontrivial MKP. An advantage of this signature is that it is not expected to be sensitive to finite temperatures. This should be contrasted with the zero bias peak, which is only quantized to $4e^2/h$ at zero temperature. The anomalous Zeeman splitting, on the other hand, while could be smeared by temperature, its dependence on $\bs{B}$ should not change.

The behavior of the MKP under the application of magnetic field is understood based on symmetry considerations. Consider a system in the TRITOPS phase with a MKP at each end of the system, and let us focus for the moment only on one end. The addition of a TRS-breaking perturbation, in this case a magnetic field, allows for a coupling between the two MBSs composing the pair. At low-energies, this is described by
\be\label{eq:H_TRS_breaking}
H^\prime = i\lambda\gamma\tilde\gamma.
\ee
where $\lambda$ is a coupling coefficient which depends on the magnetic field, $\bs{B}$, in a way that we wish to determine. While the Hamiltonian is not time-reversal symmetric, it should be invariant under a TRS which is followed by reversal of the magnetic field, namely $\mathbb{T}H'(-\bs{B})\mathbb{T}^{-1}=H'(\bs{B})$. Applying this condition to Eq.~\eqref{eq:H_TRS_breaking} results in $\lambda(-\bs{B}) = -\lambda(\bs{B})$. Indeed, in the absence of magnetic field the coupling must vanish. For small $\bs{B}$ we can expand $\lambda(\bs{B})$ in a Taylor series, which will only contain odd powers of $B$~\cite{Keselman2013inducing},
\begin{equation}\label{eq:Zeeman_splitting_expan}
\lambda(\bs{B}) = \sum_i \lambda^{(1)}_{i}B_i + \sum_{ijk} \lambda^{(3)}_{ijk}B_i B_j B_k +\dots
\end{equation}
While the first term in the expansion is linear, there is always a plane, defined by $\bs{B}\perp\bs{\lambda^{(1)}}$, in which the linear term vanishes. When the magnetic field is directed in this plane, the energy splitting of the MBSs is of order $|\bs{B}|^3$ or higher~\cite{Keselman2013inducing,Haim2014time,Dumitrescu2014magnetic}.

The absence of a linear term in the energy splitting can be demonstrated numerically. Let us consider the Hartree-Fock Hamiltonian, $\mc{H}_k^\tx{HF}$, describing the interacting proximitized wire introduced in Eq.~\eqref{eq:H_HF} of Sec.~\ref{subsubsec:realize_inter_syst_eff_of_int_R_wire}, with parameters $\tilde\mu_{\rm a}=\tilde\mu_{\rm b}=0.15$, $t_{\rm a}=t_{\rm b}=1$, $t_{\rm ab}=0.4$, $\alpha_{\rm a}=0$, $\alpha_{\rm b}=0.6$, $\tilde\Delta_{\rm a}=0.3$, and $\tilde\Delta_{\rm b}=-0.15$.  With these parameters the system is in the TRITOPS phase with a MKPs at each end. We now apply a magnetic field in the $x$ direction, by introducing a Zeeman term to the Hamiltonian, $\mc{H}_k=\mc{H}_k^\tx{HF}+\mc{H}_k^\tx{Z}$, where
\be
\mc{H}_k^\tx{Z} = B_x\sigma_x.
\ee
We attach a normal-metal lead to the end of the wire, by introducing a coupling between the modes in the lead and the first sites of the double chain [see also Fig.~\hyperref[fig:setup_and_num_phase_diagram_HF]{\ref{fig:setup_and_num_phase_diagram_HF}(a)}]. One can then calculate numerically the scattering matrix for electrons and holes in the lead, from which the conductance is obtained using Eq.~\eqref{eq:diff_conduc}.

\begin{figure}
\begin{center}
\begin{tabular}{cc}
\rlap{\hskip -0.025\tw \parbox[c]{\tw}{\vspace{-0.55\tw}(a)}}
\includegraphics[clip=true,trim =0cm -0.2cm 0cm 0.8cm,width=0.41\tw]{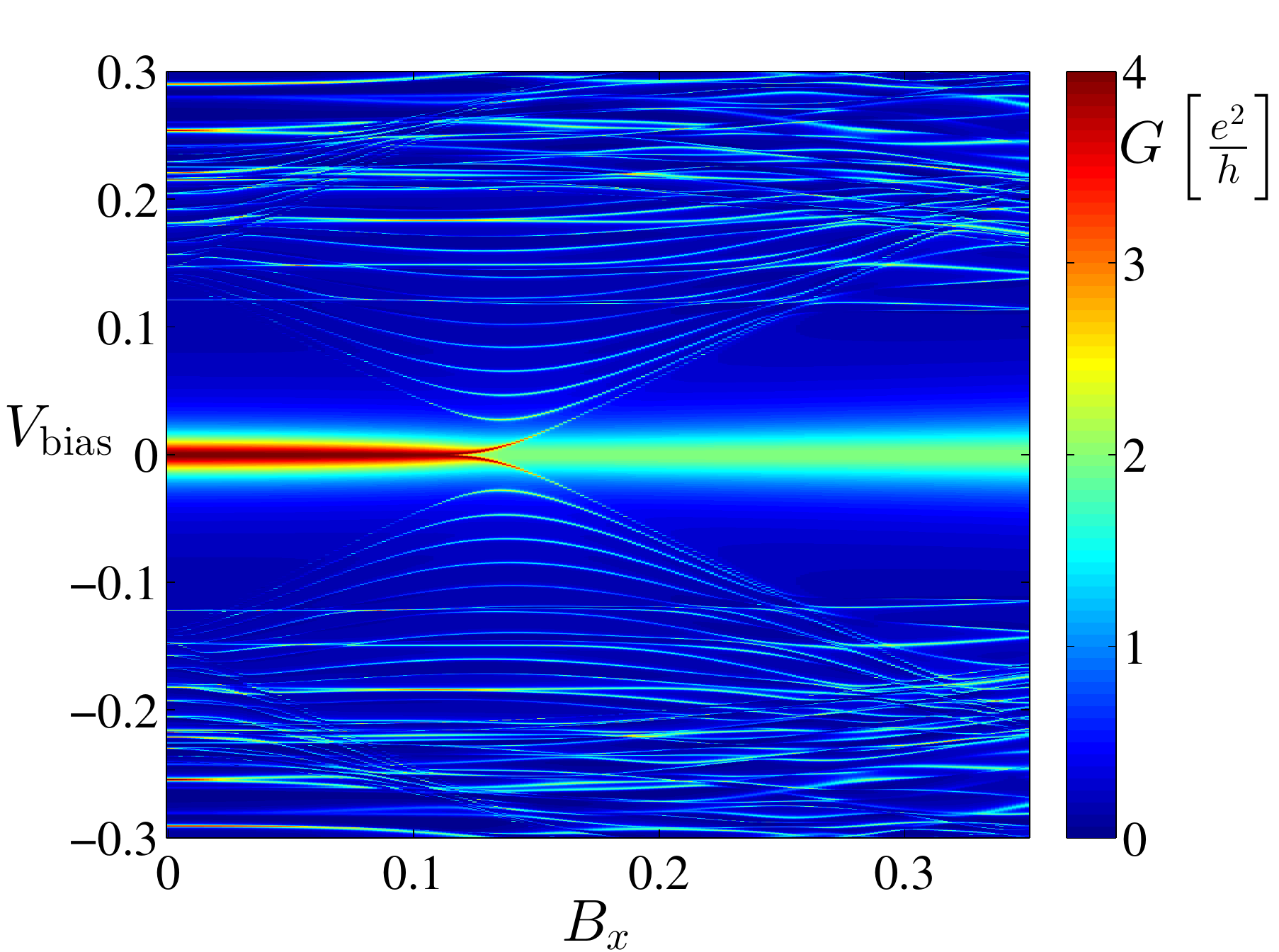}\label{fig:TRITOPS_sig_diff_cond}
\hs 5mm
&
\hs 5mm
\rlap{\hskip -0.025\tw \parbox[c]{\tw}{\vspace{-0.55\tw}(b)}}
\includegraphics[clip=true,trim =3.1cm 8.75cm 4cm 9.4cm,width=0.346\tw]{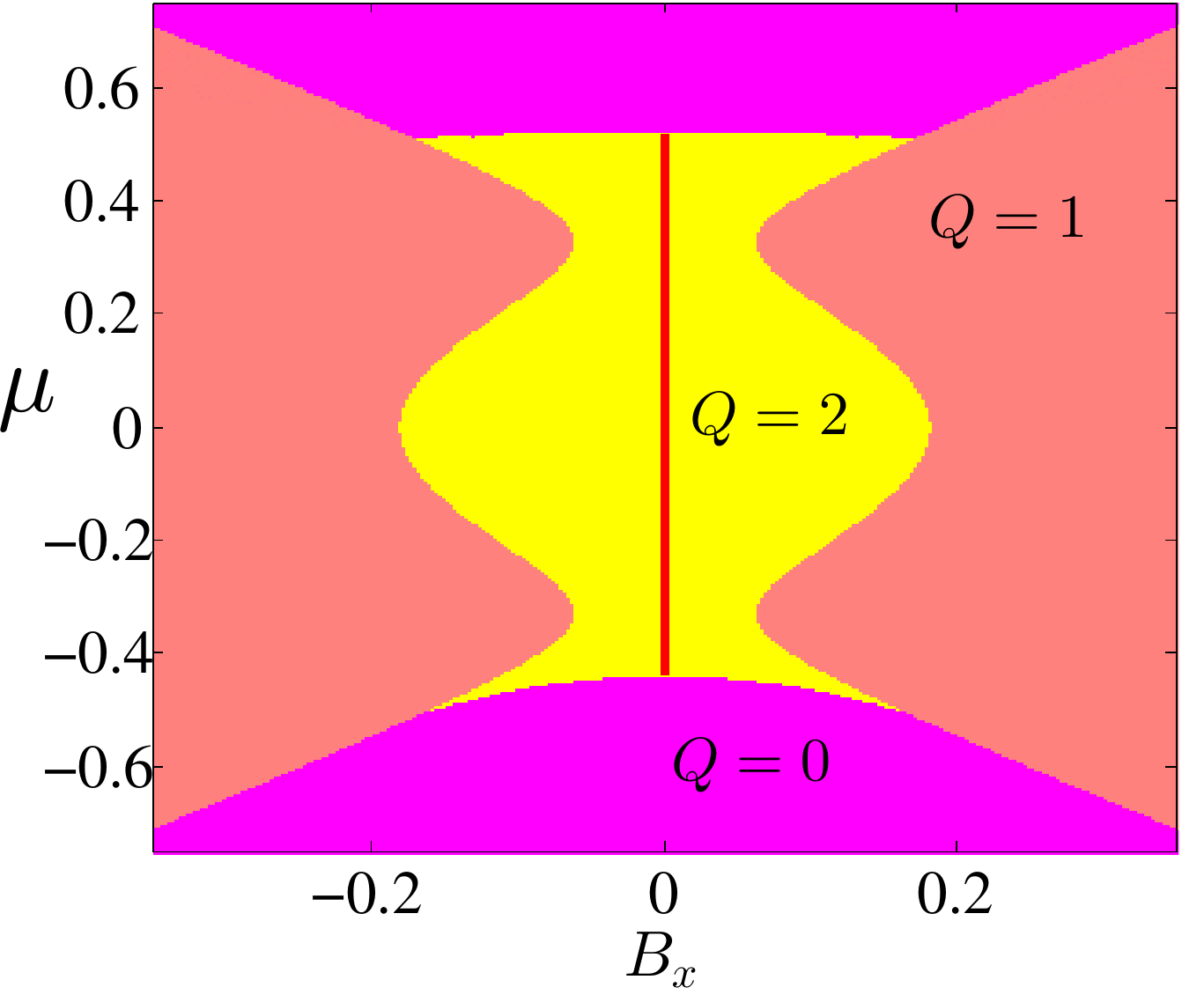}\label{fig:TRITOPS_sig_BDI_Z_invariant}
\end{tabular}
\end{center}
\caption{(a) Differential conductance through a single lead connected to the wire~\cite{Haim2014time} described in Eq.~\eqref{eq:H_HF}, as a function of bias voltage and magnetic field in the $x$ direction, for system parameters $\tilde\mu_{\rm a}=\tilde\mu_{\rm b}=0.15$, $t_{\rm a}=t_{\rm b}=1$, $t_{\rm ab}=0.4$, $\alpha_{\rm a}=0$, $\alpha_{\rm b}=0.6$, $\tilde\Delta_{\rm a}=0.3$, and $\tilde\Delta_{\rm b}=-0.15$. The zero-bias peak at $B_x=0$, originating from the MKP, does not split for small $B_x$, which is a special case of the absence of $\bs{B}$-linear splitting predicted by Eq.~\eqref{eq:Zeeman_splitting_expan} with $\bs{B}\perp\bs{\mu}^{(1)}$. (b) Phase diagram of the system as function of chemical potential $\tilde\mu_{\rm a}=\tilde\mu_{\rm b}=\mu$, and magnetic field. Since the magnetic field is applied in the $x$ direction (perpendicular to the SOC), the system is in symmetry class BDI, characterized by a $\mathbb{Z}$ topological invariant, $Q$. $Q$ equals the number of MBS at each end of the wire. The TRITOPS phase is marked by a red line.}
\end{figure}

Figure.~\hyperref[fig:TRITOPS_sig_diff_cond]{\ref{fig:TRITOPS_sig_diff_cond}(a)} presents the differential conductance as a function of $B_x$~\cite{Haim2014time}. In the absence of a magnetic field, there exists a ZBCP quantized to $4e^2/h$. Remarkably, this peak does not split upon introducing a small $B_x$~\cite{Zhang2013time,Gaidamauskas2014majorana,Haim2014time}. As the field is further increased, a topological phase transition occurs to a phase with a single MBS at each end, at which point the ZBCP peak splits to three peaks. One of them stays at zeros-bias and is quantized to $2e^2/h$, while the other two become part of the bulk spectrum. The lack of splitting for small $B_x$ is a special case of the absence of $\bs{B}$-linear splitting, as predicted by Eq.~\eqref{eq:Zeeman_splitting_expan} with $\bs{B}\perp\bs{\lambda}^{(1)}$.

In the specific system considered here,  we can understand the lack of splitting from another point of view~\cite{Gaidamauskas2014majorana,Haim2014time,Dumitrescu2014magnetic}. Even though TRS is broken by $\mc{H}_k^\tx{Z}$, the overall Hamiltonian, $\mc{H}_k=\mc{H}_k^\tx{HF}+\mc{H}_k^\tx{Z}$, still has an anti-unitary symmetry $\Lambda=\sigma^xK$, expressed by $\Lambda\mathcal{H}_{k}\Lambda^{-1}=\mathcal{H}_{-k}$, which protects the MBS from splitting~\cite{Haim2014time,Dumitrescu2014magnetic}. More specifically, due to this symmetry (together with PH symmetry) the Hamiltonian is in the BDI symmetry class~\cite{Altland1997} with a $\mathbb{Z}-$invariant, whose value determines the number of MBS at each end~\cite{schnyder2008classification,kitaev2009periodic}. In Fig.~\hyperref[fig:TRITOPS_sig_diff_cond]{\ref{fig:TRITOPS_sig_diff_cond}(b)}, we plot the number of MBS as a function of chemical potential and Zeeman field as inferred from the BDI $\mathbb{Z}-$invariant, calculated according to Ref.~\cite{tewari2012topological}. It should be noted, however, that in reality this symmetry is quite fragile, as it can be broken for instance by introducing a term $\alpha_{ab}\sigma^ys^x\tau^z$, which describes Rashba-type SOC associated with motion transverse to the wire.

\subsection{Josephson junctions}
\label{subsec:Signa_JJ}

\begin{figure}
\begin{center}
\begin{tabular}{cccc}
\hskip -\ShiftFig
\includegraphics[clip=true,trim = 0cm -5cm 0cm 0cm,width=0.24\tw]{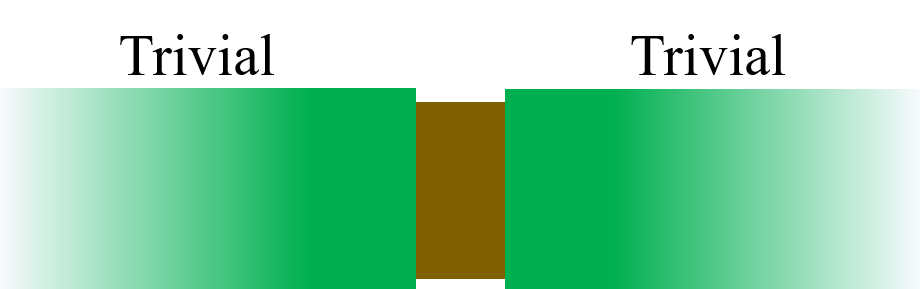}
\llap{\parbox[c]{4.1cm}{\vspace{-0.45\tw}(a)}}
&
\hskip -\ShiftFig
\includegraphics[clip=true,trim = 0cm 0cm 0cm 0cm,height=0.24\tw]{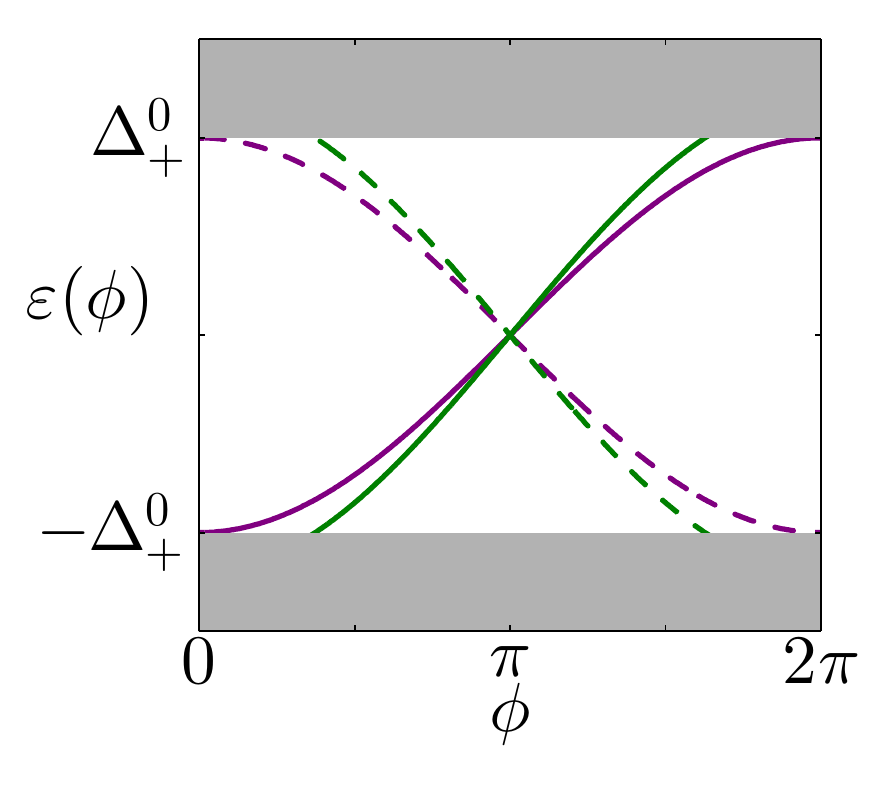}
\llap{\parbox[c]{4.3cm}{\vspace{-0.45\tw}(b)}}
&
\hskip -\ShiftFig
\includegraphics[clip=true,trim = 0cm 0cm 0cm 0cm,height=0.24\tw]{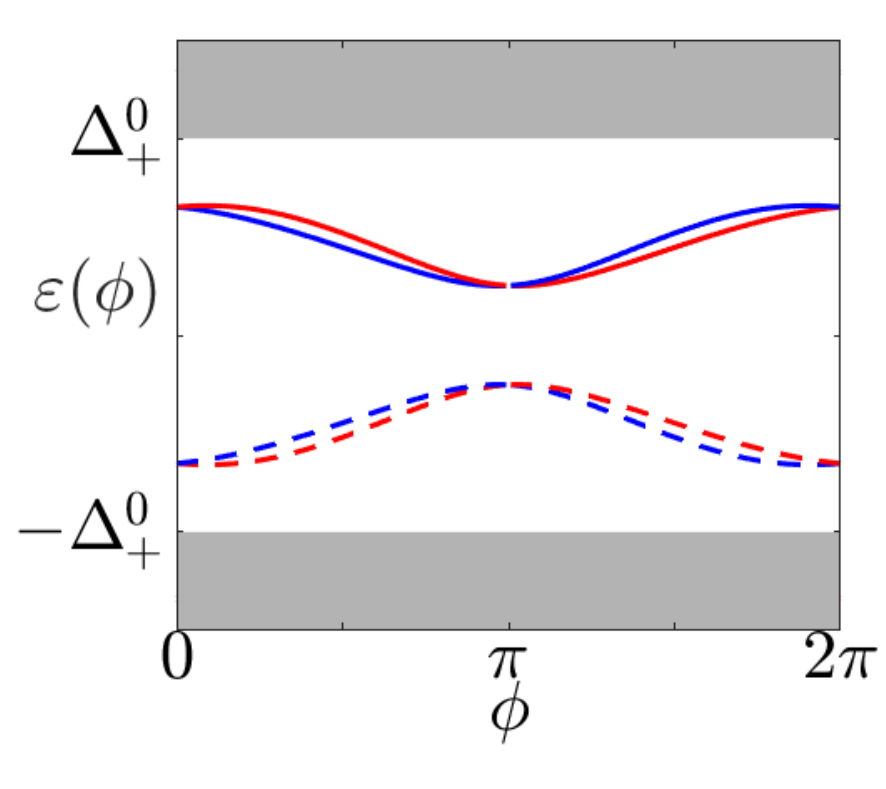}
\llap{\parbox[c]{4.3cm}{\vspace{-0.45\tw}(c)}}
&
\hskip -6mm
\includegraphics[clip=true,trim = 0cm 0cm 0cm 0cm,height=0.24\tw]{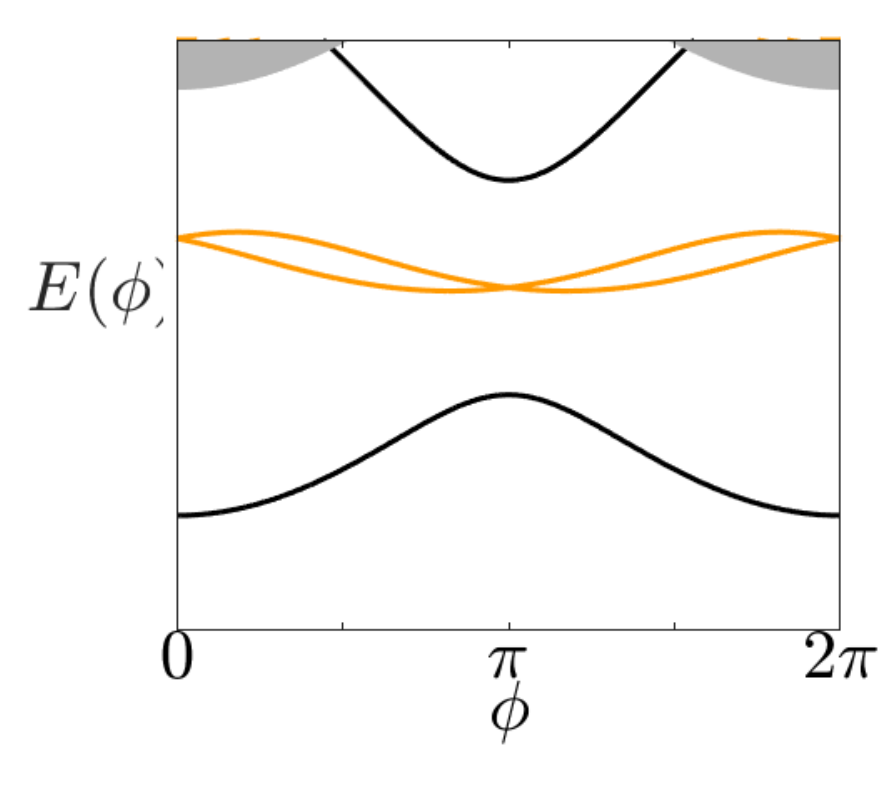}
\llap{\parbox[c]{4.3cm}{\vspace{-0.45\tw}(d)}}
\\
\hskip -\ShiftFig
\includegraphics[clip=true,trim = 0cm -5cm 0cm 0cm,width=0.24\tw]{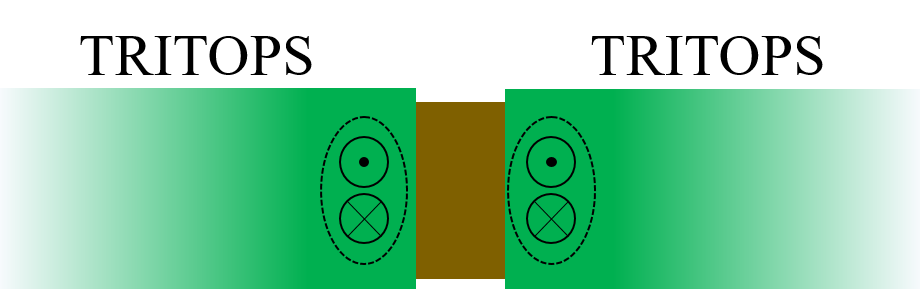}
\llap{\parbox[c]{4.1cm}{\vspace{-0.45\tw}(e)}}
&
\hskip -\ShiftFig
\includegraphics[clip=true,trim = 0cm 0cm 0cm 0cm,height=0.24\tw]{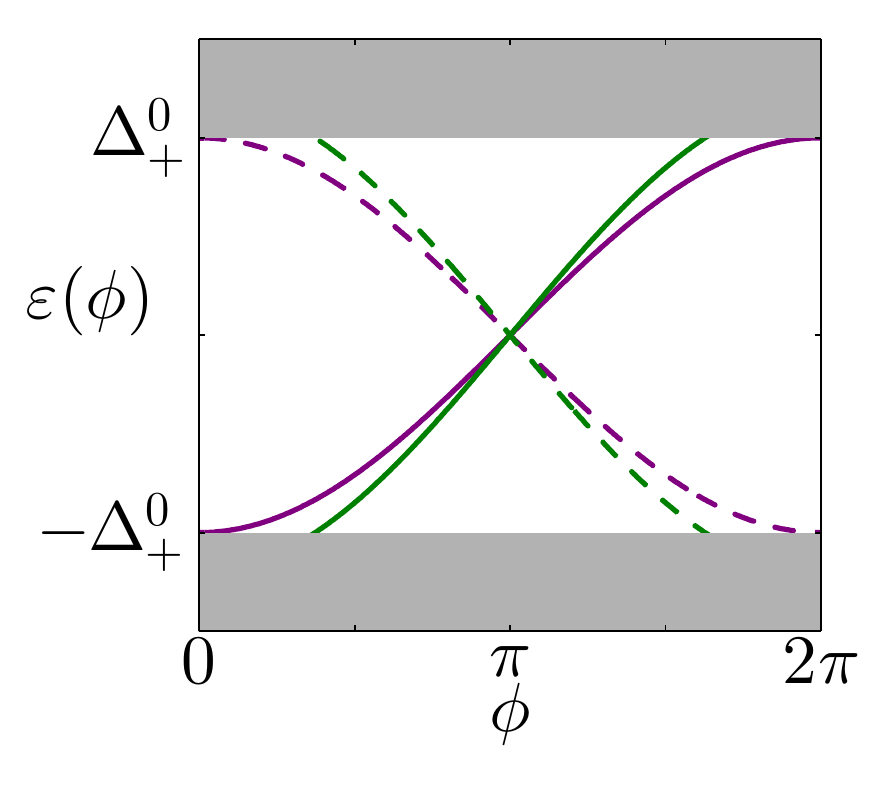}
\llap{\parbox[c]{4.3cm}{\vspace{-0.45\tw}(f)}}
&
\hskip -\ShiftFig
\includegraphics[clip=true,trim = 0cm 0cm 0cm 0cm,height=0.24\tw]{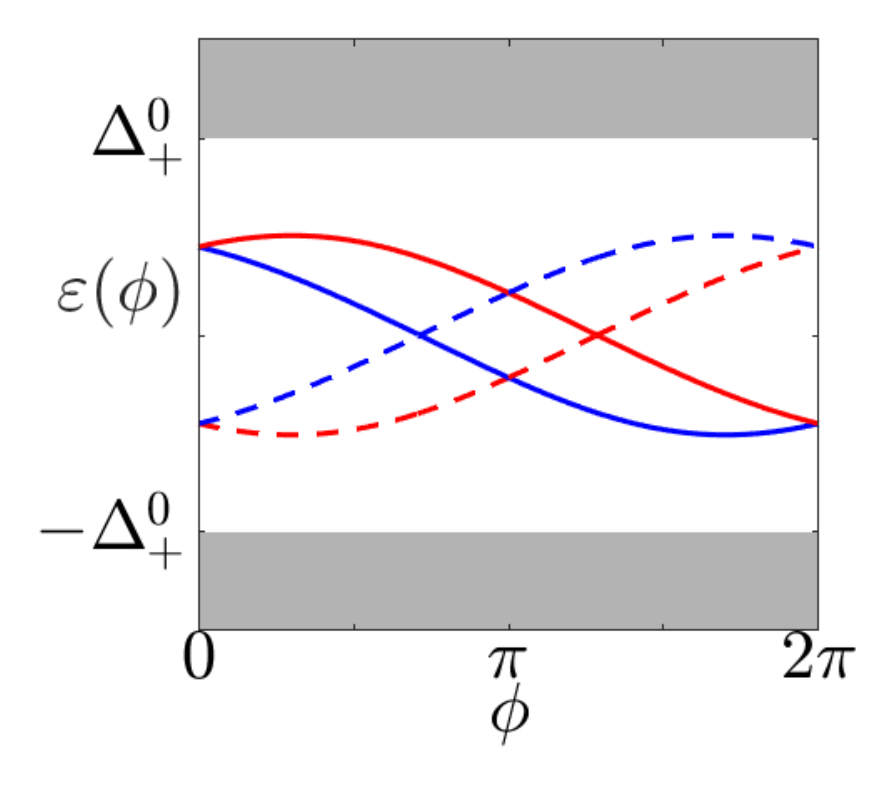}
\llap{\parbox[c]{4.3cm}{\vspace{-0.45\tw}(g)}}
&
\hskip -6mm
\includegraphics[clip=true,trim = 0cm 0cm 0cm 0cm,height=0.24\tw]{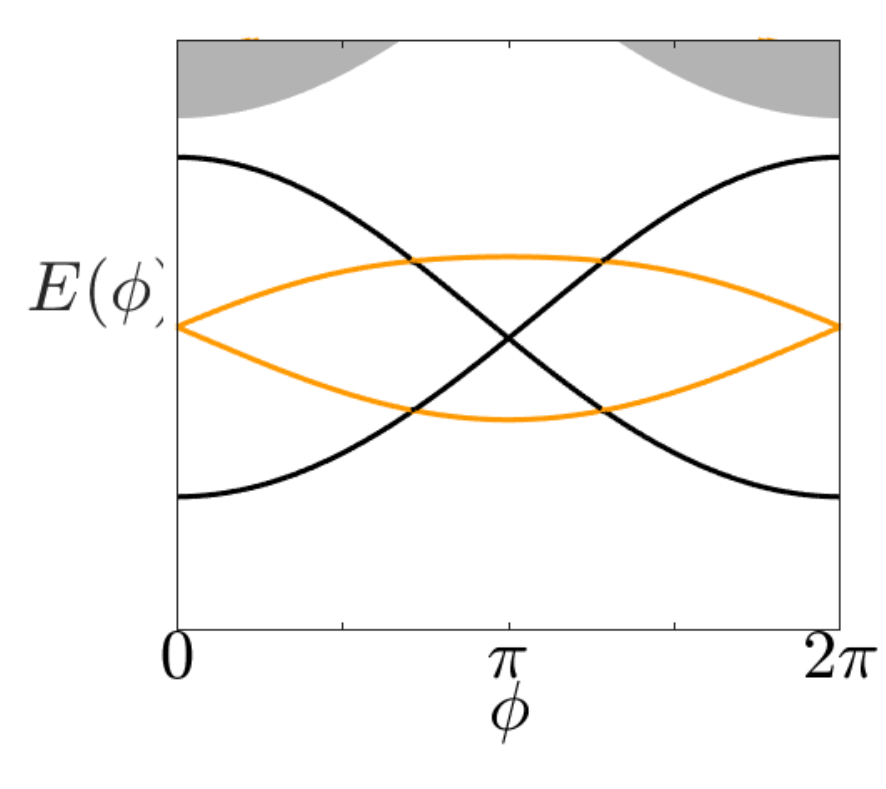}
\llap{\parbox[c]{4.3cm}{\vspace{-0.45\tw}(h)}}
\\
\hskip -\ShiftFig
\includegraphics[clip=true,trim = 0cm -5cm 0cm 0cm,width=0.24\tw]{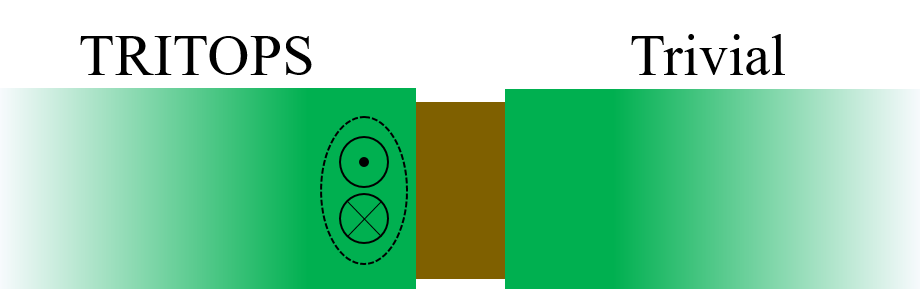}
\llap{\parbox[c]{4.1cm}{\vspace{-0.45\tw}(i)}}
&
\hskip -\ShiftFig
\includegraphics[clip=true,trim = 0cm 0cm 0cm 0cm,height=0.24\tw]{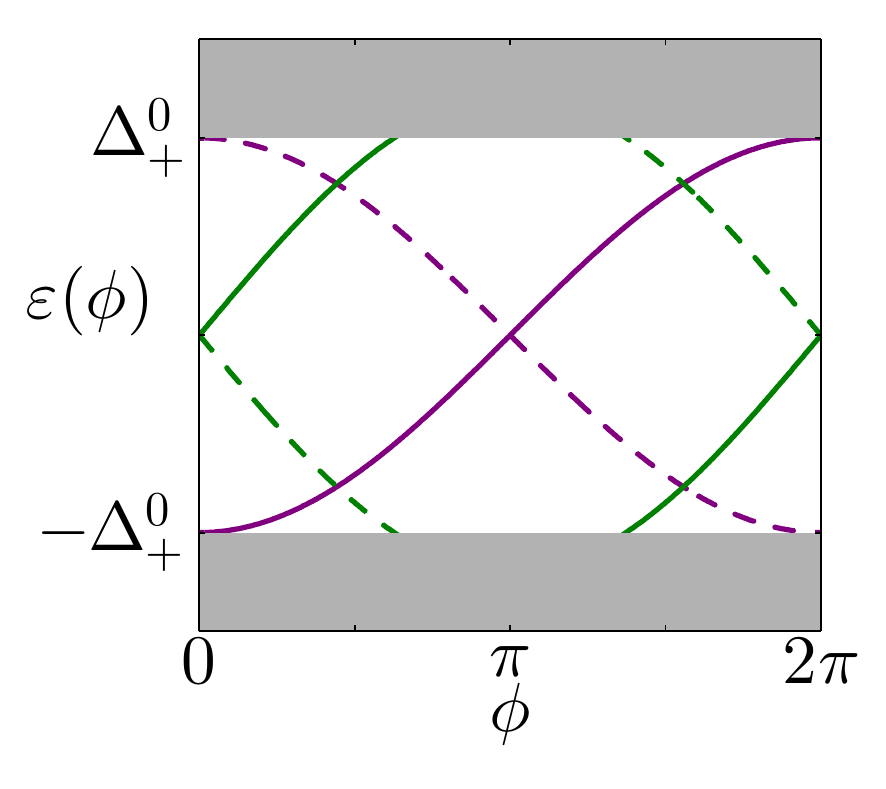}
\llap{\parbox[c]{4.3cm}{\vspace{-0.45\tw}(j)}}
&
\hskip -\ShiftFig
\includegraphics[clip=true,trim = 0cm 0cm 0cm 0cm,height=0.24\tw]{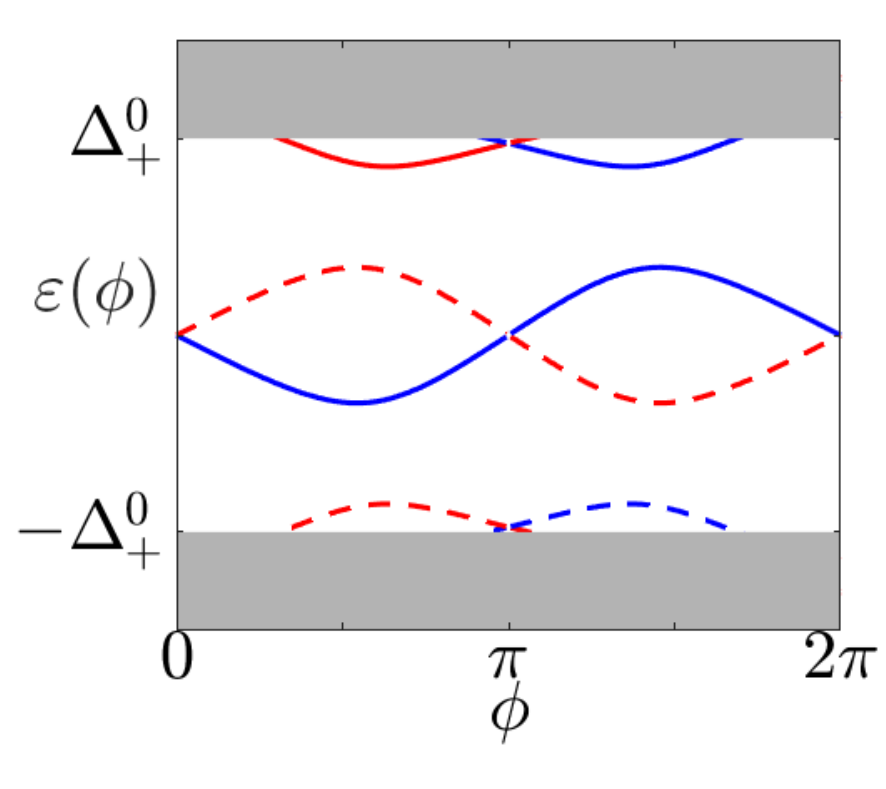}
\llap{\parbox[c]{4.3cm}{\vspace{-0.45\tw}(k)}}
&
\hskip -6mm
\includegraphics[clip=true,trim = 0cm 0cm 0cm 0cm,height=0.24\tw]{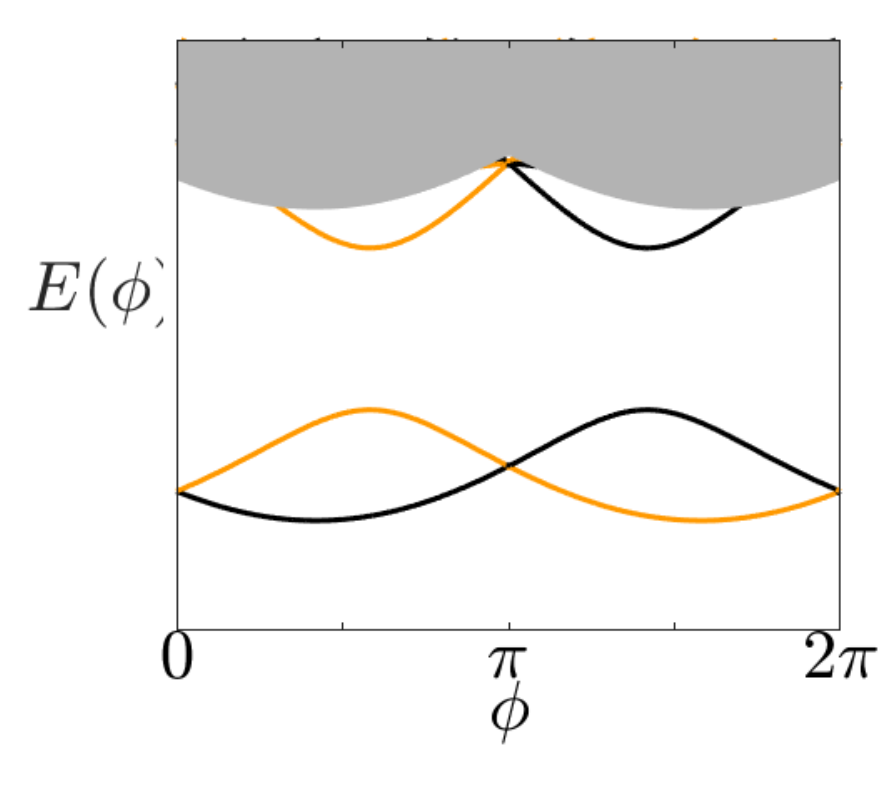}
\llap{\parbox[c]{4.3cm}{\vspace{-0.45\tw}(l)}}
\end{tabular}
\end{center}
\vskip -5mm
\caption{Three types of Josephson junctions (JJs): a junction between two trivial SCs (first row), a junction between two topological SCs (second row), and a junction between a topological SC and a trivial SC (third row).
The three types of JJs are depicted in the left column of the figure. The second column shows the single-particle excitation spectra of the junctions as a function of phase bias, $\phi$, as obtained from the simplified model, $H_0$, given in Eqs.~\eqref{eq:H_BdG} and \eqref{eq:H_0_JJ_spect}, with $\Delta^0_-=1.2\Delta^0_+$. The simplified model assumes a short junction, does not include backscattering in the junction, and possess a spin-rotation symmetry about the $z$ axis. Solid and dashed lines represent states connected by particle-hole symmetry. Notice that within this model, the excitation spectra of the trivial-trivial junction (b) and the topological-topological junction (f) are identical. The wave functions describing these excitations, however, are topologically distinct [see Eqs.~\eqref{eq:H_0_chiral_eigen_triv} and \eqref{eq:H_0_chiral_eigen_top}]. This affects how the spectra are modified by symmetry-allowed perturbations. The third column then presents the single-particle excitation spectra for a finite-length junction, and upon introducing a backscattering term and a spin-rotation symmetry-breaking term. Blue and red lines represent states related by time-reversal symmetry. The spectra are calculated using the lattice model introduced in Sec.~\ref{subsec:Intro_min_model}, as described in appendix~\ref{sec:JJ_lattice}. The difference in the spectra of the trivial-trivial junction and topological-topological junction is now apparent. In particular, the four-fold degeneracy at $\phi=\pi$ in (b) and (f) is lifted in two different ways, (c) and (g), which cannot be adiabatically connected. The spectrum for the trivial-topological junction (k) exhibits two zero-energy crossings, at $\phi=0$ and $\phi=\pi$. This are due to the Majorana Kramers pair which must exist on the boundary between a trivial and a topological system, at the time-reversal-invariant points. The forth column shows the many-body spectra of the three types of JJs, obtained by summing over the single-particle excitations energies (third column), for all possible occupations. Black and orange lines represents many-body states of even and odd fermion parity, respectively. The current-phase relation of the junction is obtained from the energy of the system using $I(\phi) = \frac{2e}{\hbar}\tx{d}E(\phi)/\tx{d}\phi$. For the trivial-trivial JJ (d) the ground state is unique for all $\phi$, and the current is therefore $2\pi$-periodic in $\phi$. For the topological-topological junction (h), on the other hand, the ground state switches as a function of $\phi$. The crossings between black and orange lines are protected by fermion-parity conservation. Upon changing $\phi$, the system goes from the ground states to an excited state. The crossing between the black lines at $\phi=0$ are protected by time-reversal symmetry. As long as $\phi$ is varied faster than the time scale, $\tau_\tx{ps}$, for processes breaking fermion-parity conservation, the current is $4\pi$ periodic in $\phi$. If $\phi$ is varied slow compared to $\tau_\tx{ps}$, the $2\pi$ periodicity is retained. For the topological-trivial junction (l) the ground state switches twice as a function of $\phi$ between states of opposite fermion-parity. Upon completing a $2\pi$ cycle the system returns to its initial states and the current is therefore $2\pi$ periodic.  Remarkably, the Josephson current is nonzero at the time-reversal-invariant points, $\phi=0$ and $\phi=\pi$, as can be seen from the nonzero slope of the many-body energies (this is true regardless of whether the system is in its lowest odd-parity state or lowest even-parity states). This effect is a manifestation of the TR anomaly of the TRITOPS phase.\label{fig:signa_JJs}}
\end{figure}

One of the ways to study the properties of a superconductor is by creating a Josephson junction, and examining the current-phase relation. In this subsection we shall consider and compare three types of Josephson junctions: (i) a junction between two trivial superconductors, (ii) a junction between two topological superconductors~\cite{liu2013non,Zhang2013time,Zhang2014anomalous,Kane2015the,gong2016influence,Mellars2016signatures,Camjayi2017fractional}, and (iii) a junction between a trivial and topological superconductor~\cite{chung2013time,gao2015josephson,Mellars2016signatures,JIANG2017Fano}. The results of this subsection are summarized in Fig.~\ref{fig:signa_JJs}.

One can study these three types of junctions within a single framework, using the linearized low-energy model introduced in Sec.~\ref{subsec:Intro_min_model}. The Hamiltonian in this case is given by
\be
\begin{split}\label{eq:H_JJ}
H_0=&-iv\int_{-\infty}^\infty \tx{d}x
\left\{\psi^\dag_{\tx{R},\ua}(x)\partial_x\psi^{\phantom{\dag}}_{\tx{R},\ua}(x)-\psi^\dag_{\tx{L},\da}(x)\partial_x\psi^{\phantom{\dag}}_{\tx{L},\da}(x)
+\left[\Delta_+(x)\psi^\dag_{\tx{R},\ua}(x)\psi^\dag_{\tx{L},\da}(x)+\tx{h.c.}\right]\right\},\\
&-iv\int_{-\infty}^\infty \tx{d}x
\left\{\psi^\dag_{\tx{R},\da}(x)\partial_x\psi^{\phantom{\dag}}_{\tx{R},\da}(x)-\psi^\dag_{\tx{L},\ua}(x)\partial_x\psi^{\phantom{\dag}}_{\tx{L},\ua}(x)
+\left[\Delta_-(x)\psi^\dag_{\tx{L},\ua}(x)\psi^\dag_{\tx{R},\da}(x)+\tx{h.c.}\right]\right\},
\end{split}
\ee
where the pairing potentials are taken to have the following spatial dependence
\be\label{eq:Delta_x_JJ}
\Delta_+(x) = \Delta^0_+\left\{
\begin{array}{lcr}
1 & , & x<0\\
e^{i\phi} & , &x>0
\end{array}\right.
\hs 4mm ; \hs 4mm
\Delta_-(x) = \Delta^0_-\left\{
\begin{array}{lcr}
s_\tx{L} & , & x<0\\
s_\tx{R}e^{i\phi} & , &x>0
\end{array}\right. ,
\ee
with $\Delta^0_\pm>0$. The parameters, $s_\tx{L}$ and $s_\tx{R}$, which can take the values $\pm1$, determine the type of the junction being studied: (i) the trivial-trivial junction corresponds to $s_\tx{R}=s_\tx{L}=1$, (ii) the topological-topological junction corresponds to $s_\tx{R}=s_\tx{L}=-1$, and (iii) the topological-trivial junction corresponds to $s_\tx{L}=-1$, $s_\tx{R}=1$ (or vice versa).

The model described by $H_0$ for the Josephson junction is somewhat oversimplified. First it assumes a very short junction (there is no normal metallic region between the SCs). More importantly, the model does not include a backscattering term, and it has a spin-rotation symmetry about the $z$ direction (namely $s_z$ is a good quantum number). Indeed one can consider a more general model for the JJ by adding the following symmetry-allowed perturbation 
\be\label{eq:H_prime}
H^\prime =
[V\psi^\dag_{\tx{R},\ua}(0)\psi^{\phantom{\dag}}_{\tx{L},\ua}(0) + V^\ast\psi^\dag_{\tx{L},\da}(0)\psi^{\phantom{\dag}}_{\tx{R},\da}(0) +  \tx{h.c.}] + [U\psi^\dag_{\tx{R},\ua}(0)\psi^{\phantom{\dag}}_{\tx{R},\da}(0) - U^\ast\psi^\dag_{\tx{L},\da}(0)\psi^{\phantom{\dag}}_{\tx{L},\ua}(0) + \tx{h.c.},
\ee
where the first term describes backscattering at the junction, and the second term accounts for spin-orbit coupling which breaks the spin-rotation symmetry about the $z$ axis of $H_0$. Nevertheless, as we shall see below, the topological properties of the junction can be inferred already from $H_0$, without accounting for $H^\prime$. In particular, the way in which the spectrum of the junction is affected by such a perturbation, such as $H^\prime$, will turn out to be determined by these properties. For a full non-perturbative analytical treatment of the three types of junctions see Ref.~\cite{Mellars2016signatures}.

To analyze the junction we look for the single-particle 
excitations of the system. To this end we write $H_0$ in a BdG 
form
\be\label{eq:H_BdG}
\begin{split}
&H_0=\half\int_{-\infty}^\infty \tx{d}x \Psi^\dag(x)\mc{H}_0(x) \Psi(x),
\\
&\mc{H}_0(x) = -iv\eta_z\tau_z\partial_x +\Real[\Delta_\tx{s}(x)+\eta_z\sigma_z\Delta_\tx{t}(x)]\tau_x + \Imag[\Delta_\tx{s}(x)+\eta_z\sigma_z\Delta_\tx{t}(x)]\tau_y,
\end{split}
\ee
where the Nambu spinor is defined as  $\Psi^\dag(x) = [\psi^\dag(x),-i\psi^\top(x)\eta_x\sigma_y]$, with $\psi^\dag = (\psi^\dag_{\tx{R},\ua},\psi^\dag_{\tx{R},\da},\psi^\dag_{\tx{L},\ua},\psi^\dag_{\tx{L},\da})$. Here, ${\tau}_{i=x,y,z}$ is a set Pauli matrices operating on the particle-hole degree of freedom, ${\sigma}_{i=x,y,z}$ are Pauli matrices operating on the spin degree of freedom, and ${\eta}_{i=x,y,z}$ are Pauli matrices operating on the right-moving/left-moving degree of freedom. In this basis, time-reversal symmetry is given by $\Theta = \eta_x\sigma_yK$, which commutes with the Hamiltonian, and particle-hole symmetry is given by $\Xi = \tau_y\sigma_y\eta_xK$, which anticommutes with the Hamiltonian. It is then immediately implied that $\mc{H}$ anticommutes with $\mc{C}=\Theta\Xi=\tau_y$, which is referred to as the chiral symmetry.

To solve the eigenvalue problem, $\mc{H}_0(x){u}(x)=\eps{u}(x)$, we first notice that $\eta_z=\pm1$ and $\sigma_z=\pm1$ are both good quantum numbers. The problem then reduces to solving $2\times2$ Hamiltonians. 
A straight-forward solution of the resulting differential equation yields the following excitation energies
\begin{subequations}
\begin{align}\label{eq:H_0_JJ_spect}
&\eps_+(\phi) = -\Delta_+^0\cos(\phi/2)
\hs 10.5mm ; \hs 4mm
\phi\in[0,2\pi),\\
&\eps_-(\phi) = -\Delta_-^0\cos(\phi/2 - \beta)
\hs 4mm ; \hs 4mm
\phi-2\beta\in[0,2\pi)
\\
&\tilde\eps_+(\phi) = -\eps_+(\phi)\\
&\tilde\eps_-(\phi) = -\eps_-(\phi)
\end{align}
\end{subequations}
with corresponding eigenvectors
\begin{subequations}
\label{eq:H_0_JJ_eigenvec}
\begin{align}
&{u}_+(x) = e^{-\frac{|x|}{\xi_+}}( 1 , -e^{-i\frac{\phi}{2}})^\top\otimes (1,0,0,0)^\top\\
&{u}_-(x) = e^{-\frac{|x|}{\xi_-}}( 1 ,-s_\tx{L} e^{-i(\frac{\phi}{2}-\beta)})^\top\otimes (0,1,0,0)^\top\\
&\tilde{{u}}_+(x) = \Xi{u}_+(x)\\
&\tilde{{u}}_-(x) = \Xi{u}_-(x),
\end{align}
\end{subequations}
where $\beta \equiv \pi(s_\tx{R} - s_\tx{L})/4$, $\xi_\pm \equiv v({\Delta^0_\pm}^2-\eps^2)^{-1/2}$, and where we focus on subgap excitation energies\footnote{Excitations with energies below the bulk gap are the ones which correspond to bound states whose wave function decays away from the junction, $x=0$.}. The subscript $\pm$ in Eqs.~(\ref{eq:H_0_JJ_spect},\ref{eq:H_0_JJ_eigenvec}) stands for the helicity of the mode, $\eta_z\sigma_z=\pm1$. Figures~\hyperref[fig:signa_JJs]{\ref{fig:signa_JJs}(b,f,j)} present the spectrum of $\mc{H}_0(x)$, as given in Eq.~\eqref{eq:H_0_JJ_spect}, for the different values of $s_\tx{R},s_\tx{L}$, corresponding to the three types of Josephson junctions. Purple and green lines mark states of positive and negative helicity, respectively, while solid and dashed lines mark states related by particle-hole symmetry.

Naturally, the spectrum of $H_0$ depends on $s_\tx{R}, s_\tx{L}$ which determined the type of the junction under study. Notice, however, that the spectrum for the trivial-trivial junction ($s_\tx{R}=s_\tx{L}=1$) is \emph{identical} to that of the topological-topological junction ($s_\tx{R}=s_\tx{L}=1$). The wave functions, on the other hand, are different for the two cases. This difference in the wave functions will determine how the spectrum is affected by a general symmetry-allowed perturbation, such as $H^\prime$. Below, we consider the effect of such a perturbation on the spectrum, separately for each of the three types of Josephson junctions. Specifically, we will be interested in the periodicity of the single- and many-body spectrum which in turn determines the periodicity of the Josephson current as a function of phase bias, $\phi$.

\subsubsection{The trivial-trivial junction}
\label{subsubsec:signa_JJ_triv-triv}

The single-particle excitation spectrum of the Josephson junction between two trivial SCs, as described by $\mc{H}_0(x)$, is presented in Fig.~\hyperref[fig:signa_JJs]{\ref{fig:signa_JJs}(b)}. Particle-hole symmetry constrains the spectrum to be symmetric about $\eps=0$, while time-reversal symmetry constrains the spectrum to be symmetric about $\phi=\pi$. Let us focus on the four-fold degenerate crossing observed at $\phi=\pi$, and let us consider how it is affected by a symmetry-allowed perturbation. Since TRS protects only a \emph{two}-fold (Kramers) degeneracy at $\phi=\pi$, we expect the four-fold crossing to split in energy. There are, however, two distinct ways in which this could happen, corresponding to Figs.~\hyperref[fig:signa_JJs]{\ref{fig:signa_JJs}(c)} and~\hyperref[fig:signa_JJs]{\ref{fig:signa_JJs}(g)}.

To understand the difference between the two, one has to examine the wave functions of the four degenerate states, and how each of them behaves under the chiral symmetry, $\mc{C}=\tau_y$. Setting $\phi=\pi$ in Eq.~\eqref{eq:H_0_JJ_eigenvec}, one can check that 
\be\label{eq:H_0_chiral_eigen_triv}
\mc{C}u_+ = u_+
\hs 4mm ; \hs 4mm
\mc{C}u_- = u_-
\hs 4mm ; \hs 4mm
\mc{C}\tilde{u}_+ = -\tilde{u}_+
\hs 4mm ; \hs 4mm
\mc{C}\tilde{u}_- = -\tilde{u}_-,
\ee
namely the eigenstates of $\mc{H}_0(x)$ are eigenstates of the chiral symmetry with eigenvalues $\pm1$. The crucial point is that any perturbation to $\mc{H}_0(x)$ which respects the chiral symmetry [such as $H^\prime$ in Eq.~\eqref{eq:H_prime}], 
cannot couple between eigenvectors of $\mc{C}$ which have the same eigenvalue\footnote{Let $|u_1\rangle$ and $|u_2\rangle$ be eigenvectors of $\mc{C}$ with the \emph{same} eigenvalues (either $+1$ or $-1$), and let $\mc{H}^\prime$ anticommutes with $\mc{C}$. Then $\langle u_1|\mc{H}^\prime|u_2\rangle=\langle u_1|\mc{C}^\dag\mc{H}^\prime\mc{C}|u_2\rangle=-\langle u_1|\mc{H}^\prime|u_2\rangle$, namely $\langle u_1|\mc{H}^\prime|u_2\rangle=0$.}. This means that any symmetry-allowed perturbation can only couple $u_+$ with $\tilde{u}_-$, and $u_-$ with $\tilde{u}_+$\footnote{Notice that even though $u_+$ and $\tilde{u}_+=\Xi u_+$ have opposite eigenvalues under $\mc{C}$, they cannot be coupled because they are related by particle-hole symmetry (and similarly for $u_-$ and $\tilde{u}_-=\Xi u_-$). This can be inferred from $\langle \Xi u_+|\mc{H}^\prime|u_+\rangle={\langle \Xi^2 u_+|\Xi\mc{H}^\prime| u_+\rangle}^\ast = \langle u_+|-\mc{H}^\prime\Xi|u_+\rangle^\ast = -\langle \Xi u_+|\mc{H}^\prime|u_+\rangle=0$, where in the first step we have used the fact that operating with an anti-unitary transformations on both vectors in an inner product amounts to complex conjugation, and in the second step we used the fact that $\mc{H}^\prime$ anticommutes with $\Xi$.}. Since the energy of $u_+$ [purple solid line Fig.~\hyperref[fig:signa_JJs]{\ref{fig:signa_JJs}(b)}] increases with $\phi$, while the energy of $\tilde{u}_-$ [green solid line in Fig.~\hyperref[fig:signa_JJs]{\ref{fig:signa_JJs}(b)}] decreases with $\phi$, the coupling between these states results in an avoided crossing, and the resulting spectrum is of the form shown in Fig.~\hyperref[fig:signa_JJs]{\ref{fig:signa_JJs}(c)}. The blue and the red line colors correspond to states related by TRS, while the solid and dashed line shapes, as before, represent states related by PHS. The spectrum in Fig.~\hyperref[fig:signa_JJs]{\ref{fig:signa_JJs}(c)} is calculated from a lattice model for the JJ which includes backscattering and breaks all spin-rotation symmetries (see appendix~\ref{sec:JJ_lattice}). This behavior can also be verified by explicitly treating $H^\prime$ using degenerate perturbation theory.

The spectrum of the many-body states can be constructed from the single-particle excitation spectrum by summing over the excitation energies, for all possible occupations. The result is shown in Fig.~\hyperref[fig:signa_JJs]{\ref{fig:signa_JJs}(d)}. States having an even fermion-number parity are shown in black, while states having an odd fermion-number parity are shown in orange. Since the excitation spectrum is gapped for all $\phi$, the ground state is unique and the system returns to the same state upon a cycle, $\phi:0\rightarrow2\pi$. As a result, the current, which is related to the energy of the system through $I(\phi) = \frac{2e}{\hbar}\tx{d}E(\phi)/\tx{d}\phi$, is $2\pi$-periodic in $\phi$, as is expected from a Josephson junction between two trivial SCs~\cite{Beenakker1991universal}. This will now be contrasted with the case of a Josephson junction between two topological superconductors.

\subsubsection{The topological-topological junction}
\label{subsubsec:signa_JJ_top-top}

We now move on to the case of a junction between two topological SCs~\cite{liu2013non,Zhang2013time,Zhang2014anomalous,Kane2015the,gong2016influence,Mellars2016signatures,Camjayi2017fractional}. This case is obtained by taking $s_\tx{R}=s_\tx{L}=-1$ in Eq.~\eqref{eq:Delta_x_JJ}. The spectrum of $\mc{H}_0$ in this case is presented in Fig.~\hyperref[fig:signa_JJs]{\ref{fig:signa_JJs}(e)}, and as noted above is identical to that of the trivial-trivial junction. While the spectra in these two cases are the same, the wave functions are not, as can be seen from Eq.~\eqref{eq:H_0_JJ_eigenvec}. Importantly, at the four-fold crossing point, $\phi=\pi$, the wave functions in the present case obey
\be\label{eq:H_0_chiral_eigen_top}
\mc{C}u_+ = u_+
\hs 4mm ; \hs 4mm
\mc{C}u_- = -u_-
\hs 4mm ; \hs 4mm
\mc{C}\tilde{u}_+ = -\tilde{u}_+
\hs 4mm ; \hs 4mm
\mc{C}\tilde{u}_- = \tilde{u}_-.
\ee
Now these are $u_+$ and $\tilde{u}_-$ which have eigenvalue $+1$ under $\mc{C}$, while $u_-$ and $\tilde{u}_+$ have eigenvalue $-1$. As a result, any symmetry-allowed perturbation can only couple $u_+$ with $u_-$ and $\tilde{u}_+$ with $\tilde{u}_-$. Since the energies of $u_+$ and $u_-$ both increase as a function of $\phi$ [solid purple and green lines in Fig.~\hyperref[fig:signa_JJs]{\ref{fig:signa_JJs}(f)}, respectively], the coupling between them does not result in an avoided crossing. The same is true for the coupling between $\tilde{u}_+$ and $\tilde{u}_-$, whose energies both decrease with $\phi$. This yields a spectrum of the form shown in Fig.~\hyperref[fig:signa_JJs]{\ref{fig:signa_JJs}(g)}, which is a result of a numerical calculation, based on a lattice mode which includes, on top of $\mc{H}_0$ a backscattering term and a spin-orbit coupling term (see Appendix~\ref{sec:JJ_lattice} for details). Unlike the case of the trivial-trivial junction, the four-fold degeneracy at $\phi=\pi$ is now lifted in a way which does not leave the spectrum gapped. Notice that the crossings between states marked by solid and dashed lines are protected by PHS, and the two-fold degeneracies at $\phi=0$ and $\phi=\pi$ are protected by TRS. Indeed, this spectrum cannot be smoothly connected with that of the trivial-trivial junction [compare Fig.~\hyperref[fig:signa_JJs]{\ref{fig:signa_JJs}(g)} with Fig.~\hyperref[fig:signa_JJs]{\ref{fig:signa_JJs}(c)}]. In fact, Zhang and Kane~\cite{Zhang2014anomalous} have shown that these two spectra correspond to two topologically distinct adiabatic pumping cycles.

As before, the many-body states can be constructed from the single-particle excitation spectrum, and is shown in Fig.~\hyperref[fig:signa_JJs]{\ref{fig:signa_JJs}(h)}, with states of even (odd) fermion-number parity shown in black (orange). Unlike the case of the trivial-trivial junction, here the ground state as a function of $\phi$ is not unique. If the system starts at the ground states at $\phi=0$, it finishes at an excited states upon completing an adiabatic cycle, $\phi:0\rightarrow 2\pi$. The crossing at $\phi=\pi$ is protected by TRS, while the other two crossings are protected by conservation of Fermion-number parity. As a result, it takes \emph{two} cycles for the system to return to its initial state. The current, $I(\phi) = \frac{2e}{\hbar}\tx{d}E(\phi)/\tx{d}\phi$, is therefore $4\pi$-periodic in $\phi$~\cite{Kane2015the,gong2016influence,Mellars2016signatures,Camjayi2017fractional}.

In practice, there are processes which can effectively change the Fermion parity. For example, at finite temperature, thermally-excited quasiparticles can be present in the system, and these can relax by switching the Fermion-number parity in the junction. The behavior of the junction therefore depends on the characteristic time scale for these parity-switching processes, $\tau_\tx{ps}$. If the rate at which $\phi$ is varied, $\omega_\tx{J}$, is fast compared with $\tau_\tx{ps}^{-1}$ (but still adiabatic as defined by the system's bulk energy gap), then the current is $4\pi$ periodic. If on the other hand $\omega_\tx{J}\ll\tau_\tx{ps}^{-1}$, then the system stays in its true (thermodynamic) ground state, and the current is $2\pi$ periodic~\cite{Kane2015the,Mellars2016signatures}.

Another signature of the topological-topological junction can be obtained by incorporating an interacting quantum dot (QD) into the junction and examining the dependence of the current-phase relation on the coupling strength between the dot and the SCs~\cite{gong2016influence,Camjayi2017fractional}. In a trivial SC-QD-SC junction, the behavior of the current depends strongly on the coupling strength between the dot and the SCs~\cite{Franceschi2010hybrid,Rodero2011Josephson}. At weak coupling, and when the dot is singly occupied (therefore serving as an impurity spin), the system minimizes its energy when the phase difference is $\pi$. This realizes a so called $\pi$ junction. At strong coupling, on the other hand, the impurity spin is screened by the (high-energy) electron states in the SCs, forming a singlet ground stats, and retaining the $0$-junction behavior. This transition between a $\pi$-junction behavior and a $0$-junction behavior as a function of coupling strength is predicted to be absent in the topological case (topological-QD-topological junction)~\cite{Camjayi2017fractional}. The two MKPs at the junction (one MKP pair from each SC) together form a spin which can form a singlet with the impurity spin, thereby screening it. Since the MKPs are zero energy-states (unlike the electronic states whose energies are larger than the bulk gap), this can happen even at small coupling to the QD. The system therefore exhibits $0$-junction behavior independent of the coupling, qualitatively different than a SC-QD-SC system.

\subsubsection{The topological-trivial junction}
\label{subsubsec:signa_JJ_top-triv}

Finally, we discuss the Josephson junction between a topological and a trivial SC~\cite{chung2013time,gao2015josephson,Mellars2016signatures,JIANG2017Fano}. The results are summarized in the bottom row in Fig.~\ref{fig:signa_JJs}. The single-particle spectrum of the unperturbed Hamiltonian, $\mc{H}_0$, is presented in Fig.~\hyperref[fig:signa_JJs]{\ref{fig:signa_JJs}(j)}, which is obtained by setting $s_\tx{R}=1$, $s_\tx{L}=-1$, in Eq.~\eqref{eq:H_0_JJ_spect}.

Notice the zero-energy crossings at the TR-invariant points, $\phi=0$, and $\phi=\pi$. These crossings are due to the MKP at the junction, and are protected by both particle-hole and time-reversal symmetry. In contrast, the other crossing points in the spectrum are not protected. Upon adding a symmetry-allowed perturbation, such as $H^\prime$ in Eq.~\eqref{eq:H_prime}, these crossings become avoided crossings, resulting in a spectrum of the form shown in Fig.~\hyperref[fig:signa_JJs]{\ref{fig:signa_JJs}(k)}, calculated using the lattice model of Appendix~\ref{sec:JJ_lattice}.

The many-body spectrum, constructed by summing over the excitation energies for all possible occupations, is presented in Fig.~\hyperref[fig:signa_JJs]{\ref{fig:signa_JJs}(l)}. The time-reversal anomaly~\cite{Qi2009time}, described in Sec.~\ref{subsec:Intro_top_prot_TRS_anom}, is manifested in the spectrum in two ways. First, the Kramers degeneracies at $\phi=0,\pi$ are between states of \emph{opposite} Fermion parity (and not between two states of \emph{odd} parity as in the usual case of Kramers degeneracy). This is because in systems with TR anomaly, TRS locally anticommutes with the Fermion parity. This is resolved by accounting for the other MKP which must exists at the left end of the system [the energies due to this MKP are shown in Fig.~\hyperref[fig:signa_JJs]{\ref{fig:signa_JJs}(i,j,k)}].

The second manifestation of the TR anomaly is the remarkable fact that the current through the junction, $I(\phi) = \frac{2e}{\hbar}\tx{d}E(\phi)/\tx{d}\phi$, is nonzero even at the TR-invariant points $\phi=0,\pi$~\cite{chung2013time,gao2015josephson,JIANG2017Fano}. This is regardless of whether the system is in its even-parity ground state (black line) or its odd-parity ground state (orange line), as can be seen from the nonzero slope of the energies of these states at $\phi=0,\pi$. This phenomena is possible due to the TR anomaly; since TRS anticommutes with fermion parity~\cite{Qi2009time}, states of definite fermion parity are \emph{not} time-reversal invariant. Essentially, by choosing a state of a given parity, the system spontaneously breaks TRS~~\cite{chung2013time}.

The behavior of current-phase relation is again dependent on the rate at which $\phi$ is varied, $\omega_\tx{J}$, relative to the rate at which fermion-parity switches, $\tau_\tx{ps}^{-1}$. For $\omega_\tx{J}\gg\tau_\tx{ps}^{-1}$, the system remains in a state of given parity, and returns to the same state upon varying $\phi$ from $0$ to $2\pi$, as can be seen from Fig.~\hyperref[fig:signa_JJs]{\ref{fig:signa_JJs}(l)}. Unlike the topological-topological junction, the current in this case is therefore $2\pi$ periodic. If on the other hand, $\omega_\tx{J}\ll\tau_\tx{ps}^{-1}$, the system always remains in its thermodynamic ground state, namely it switches parity at integer multiples of $\pi$. The current in this case is approximately $\pi$ periodic.

\section{Braiding properties of Majorana Kramers pairs}
\label{sec:Braid}

One of the interesting questions one can ask about the TRITOPS phase is concerned with the braiding of two Kramers pairs of MBSs. Indeed, since the two MBSs in a Kramers pair overlap in space~\cite{leijnse2012introduction}, it is impossible to braid single MBSs in a TRS system, and one is therefore forced to consider the braiding of one MKP with another MKP.

It is established that single MBSs (in a TRS-broken system) obey non-abelian exchange statistics~\cite{Read2000paired,Ivanov2001non,Stern2004geometric,Alicea2011non,Sau2011controlling}. Namely, the operation of adiabatically exchanging two MBSs (keeping the two spatially separated at all times) results in a non-trivial unitary operation on the ground-state manifold, which is universal and is independent of the exact path taken during the braiding. It turns out that, unlike the case of single MBSs, braiding of MKPs results in a \emph{non-universal} operation, which depends on the microscopic details of the exchange operation~\cite{Wolms2014local,Wolms2015braiding}. This has to do with the fact that the fermion composed out of the two TR-related Majoranas is a \emph{local} fermion, and is therefore sensitive to local perturbations.

An exception to this is the case where an additional symmetry exists in the system (in addition to TRS). For example, if the system obeys a $U(1)$ spin-rotation symmetry such that one spin component (say $s_z$) is a good quantum number, then the system can be decomposed into two spin-opposite copies ($s_z=\ua.\da$) which are completely decoupled from each other throughout the braiding process. Each copy describes a TRS-broken TSC, such that the braiding of MKPs can be described as the two braiding processes of single isolated MBSs, performed separately in each sector~\cite{Liu2014non,Wolms2015braiding}. If such a symmetry is absent, on the other hand, the system can still be thought of as composed of two such copies, however, the basis in which the system is decoupled depends on the microscopic parameters, and can change during the braiding process. It is this path-dependent change of basis which ultimately mixes the MBSs in the MKP and is responsible for the path dependence of the braiding.

We start by briefly reviewing the subject of non-abelian berry phase in the adiabatic theory of degenerate states. We then consider the case of braiding single MBSs in a TRS-broken systems, showing the existence of a path-independent non-abelian berry phase. Finally, we examine braiding of MKPs in TRS systems, and show that local mixing of the MBSs composing a pair produces a path-dependent contribution, rendering the resulting unitary operation non universal.

\subsection{Non-abelian Berry phase}
\label{subsec:Braid_NABP}

Consider a Hamiltonian, $H[\R]$, which depends on time through a set of parameters, $\R$. At any time, $t$, the Hamiltonian $H[\R]$ has a set of instantaneous eigenstates. We concentrate on situations where there is a subset of instantaneous eigenstates, $\{\ket{\eta_i[\R]}\}_{i=1}^g$, which are degenerate for all $t$,
\begin{equation}\label{eq:instant_es}
H[\R]\ket{\eta_i[\R]}=E[\R]\ket{\eta_i[\R]},
\end{equation}
and are separated from the rest of the states by energy gaps\footnote{For our purposes, $\{\ket{\eta_i[\R]}\}_{i=1}^g$ will be the degenerate ground-state manifold, and the degeneracy, $g$, will be either $2$ or $4$, depending on whether there are two isolated MBSs or two Kramers' pairs of MBSs.}.

According to the adiabatic theorem, if the parameters, $\R$, are changed slowly enough in time (compared with the energy gaps separating $\{\ket{\eta_i[\R]}\}_i$ from the rest of the states), then the unitary time evolution of the system does not take the system out of the degenerate manifold. Namely, if at time $t=0$ the system is in the state $\ket{\psi(0)}=\ket{\eta_i[\bs{R}(0)]}$, then at later times it will evolve according to
\begin{equation}\label{eq:U_evolution}
\ket{\psi(t)}=e^{-i\int_0^t \tx{d}t'E[\bs{R}(t')]}\hat{U}(t)\ket{\eta_i[\bs{R}(0)]}=
e^{-i\int_0^t \tx{d}t'E[\bs{R}(t')]}\sum_j U_{ij}(t)\ket{\eta_j[\R]},
\end{equation}
with the matrix $U(t)$ defined by
\begin{equation}\label{eq:U_def}
U_{ij}(t)=\langle \eta_j[\R] |\hat{U}(t)| \eta_i[\bs{R}(0)] \rangle,
\end{equation}
where we have artificially separated the dynamical phase, $\exp \{-i\int_0^t \tx{d}t'E[\bs{R}(t')]\}$, from the rest of the time-evolution operator as it simply contributes an abelian phase which will be of no interest to us here.

By plugging $\ket{\psi(t)}$ into the Schr\"odinger equation, we arrive at~\cite{Wilczek1984apperance}
\begin{equation}
\partial_t{U_{ij}(t)}=U_{ij}(t)\langle \eta_j[\R] |\partial_t| \eta_i[\R] \rangle,
\end{equation}
which is formally solved by
\begin{equation}\label{eq:NABP}
U_C = \mc{P} \exp \left[-i\int_C \tx{d}\bs{R}\cdot\bs{A}(\bs{R})\right] \hs 3mm ; \hs 3mm
\bs{A}_{ij}(\bs{R})\equiv i\langle \eta_j(\bs{R}) |\nabR| \eta_i(\bs{R}) \rangle,
\end{equation}
where $C$ denotes the path taken in parameter space and $\mc{P}$ is the path-ordering operator. This result generalizes Berry's geometric phase~\cite{berry1984quantal} to the case where there is a degenerate subspace of states, rather than a \emph{single} state separated by a gap from the rest. Here, $\bs{A}_{ij}(\bs{R})$ has a vector structure in the space of parameters, $\bs{R}$, and a matrix structure in the space of degenerate eigenstates [the expression on the left of Eq.~\eqref{eq:NABP} involves the exponentiation of a matrix].

We note that $U_C$ depends on the basis. Under a path-dependent basis transformation of the degenerate subspace,
\begin{equation}
\ket{\eta_i(\bs{R})}\rightarrow \sum_j W_{ij}(\bs{R})\ket{\eta_i(\bs{R})},
\end{equation}
the matrix $U_C$ transforms as
\begin{equation}\label{eq:U_under_base_trans}
U_C \rightarrow W(\bs{R}_\tx{i})U_C W^\dag(\bs{R}_\tx{f}),
\end{equation}
which can be inferred from Eq.~\eqref{eq:U_def}, with $\bs{R}_\tx{i}=\bs{R}(0)$ and $\bs{R}_\tx{f}=\R$ being the parameters at the beginning and end of the path, $C$, respectively.

\subsection{Braiding in the TRS-broken case}
\label{subsec:Braid_TRS-broken}

\begin{figure}
\begin{center}
\begin{tabular}{cc}
\rlap{\hskip -0.025\tw \parbox[c]{\tw}{\vspace{-0.28\tw}(a)}}
\includegraphics[clip=true,trim =0cm 0cm 0cm 0cm,width=0.28\tw]{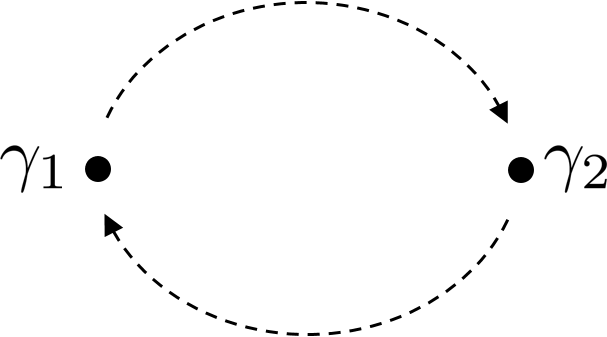}
\hs 5mm
&
\hs 5mm
\rlap{\hskip -0.025\tw \parbox[c]{\tw}{\vspace{-0.28\tw}(b)}}
\includegraphics[clip=true,trim =0cm 0cm 0cm 0cm,width=0.28\tw]{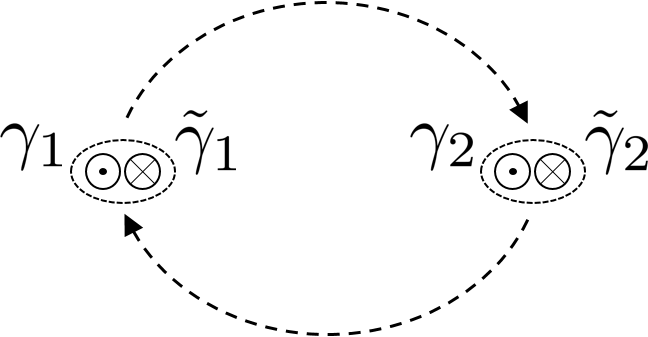}
\end{tabular}
\end{center}
\caption{(a) Braiding of two single Majorana bound states. (b) Braiding of two Majorana Kramers pairs.\label{fig:signa_Braiding}}
\end{figure}

We consider a system in class D with two spatially-separated MBSs, described by $\g_1$ and $\g_2$. The Hamiltonian, $H(\bs{R})$, is varied adiabatically along a path in parameter space in a way that makes $\g_1$ and $\g_2$ switch places, as depicted in Fig.~\hyperref[fig:signa_Braiding]{\ref{fig:signa_Braiding}(a)}. At any point along the path the systems has a pair of zero-energy MBSs, namely $[H(\bs{R}),\c_1(\bs{R})]=[H(\bs{R}),\c_2(\bs{R})]=0$, where $\c_1(\bs{R}_\tx{i})=\g_1$ and $\c_2(\bs{R}_\tx{i})=\g_2$. We choose the basis for the instantaneous ground-states manifold to be the two states having a given occupancy of the fermion $f(\bs{R})=[\c_1(\bs{R})+i\c_2(\bs{R})]/2$,
\begin{equation}\label{eq:signa_braid_bais}
\begin{split}
&\ket{0},\\
&\ket{1}=f^\dag(\bs{R}) \ket{0}.
\end{split}
\end{equation}

Now we are in the position to calculate $\bs{A}_{ij}(\bs{R})$, with $i,j\in\{0,1\}$. First, since the Hamiltonian conserves fermion parity, and since $\ket{0}$ and $\ket{1}$ have different parity, one has
\begin{equation}
\langle 0|\nabR|1 \rangle=\langle 1|\nabR|0 \rangle =0,
\end{equation}
namely $\bs{A}_{01}(\bs{R})=\bs{A}_{10}(\bs{R})=0$. The diagonal elements, $\langle 0|\nabR|0 \rangle$ and $\langle 1|\nabR|1 \rangle$ can be related through
\begin{equation}
\begin{split}
\langle 1|\nabR|1 \rangle = \langle 0|f(\bs{R})\nabR\left[f^\dag(\bs{R})|0 \rangle \right] =
\langle 0|\nabR|0 \rangle + \left\{f(\bs{R}),\nabR f^\dag(\bs{R})\right\}= \langle 0|\nabR|0 \rangle -\frac{i}{2}\left\{\chi_1(\bs{R}),\nabR \chi_2(\bs{R})\right\},
\end{split}
\end{equation}
where in the last step we have used the fact that $\left\{\chi_2(\bs{R}),\nabR \chi_1(\bs{R})\right\}=-\left\{\chi_1(\bs{R}),\nabR \chi_2(\bs{R})\right\}$, as well as $\left\{\chi_1(\bs{R}),\nabR \chi_1(\bs{R})\right\}=\nabR(\chi_1^2)=0$, and similarly for $\chi_2$. Ignoring the part of $\bs{A}(\bs{R})$ which is proportional to the identity matrix (as it only gives rise to an overall abelian phase), we are left with
\begin{equation}
\bs{A}(\bs{R})=\frac{1}{4}\begin{pmatrix}1&0\\0&-1\end{pmatrix}\left\{\chi_1(\bs{R}),\nabR \chi_2(\bs{R})\right\}.
\end{equation}
We now notice that since $\c_1$ and $\c_2$ are spatially separated at all instances during the braiding procedure, one necessarily has $\left\{\chi_1(\bs{R}),\nabR \chi_2(\bs{R})\right\}=0$, namely $\bs{A}(\bs{R})=0$ for all $\bs{R}$\footnote{Importantly, this will not be the case when we deal below with a Kramers' pair of MBSs, since there the two Majorana operators overlap in space.}.

Naively, this would mean that the braiding operation is trivial, however, one must bear in mind that at the end of the braiding procedure the states $\ket{0}$ and $\ket{1}$ return to themselves only up to a basis transformation. What determines this basis transformation is the definition of the braiding operation, $\c_1(\bs{R}_\tx{f})\propto\g_2$, $\c_2(\bs{R}_\tx{f})\propto\g_1$ together with Eq.~\eqref{eq:signa_braid_bais}. Moreover, the proportionality constants are not arbitrary; they are determined by conservation of Fermion parity, $P=i\c_1\c_2$, yielding either $\c_1(\bs{R}_\tx{f})=\g_2$, $\c_2(\bs{R}_\tx{f})=-\g_1$ or $\c_1(\bs{R}_\tx{f})=-\g_2$, $\c_2(\bs{R}_\tx{f})=\g_1$. When applied to the many-body states, these two options imply that the basis transformation is given by $|0\rangle\rightarrow|0\rangle ; |1\rangle\rightarrow\mp i|1\rangle$ , namely
\begin{equation}
W(\bs{R}_\tx{f})=\bmat 1&0\\0&\mp i\emat.
\end{equation}
Overall, Eq.~\eqref{eq:U_under_base_trans} implies that the braiding operation is given by $U=W^\dag(\bs{R}_\tx{f})$. Finally, one can check that this operation can also be written (up to an abelian phase) in the following basis-independent form
\begin{equation}
\hat{U}=e^{\mp\frac{\pi}{4}\g_1\g_2}.
\end{equation}

To summarize, the operator describing the braiding can be thought of as composed of two parts; a local part described by Eq.~\eqref{eq:NABP} and a part which comes from a nonlocal basis transformation - that is the difference between the initial and final basis. Since the two MBSs are kept far apart throughout the braiding process, the former part did not contribute. As we shall see below, this would no longer be the case when dealing with MKPs, since now the two MBSs composing the Kramers pair can locally mix with each other.

\subsection{Braiding in the TRS case}
\label{subsec:Braid_TRS}

We now consider a system in class DIII with two separated MKPs~\cite{Liu2014non,Wolms2014local,Wolms2015braiding,Gao2017symmetry}. The first Kramers pair is described by $\gamma_1,\tilde\gamma_1$ and the second by $\gamma_2,\tilde\gamma_2$. We vary the Hamiltonian adiabatically such that the two pairs are exchanged while keeping them spatially separated at all times, as depicted in Fig.~\hyperref[fig:signa_Braiding]{\ref{fig:signa_Braiding}(b)}. As before, we are interested in the unitary operation on the ground-state manifold as a result of the braiding process.

At any instance during the braiding process, there are two pairs of zero-energy Majorana bound states, namely $[H(\bs{R}),\c_1(\bs{R})]=[H(\bs{R}),\tilde\c_1(\bs{R})]=[H(\bs{R}),\c_2(\bs{R})]=[H(\bs{R}),\tilde\c_2(\bs{R})]=0$.
Out of these four Majorana operators we construct two fermions (related by TRS),
\begin{align}
f(\bs{R})=\chi_1(\bs{R})+i\chi_2(\bs{R}) \hs 4mm ;\hs 4mm
\tilde f(\bs{R})=\tilde\chi_1(\bs{R})-i\tilde\chi_2(\bs{R})
\end{align}
and use them to define a basis for the ground-state manifold,
\begin{equation}
\begin{split}
&\ket{00}, \\
&\ket{10}=f^\dag(\bs{R}) \ket{00}, \\
&\ket{01}=\tilde{f}^\dag(\bs{R}) \ket{00}, \\
&\ket{11}=f^\dag(\bs{R}) \tilde{f}^\dag(\bs{R}) \ket{00}.
\end{split}
\end{equation}

When considering the operation of braiding on the ground-state manifold, we can use the fact the total parity of the system is conserved and therefore examine separately the even- and odd-parity sectors, namely $\{\ket{00},\ket{11}\}$ and $\{\ket{10},\ket{01}\}$, respectively. Focusing, for example, on the odd-parity sector, one arrives at
\begin{equation}
\begin{split}
&\langle10|\nabR|10\rangle = \langle 00|\nabR|00 \rangle - \frac{i}{2}\left\{\chi_1(\bs{R}),\nabR \chi_2(\bs{R})\right\},\\
&\langle01|\nabR|01\rangle = \langle 00|\nabR|00 \rangle + \frac{i}{2}\left\{\tilde\chi_1(\bs{R}),\nabR \tilde\chi_2(\bs{R})\right\},\\
&\langle10|\nabR|01\rangle = \frac{1}{4}\left[\left\{\chi_1(\bs{R}),\nabR \tilde\chi_1(\bs{R})\right\} -\left\{\chi_2(\bs{R}),\nabR \tilde\chi_2(\bs{R})\right\}\right] +
\frac{i}{4}\left[\left\{\chi_1(\bs{R}),\nabR \tilde\chi_2(\bs{R})\right\} + \left\{\chi_2(\bs{R}),\nabR \tilde\chi_1(\bs{R})\right\}\right]\\
&\langle10|\nabR|01\rangle = -\langle01|\nabR|10\rangle.
\end{split}
\end{equation}
As in before, since $\chi_1(\bs{R})$ and $\chi_2(\bs{R})$ are kept spatially separated (and similarly $\tilde\chi_1(\bs{R})$ and $\tilde\chi_2(\bs{R})$), we have $\left\{\chi_1(\bs{R}),\nabR \chi_2(\bs{R})\right\}=\left\{\tilde\chi_1(\bs{R}),\nabR \tilde\chi_2(\bs{R})\right\}=\left\{\chi_1(\bs{R}),\nabR \tilde\chi_2(\bs{R})\right\}=\left\{\chi_2(\bs{R}),\nabR \tilde\chi_1(\bs{R})\right\}=0$. Up to an abelian phase one then has for the non-abelian Berry phase in the odd-parity sector~\cite{Wolms2015braiding}
\begin{equation}
\bs{A}^\tx{odd}(\bs{R}) = \frac{i}{4}\bmat 0&-1\\1&0 \emat \left[\left\{\chi_1(\bs{R}),\nabR \tilde\chi_1(\bs{R})\right\} - \left\{\chi_2(\bs{R}),\nabR \tilde\chi_2(\bs{R})\right\}\right].
\end{equation}
A similar calculation for the even-parity subspace yields
\begin{equation}
\bs{A}^\tx{even}(\bs{R}) = \frac{i}{4}\bmat 0&-1\\1&0 \emat \left[\left\{\chi_1(\bs{R}),\nabR \tilde\chi_1(\bs{R})\right\} + \left\{\chi_2(\bs{R}),\nabR \tilde\chi_2(\bs{R})\right\}\right].
\end{equation}

Importantly, unlike in the TRS-broken case where $\bs{A}(\bs{R})$ vanished due to the two MBSs being spatially separated, here the commutator $\left\{\chi_n(\bs{R}),\nabR \tilde\chi_n(\bs{R})\right\}$ involves two spatially-overlapping MBSs and is therefore generally nonzero. Moreover, it depends on the microscopic local properties of the Hamiltonian during the braiding process\footnote{For an example of a model yielding a nonzero $\left\{\chi_n(\bs{R}),\nabR \tilde\chi_n(\bs{R})\right\}$ see Ref.~\cite{Wolms2014local}.}. We conclude that braiding of MKPs results in a nonuniversal (path-dependent) non-abelian Berry phase.

An exception to this is the case where the system has an additional symmetry~\cite{Wolms2014local}. For example, consider a system which has, in addition to TRS, also a U(1) spin-rotation symmetry (say if the system conserves spin in the $z$ direction). In this case each MBS composing a Kramers pair belongs to a different spin sector (one belongs to the spin-up sector and the other to the spin-down sector), resulting in $\left\{\chi_n(\bs{R}),\nabR \tilde\chi_n(\bs{R})\right\}=0$. Like in the TRS-broken case, this leaves us with the contribution coming from the non-local basis transformation which is described by~\cite{Liu2014non,Wolms2015braiding}
\begin{equation}
\hat{U}=e^{\frac{\pi}{4}\gamma_1\gamma_2} e^{\frac{\pi}{4}\tilde\gamma_1\tilde\gamma_2},
\end{equation}
namely it is equivalent to two separate exchanges of isolated MBSs. Recently it was suggested that local mixing between MBSs in a Kramers' pair can also be somewhat suppressed by charging energy~\cite{Schrade2018quantum}, which exists when the (topological) superconductor hosting the MKPs is floating~\cite{Bao2017topological}.

\section{Summary and outlook}
\label{sec:Summary}

Time-reversal-invariant topological superconductivity (or 
TRITOPS) has been 
discussed in many (mostly theoretical) studies in recent 
years. In this article we pedagogically reviewed the progress made towards 
realizing the TRITOPS 
phase in one and two dimensions. The main emerging feature of 
this phase is the 
Kramers pair of Majorana boundary modes; in 1d this is a pair 
of Majorana bound 
states at each end, while in 2d it is a pair of 
counter-propagating gapless helical Majorana modes at each edge. The properties of the 
Majorana Kramers 
pair (MKP) were discusses, and in particular their symmetry and 
topological 
protection (see Sec.~\ref{sec:Intro} and Sec.~\ref{sec:TopInv}).

We have discussed possible mechanisms for experimentally realizing this phase. We focused on three main routs: (i) proximity 
to a pair of superconductors with an externally-tuned $\pi$ phase difference (Sec.~\ref{subsec:realize_pi_junc}), (ii) 
proximity-coupled interacting semiconductors (Sec.~\ref{subsec:realize_inter_syst}), and (iii) proximity to 
unconventional superconductors (Sec.~\ref{subsec:realize_unconv_sc}). The key for understanding these mechanisms lies in 
the sign change they induce in the pairing potential as a function of momentum. We emphasized the role of the interplay 
between, spin-orbit coupling, superconducting proximity and repulsive interactions, in the stabilization of the TRITOPS 
phase.

We then reviewed, in Sec.~\ref{sec:Signa}, various ways to 
experimentally detect the TRITOPS phase. Specifically, we discussed 
how the presence of the 
Majorana Kramers pair should effect various transport properties. 
These include, in particular, the anomalous Zeeman splitting of the 
zero-bias conductance 
peak, and a unique current-phase relation in Josephson junctions. 
In brief, the conductance into the system's end from a 
normal-metal lead should exhibit a zero 
bias conductance peak, quantized to $4 e^2/h$, which splits in a 
non-linear way as a function of magnetic field in certain 
directions 
(Sec.~\ref{subsec:Signa_AnomZeemSplit}). Signatures in Josephson junctions 
include 4$\pi$ periodicity of the Josephson current in a 
junction between a TRITOPS 
and a trivial SC (Sec.~\ref{subsubsec:signa_JJ_top-triv}), and a finite current 
at a phase difference of $\pi$ in a junction between two 
TRITOPSs 
(Sec.~\ref{subsubsec:signa_JJ_top-top}). The latter property is a consequence of 
the time-reversal anomaly (Sec.~\ref{subsec:Intro_top_prot_TRS_anom}).

The behavior of MKPs under exchange was discussed in Sec.~\ref{sec:Braid}. While exchanging two MKPs results in a non-trivial operation on the ground-state manifold, since each MKP comprises of two \emph{overlapping} Majorana bound states, this operation is rendered non-universal; it depends on the particular adiabatic path taken in parameter space. Nevertheless, if additional symmetries are present, for example a $U(1)$ spin-rotation symmetry, the exchange process becomes universal. Protected quantum-information processing is then possible using non-abelian Braiding of MKPs, similar to the time-reversal-broken case.

While they have been by now a large body of theoretical works proposing realizations of the TRITOPS phase, only a few 
experimental attempts have been made in this direction (see for example Ref.~\cite{Williams2012unconventional}). This comes in 
contrast to the time-reversal broken case (class D), where evidence of topological superconductivity have been reported in 
several studies~\cite{mourik2012signatures,deng2012anomalous,Das2012zero,churchill2013superconductor,Finck2013anomalous,Nadj-Perge2014observation,Pawlak2016probing,Ruby2015end,Albrecht2016exponential,Deng2016Majorana,Chen2017Experimental,gul2018ballistic,Suominen2017zero,Nichele2017scaling,zhang2017quantized}. The key there was to combine spin-orbit coupling, superconducting proximity, and magnetism. In order to stabilize the TRITOPS phase (without the need for fine tuning), it seems that the magnetism ingredient is replaced by repulsive e-e interactions. This suggests that the same experimental setups already being studied, such as proximitized nanowires and TIs, should be adequate for realizing the TRITOPS phase. The lack of need for magnetism is an advantage, as it generally tends to suppress superconductivity. In fact, some of the currently studied experimental setups already closely resemble the proposals described in Sec~\ref{sec:realizations} - see for example Ref.~\cite{Hart2014Induced}. The challenge of controlling the strength of the e-e interaction can be addressed, for example, by controlling the electron density in the semiconductor. 
In this regard, it might be beneficial in the future to also consider quantum dot chains as a possible platform in which one can control the on-site Coulomb interaction, similar to the proposals for the TRS-broken TSC~\cite{Sau2012realizing,Fulga2013adaptive,Zhang2016Majorana,Levine2017realizing}.

From the theory point of view, it would be interesting to explore phases related to TRITOPS which cannot be described by quadratic Hamiltonians, analogous to the generalization of the integer quantum Hall effect to the fractional one. What are the possible fractional TRITOPS phases and what are the systems most likely to realize them? Some progress in this direction was made~\cite{Neupert2014wire} by using a coupled-wires approach~\cite{Kane2002Fractional,Klinovaja2013topologicalWireConstruc,Teo2014From,Seroussi2014topological,Sagi2014nonabelian,Klinovaja2014Quantum,Meng2014Time}. Another possible direction can be inspired by a recent proposal to realize a two-dimensional Ising Spin Liquid in a network of interacting Majorana Cooper Boxes~\cite{Barkeshli2015physical,Sagi2018spin}. What phases can be realized in a network of \emph{TRITOPS} Cooper Boxes?  
Recently, a mapping was found between a topological phase of $\mathbb{Z}_4$ parafermions and the TRITOPS phase in 1d~\cite{Chew2018fermionized}. It will be interesting to explore this connection further; for example, how are the various physical signatures of the TRITOPS phase manifest themselves in the parafermionic system? It will certainly be exciting to witness the future theoretical and experimental fruits from the study of TRITOPS.

\section*{Acknowledgments}
\label{sec:Acknowledgments}
\addcontentsline{toc}{section}{Acknowledgements}

Our research of time-reversal-invariant topological superconductivity was conducted in collaboration with E. Berg, K. Flensberg, A Keselman, and K. W\"olms. We have also benefited from discussions with I. C. Fulga, C. M. Marcus, K. Michaeli, F. von Oppen, M.-T. Rieder, Y. Schattner, E. Sela and A. Stern. A. H acknowledges support from the Walter Burke Institute for theoretical physics at Caltech. Y. O. acknowledges support from the Israeli Science Foundation (ISF), the Minerva Foundation, the Binational Science Foundation (BSF), and the European Research Council under the European Union's Seventh Framework Programme (FP7/2007-2013)/ERC Grant agreement MUNATOP No. 340210.

\begin{appendices}

\section{Necessity of electron-electron interactions}
\label{subsec:realize_inter_syst_necess_int}

In this appendix we follow Ref.~\cite{Haim2016NoGo} and show that the topological phase of class DIII superconductors (in 1d and 2d) cannot be realized using proximity of a single conventional $s$-wave superconductor to a system of \emph{noninteracting} electrons. We begin by considering a model which consist of both the parent superconductor and the system. By integrating out the superconductor's degrees of freedom, we obtain the Green's function of the system alone. From that Green function, we then show that, for a conventional $s$-wave SC, the topological invariant (as derived in Sec.~\ref{sec:TopInv}), always takes its trivial value. We perform this procedure for a clean 1d system, and then generalize the result to the case of 2d and to the case of system lacking translation invariance.

\paragraph{The model}
\label{parag:realize_inter_syst_necess_int_model}

We consider a quasi--1d system of noninteracting electrons (hereafter referred to as a ``wire''), coupled to a bulk SC. The Hamiltonian describing the combined system reads $H = H_\w + H_\s + H_\tx{c}$, with
\begin{equation}\label{eq:H}
\begin{split}
&H_\w = \sum_{\k} \psi_\k^\dag h^\tx{w}_\k \psi_\k,
\hs 5mm ; \hs 5mm
H_\s = \sum_\k \left[ \eta_\k^\dag h^\tx{sc}_\k \eta_\k + \frac{1}{2}(\eta_\k^\dag \Delta_\k \eta_\mk^{\dag\top} + \tx{h. c.})\right],\\
&H_\tx{c} = \sum_\k ( \eta_\k^\dag t_\k \psi_\k+\tx{h. c.} ),
\end{split}
\end{equation}
where $k$ is the momentum along the wire's axis. $H_\w$ and $H_\s$ are the Hamiltonians describing the wire and the SC, respectively, and $H_\tx{c}$ describes the coupling between them. As in Sec.~\ref{subsec:TopInv_wind_num}, for every $\k$, $\psi^\dag_\k$ is a $M_\w$--dimensional row vector of fermionic creation operators of states in the wire. Similarly, $\eta^\dag_\k$ is a $M_\s$--dimensional row vector of fermionic creation operators of states in the SC, respectively. These states include all degrees of freedom within a unit cell including spin, transverse modes, sublattice sites, atomic orbitals etc. Correspondingly, $h^\w_\k$ is a $M_\w\times M_\w$ matrix, $t_\k$ is a $M_\s\times M_\w$ matrix, and $h^\s_\k$, $\Delta_\k$, are $M_\s\times M_\s$ matrices, operating on these internal degrees of freedom. Notice that as before $\Delta^\top_\mk=-\Delta_\k$. due to fermionic statistics. The purpose of this appendix is to show that as long as $\Delta_\k$ corresponds to a pairing of a conventional $s$-wave SC (to be defined more precisely below), the system cannot be in the topological phase of class DIII\footnote{In writing Eq.~\eqref{eq:H}, we have assumed that the interactions in the SC are adequately described within mean-field theory through the pairing potential matrix, $\Delta_\k$. In principle, the coupling of the SC to the wire can itself affect $\Delta_\k$, through what is known as the reverse proximity effect. this can reduce $\Delta_k$ near the interface with the wire, compared to the bulk value. This effect should be small when the density of states in the SC (within a coherence length away from the interface) is large compared to the density of states in the wire, or alternatively, when the coupling to the wire is weak. In any case, this effect is not expected to turn the pairing matrix $\Delta_\k$ to that of an unconventional SC.}.

As usual, we define the TR operation by its form when acting on the fermionic operators,
\begin{equation}\label{eq:TR}
\mathbb{T}\psi_\k \mathbb{T}^{-1}=\mc{T}_\w \psi_\mk \hskip 5mm ; \hskip 5mm
\mathbb{T}\eta_\k \mathbb{T}^{-1}=\mc{T}_\s \eta_\mk  \hskip 5mm ; \hskip 5mm
\mathbb{T} i \mathbb{T}^{-1}=-i,
\end{equation}
where $\mc{T}_\w$ and $\mc{T}_\s$ are unitary matrices operating 
in the spaces 
of states in the wire and the superconductor, respectively. For the TR 
operation to square to $-1$, we further require 
$\mc{T}^{\phantom{\ast}}_{\w(\s)}\mc{T}_{\w(\s)}^\ast=-1$. Enforcing TRS on 
the 
system, $\mathbb{T}H\mathbb{T}^{-1}=H$, amounts to the following 
conditions
\begin{equation}\label{eq:TR_on_matrices}
\begin{split}
\mc{T}_\w^\dag h^{\w\ast}_\mk \mc{T}_\w = h^\w_\k \hskip 4mm ; \hskip 4mm
\mc{T}_\s^\dag h^{\s\ast}_\mk \mc{T}_\s = h^\s_\k \hskip 4mm ; \hskip 4mm
\mc{T}_\s^\dag t^\ast_\mk \mc{T}_\w = t_\k \hskip 4mm ; \hskip 4mm
\mc{T}_\s^\dag \Delta^\ast_\mk \mc{T}_\s^\ast = \Delta_\k.
\end{split}
\end{equation}
The last equality, together with the property $\Delta_\mk^\top=-\Delta_\k$, guarantee that $\Delta_\k\mc{T}_\s$ is a Hermitian matrix.

We are now in a position to define more precisely the statement we wish to prove: The system will always be in the topologically-trivial phase if $\Delta_\k\mc{T}_\s$ is a positive semi-definite (PSD) matrix\footnote{More generally, $\Delta_\k\mc{T}_\s$ can be a PSD matrix times some complex number that does not depend on $k$. The phase of this complex number can always be absorbed in the definition of $\mc{T}_\s$, rendering $\Delta_\k\mc{T}_\s$ PSD.}. Namely, we shall assume that $\langle u| \Delta_\k \mc{T}_\s |u\rangle \ge 0$ for all vectors $|u\rangle$, and all momenta $k$, and show that the topological invariant of the system is always trivial. Importantly, the case of a PSD $\Delta_\k\mc{T}_\s$ includes, in particular, the case of a \emph{conventional s-wave superconductor}, in which the order parameter has a uniform phase on all the bands, and there are no inter-band pairing\footnote{Note also that this condition excludes the case considered in Sec.~\ref{subsec:realize_pi_junc}, in which case $\Delta_k$ describes \emph{two} superconducting leads in a $\pi$ junction}$^,$\footnote{An interesting example is the system considered in Ref.~\cite{Reeg2017DIII}, where a single $s$-wave SC is coupled to two wires experiencing opposite magnetic fields. While the usual TRS is not preserved, the system obeys the product of TRS and mirror, which is an anti-unitary symmetry, $\Theta'=\mc{T}'\mc{K}$, squaring to $-1$. Interestingly, in this case $\Delta\mc{T}'$ is \emph{not} PSD, allowing for realization of a class DIII TSC in the absence of interactions, protected by TRS$\times$Mirror.}.

\paragraph{Integrating out the superconductor}
\label{parag:realize_inter_syst_necess_int_integ_out}
We wish to obtain the Green's function describing the wire, from which one can then extract the topological invariant of the system. 
To this end we first write the Hamiltonian in a BdG form
\begin{equation}\label{eq:H_BdG_no_go}
H=\frac{1}{2}\sum_\k \Psi_\k^\dag \begin{pmatrix} \mc{H}^\w_\k & V^\dag_\k \\ V_\k & \mc{H}^\s_\k \end{pmatrix} \Psi_\k,
\end{equation}
using the Nambu spinor $\Psi^\dag_\k =(\psi^\dag_\k , \psi^\top_\mk \mc{T}_\w , \eta^\dag_\k , \eta^\top_\mk \mc{T}_\s)$, where
\begin{subequations}\label{eq:BDG_matrices}
	\begin{align}
	\mc{H}^\w_\k &= \tau^z \otimes h^\w_\k \label{eq:H_wire}, \\
	\mc{H}^\s_\k &= \tau^z \otimes h^\s_\k + \tau^x \otimes \Delta_\k \mc{T}_\s, \label{eq:H_sc}, \\
	V_\k &= \tau^z \otimes t_\k, \label{eq:T_w_sc}
	\end{align}
\end{subequations}
and where as usual $\{\tau^\alpha\}_{\alpha=x,y,z}$ are Pauli matrices in particle-hole space. In writing Eqs.~(\ref{eq:H_BdG_no_go}, \ref{eq:BDG_matrices}), we have used the relations given in Eq.~(\ref{eq:TR_on_matrices}).

The Green's function of the wire, $\mc{G}^\w_\k(\omega)$, is obtained by integrating out the SC,
\begin{subequations}\label{eq:Green_function}
	\begin{align}
	& \mc{G}^\w_\k(\omega) = [i\omega - \mc{H}^\w_\k - \Sigma_\k(\omega)]^{-1},\label{eq:G_w}  \\
	& \Sigma_\k(\omega) = V^\dag_\k g^\s_\k(\omega) V_\k,\label{eq:self_E} \\
	& g^\s_\k(\omega) = (i\omega - \mc{H}^\s_\k)^{-1},\label{eq:g_sc}
	\end{align}
\end{subequations}
where $\Sigma_\k(\omega)$ is the self energy, and $g^\s_\k(\omega)$ is the Green's function of the parent SC in the absence of coupling to the wire.

In the next step, we wish to relate the properties of $\Delta_\k$ to the properties of $\Sigma_\k(0)$. Using Eqs.~\eqref{eq:H_sc} and~\eqref{eq:g_sc}, one can check that $g^\s_k(0)$ is Hermitian and obeys $\tau^y g^\s_\k(0) \tau^y= -g^\s_\k(0)$. It therefore has the following structure:
\begin{equation}\label{eq:g_sc_struct}
g^\s_\k(0) = \tau^z \otimes g^\tx{N}_\k + \tau^x \otimes g^\tx{A}_\k,
\end{equation}
where $g^\tx{N}_\k$ and $g^\tx{A}_\k$ are Hermitian matrices. This also means that the zero-frequency self energy has the same structure, $\Sigma_\k(0) = \tau^z \otimes \Sigma^\tx{N}_\k + \tau^x \otimes \Sigma^\tx{A}_\k$, with $\Sigma^\tx{N}_\k=t^\dag_\k g^\tx{N}_\k t_k$ and $\Sigma^\tx{A}_\k=-t^\dag_\k g^\tx{A}_\k t_k$ being the normal and anomalous parts, respectively. Upon rotating $g^\s_\k(0)$ in Eq.~\eqref{eq:g_sc} by the unitary transformation $\exp(i\pi\tau^x/4)$, and using Eqs.~\eqref{eq:H_sc} and~\eqref{eq:g_sc_struct}, it follows that
\begin{equation}\label{eq:Sigma_times_H}
(\Delta_\k \mc{T}_\s - ih^\s_\k)(g^\tx{A}_k + ig^\tx{N}_\k)=-\mathbbm{1}.
\end{equation}
One then arrives at
\begin{equation}\label{eq:Sigma_A_PSD}
\begin{split}
\langle u | \Sigma^\tx{A}_\k | u \rangle &= -\langle u | t^\dag_\k g^\tx{A}_\k t_\k | u \rangle =  -\Real \langle u | t^\dag_\k(g^\tx{A}_\k - i g^\tx{N}_\k) t_\k | u \rangle \\
&= \Real \langle u | t^\dag_\k(g^\tx{A}_\k - i g^\tx{N}_\k) (\Delta_\k \mc{T}_\s  - ih^\s_\k) (g^\tx{A}_\k + i g^\tx{N}_\k) t_\k| u \rangle = \Real \langle v |\Delta_\k \mc{T}_\s  - ih^\s_\k| v \rangle \\
&= \langle v |\Delta_\k \mc{T}_\s| v \rangle \ge 0, 
\end{split}
\end{equation}
where $\vert u \rangle$ is an arbitrary vector, $| v \rangle \equiv (g^\tx{A}_\k + i g^\tx{N}_\k) t_\k| u \rangle$, and we have used the fact that $g^\tx{N}_\k$, $g^\tx{A}_\k$, and $h^\s_\k$ are Hermitian. Namely, we have proved that $\Sigma^\tx{A}_\k$ is PSD.

We now use this property of $\Sigma^\tx{A}_\k$ to prove the topological invariant is trivial. This can be obtained from the Green function by defining~\cite{manmana2012topological}
\begin{equation}\label{eq:H_eff}
\mc{H}^\tx{eff}_\k \equiv -[\mc{G}^\w_\k(0)]^{-1} = \tau^z\otimes(h^\w_\k + \Sigma^{N}_\k) + \tau^x\otimes\Sigma^\tx{A}_\k.
\end{equation}
Indeed, this Hamiltonian obeys a time-reversal symmetry, $\mc{T}^\dag_\w \mc{H}^\tx{eff\ast}_\mk \mc{T}_\w=\mc{H}^\tx{eff}_\k$, as well as a chiral symmetry, $\tau^y \mc{H}^\tx{eff}_\k \tau^y=-\mc{H}^\tx{eff}_\k$, and is therefore in class DIII. It was shown in Sec.~\ref{subsec:TopInv_wind_num} how the invariant can be obtained from the matrix $Q_k=U^\dag_kD_kV_k$, defined in Eq.~\eqref{eq:H_rot}, by calculating the windings of the eigenvalues $\tilde{Q}_k=U^\dag_kV_k$ [see Eq.~\eqref{eq:Winding_parity}]. 
Inserting Eq.~\eqref{eq:H_eff} in Eq.~\eqref{eq:H_rot}, one arrives at $Q_\k =\Sigma^\tx{A}_\k - i (h^\w_\k+\Sigma^\tx{N}_\k)$. From the positivity of $\Sigma^\tx{A}_\k$, and the fact that $h^\w_\k$ and $\Sigma^\tx{N}_\k$ are Hermitian, it follows that
\begin{equation}\label{eq:zero_winding}
\begin{split}
0 \le & \langle \alpha_{n,\k} | \Sigma^\tx{A}_\k | \alpha_{n,\k} \rangle
= \Real \langle \alpha_{n,\k} | Q_\k | \alpha_{n,\k} \rangle
= \Real \langle \alpha_{n,\k} | U^\dag_\k D_\k V_\k  | \alpha_{n,\k} \rangle  \\
= &\Real \langle \alpha_{n,\k} | \tilde{Q}_\k V^\dag_\k D_\k V_\k  | \alpha_{n,\k} \rangle
= 2\cos\theta_{n,k} \cdot \langle V_\k \alpha_{n,\k} | D_\k | V_\k \alpha_{n,\k} \rangle,
\end{split}
\end{equation}
where as define in Sec.~\ref{subsec:TopInv_wind_num}, $| \alpha_{n,\k} \rangle$ are the eigenvectors of $\tilde{Q}_k$ with corresponding eigenvalues $\exp(i\theta_{n,k})$. Since $D_\k$ is positive definite (see Sec.~\ref{subsec:TopInv_wind_num}), we conclude that $\cos\theta_{n,k} \ge 0$ for all $n$ and $k$. Namely none of the phases $\theta_{n,k}$ can have a non-zero winding number as $k$ changes from $-\pi$ to $\pi$, which in particular means that the topological invariant, Eq.~(\ref{eq:Winding_parity}), is always trivial, $\nu=1$.

The results of this section immediately generalizes to the case a proximity-coupled 2d system. The combined system in this case is described by the Hamiltonian of Eq.~\eqref{eq:H}, with $k\rightarrow \bs{k}=(k_x,k_y)$. All the above results are still valid in the 2d case under this substitution. As show in Sec.~\ref{subsec:TopInv_1dto2d}, the $\mathbb{Z}_2$ two-dimensional topological invariant can be obtained from the 1d invariant by
\begin{equation}\label{eq:2d_top_inv_2nd}
\nu_\tx{2d} = \nu[\mc{H}^\tx{eff}_{k_x=0,k_y}]\cdot\nu[\mc{H}^\tx{eff}_{k_x=\pi,k_y}].
\end{equation}
The Hamiltonians $\mc{H}^\tx{eff}_{k_x=0,k_y}$ and $\mc{H}^\tx{eff}_{k_x=\pi,k_y}$ both belong to class DIII in 1d, and are of the form of Eq.~\eqref{eq:H_eff} with a PSD anomalous part. Consequently, as we proved above, both $\mc{H}^\tx{eff}_{k_x=0,k_y}$ and $\mc{H}^\tx{eff}_{k_x=\pi,k_y}$ are topologically trivial. From Eq.~\eqref{eq:2d_top_inv_2nd} it then follows that the 2d Hamiltonian $\mc{H}^\tx{eff}_{k_x,k_y}$ is trivial as well\footnote{We note that since \emph{both} $\mc{H}^\tx{eff}_{k_x=0,k_y}$ and $\mc{H}^\tx{eff}_{k_x=\pi,k_y}$ (and similarly $\mc{H}^\tx{eff}_{k_x,k_y=0}$ and $\mc{H}^\tx{eff}_{k_x,k_y=\pi}$) are trivial, the weak topological indices are trivial as well.}.

\paragraph{Extension to non-translationally-invariant systems}
\label{parag:realize_inter_syst_necess_int_dis}
So far, we assumed that the system is translationally invariant along the direction of the wire in the 1d case, or in the plane of the system in the 2d case. However, our results holds even without translational symmetry, e.g., in the presence of disorder.

To see this, consider a disordered system in either 1d or 2d, coupled to a superconductor. Imagine a disorder realization which is periodic in space, with a period that is much larger than any microscopic length scale (in particular, the induced superconducting coherence length).
By the arguments presented in the preceding sections, the resulting translationally invariant system is topologically trivial. Hence, at its boundary there are no topologically non-trivial edge states. Since the size of the unit cell is much larger than the coherence length, the periodicity of the system cannot matter for the existence or the lack of edge states. Therefore, a \emph{single} unit cell corresponds to a finite disordered system, which (as its size tends to infinity) is in the topologically trivial phase, as well.

\section{Lattice model for Josephson junctions}
\label{sec:JJ_lattice}

In Sec.~\ref{subsec:Signa_JJ} the spectra of several Josephson junctions were studied within a simplified model, Eq.~\eqref{eq:H_JJ}, which is based on the low-energy Hamiltonian introduced in Sec.~\ref{subsec:Intro_min_model}. This model, however, does not include effects that arises from backscattering in the junction or from completely breaking of spin-rotation symmetry. The effect of such perturbations were discussed in a qualitative manner in Sec.~\ref{subsec:Signa_JJ}, based on degenerate perturbation theory. In this appendix we present a lattice model for the studied Josephson junctions, which includes backscattering and which completely breaks spin-rotation symmetry. This model was used to produce the spectra presented in the third and forth columns of Fig.~\ref{fig:signa_JJs}.

The Hamiltonian describing the Josephson junction is based on the lattice model of Sec.~\ref{subsec:Lattice_model}, and is given by 
\be\label{eq:H_min-latt}
\begin{split}
	&H^\tx{JJ}_\tx{Latt} = \sum_{n=1}^{2N+N_\tx{J}}\left\{-\mu \bs{c}^\dag_{n} \bs{c}_{n} + \left[\bs{c}^\dag_n(-t_n + iu_n\sigma^x)\bs{c}_{n+1}
	+\half\Delta_{0,n} \bs{c}^\dag_n i\sigma^y \bs{c}^{\dag\top}_{n} + \half \Delta''_{1,n} \bs{c}^\dag_n \sigma^x \bs{c}^{\dag\top}_{n+1} +\tx{h.c.}\right]\right\},
\end{split}
\ee
where 
\begin{equation}
\Delta_{0,n} =\left\{
\begin{array}{lll}
\Delta_{0,\tx{L}} & , & n\le N\\
0 & , & N<n\le N_\tx{J}\\
\Delta_{0,\tx{R}}e^{i\phi} & , & n> N+N_\tx{J}
\end{array}
\right.
\hs 4mm ; \hs 4mm
\Delta''_{1,n} =\left\{
\begin{array}{lll}
\Delta''_{1,\tx{L}} & , & n\le N\\
0 & , & N<n\le N_\tx{J}\\
\Delta''_{1,\tx{R}}e^{i\phi} & , &n> N+N_\tx{J}
\end{array}
\right.,
\end{equation}
and
\begin{equation}
t_n =\left\{
\begin{array}{lll}
t_0 & , & n\neq N,N+N_\tx{J}\\
t' & , & n=N, N+N_\tx{J}
\end{array}
\right.
\hs 4mm ; \hs 4mm
u_n =\left\{
\begin{array}{lll}
0 & , & N\ge n,\hs 1mm \tx{or} \hs 1mm n> N+N_\tx{J}\\
u' & , & N<n\le N+N_\tx{J}
\end{array}
\right..
\end{equation}
It describes two superconductors, each of length $Na$, where $a$ is the lattice constant. The length of the junction is $N_\tx{J}a$.

By adjusting $\Delta_{0,\tx{L,R}}$ and $\Delta_{1,\tx{L,R}}$ we can go between the three different Josephson junctions studied in Sec.~\ref{subsec:Signa_JJ}: (i) trivial-trivial junction, (ii) topological-topological junction, and (iii) topological-trivial junction. The left (right) superconductor is in the topological (TRITOPS) phase for $|\Delta''_{1,\tx{L(R)}}|>|\Delta_{0,\tx{L(R)}}|$ (see Sec.~\ref{subsec:Lattice_model}). Notice the spin-orbit coupling term, $u'$ is in the $\sigma^x$ direction, therefore breaking the symmetry under $\bs{c}_n\rightarrow \exp(i\theta\sigma^z) \bs{c}_n$. Furthermore, the coupling $t'$ at the edges of the junction controls the transparency of the junctions, and introduce backscattering.

The parameters used in the simulations shown in Fig.~\ref{fig:signa_JJs} are as follows. (i) For the trivial-trivial junction we take $\Delta_{0,\tx{L}}=\Delta_{0,\tx{R}}=1$, $\Delta''_{1,\tx{L}}=\Delta''_{1,\tx{R}}=0.3$, $t=10$, $\mu=-10$, $t'=5$, $u'=0.5$, $N=125$, $N_\tx{J}=8$. (ii) For the topological-topological junction we take $\Delta_{0,\tx{L}}=\Delta_{0,\tx{R}}=0.3$, $\Delta''_{1,\tx{L}}=\Delta''_{1,\tx{R}}=1$, $t=10$, $\mu=-10$, $t'=5$, $u'=0.5$, $N=125$, $N_\tx{J}=8$. (iii) For the topological-trivial junction we take $\Delta_{0,\tx{L}}=0.3$, $\Delta_{0,\tx{R}}=1$, $\Delta''_{1,\tx{L}}=1$, $\Delta''_{1,\tx{R}}=0.3$, $t=10$, $\mu=-10$, $t'=5$, $u'=0.5$, $N=125$, $N_\tx{J}=10$.

\end{appendices}

\section*{References}
\addcontentsline{toc}{section}{References}

\bibliography{Refs_review1}

\begin{thebibliography}{100}
\expandafter\ifx\csname url\endcsname\relax
  \def\url#1{\texttt{#1}}\fi
\expandafter\ifx\csname urlprefix\endcsname\relax\def\urlprefix{URL }\fi
\expandafter\ifx\csname href\endcsname\relax
  \def\href#1#2{#2} \def\path#1{#1}\fi

\bibitem{Klitzing1980new}
K.~v. Klitzing, G.~Dorda, M.~Pepper,
  \href{http://link.aps.org/doi/10.1103/PhysRevLett.45.494}{New method for
  high-accuracy determination of the fine-structure constant based on quantized
  hall resistance}, Phys. Rev. Lett. 45 (1980) 494--497.
\newblock \href {http://dx.doi.org/10.1103/PhysRevLett.45.494}
  {\path{doi:10.1103/PhysRevLett.45.494}}.
\newline\urlprefix\url{http://link.aps.org/doi/10.1103/PhysRevLett.45.494}

\bibitem{Laughlin1981quantized}
R.~B. Laughlin,
  \href{http://link.aps.org/doi/10.1103/PhysRevB.23.5632}{Quantized hall
  conductivity in two dimensions}, Phys. Rev. B 23 (1981) 5632--5633.
\newblock \href {http://dx.doi.org/10.1103/PhysRevB.23.5632}
  {\path{doi:10.1103/PhysRevB.23.5632}}.
\newline\urlprefix\url{http://link.aps.org/doi/10.1103/PhysRevB.23.5632}

\bibitem{Thouless1982quantized}
D.~J. Thouless, M.~Kohmoto, M.~P. Nightingale, M.~den Nijs,
  \href{http://link.aps.org/doi/10.1103/PhysRevLett.49.405}{Quantized hall
  conductance in a two-dimensional periodic potential}, Phys. Rev. Lett. 49
  (1982) 405--408.
\newblock \href {http://dx.doi.org/10.1103/PhysRevLett.49.405}
  {\path{doi:10.1103/PhysRevLett.49.405}}.
\newline\urlprefix\url{http://link.aps.org/doi/10.1103/PhysRevLett.49.405}

\bibitem{Wen04}
X.-G. Wen, Quantum Field Theory of Many-Body Systems, Oxford, University Press,
  2004.

\bibitem{schnyder2008classification}
A.~P. Schnyder, S.~Ryu, A.~Furusaki, A.~W.~W. Ludwig,
  \href{http://link.aps.org/doi/10.1103/PhysRevB.78.195125}{Classification of
  topological insulators and superconductors in three spatial dimensions},
  Phys. Rev. B 78 (2008) 195125.
\newblock \href {http://dx.doi.org/10.1103/PhysRevB.78.195125}
  {\path{doi:10.1103/PhysRevB.78.195125}}.
\newline\urlprefix\url{http://link.aps.org/doi/10.1103/PhysRevB.78.195125}

\bibitem{kitaev2009periodic}
A.~Kitaev, \href{http://dx.doi.org/10.1063/1.3149495}{Periodic table for
  topological insulators and superconductors}, in: AIP Conf. Proc., Vol. 1134,
  2009, p.~22.
\newline\urlprefix\url{http://dx.doi.org/10.1063/1.3149495}

\bibitem{Ryu2010a}
S.~Ryu, A.~P. Schnyder, A.~Furusaki, A.~W.~W. Ludwig, {Topological insulators
  and superconductors: tenfold way and dimensional hierarchy}, New J. Phys.
  12~(6) (2010) 065010.
\newblock \href {http://dx.doi.org/10.1088/1367-2630/12/6/065010}
  {\path{doi:10.1088/1367-2630/12/6/065010}}.

\bibitem{Qi2011topological}
X.-L. Qi, S.-C. Zhang,
  \href{http://link.aps.org/doi/10.1103/RevModPhys.83.1057}{Topological
  insulators and superconductors}, Rev. Mod. Phys. 83 (2011) 1057--1110.
\newblock \href {http://dx.doi.org/10.1103/RevModPhys.83.1057}
  {\path{doi:10.1103/RevModPhys.83.1057}}.
\newline\urlprefix\url{http://link.aps.org/doi/10.1103/RevModPhys.83.1057}

\bibitem{Kane2005quantum}
C.~L. Kane, E.~J. Mele,
  \href{http://link.aps.org/doi/10.1103/PhysRevLett.95.226801}{Quantum spin
  hall effect in graphene}, Phys. Rev. Lett. 95 (2005) 226801.
\newblock \href {http://dx.doi.org/10.1103/PhysRevLett.95.226801}
  {\path{doi:10.1103/PhysRevLett.95.226801}}.
\newline\urlprefix\url{http://link.aps.org/doi/10.1103/PhysRevLett.95.226801}

\bibitem{Bernevig2006quantum}
B.~A. Bernevig, T.~L. Hughes, S.-C. Zhang,
  \href{http://www.sciencemag.org/content/314/5806/1757.abstract}{Quantum spin
  hall effect and topological phase transition in hgte quantum wells}, Science
  314~(5806) (2006) 1757--1761.
\newblock \href {http://dx.doi.org/10.1126/science.1133734}
  {\path{doi:10.1126/science.1133734}}.
\newline\urlprefix\url{http://www.sciencemag.org/content/314/5806/1757.abstract}

\bibitem{konig2007quantum}
M.~K{\"o}nig, S.~Wiedmann, C.~Br{\"u}ne, A.~Roth, H.~Buhmann, L.~W. Molenkamp,
  X.-L. Qi, S.-C. Zhang,
  \href{http://www.sciencemag.org/content/318/5851/766.short}{Quantum spin hall
  insulator state in hgte quantum wells}, Science 318~(5851) (2007) 766--770.
\newline\urlprefix\url{http://www.sciencemag.org/content/318/5851/766.short}

\bibitem{Fu2007topological}
L.~Fu, C.~L. Kane, E.~J. Mele,
  \href{https://link.aps.org/doi/10.1103/PhysRevLett.98.106803}{Topological
  insulators in three dimensions}, Phys. Rev. Lett. 98 (2007) 106803.
\newblock \href {http://dx.doi.org/10.1103/PhysRevLett.98.106803}
  {\path{doi:10.1103/PhysRevLett.98.106803}}.
\newline\urlprefix\url{https://link.aps.org/doi/10.1103/PhysRevLett.98.106803}

\bibitem{Moore2007topological}
J.~E. Moore, L.~Balents,
  \href{https://link.aps.org/doi/10.1103/PhysRevB.75.121306}{Topological
  invariants of time-reversal-invariant band structures}, Phys. Rev. B 75
  (2007) 121306.
\newblock \href {http://dx.doi.org/10.1103/PhysRevB.75.121306}
  {\path{doi:10.1103/PhysRevB.75.121306}}.
\newline\urlprefix\url{https://link.aps.org/doi/10.1103/PhysRevB.75.121306}

\bibitem{hsieh2008topological}
D.~Hsieh, D.~Qian, L.~Wray, Y.~Xia, Y.~S. Hor, R.~J. Cava, M.~Z. Hasan, A
  topological dirac insulator in a quantum spin hall phase, Nature 452~(7190)
  (2008) 970.

\bibitem{Hasan2010}
M.~Hasan, C.~Kane,
  \href{http://link.aps.org/doi/10.1103/RevModPhys.82.3045}{{Colloquium:
  Topological insulators}}, Rev. Mod. Phys. 82~(4) (2010) 3045--3067.
\newblock \href {http://dx.doi.org/10.1103/RevModPhys.82.3045}
  {\path{doi:10.1103/RevModPhys.82.3045}}.
\newline\urlprefix\url{http://link.aps.org/doi/10.1103/RevModPhys.82.3045}

\bibitem{Altland1997}
A.~Altland, M.~R. Zirnbauer,
  \href{http://link.aps.org/doi/10.1103/PhysRevB.55.1142}{{Nonstandard symmetry
  classes in mesoscopic normal-superconducting hybrid structures}}, Phys. Rev.
  B 55~(2) (1997) 1142--1161.
\newblock \href {http://dx.doi.org/10.1103/PhysRevB.55.1142}
  {\path{doi:10.1103/PhysRevB.55.1142}}.
\newline\urlprefix\url{http://link.aps.org/doi/10.1103/PhysRevB.55.1142}

\bibitem{Read2000paired}
N.~Read, D.~Green,
  \href{http://link.aps.org/doi/10.1103/PhysRevB.61.10267}{Paired states of
  fermions in two dimensions with breaking of parity and time-reversal
  symmetries and the fractional quantum hall effect}, Phys. Rev. B 61~(15)
  (2000) 10267.
\newline\urlprefix\url{http://link.aps.org/doi/10.1103/PhysRevB.61.10267}

\bibitem{Kitaev2001unpaired}
A.~Kitaev,
  \href{http://iopscience.iop.org/article/10.1070/1063-7869/44/10S/S29/meta}{Unpaired
  majorana fermions in quantum wires}, Phys. Usp. 44~(10S) (2001) 131.
\newline\urlprefix\url{http://iopscience.iop.org/article/10.1070/1063-7869/44/10S/S29/meta}

\bibitem{Alicea2012}
J.~Alicea, \href{http://www.ncbi.nlm.nih.gov/pubmed/22790778}{{New directions
  in the pursuit of Majorana fermions in solid state systems.}}, Rep. Prog.
  Phys. 75~(7) (2012) 076501.
\newblock \href {http://dx.doi.org/10.1088/0034-4885/75/7/076501}
  {\path{doi:10.1088/0034-4885/75/7/076501}}.
\newline\urlprefix\url{http://www.ncbi.nlm.nih.gov/pubmed/22790778}

\bibitem{Beenakker2013}
C.~W.~J. Beenakker,
  \href{http://www.annualreviews.org/doi/abs/10.1146/annurev-conmatphys-030212-184337}{{Search
  for Majorana Fermions in Superconductors}}, Ann. Rev. Condens. Matt. Phys.
  4~(1) (2013) 113--136.
\newblock \href {http://dx.doi.org/10.1146/annurev-conmatphys-030212-184337}
  {\path{doi:10.1146/annurev-conmatphys-030212-184337}}.
\newline\urlprefix\url{http://www.annualreviews.org/doi/abs/10.1146/annurev-conmatphys-030212-184337}

\bibitem{Kopnin1991mutual}
N.~B. Kopnin, M.~M. Salomaa,
  \href{https://link.aps.org/doi/10.1103/PhysRevB.44.9667}{Mutual friction in
  superfluid $^{3}\mathrm{He}$: Effects of bound states in the vortex core},
  Phys. Rev. B 44 (1991) 9667--9677.
\newblock \href {http://dx.doi.org/10.1103/PhysRevB.44.9667}
  {\path{doi:10.1103/PhysRevB.44.9667}}.
\newline\urlprefix\url{https://link.aps.org/doi/10.1103/PhysRevB.44.9667}

\bibitem{Ivanov2001non}
D.~A. Ivanov,
  \href{http://link.aps.org/doi/10.1103/PhysRevLett.86.268}{Non-abelian
  statistics of half-quantum vortices in $\mathit{p}$-wave superconductors},
  Phys. Rev. Lett. 86 (2001) 268--271.
\newblock \href {http://dx.doi.org/10.1103/PhysRevLett.86.268}
  {\path{doi:10.1103/PhysRevLett.86.268}}.
\newline\urlprefix\url{http://link.aps.org/doi/10.1103/PhysRevLett.86.268}

\bibitem{Freedman2003topological}
M.~Freedman, A.~Kitaev, M.~Larsen, Z.~Wang, Topological quantum computation,
  Bulletin of the American Mathematical Society 40~(1) (2003) 31--38.

\bibitem{Kitaev2003}
A.~Y. Kitaev,
  \href{http://www.sciencedirect.com/science/article/pii/S0003491602000180}{Fault-tolerant
  quantum computation by anyons}, Ann. Phys. 303 (2003) 2.
\newline\urlprefix\url{http://www.sciencedirect.com/science/article/pii/S0003491602000180}

\bibitem{nayak2008non}
C.~Nayak, S.~Simon, A.~Stern, M.~Freedman, S.~Sarma,
  \href{http://link.aps.org/doi/10.1103/RevModPhys.80.1083}{Non-abelian anyons
  and topological quantum computation}, Rev. Mod. Phys. 80~(3) (2008) 1083.
\newline\urlprefix\url{http://link.aps.org/doi/10.1103/RevModPhys.80.1083}

\bibitem{Bonderson2008measurement}
P.~Bonderson, M.~Freedman, C.~Nayak,
  \href{https://link.aps.org/doi/10.1103/PhysRevLett.101.010501}{Measurement-only
  topological quantum computation}, Phys. Rev. Lett. 101 (2008) 010501.
\newblock \href {http://dx.doi.org/10.1103/PhysRevLett.101.010501}
  {\path{doi:10.1103/PhysRevLett.101.010501}}.
\newline\urlprefix\url{https://link.aps.org/doi/10.1103/PhysRevLett.101.010501}

\bibitem{Karzig2017Scalable}
T.~Karzig, C.~Knapp, R.~M. Lutchyn, P.~Bonderson, M.~B. Hastings, C.~Nayak,
  J.~Alicea, K.~Flensberg, S.~Plugge, Y.~Oreg, C.~M. Marcus, M.~H. Freedman,
  \href{https://link.aps.org/doi/10.1103/PhysRevB.95.235305}{Scalable designs
  for quasiparticle-poisoning-protected topological quantum computation with
  majorana zero modes}, Phys. Rev. B 95 (2017) 235305.
\newblock \href {http://dx.doi.org/10.1103/PhysRevB.95.235305}
  {\path{doi:10.1103/PhysRevB.95.235305}}.
\newline\urlprefix\url{https://link.aps.org/doi/10.1103/PhysRevB.95.235305}

\bibitem{Qi2009time}
X.-L. Qi, T.~L. Hughes, S.~Raghu, S.-C. Zhang,
  \href{http://link.aps.org/doi/10.1103/PhysRevLett.102.187001}{Time-reversal-invariant
  topological superconductors and superfluids in two and three dimensions},
  Phys. Rev. Lett. 102 (2009) 187001.
\newblock \href {http://dx.doi.org/10.1103/PhysRevLett.102.187001}
  {\path{doi:10.1103/PhysRevLett.102.187001}}.
\newline\urlprefix\url{http://link.aps.org/doi/10.1103/PhysRevLett.102.187001}

\bibitem{Fu2010}
L.~Fu, E.~Berg,
  \href{http://link.aps.org/doi/10.1103/PhysRevLett.105.097001}{Odd-parity
  topological superconductors: Theory and application to
  ${\mathrm{cu}}_{x}{\mathrm{bi}}_{2}{\mathrm{se}}_{3}$}, Phys. Rev. Lett. 105
  (2010) 097001.
\newblock \href {http://dx.doi.org/10.1103/PhysRevLett.105.097001}
  {\path{doi:10.1103/PhysRevLett.105.097001}}.
\newline\urlprefix\url{http://link.aps.org/doi/10.1103/PhysRevLett.105.097001}

\bibitem{volovik2003universe}
G.~E. Volovik, The Universe in a Helium Droplet, Oxford, 2003.

\bibitem{vollhardt2013superfluid}
D.~Vollhardt, P.~W{\"o}lfle, The superfluid phases of helium 3, Courier
  Corporation, 2013.

\bibitem{Fu2008superconducting}
L.~Fu, C.~L. Kane,
  \href{http://link.aps.org/doi/10.1103/PhysRevLett.100.096407}{Superconducting
  proximity effect and majorana fermions at the surface of a topological
  insulator}, Phys. Rev. Lett. 100 (2008) 096407.
\newblock \href {http://dx.doi.org/10.1103/PhysRevLett.100.096407}
  {\path{doi:10.1103/PhysRevLett.100.096407}}.
\newline\urlprefix\url{http://link.aps.org/doi/10.1103/PhysRevLett.100.096407}

\bibitem{Fu2009josephson}
L.~Fu, C.~L. Kane,
  \href{http://link.aps.org/doi/10.1103/PhysRevB.79.161408}{Josephson current
  and noise at a superconductor/quantum-spin-hall-insulator/superconductor
  junction}, Phys. Rev. B 79 (2009) 161408.
\newblock \href {http://dx.doi.org/10.1103/PhysRevB.79.161408}
  {\path{doi:10.1103/PhysRevB.79.161408}}.
\newline\urlprefix\url{http://link.aps.org/doi/10.1103/PhysRevB.79.161408}

\bibitem{Sau2010generic}
J.~D. Sau, R.~M. Lutchyn, S.~Tewari, S.~Das~Sarma,
  \href{http://link.aps.org/doi/10.1103/PhysRevLett.104.040502}{Generic new
  platform for topological quantum computation using semiconductor
  heterostructures}, Phys. Rev. Lett. 104 (2010) 040502.
\newblock \href {http://dx.doi.org/10.1103/PhysRevLett.104.040502}
  {\path{doi:10.1103/PhysRevLett.104.040502}}.
\newline\urlprefix\url{http://link.aps.org/doi/10.1103/PhysRevLett.104.040502}

\bibitem{Alicea2010Majorana}
J.~Alicea, \href{http://link.aps.org/doi/10.1103/PhysRevB.81.125318}{Majorana
  fermions in a tunable semiconductor device}, Phys. Rev. B 81 (2010) 125318.
\newblock \href {http://dx.doi.org/10.1103/PhysRevB.81.125318}
  {\path{doi:10.1103/PhysRevB.81.125318}}.
\newline\urlprefix\url{http://link.aps.org/doi/10.1103/PhysRevB.81.125318}

\bibitem{Lutchyn2010majorana}
R.~M. Lutchyn, J.~D. Sau, S.~Das~Sarma,
  \href{http://link.aps.org/doi/10.1103/PhysRevLett.105.077001}{Majorana
  fermions and a topological phase transition in semiconductor-superconductor
  heterostructures}, Phys. Rev. Lett. 105 (2010) 077001.
\newblock \href {http://dx.doi.org/10.1103/PhysRevLett.105.077001}
  {\path{doi:10.1103/PhysRevLett.105.077001}}.
\newline\urlprefix\url{http://link.aps.org/doi/10.1103/PhysRevLett.105.077001}

\bibitem{Oreg2010helical}
Y.~Oreg, G.~Refael, F.~von Oppen,
  \href{http://link.aps.org/doi/10.1103/PhysRevLett.105.177002}{Helical liquids
  and majorana bound states in quantum wires}, Phys. Rev. Lett. 105 (2010)
  177002.
\newblock \href {http://dx.doi.org/10.1103/PhysRevLett.105.177002}
  {\path{doi:10.1103/PhysRevLett.105.177002}}.
\newline\urlprefix\url{http://link.aps.org/doi/10.1103/PhysRevLett.105.177002}

\bibitem{Duckheim2010Andreev}
M.~Duckheim, P.~W. Brouwer,
  \href{http://link.aps.org/doi/10.1103/PhysRevB.83.054513}{Andreev reflection
  from noncentrosymmetric superconductors and majorana bound-state generation
  in half-metallic ferromagnets}, Phys. Rev. B 83 (2011) 054513.
\newblock \href {http://dx.doi.org/10.1103/PhysRevB.83.054513}
  {\path{doi:10.1103/PhysRevB.83.054513}}.
\newline\urlprefix\url{http://link.aps.org/doi/10.1103/PhysRevB.83.054513}

\bibitem{Pientka2013Topological}
F.~Pientka, L.~I. Glazman, F.~von Oppen,
  \href{http://link.aps.org/doi/10.1103/PhysRevB.88.155420}{Topological
  superconducting phase in helical shiba chains}, Phys. Rev. B 88 (2013)
  155420.
\newblock \href {http://dx.doi.org/10.1103/PhysRevB.88.155420}
  {\path{doi:10.1103/PhysRevB.88.155420}}.
\newline\urlprefix\url{http://link.aps.org/doi/10.1103/PhysRevB.88.155420}

\bibitem{Nadj-Perge2013proposal}
S.~Nadj-Perge, I.~K. Drozdov, B.~A. Bernevig, A.~Yazdani,
  \href{http://link.aps.org/doi/10.1103/PhysRevB.88.020407}{Proposal for
  realizing majorana fermions in chains of magnetic atoms on a superconductor},
  Phys. Rev. B 88 (2013) 020407.
\newblock \href {http://dx.doi.org/10.1103/PhysRevB.88.020407}
  {\path{doi:10.1103/PhysRevB.88.020407}}.
\newline\urlprefix\url{http://link.aps.org/doi/10.1103/PhysRevB.88.020407}

\bibitem{Klinovaja2013topological}
J.~Klinovaja, P.~Stano, A.~Yazdani, D.~Loss,
  \href{http://link.aps.org/doi/10.1103/PhysRevLett.111.186805}{Topological
  superconductivity and majorana fermions in rkky systems}, Phys. Rev. Lett.
  111 (2013) 186805.
\newblock \href {http://dx.doi.org/10.1103/PhysRevLett.111.186805}
  {\path{doi:10.1103/PhysRevLett.111.186805}}.
\newline\urlprefix\url{http://link.aps.org/doi/10.1103/PhysRevLett.111.186805}

\bibitem{Braunecker2013interplay}
B.~Braunecker, P.~Simon,
  \href{http://link.aps.org/doi/10.1103/PhysRevLett.111.147202}{Interplay
  between classical magnetic moments and superconductivity in quantum
  one-dimensional conductors: Toward a self-sustained topological majorana
  phase}, Phys. Rev. Lett. 111 (2013) 147202.
\newblock \href {http://dx.doi.org/10.1103/PhysRevLett.111.147202}
  {\path{doi:10.1103/PhysRevLett.111.147202}}.
\newline\urlprefix\url{http://link.aps.org/doi/10.1103/PhysRevLett.111.147202}

\bibitem{Vazifeh2013self}
M.~M. Vazifeh, M.~Franz,
  \href{http://link.aps.org/doi/10.1103/PhysRevLett.111.206802}{Self-organized
  topological state with majorana fermions}, Phys. Rev. Lett. 111 (2013)
  206802.
\newblock \href {http://dx.doi.org/10.1103/PhysRevLett.111.206802}
  {\path{doi:10.1103/PhysRevLett.111.206802}}.
\newline\urlprefix\url{http://link.aps.org/doi/10.1103/PhysRevLett.111.206802}

\bibitem{Lutchyn2018majorana}
R.~Lutchyn, E.~Bakkers, L.~Kouwenhoven, P.~Krogstrup, C.~Marcus, Y.~Oreg,
  \href{https://www.nature.com/articles/s41578-018-0003-1}{Majorana zero modes
  in superconductor--semiconductor heterostructures}, Nat. Rev. Mat. 3 (2018)
  52--68.
\newline\urlprefix\url{https://www.nature.com/articles/s41578-018-0003-1}

\bibitem{mourik2012signatures}
V.~Mourik, K.~Zuo, S.~Frolov, S.~Plissard, E.~Bakkers, L.~Kouwenhoven,
  \href{http://www.sciencemag.org/content/336/6084/1003}{Signatures of majorana
  fermions in hybrid superconductor-semiconductor nanowire devices}, Science
  336~(6084) (2012) 1003--1007.
\newline\urlprefix\url{http://www.sciencemag.org/content/336/6084/1003}

\bibitem{deng2012anomalous}
M.~Deng, C.~Yu, G.~Huang, M.~Larsson, P.~Caroff, H.~Xu,
  \href{http://pubs.acs.org/doi/abs/10.1021/nl303758w}{Anomalous zero-bias
  conductance peak in a nb--insb nanowire--nb hybrid device}, Nano Lett.
  12~(12) (2012) 6414--6419.
\newline\urlprefix\url{http://pubs.acs.org/doi/abs/10.1021/nl303758w}

\bibitem{Das2012zero}
A.~Das, Y.~Ronen, Y.~Most, Y.~Oreg, M.~Heiblum, H.~Shtrikman,
  \href{http://www.nature.com/doifinder/10.1038/nphys2479}{{Zero-bias peaks and
  splitting in an AlגAlInAs nanowire topological superconductor as a signature
  of Majorana fermions}}, Nat. Phys. 8 (2012) 887.
\newblock \href {http://dx.doi.org/10.1038/nphys2479}
  {\path{doi:10.1038/nphys2479}}.
\newline\urlprefix\url{http://www.nature.com/doifinder/10.1038/nphys2479}

\bibitem{churchill2013superconductor}
H.~O.~H. Churchill, V.~Fatemi, K.~Grove-Rasmussen, M.~T. Deng, P.~Caroff, H.~Q.
  Xu, C.~M. Marcus,
  \href{http://link.aps.org/doi/10.1103/PhysRevB.87.241401}{Superconductor-nanowire
  devices from tunneling to the multichannel regime: Zero-bias oscillations and
  magnetoconductance crossover}, Phys. Rev. B 87 (2013) 241401.
\newblock \href {http://dx.doi.org/10.1103/PhysRevB.87.241401}
  {\path{doi:10.1103/PhysRevB.87.241401}}.
\newline\urlprefix\url{http://link.aps.org/doi/10.1103/PhysRevB.87.241401}

\bibitem{Finck2013anomalous}
A.~D.~K. Finck, D.~J. Van~Harlingen, P.~K. Mohseni, K.~Jung, X.~Li,
  \href{http://link.aps.org/doi/10.1103/PhysRevLett.110.126406}{Anomalous
  modulation of a zero-bias peak in a hybrid nanowire-superconductor device},
  Phys. Rev. Lett. 110 (2013) 126406.
\newblock \href {http://dx.doi.org/10.1103/PhysRevLett.110.126406}
  {\path{doi:10.1103/PhysRevLett.110.126406}}.
\newline\urlprefix\url{http://link.aps.org/doi/10.1103/PhysRevLett.110.126406}

\bibitem{Nadj-Perge2014observation}
S.~Nadj-Perge, I.~K. Drozdov, J.~Li, H.~Chen, S.~Jeon, J.~Seo, A.~H. MacDonald,
  B.~A. Bernevig, A.~Yazdani,
  \href{http://www.sciencemag.org/content/early/2014/10/01/science.1259327.abstract}{Observation
  of majorana fermions in ferromagnetic atomic chains on a superconductor},
  Science 346 (2014) 602.
\newline\urlprefix\url{http://www.sciencemag.org/content/early/2014/10/01/science.1259327.abstract}

\bibitem{Pawlak2016probing}
R.~Pawlak, M.~Kisiel, J.~Klinovaja, T.~Meier, S.~Kawai, T.~Glatzel, D.~Loss,
  E.~Meyer, \href{https://www.nature.com/articles/npjqi201635}{Probing atomic
  structure and majorana wavefunctions in mono-atomic fe chains on
  superconducting pb surface}, NPJ Quantum Information 2 (2016) 16035.
\newline\urlprefix\url{https://www.nature.com/articles/npjqi201635}

\bibitem{Ruby2015end}
M.~Ruby, F.~Pientka, Y.~Peng, F.~von Oppen, B.~W. Heinrich, K.~J. Franke,
  \href{http://link.aps.org/doi/10.1103/PhysRevLett.115.197204}{End states and
  subgap structure in proximity-coupled chains of magnetic adatoms}, Phys. Rev.
  Lett. 115 (2015) 197204.
\newblock \href {http://dx.doi.org/10.1103/PhysRevLett.115.197204}
  {\path{doi:10.1103/PhysRevLett.115.197204}}.
\newline\urlprefix\url{http://link.aps.org/doi/10.1103/PhysRevLett.115.197204}

\bibitem{Albrecht2016exponential}
S.~Albrecht, A.~Higginbotham, M.~Madsen, F.~Kuemmeth, T.~Jespersen,
  J.~Nyg{\aa}rd, P.~Krogstrup, C.~Marcus,
  \href{http://www.nature.com/nature/journal/v531/n7593/abs/nature17162.html}{Exponential
  protection of zero modes in majorana islands}, Nature 531~(7593) (2016)
  206--209.
\newline\urlprefix\url{http://www.nature.com/nature/journal/v531/n7593/abs/nature17162.html}

\bibitem{Deng2016Majorana}
M.~T. Deng, S.~Vaitiekenas, E.~B. Hansen, J.~Danon, M.~Leijnse, K.~Flensberg,
  J.~Nyg{\r a}rd, P.~Krogstrup, C.~M. Marcus,
  \href{http://science.sciencemag.org/content/354/6319/1557}{Majorana bound
  state in a coupled quantum-dot hybrid-nanowire system}, Science 354~(6319)
  (2016) 1557--1562.
\newblock \href
  {http://arxiv.org/abs/http://science.sciencemag.org/content/354/6319/1557.full.pdf}
  {\path{arXiv:http://science.sciencemag.org/content/354/6319/1557.full.pdf}},
  \href {http://dx.doi.org/10.1126/science.aaf3961}
  {\path{doi:10.1126/science.aaf3961}}.
\newline\urlprefix\url{http://science.sciencemag.org/content/354/6319/1557}

\bibitem{Chen2017Experimental}
J.~Chen, P.~Yu, J.~Stenger, M.~Hocevar, D.~Car, S.~R. Plissard, E.~P. A.~M.
  Bakkers, T.~D. Stanescu, S.~M. Frolov,
  \href{http://advances.sciencemag.org/content/3/9/e1701476}{Experimental phase
  diagram of zero-bias conductance peaks in superconductor/semiconductor
  nanowire devices}, Science Advances 3~(9).
\newblock \href
  {http://arxiv.org/abs/http://advances.sciencemag.org/content/3/9/e1701476.full.pdf}
  {\path{arXiv:http://advances.sciencemag.org/content/3/9/e1701476.full.pdf}},
  \href {http://dx.doi.org/10.1126/sciadv.1701476}
  {\path{doi:10.1126/sciadv.1701476}}.
\newline\urlprefix\url{http://advances.sciencemag.org/content/3/9/e1701476}

\bibitem{gul2018ballistic}
{\"O}.~G{\"u}l, H.~Zhang, J.~D. Bommer, M.~W. de~Moor, D.~Car, S.~R. Plissard,
  E.~P. Bakkers, A.~Geresdi, K.~Watanabe, T.~Taniguchi, et~al., Ballistic
  majorana nanowire devices, Nature nanotechnology (2018) 1.

\bibitem{Suominen2017zero}
H.~J. Suominen, M.~Kjaergaard, A.~R. Hamilton, J.~Shabani, C.~J. Palmstr\o{}m,
  C.~M. Marcus, F.~Nichele,
  \href{https://link.aps.org/doi/10.1103/PhysRevLett.119.176805}{Zero-energy
  modes from coalescing andreev states in a two-dimensional
  semiconductor-superconductor hybrid platform}, Phys. Rev. Lett. 119 (2017)
  176805.
\newblock \href {http://dx.doi.org/10.1103/PhysRevLett.119.176805}
  {\path{doi:10.1103/PhysRevLett.119.176805}}.
\newline\urlprefix\url{https://link.aps.org/doi/10.1103/PhysRevLett.119.176805}

\bibitem{Nichele2017scaling}
F.~Nichele, A.~C.~C. Drachmann, A.~M. Whiticar, E.~C.~T. O'Farrell, H.~J.
  Suominen, A.~Fornieri, T.~Wang, G.~C. Gardner, C.~Thomas, A.~T. Hatke,
  P.~Krogstrup, M.~J. Manfra, K.~Flensberg, C.~M. Marcus,
  \href{https://link.aps.org/doi/10.1103/PhysRevLett.119.136803}{Scaling of
  majorana zero-bias conductance peaks}, Phys. Rev. Lett. 119 (2017) 136803.
\newblock \href {http://dx.doi.org/10.1103/PhysRevLett.119.136803}
  {\path{doi:10.1103/PhysRevLett.119.136803}}.
\newline\urlprefix\url{https://link.aps.org/doi/10.1103/PhysRevLett.119.136803}

\bibitem{zhang2017quantized}
H.~Zhang, C.-X. Liu, S.~Gazibegovic, D.~Xu, J.~A. Logan, G.~Wang, N.~van Loo,
  J.~D. Bommer, M.~W. de~Moor, D.~Car, et~al., Quantized majorana conductance,
  arXiv preprint arXiv:1710.10701.

\bibitem{Schrade2015proximity}
C.~Schrade, A.~A. Zyuzin, J.~Klinovaja, D.~Loss,
  \href{http://link.aps.org/doi/10.1103/PhysRevLett.115.237001}{Proximity-induced
  $\ensuremath{\pi}$ josephson junctions in topological insulators and kramers
  pairs of majorana fermions}, Phys. Rev. Lett. 115 (2015) 237001.
\newblock \href {http://dx.doi.org/10.1103/PhysRevLett.115.237001}
  {\path{doi:10.1103/PhysRevLett.115.237001}}.
\newline\urlprefix\url{http://link.aps.org/doi/10.1103/PhysRevLett.115.237001}

\bibitem{Hor2010}
Y.~S. Hor, A.~J. Williams, J.~G. Checkelsky, P.~Roushan, J.~Seo, Q.~Xu, H.~W.
  Zandbergen, A.~Yazdani, N.~P. Ong, R.~J. Cava,
  \href{http://link.aps.org/doi/10.1103/PhysRevLett.104.057001}{{Superconductivity
  in Cu$_{x}$Bi$_{2}$Se$_{3}$ and its Implications for Pairing in the Undoped
  Topological Insulator}}, Phys. Rev. Lett. 104~(5) (2010) 057001.
\newblock \href {http://dx.doi.org/10.1103/PhysRevLett.104.057001}
  {\path{doi:10.1103/PhysRevLett.104.057001}}.
\newline\urlprefix\url{http://link.aps.org/doi/10.1103/PhysRevLett.104.057001}

\bibitem{Brydon2014Odd}
P.~M.~R. Brydon, S.~Das~Sarma, H.-Y. Hui, J.~D. Sau,
  \href{http://link.aps.org/doi/10.1103/PhysRevB.90.184512}{Odd-parity
  superconductivity from phonon-mediated pairing: Application to
  ${\mathrm{cu}}_{x}{\mathrm{bi}}_{2}{\mathrm{se}}_{3}$}, Phys. Rev. B 90
  (2014) 184512.
\newblock \href {http://dx.doi.org/10.1103/PhysRevB.90.184512}
  {\path{doi:10.1103/PhysRevB.90.184512}}.
\newline\urlprefix\url{http://link.aps.org/doi/10.1103/PhysRevB.90.184512}

\bibitem{Kriener2011bulk}
M.~Kriener, K.~Segawa, Z.~Ren, S.~Sasaki, Y.~Ando,
  \href{https://link.aps.org/doi/10.1103/PhysRevLett.106.127004}{Bulk
  superconducting phase with a full energy gap in the doped topological
  insulator ${\mathrm{cu}}_{x}{\mathrm{bi}}_{2}{\mathrm{se}}_{3}$}, Phys. Rev.
  Lett. 106 (2011) 127004.
\newblock \href {http://dx.doi.org/10.1103/PhysRevLett.106.127004}
  {\path{doi:10.1103/PhysRevLett.106.127004}}.
\newline\urlprefix\url{https://link.aps.org/doi/10.1103/PhysRevLett.106.127004}

\bibitem{Bay2012superconductivity}
T.~V. Bay, T.~Naka, Y.~K. Huang, H.~Luigjes, M.~S. Golden, A.~de~Visser,
  \href{https://link.aps.org/doi/10.1103/PhysRevLett.108.057001}{Superconductivity
  in the doped topological insulator
  ${\mathrm{cu}}_{x}{\mathrm{bi}}_{2}{\mathrm{se}}_{3}$ under high pressure},
  Phys. Rev. Lett. 108 (2012) 057001.
\newblock \href {http://dx.doi.org/10.1103/PhysRevLett.108.057001}
  {\path{doi:10.1103/PhysRevLett.108.057001}}.
\newline\urlprefix\url{https://link.aps.org/doi/10.1103/PhysRevLett.108.057001}

\bibitem{matano2016spin}
K.~Matano, M.~Kriener, K.~Segawa, Y.~Ando, G.-q. Zheng, Spin-rotation symmetry
  breaking in the superconducting state of cuxbi2se3, Nature Physics 12~(9)
  (2016) 852--854.

\bibitem{Fu2014Odd}
L.~Fu, \href{https://link.aps.org/doi/10.1103/PhysRevB.90.100509}{Odd-parity
  topological superconductor with nematic order: Application to
  ${\mathrm{cu}}_{x}{\mathrm{bi}}_{2}{\mathrm{se}}_{3}$}, Phys. Rev. B 90
  (2014) 100509.
\newblock \href {http://dx.doi.org/10.1103/PhysRevB.90.100509}
  {\path{doi:10.1103/PhysRevB.90.100509}}.
\newline\urlprefix\url{https://link.aps.org/doi/10.1103/PhysRevB.90.100509}

\bibitem{Venderbos2016Odd}
J.~W.~F. Venderbos, V.~Kozii, L.~Fu,
  \href{https://link.aps.org/doi/10.1103/PhysRevB.94.180504}{Odd-parity
  superconductors with two-component order parameters: Nematic and chiral, full
  gap, and majorana node}, Phys. Rev. B 94 (2016) 180504.
\newblock \href {http://dx.doi.org/10.1103/PhysRevB.94.180504}
  {\path{doi:10.1103/PhysRevB.94.180504}}.
\newline\urlprefix\url{https://link.aps.org/doi/10.1103/PhysRevB.94.180504}

\bibitem{fu2016superconductivity}
L.~Fu, Superconductivity: Finding a direction, Nature Physics 12~(9) (2016)
  822.

\bibitem{Keselman2015gapless}
A.~Keselman, E.~Berg,
  \href{http://link.aps.org/doi/10.1103/PhysRevB.91.235309}{Gapless
  symmetry-protected topological phase of fermions in one dimension}, Phys.
  Rev. B 91 (2015) 235309.
\newblock \href {http://dx.doi.org/10.1103/PhysRevB.91.235309}
  {\path{doi:10.1103/PhysRevB.91.235309}}.
\newline\urlprefix\url{http://link.aps.org/doi/10.1103/PhysRevB.91.235309}

\bibitem{Kainaris2015Emergent}
N.~Kainaris, S.~T. Carr,
  \href{http://link.aps.org/doi/10.1103/PhysRevB.92.035139}{Emergent
  topological properties in interacting one-dimensional systems with spin-orbit
  coupling}, Phys. Rev. B 92 (2015) 035139.
\newblock \href {http://dx.doi.org/10.1103/PhysRevB.92.035139}
  {\path{doi:10.1103/PhysRevB.92.035139}}.
\newline\urlprefix\url{http://link.aps.org/doi/10.1103/PhysRevB.92.035139}

\bibitem{kainaris2017interaction}
N.~Kainaris, R.~A. Santos, D.~Gutman, S.~T. Carr, Interaction induced
  topological protection in one-dimensional conductors, Fortschritte der
  Physik.

\bibitem{Qi2010topological}
X.-L. Qi, T.~L. Hughes, S.-C. Zhang,
  \href{http://link.aps.org/doi/10.1103/PhysRevB.81.134508}{Topological
  invariants for the fermi surface of a time-reversal-invariant
  superconductor}, Phys. Rev. B 81 (2010) 134508.
\newblock \href {http://dx.doi.org/10.1103/PhysRevB.81.134508}
  {\path{doi:10.1103/PhysRevB.81.134508}}.
\newline\urlprefix\url{http://link.aps.org/doi/10.1103/PhysRevB.81.134508}

\bibitem{Fulga2011scattering}
I.~C. Fulga, F.~Hassler, A.~R. Akhmerov, C.~W.~J. Beenakker,
  \href{http://link.aps.org/doi/10.1103/PhysRevB.83.155429}{Scattering formula
  for the topological quantum number of a disordered multimode wire}, Phys.
  Rev. B 83 (2011) 155429.
\newblock \href {http://dx.doi.org/10.1103/PhysRevB.83.155429}
  {\path{doi:10.1103/PhysRevB.83.155429}}.
\newline\urlprefix\url{http://link.aps.org/doi/10.1103/PhysRevB.83.155429}

\bibitem{andreev1964thermal}
A.~Andreev,
  \href{http://www.jetp.ac.ru/cgi-bin/e/index/r/46/5/p1823?a=list}{Thermal
  conductivity of the intermediate state of superconductors}, Zh. Eksp. Teor.
  Fiz. 46 (1964) 1823.
\newline\urlprefix\url{http://www.jetp.ac.ru/cgi-bin/e/index/r/46/5/p1823?a=list}

\bibitem{Dumitrescu2013topological}
E.~Dumitrescu, S.~Tewari,
  \href{http://link.aps.org/doi/10.1103/PhysRevB.88.220505}{Topological
  properties of the time-reversal-symmetric kitaev chain and applications to
  organic superconductors}, Phys. Rev. B 88 (2013) 220505.
\newblock \href {http://dx.doi.org/10.1103/PhysRevB.88.220505}
  {\path{doi:10.1103/PhysRevB.88.220505}}.
\newline\urlprefix\url{http://link.aps.org/doi/10.1103/PhysRevB.88.220505}

\bibitem{Wong2012majorana}
C.~L.~M. Wong, K.~T. Law,
  \href{http://link.aps.org/doi/10.1103/PhysRevB.86.184516}{Majorana kramers
  doublets in $d_{x^2-y^2}$-wave superconductors with rashba spin-orbit
  coupling}, Phys. Rev. B 86 (2012) 184516.
\newblock \href {http://dx.doi.org/10.1103/PhysRevB.86.184516}
  {\path{doi:10.1103/PhysRevB.86.184516}}.
\newline\urlprefix\url{http://link.aps.org/doi/10.1103/PhysRevB.86.184516}

\bibitem{Zhang2013time}
F.~Zhang, C.~L. Kane, E.~J. Mele,
  \href{http://link.aps.org/doi/10.1103/PhysRevLett.111.056402}{Time-reversal-invariant
  topological superconductivity and majorana kramers pairs}, Phys. Rev. Lett.
  111 (2013) 056402.
\newblock \href {http://dx.doi.org/10.1103/PhysRevLett.111.056402}
  {\path{doi:10.1103/PhysRevLett.111.056402}}.
\newline\urlprefix\url{http://link.aps.org/doi/10.1103/PhysRevLett.111.056402}

\bibitem{chung2013time}
S.~B. Chung, J.~Horowitz, X.-L. Qi,
  \href{http://link.aps.org/doi/10.1103/PhysRevB.88.214514}{Time-reversal
  anomaly and josephson effect in time-reversal-invariant topological
  superconductors}, Phys. Rev. B 88 (2013) 214514.
\newblock \href {http://dx.doi.org/10.1103/PhysRevB.88.214514}
  {\path{doi:10.1103/PhysRevB.88.214514}}.
\newline\urlprefix\url{http://link.aps.org/doi/10.1103/PhysRevB.88.214514}

\bibitem{Nakosai2013majorana}
S.~Nakosai, J.~C. Budich, Y.~Tanaka, B.~Trauzettel, N.~Nagaosa,
  \href{http://link.aps.org/doi/10.1103/PhysRevLett.110.117002}{Majorana bound
  states and nonlocal spin correlations in a quantum wire on an unconventional
  superconductor}, Phys. Rev. Lett. 110 (2013) 117002.
\newblock \href {http://dx.doi.org/10.1103/PhysRevLett.110.117002}
  {\path{doi:10.1103/PhysRevLett.110.117002}}.
\newline\urlprefix\url{http://link.aps.org/doi/10.1103/PhysRevLett.110.117002}

\bibitem{Keselman2013inducing}
A.~Keselman, L.~Fu, A.~Stern, E.~Berg,
  \href{http://link.aps.org/doi/10.1103/PhysRevLett.111.116402}{Inducing
  time-reversal-invariant topological superconductivity and fermion parity
  pumping in quantum wires}, Phys. Rev. Lett. 111 (2013) 116402.
\newblock \href {http://dx.doi.org/10.1103/PhysRevLett.111.116402}
  {\path{doi:10.1103/PhysRevLett.111.116402}}.
\newline\urlprefix\url{http://link.aps.org/doi/10.1103/PhysRevLett.111.116402}

\bibitem{Nakosai2014theoretical}
S.~Nakosai, Y.~Tanaka, N.~Nagaosa,
  \href{https://www.sciencedirect.com/science/article/pii/S1386947713002816}{Theoretical
  modeling and properties of class diii topological superconductors}, Physica
  E: Low-dimensional Systems and Nanostructures 55 (2014) 37--41.
\newline\urlprefix\url{https://www.sciencedirect.com/science/article/pii/S1386947713002816}

\bibitem{Haim2016NoGo}
A.~Haim, E.~Berg, K.~Flensberg, Y.~Oreg,
  \href{https://link.aps.org/doi/10.1103/PhysRevB.94.161110}{No-go theorem for
  a time-reversal invariant topological phase in noninteracting systems coupled
  to conventional superconductors}, Phys. Rev. B 94 (2016) 161110.
\newblock \href {http://dx.doi.org/10.1103/PhysRevB.94.161110}
  {\path{doi:10.1103/PhysRevB.94.161110}}.
\newline\urlprefix\url{https://link.aps.org/doi/10.1103/PhysRevB.94.161110}

\bibitem{budich2013topological}
J.~C. Budich, E.~Ardonne,
  \href{http://link.aps.org/doi/10.1103/PhysRevB.88.134523}{Topological
  invariant for generic one-dimensional time-reversal-symmetric superconductors
  in class diii}, Phys. Rev. B 88 (2013) 134523.
\newblock \href {http://dx.doi.org/10.1103/PhysRevB.88.134523}
  {\path{doi:10.1103/PhysRevB.88.134523}}.
\newline\urlprefix\url{http://link.aps.org/doi/10.1103/PhysRevB.88.134523}

\bibitem{Gaidamauskas2014majorana}
E.~Gaidamauskas, J.~Paaske, K.~Flensberg,
  \href{http://link.aps.org/doi/10.1103/PhysRevLett.112.126402}{Majorana bound
  states in two-channel time-reversal-symmetric nanowire systems}, Phys. Rev.
  Lett. 112 (2014) 126402.
\newblock \href {http://dx.doi.org/10.1103/PhysRevLett.112.126402}
  {\path{doi:10.1103/PhysRevLett.112.126402}}.
\newline\urlprefix\url{http://link.aps.org/doi/10.1103/PhysRevLett.112.126402}

\bibitem{Mandal2015counting}
I.~Mandal, \href{http://arxiv.org/abs/1503.06804}{Counting majorana bound
  states using complex momenta}, arXiv preprint arXiv:1503.06804.
\newline\urlprefix\url{http://arxiv.org/abs/1503.06804}

\bibitem{Ringel2012strong}
Z.~Ringel, Y.~E. Kraus, A.~Stern,
  \href{https://link.aps.org/doi/10.1103/PhysRevB.86.045102}{Strong side of
  weak topological insulators}, Phys. Rev. B 86 (2012) 045102.
\newblock \href {http://dx.doi.org/10.1103/PhysRevB.86.045102}
  {\path{doi:10.1103/PhysRevB.86.045102}}.
\newline\urlprefix\url{https://link.aps.org/doi/10.1103/PhysRevB.86.045102}

\bibitem{Fu2012Topology}
L.~Fu, C.~L. Kane,
  \href{https://link.aps.org/doi/10.1103/PhysRevLett.109.246605}{Topology,
  delocalization via average symmetry and the symplectic anderson transition},
  Phys. Rev. Lett. 109 (2012) 246605.
\newblock \href {http://dx.doi.org/10.1103/PhysRevLett.109.246605}
  {\path{doi:10.1103/PhysRevLett.109.246605}}.
\newline\urlprefix\url{https://link.aps.org/doi/10.1103/PhysRevLett.109.246605}

\bibitem{Fulga2014statistical}
I.~C. Fulga, B.~van Heck, J.~M. Edge, A.~R. Akhmerov,
  \href{https://link.aps.org/doi/10.1103/PhysRevB.89.155424}{Statistical
  topological insulators}, Phys. Rev. B 89 (2014) 155424.
\newblock \href {http://dx.doi.org/10.1103/PhysRevB.89.155424}
  {\path{doi:10.1103/PhysRevB.89.155424}}.
\newline\urlprefix\url{https://link.aps.org/doi/10.1103/PhysRevB.89.155424}

\bibitem{Dahlhaus2010}
J.~P. Dahlhaus, B.~B\'eri, C.~W.~J. Beenakker,
  \href{http://link.aps.org/doi/10.1103/PhysRevB.82.014536}{Random-matrix
  theory of thermal conduction in superconducting quantum dots}, Phys. Rev. B
  82 (2010) 014536.
\newblock \href {http://dx.doi.org/10.1103/PhysRevB.82.014536}
  {\path{doi:10.1103/PhysRevB.82.014536}}.
\newline\urlprefix\url{http://link.aps.org/doi/10.1103/PhysRevB.82.014536}

\bibitem{Seradjeh2012majorana}
B.~Seradjeh, \href{http://link.aps.org/doi/10.1103/PhysRevB.86.121101}{Majorana
  edge modes of topological exciton condensate with superconductors}, Phys.
  Rev. B 86 (2012) 121101.
\newblock \href {http://dx.doi.org/10.1103/PhysRevB.86.121101}
  {\path{doi:10.1103/PhysRevB.86.121101}}.
\newline\urlprefix\url{http://link.aps.org/doi/10.1103/PhysRevB.86.121101}

\bibitem{Haim2014time}
A.~Haim, A.~Keselman, E.~Berg, Y.~Oreg,
  \href{http://link.aps.org/doi/10.1103/PhysRevB.89.220504}{Time-reversal-invariant
  topological superconductivity induced by repulsive interactions in quantum
  wires}, Phys. Rev. B 89 (2014) 220504.
\newblock \href {http://dx.doi.org/10.1103/PhysRevB.89.220504}
  {\path{doi:10.1103/PhysRevB.89.220504}}.
\newline\urlprefix\url{http://link.aps.org/doi/10.1103/PhysRevB.89.220504}

\bibitem{Klinovaja2014Kramers}
J.~Klinovaja, A.~Yacoby, D.~Loss,
  \href{http://link.aps.org/doi/10.1103/PhysRevB.90.155447}{Kramers pairs of
  majorana fermions and parafermions in fractional topological insulators},
  Phys. Rev. B 90 (2014) 155447.
\newblock \href {http://dx.doi.org/10.1103/PhysRevB.90.155447}
  {\path{doi:10.1103/PhysRevB.90.155447}}.
\newline\urlprefix\url{http://link.aps.org/doi/10.1103/PhysRevB.90.155447}

\bibitem{Klinovaja2014time}
J.~Klinovaja, D.~Loss,
  \href{http://link.aps.org/doi/10.1103/PhysRevB.90.045118}{Time-reversal
  invariant parafermions in interacting rashba nanowires}, Phys. Rev. B 90
  (2014) 045118.
\newblock \href {http://dx.doi.org/10.1103/PhysRevB.90.045118}
  {\path{doi:10.1103/PhysRevB.90.045118}}.
\newline\urlprefix\url{http://link.aps.org/doi/10.1103/PhysRevB.90.045118}

\bibitem{Danon2015interaction}
J.~Danon, K.~Flensberg,
  \href{http://link.aps.org/doi/10.1103/PhysRevB.91.165425}{Interaction effects
  on proximity-induced superconductivity in semiconducting nanowires}, Phys.
  Rev. B 91 (2015) 165425.
\newblock \href {http://dx.doi.org/10.1103/PhysRevB.91.165425}
  {\path{doi:10.1103/PhysRevB.91.165425}}.
\newline\urlprefix\url{http://link.aps.org/doi/10.1103/PhysRevB.91.165425}

\bibitem{Thakurathi2018majorana}
M.~Thakurathi, P.~Simon, I.~Mandal, J.~Klinovaja, D.~Loss,
  \href{https://link.aps.org/doi/10.1103/PhysRevB.97.045415}{Majorana kramers
  pairs in rashba double nanowires with interactions and disorder}, Phys. Rev.
  B 97 (2018) 045415.
\newblock \href {http://dx.doi.org/10.1103/PhysRevB.97.045415}
  {\path{doi:10.1103/PhysRevB.97.045415}}.
\newline\urlprefix\url{https://link.aps.org/doi/10.1103/PhysRevB.97.045415}

\bibitem{Nakosai2012topological}
S.~Nakosai, Y.~Tanaka, N.~Nagaosa,
  \href{http://link.aps.org/doi/10.1103/PhysRevLett.108.147003}{Topological
  superconductivity in bilayer rashba system}, Phys. Rev. Lett. 108 (2012)
  147003.
\newblock \href {http://dx.doi.org/10.1103/PhysRevLett.108.147003}
  {\path{doi:10.1103/PhysRevLett.108.147003}}.
\newline\urlprefix\url{http://link.aps.org/doi/10.1103/PhysRevLett.108.147003}

\bibitem{Chen2018helical}
Y.~Chen, H.-Y. Kee,
  \href{https://link.aps.org/doi/10.1103/PhysRevB.97.085155}{Helical majorana
  fermions and flat edge states in the heterostructures of iridates and
  high-${\mathrm{t}}_{C}$ cuprates}, Phys. Rev. B 97 (2018) 085155.
\newblock \href {http://dx.doi.org/10.1103/PhysRevB.97.085155}
  {\path{doi:10.1103/PhysRevB.97.085155}}.
\newline\urlprefix\url{https://link.aps.org/doi/10.1103/PhysRevB.97.085155}

\bibitem{Kotetes2015topological}
P.~Kotetes,
  \href{http://link.aps.org/doi/10.1103/PhysRevB.92.014514}{Topological
  superconductivity in rashba semiconductors without a zeeman field}, Phys.
  Rev. B 92 (2015) 014514.
\newblock \href {http://dx.doi.org/10.1103/PhysRevB.92.014514}
  {\path{doi:10.1103/PhysRevB.92.014514}}.
\newline\urlprefix\url{http://link.aps.org/doi/10.1103/PhysRevB.92.014514}

\bibitem{Wang2016Electrically}
J.~Wang,
  \href{https://link.aps.org/doi/10.1103/PhysRevB.94.214502}{Electrically
  tunable topological superconductivity and majorana fermions in two
  dimensions}, Phys. Rev. B 94 (2016) 214502.
\newblock \href {http://dx.doi.org/10.1103/PhysRevB.94.214502}
  {\path{doi:10.1103/PhysRevB.94.214502}}.
\newline\urlprefix\url{https://link.aps.org/doi/10.1103/PhysRevB.94.214502}

\bibitem{Parhizgar2017highly}
F.~Parhizgar, A.~M. Black-Schaffer,
  \href{https://www.nature.com/articles/s41598-017-10510-y}{Highly tunable
  time-reversal-invariant topological superconductivity in topological
  insulator thin films}, Scientific reports 7~(1) (2017) 9817.
\newline\urlprefix\url{https://www.nature.com/articles/s41598-017-10510-y}

\bibitem{Parhizgar2014unconventional}
F.~Parhizgar, A.~M. Black-Schaffer,
  \href{https://link.aps.org/doi/10.1103/PhysRevB.90.184517}{Unconventional
  proximity-induced superconductivity in bilayer systems}, Phys. Rev. B 90
  (2014) 184517.
\newblock \href {http://dx.doi.org/10.1103/PhysRevB.90.184517}
  {\path{doi:10.1103/PhysRevB.90.184517}}.
\newline\urlprefix\url{https://link.aps.org/doi/10.1103/PhysRevB.90.184517}

\bibitem{Deng2012majorana}
S.~Deng, L.~Viola, G.~Ortiz,
  \href{http://link.aps.org/doi/10.1103/PhysRevLett.108.036803}{Majorana modes
  in time-reversal invariant $s$-wave topological superconductors}, Phys. Rev.
  Lett. 108 (2012) 036803.
\newblock \href {http://dx.doi.org/10.1103/PhysRevLett.108.036803}
  {\path{doi:10.1103/PhysRevLett.108.036803}}.
\newline\urlprefix\url{http://link.aps.org/doi/10.1103/PhysRevLett.108.036803}

\bibitem{Deng2013Multiband}
S.~Deng, G.~Ortiz, L.~Viola,
  \href{https://link.aps.org/doi/10.1103/PhysRevB.87.205414}{Multiband $s$-wave
  topological superconductors: Role of dimensionality and magnetic field
  response}, Phys. Rev. B 87 (2013) 205414.
\newblock \href {http://dx.doi.org/10.1103/PhysRevB.87.205414}
  {\path{doi:10.1103/PhysRevB.87.205414}}.
\newline\urlprefix\url{https://link.aps.org/doi/10.1103/PhysRevB.87.205414}

\bibitem{Kwon2004fractional}
H.-J. Kwon, K.~Sengupta, V.~M. Yakovenko, Fractional ac josephson effect in
  p-and d-wave superconductors, The European Physical Journal B-Condensed
  Matter and Complex Systems 37~(3) (2004) 349--361.

\bibitem{Haim2016interaction}
A.~Haim, K.~W\"olms, E.~Berg, Y.~Oreg, K.~Flensberg,
  \href{https://link.aps.org/doi/10.1103/PhysRevB.94.115124}{Interaction-driven
  topological superconductivity in one dimension}, Phys. Rev. B 94 (2016)
  115124.
\newblock \href {http://dx.doi.org/10.1103/PhysRevB.94.115124}
  {\path{doi:10.1103/PhysRevB.94.115124}}.
\newline\urlprefix\url{https://link.aps.org/doi/10.1103/PhysRevB.94.115124}

\bibitem{Moroz1999Effect}
A.~V. Moroz, C.~H.~W. Barnes,
  \href{http://link.aps.org/doi/10.1103/PhysRevB.60.14272}{Effect of the
  spin-orbit interaction on the band structure and conductance of
  quasi-one-dimensional systems}, Phys. Rev. B 60 (1999) 14272--14285.
\newblock \href {http://dx.doi.org/10.1103/PhysRevB.60.14272}
  {\path{doi:10.1103/PhysRevB.60.14272}}.
\newline\urlprefix\url{http://link.aps.org/doi/10.1103/PhysRevB.60.14272}

\bibitem{DeGennes1964Boundary}
P.~G. De~Gennes,
  \href{http://link.aps.org/doi/10.1103/RevModPhys.36.225}{Boundary effects in
  superconductors}, Rev. Mod. Phys. 36 (1964) 225--237.
\newblock \href {http://dx.doi.org/10.1103/RevModPhys.36.225}
  {\path{doi:10.1103/RevModPhys.36.225}}.
\newline\urlprefix\url{http://link.aps.org/doi/10.1103/RevModPhys.36.225}

\bibitem{Reeg2017destructive}
C.~Reeg, J.~Klinovaja, D.~Loss,
  \href{https://link.aps.org/doi/10.1103/PhysRevB.96.081301}{Destructive
  interference of direct and crossed andreev pairing in a system of two
  nanowires coupled via an $s$-wave superconductor}, Phys. Rev. B 96 (2017)
  081301.
\newblock \href {http://dx.doi.org/10.1103/PhysRevB.96.081301}
  {\path{doi:10.1103/PhysRevB.96.081301}}.
\newline\urlprefix\url{https://link.aps.org/doi/10.1103/PhysRevB.96.081301}

\bibitem{Ebisu2016theory}
H.~Ebisu, B.~Lu, J.~Klinovaja, Y.~Tanaka,
  \href{http://dx.doi.org/10.1093/ptep/ptw094}{Theory of time-reversal
  topological superconductivity in double rashba wires: symmetries of cooper
  pairs and andreev bound states}, Progress of Theoretical and Experimental
  Physics 2016~(8) (2016) 083I01.
\newblock \href
  {http://arxiv.org/abs//oup/backfile/content_public/journal/ptep/2016/8/10.1093_ptep_ptw094/3/ptw094.pdf}
  {\path{arXiv:/oup/backfile/content_public/journal/ptep/2016/8/10.1093_ptep_ptw094/3/ptw094.pdf}},
  \href {http://dx.doi.org/10.1093/ptep/ptw094}
  {\path{doi:10.1093/ptep/ptw094}}.
\newline\urlprefix\url{http://dx.doi.org/10.1093/ptep/ptw094}

\bibitem{Wang2014two}
J.~Wang, Y.~Xu, S.-C. Zhang,
  \href{http://link.aps.org/doi/10.1103/PhysRevB.90.054503}{Two-dimensional
  time-reversal-invariant topological superconductivity in a doped quantum
  spin-hall insulator}, Phys. Rev. B 90 (2014) 054503.
\newblock \href {http://dx.doi.org/10.1103/PhysRevB.90.054503}
  {\path{doi:10.1103/PhysRevB.90.054503}}.
\newline\urlprefix\url{http://link.aps.org/doi/10.1103/PhysRevB.90.054503}

\bibitem{Yang2015time}
F.~Yang, C.-C. Liu, Y.-Z. Zhang, Y.~Yao, D.-H. Lee,
  \href{https://link.aps.org/doi/10.1103/PhysRevB.91.134514}{Time-reversal-invariant
  topological superconductivity in $n$-doped bih}, Phys. Rev. B 91 (2015)
  134514.
\newblock \href {http://dx.doi.org/10.1103/PhysRevB.91.134514}
  {\path{doi:10.1103/PhysRevB.91.134514}}.
\newline\urlprefix\url{https://link.aps.org/doi/10.1103/PhysRevB.91.134514}

\bibitem{Shankar1994Renormalization}
R.~Shankar,
  \href{http://link.aps.org/doi/10.1103/RevModPhys.66.129}{Renormalization-group
  approach to interacting fermions}, Rev. Mod. Phys. 66 (1994) 129--192.
\newblock \href {http://dx.doi.org/10.1103/RevModPhys.66.129}
  {\path{doi:10.1103/RevModPhys.66.129}}.
\newline\urlprefix\url{http://link.aps.org/doi/10.1103/RevModPhys.66.129}

\bibitem{Yu2016Gapped}
T.~Yu, M.~W. Wu,
  \href{https://link.aps.org/doi/10.1103/PhysRevB.93.195308}{Gapped triplet
  $p$-wave superconductivity in strong spin-orbit-coupled semiconductor quantum
  wells in proximity to $s$-wave superconductor}, Phys. Rev. B 93 (2016)
  195308.
\newblock \href {http://dx.doi.org/10.1103/PhysRevB.93.195308}
  {\path{doi:10.1103/PhysRevB.93.195308}}.
\newline\urlprefix\url{https://link.aps.org/doi/10.1103/PhysRevB.93.195308}

\bibitem{Sau2010robustness}
J.~D. Sau, R.~M. Lutchyn, S.~Tewari, S.~Das~Sarma,
  \href{http://link.aps.org/doi/10.1103/PhysRevB.82.094522}{Robustness of
  majorana fermions in proximity-induced superconductors}, Phys. Rev. B 82
  (2010) 094522.
\newblock \href {http://dx.doi.org/10.1103/PhysRevB.82.094522}
  {\path{doi:10.1103/PhysRevB.82.094522}}.
\newline\urlprefix\url{http://link.aps.org/doi/10.1103/PhysRevB.82.094522}

\bibitem{Bolech2007Observing}
C.~J. Bolech, E.~Demler,
  \href{http://link.aps.org/doi/10.1103/PhysRevLett.98.237002}{Observing
  majorana bound states in $p$-wave superconductors using noise measurements in
  tunneling experiments}, Phys. Rev. Lett. 98 (2007) 237002.
\newblock \href {http://dx.doi.org/10.1103/PhysRevLett.98.237002}
  {\path{doi:10.1103/PhysRevLett.98.237002}}.
\newline\urlprefix\url{http://link.aps.org/doi/10.1103/PhysRevLett.98.237002}

\bibitem{Law2009majorana}
K.~T. Law, P.~A. Lee, T.~K. Ng,
  \href{http://link.aps.org/doi/10.1103/PhysRevLett.103.237001}{Majorana
  fermion induced resonant andreev reflection}, Phys. Rev. Lett. 103 (2009)
  237001.
\newblock \href {http://dx.doi.org/10.1103/PhysRevLett.103.237001}
  {\path{doi:10.1103/PhysRevLett.103.237001}}.
\newline\urlprefix\url{http://link.aps.org/doi/10.1103/PhysRevLett.103.237001}

\bibitem{Fidkowski2012universal}
L.~Fidkowski, J.~Alicea, N.~H. Lindner, R.~M. Lutchyn, M.~P.~A. Fisher,
  \href{http://link.aps.org/doi/10.1103/PhysRevB.85.245121}{Universal transport
  signatures of majorana fermions in superconductor-luttinger liquid
  junctions}, Phys. Rev. B 85 (2012) 245121.
\newblock \href {http://dx.doi.org/10.1103/PhysRevB.85.245121}
  {\path{doi:10.1103/PhysRevB.85.245121}}.
\newline\urlprefix\url{http://link.aps.org/doi/10.1103/PhysRevB.85.245121}

\bibitem{flensberg2010tunneling}
K.~Flensberg,
  \href{http://link.aps.org/doi/10.1103/PhysRevB.82.180516}{Tunneling
  characteristics of a chain of majorana bound states}, Phys. Rev. B 82 (2010)
  180516.
\newblock \href {http://dx.doi.org/10.1103/PhysRevB.82.180516}
  {\path{doi:10.1103/PhysRevB.82.180516}}.
\newline\urlprefix\url{http://link.aps.org/doi/10.1103/PhysRevB.82.180516}

\bibitem{Pikulin2016Luttinger}
D.~I. Pikulin, Y.~Komijani, I.~Affleck,
  \href{https://link.aps.org/doi/10.1103/PhysRevB.93.205430}{Luttinger liquid
  in contact with a kramers pair of majorana bound states}, Phys. Rev. B 93
  (2016) 205430.
\newblock \href {http://dx.doi.org/10.1103/PhysRevB.93.205430}
  {\path{doi:10.1103/PhysRevB.93.205430}}.
\newline\urlprefix\url{https://link.aps.org/doi/10.1103/PhysRevB.93.205430}

\bibitem{Li2016detection}
J.~Li, W.~Pan, B.~A. Bernevig, R.~M. Lutchyn,
  \href{http://link.aps.org/doi/10.1103/PhysRevLett.117.046804}{Detection of
  majorana kramers pairs using a quantum point contact}, Phys. Rev. Lett. 117
  (2016) 046804.
\newblock \href {http://dx.doi.org/10.1103/PhysRevLett.117.046804}
  {\path{doi:10.1103/PhysRevLett.117.046804}}.
\newline\urlprefix\url{http://link.aps.org/doi/10.1103/PhysRevLett.117.046804}

\bibitem{shelankov1980resistance}
A.~Shelankov,
  \href{http://www.jetpletters.ac.ru/ps/1424/article_21649.shtml}{Resistance of
  n/s interface at low temperatures}, JETP Lett., 32~(2) (1980) 111--114.
\newline\urlprefix\url{http://www.jetpletters.ac.ru/ps/1424/article_21649.shtml}

\bibitem{blonder1982transition}
G.~E. Blonder, M.~Tinkham, T.~M. Klapwijk,
  \href{http://link.aps.org/doi/10.1103/PhysRevB.25.4515}{Transition from
  metallic to tunneling regimes in superconducting microconstrictions: Excess
  current, charge imbalance, and supercurrent conversion}, Phys. Rev. B 25
  (1982) 4515--4532.
\newblock \href {http://dx.doi.org/10.1103/PhysRevB.25.4515}
  {\path{doi:10.1103/PhysRevB.25.4515}}.
\newline\urlprefix\url{http://link.aps.org/doi/10.1103/PhysRevB.25.4515}

\bibitem{lesovik2011scattering}
G.~B. Lesovik, I.~A. Sadovskyy, Scattering matrix approach to the description
  of quantum electron transport, Physics-Uspekhi 54~(10) (2011) 1007--1059.

\bibitem{Golub2011shot}
A.~Golub, B.~Horovitz,
  \href{http://link.aps.org/doi/10.1103/PhysRevB.83.153415}{Shot noise in a
  majorana fermion chain}, Phys. Rev. B 83 (2011) 153415.
\newblock \href {http://dx.doi.org/10.1103/PhysRevB.83.153415}
  {\path{doi:10.1103/PhysRevB.83.153415}}.
\newline\urlprefix\url{http://link.aps.org/doi/10.1103/PhysRevB.83.153415}

\bibitem{Haim2015current}
A.~Haim, E.~Berg, F.~von Oppen, Y.~Oreg,
  \href{http://link.aps.org/doi/10.1103/PhysRevB.92.245112}{Current
  correlations in a majorana beam splitter}, Phys. Rev. B 92 (2015) 245112.
\newblock \href {http://dx.doi.org/10.1103/PhysRevB.92.245112}
  {\path{doi:10.1103/PhysRevB.92.245112}}.
\newline\urlprefix\url{http://link.aps.org/doi/10.1103/PhysRevB.92.245112}

\bibitem{Haim2015signatures}
A.~Haim, E.~Berg, F.~von Oppen, Y.~Oreg,
  \href{http://link.aps.org/doi/10.1103/PhysRevLett.114.166406}{Signatures of
  majorana zero modes in spin-resolved current correlations}, Phys. Rev. Lett.
  114 (2015) 166406.
\newblock \href {http://dx.doi.org/10.1103/PhysRevLett.114.166406}
  {\path{doi:10.1103/PhysRevLett.114.166406}}.
\newline\urlprefix\url{http://link.aps.org/doi/10.1103/PhysRevLett.114.166406}

\bibitem{Aasen2016Interaction}
D.~Aasen, S.-P. Lee, T.~Karzig, J.~Alicea,
  \href{https://link.aps.org/doi/10.1103/PhysRevB.94.165113}{Interaction
  effects in superconductor/quantum spin hall devices: Universal transport
  signatures and fractional coulomb blockade}, Phys. Rev. B 94 (2016) 165113.
\newblock \href {http://dx.doi.org/10.1103/PhysRevB.94.165113}
  {\path{doi:10.1103/PhysRevB.94.165113}}.
\newline\urlprefix\url{https://link.aps.org/doi/10.1103/PhysRevB.94.165113}

\bibitem{Dumitrescu2014magnetic}
E.~Dumitrescu, J.~D. Sau, S.~Tewari,
  \href{http://link.aps.org/doi/10.1103/PhysRevB.90.245438}{Magnetic field
  response and chiral symmetry of time-reversal-invariant topological
  superconductors}, Phys. Rev. B 90 (2014) 245438.
\newblock \href {http://dx.doi.org/10.1103/PhysRevB.90.245438}
  {\path{doi:10.1103/PhysRevB.90.245438}}.
\newline\urlprefix\url{http://link.aps.org/doi/10.1103/PhysRevB.90.245438}

\bibitem{Aligia2018entangled}
A.~Aligia, L.~Arrachea, \href{https://arxiv.org/abs/1806.06104}{Entangled end
  states with fractionalized spin projection in a time-reversal-invariant
  topological superconducting wire}, arXiv preprint arXiv:1806.06104.
\newline\urlprefix\url{https://arxiv.org/abs/1806.06104}

\bibitem{tewari2012topological}
S.~Tewari, J.~D. Sau,
  \href{http://link.aps.org/doi/10.1103/PhysRevLett.109.150408}{Topological
  invariants for spin-orbit coupled superconductor nanowires}, Phys. Rev. Lett.
  109~(15) (2012) 150408.
\newline\urlprefix\url{http://link.aps.org/doi/10.1103/PhysRevLett.109.150408}

\bibitem{liu2013non}
X.-J. Liu, C.~L. Wong, K.~Law,
  \href{http://arxiv.org/abs/1304.3765}{Non-abelian majorana doublets in
  time-reversal invariant topological superconductor}, arXiv preprint
  arXiv:1304.3765.
\newline\urlprefix\url{http://arxiv.org/abs/1304.3765}

\bibitem{Zhang2014anomalous}
F.~Zhang, C.~L. Kane,
  \href{http://link.aps.org/doi/10.1103/PhysRevB.90.020501}{Anomalous
  topological pumps and fractional josephson effects}, Phys. Rev. B 90 (2014)
  020501.
\newblock \href {http://dx.doi.org/10.1103/PhysRevB.90.020501}
  {\path{doi:10.1103/PhysRevB.90.020501}}.
\newline\urlprefix\url{http://link.aps.org/doi/10.1103/PhysRevB.90.020501}

\bibitem{Kane2015the}
C.~L. Kane, F.~Zhang,
  \href{http://stacks.iop.org/1402-4896/2015/i=T164/a=014011}{The time reversal
  invariant fractional josephson effect}, Physica Scrip. 2015~(T164) (2015)
  014011.
\newline\urlprefix\url{http://stacks.iop.org/1402-4896/2015/i=T164/a=014011}

\bibitem{gong2016influence}
W.-J. Gong, Z.~Gao, W.-F. Shan, G.-Y. Yi, Influence of an embedded quantum dot
  on the josephson effect in the topological superconducting junction with
  majorana doublets, Scientific reports 6.

\bibitem{Mellars2016signatures}
E.~Mellars, B.~B\'eri,
  \href{https://link.aps.org/doi/10.1103/PhysRevB.94.174508}{Signatures of
  time-reversal-invariant topological superconductivity in the josephson
  effect}, Phys. Rev. B 94 (2016) 174508.
\newblock \href {http://dx.doi.org/10.1103/PhysRevB.94.174508}
  {\path{doi:10.1103/PhysRevB.94.174508}}.
\newline\urlprefix\url{https://link.aps.org/doi/10.1103/PhysRevB.94.174508}

\bibitem{Camjayi2017fractional}
A.~Camjayi, L.~Arrachea, A.~Aligia, F.~von Oppen,
  \href{https://link.aps.org/doi/10.1103/PhysRevLett.119.046801}{Fractional
  spin and josephson effect in time-reversal-invariant topological
  superconductors}, Phys. Rev. Lett. 119 (2017) 046801.
\newblock \href {http://dx.doi.org/10.1103/PhysRevLett.119.046801}
  {\path{doi:10.1103/PhysRevLett.119.046801}}.
\newline\urlprefix\url{https://link.aps.org/doi/10.1103/PhysRevLett.119.046801}

\bibitem{gao2015josephson}
Z.~Gao, W.-F. Shan, W.-J. Gong, Josephson effects in the junction formed
  by$\backslash$emph $\{$DIII$\}$-class topological and $ s $-wave
  superconductors with an embedded quantum dot, arXiv preprint
  arXiv:1506.05188.

\bibitem{JIANG2017Fano}
C.~Jiang, G.-Y. Yi, G.-Y. Meng, W.-J. Gong,
  \href{http://www.sciencedirect.com/science/article/pii/S0749603616317074}{Fano-josephson
  effect in the junction with diii-class topological and s-wave
  superconductors}, Superlattices and Microstructures 104~(Supplement C) (2017)
  382 -- 389.
\newblock \href {http://dx.doi.org/https://doi.org/10.1016/j.spmi.2017.02.047}
  {\path{doi:https://doi.org/10.1016/j.spmi.2017.02.047}}.
\newline\urlprefix\url{http://www.sciencedirect.com/science/article/pii/S0749603616317074}

\bibitem{Beenakker1991universal}
C.~W.~J. Beenakker,
  \href{http://link.aps.org/doi/10.1103/PhysRevLett.67.3836}{Universal limit of
  critical-current fluctuations in mesoscopic josephson junctions}, Phys. Rev.
  Lett. 67 (1991) 3836--3839.
\newblock \href {http://dx.doi.org/10.1103/PhysRevLett.67.3836}
  {\path{doi:10.1103/PhysRevLett.67.3836}}.
\newline\urlprefix\url{http://link.aps.org/doi/10.1103/PhysRevLett.67.3836}

\bibitem{Franceschi2010hybrid}
S.~De~Franceschi, L.~Kouwenhoven, C.~Sch{\"o}nenberger, W.~Wernsdorfer, Hybrid
  superconductor-quantum dot devices, Nature Nanotechnology 5~(10) (2010)
  703--711.

\bibitem{Rodero2011Josephson}
A.~Martín-Rodero, A.~L. Yeyati,
  \href{https://doi.org/10.1080/00018732.2011.624266}{Josephson and andreev
  transport through quantum dots}, Advances in Physics 60~(6) (2011) 899--958.
\newblock \href
  {http://arxiv.org/abs/https://doi.org/10.1080/00018732.2011.624266}
  {\path{arXiv:https://doi.org/10.1080/00018732.2011.624266}}, \href
  {http://dx.doi.org/10.1080/00018732.2011.624266}
  {\path{doi:10.1080/00018732.2011.624266}}.
\newline\urlprefix\url{https://doi.org/10.1080/00018732.2011.624266}

\bibitem{leijnse2012introduction}
M.~Leijnse, K.~Flensberg, Introduction to topological superconductivity and
  majorana fermions, arXiv preprint arXiv:1206.1736.

\bibitem{Stern2004geometric}
A.~Stern, F.~von Oppen, E.~Mariani,
  \href{http://link.aps.org/doi/10.1103/PhysRevB.70.205338}{Geometric phases
  and quantum entanglement as building blocks for non-abelian quasiparticle
  statistics}, Phys. Rev. B 70~(20) (2004) 205338.
\newline\urlprefix\url{http://link.aps.org/doi/10.1103/PhysRevB.70.205338}

\bibitem{Alicea2011non}
J.~Alicea, Y.~Oreg, G.~Refael, F.~von Oppen, M.~P. Fisher,
  \href{http://www.nature.com/nphys/journal/v7/n5/abs/nphys1915.html}{Non-abelian
  statistics and topological quantum information processing in 1d wire
  networks}, Nat. Phys. 7~(5) (2011) 412--417.
\newline\urlprefix\url{http://www.nature.com/nphys/journal/v7/n5/abs/nphys1915.html}

\bibitem{Sau2011controlling}
J.~D. Sau, D.~J. Clarke, S.~Tewari,
  \href{http://link.aps.org/doi/10.1103/PhysRevB.84.094505}{Controlling
  non-abelian statistics of majorana fermions in semiconductor nanowires},
  Phys. Rev. B 84 (2011) 094505.
\newblock \href {http://dx.doi.org/10.1103/PhysRevB.84.094505}
  {\path{doi:10.1103/PhysRevB.84.094505}}.
\newline\urlprefix\url{http://link.aps.org/doi/10.1103/PhysRevB.84.094505}

\bibitem{Wolms2014local}
K.~W\"olms, A.~Stern, K.~Flensberg,
  \href{http://link.aps.org/doi/10.1103/PhysRevLett.113.246401}{Local adiabatic
  mixing of kramers pairs of majorana bound states}, Phys. Rev. Lett. 113
  (2014) 246401.
\newblock \href {http://dx.doi.org/10.1103/PhysRevLett.113.246401}
  {\path{doi:10.1103/PhysRevLett.113.246401}}.
\newline\urlprefix\url{http://link.aps.org/doi/10.1103/PhysRevLett.113.246401}

\bibitem{Wolms2015braiding}
K.~W{\"o}lms, A.~Stern, K.~Flensberg,
  \href{http://arxiv.org/abs/1507.02881}{Braiding properties of majorana
  kramers pairs}, arXiv preprint arXiv:1507.02881.
\newline\urlprefix\url{http://arxiv.org/abs/1507.02881}

\bibitem{Liu2014non}
X.-J. Liu, C.~L.~M. Wong, K.~T. Law,
  \href{http://link.aps.org/doi/10.1103/PhysRevX.4.021018}{Non-abelian majorana
  doublets in time-reversal-invariant topological superconductors}, Phys. Rev.
  X 4 (2014) 021018.
\newblock \href {http://dx.doi.org/10.1103/PhysRevX.4.021018}
  {\path{doi:10.1103/PhysRevX.4.021018}}.
\newline\urlprefix\url{http://link.aps.org/doi/10.1103/PhysRevX.4.021018}

\bibitem{Wilczek1984apperance}
F.~Wilczek, A.~Zee,
  \href{https://link.aps.org/doi/10.1103/PhysRevLett.52.2111}{Appearance of
  gauge structure in simple dynamical systems}, Phys. Rev. Lett. 52 (1984)
  2111--2114.
\newblock \href {http://dx.doi.org/10.1103/PhysRevLett.52.2111}
  {\path{doi:10.1103/PhysRevLett.52.2111}}.
\newline\urlprefix\url{https://link.aps.org/doi/10.1103/PhysRevLett.52.2111}

\bibitem{berry1984quantal}
M.~V. Berry,
  \href{http://rspa.royalsocietypublishing.org/content/392/1802/45}{Quantal
  phase factors accompanying adiabatic changes}, Proceedings of the Royal
  Society of London A: Mathematical, Physical and Engineering Sciences
  392~(1802) (1984) 45--57.
\newblock \href
  {http://arxiv.org/abs/http://rspa.royalsocietypublishing.org/content/392/1802/45.full.pdf}
  {\path{arXiv:http://rspa.royalsocietypublishing.org/content/392/1802/45.full.pdf}},
  \href {http://dx.doi.org/10.1098/rspa.1984.0023}
  {\path{doi:10.1098/rspa.1984.0023}}.
\newline\urlprefix\url{http://rspa.royalsocietypublishing.org/content/392/1802/45}

\bibitem{Gao2017symmetry}
P.~Gao, Y.-P. He, X.-J. Liu,
  \href{https://link.aps.org/doi/10.1103/PhysRevB.94.224509}{Symmetry-protected
  non-abelian braiding of majorana kramers pairs}, Phys. Rev. B 94 (2016)
  224509.
\newblock \href {http://dx.doi.org/10.1103/PhysRevB.94.224509}
  {\path{doi:10.1103/PhysRevB.94.224509}}.
\newline\urlprefix\url{https://link.aps.org/doi/10.1103/PhysRevB.94.224509}

\bibitem{Schrade2018quantum}
C.~Schrade, L.~Fu, \href{https://arxiv.org/abs/1807.06620}{Quantum computing
  with majorana kramers pairs}, arXiv preprint arXiv:1807.06620.
\newline\urlprefix\url{https://arxiv.org/abs/1807.06620}

\bibitem{Bao2017topological}
Z.-q. Bao, F.~Zhang,
  \href{https://link.aps.org/doi/10.1103/PhysRevLett.119.187701}{Topological
  majorana two-channel kondo effect}, Phys. Rev. Lett. 119 (2017) 187701.
\newblock \href {http://dx.doi.org/10.1103/PhysRevLett.119.187701}
  {\path{doi:10.1103/PhysRevLett.119.187701}}.
\newline\urlprefix\url{https://link.aps.org/doi/10.1103/PhysRevLett.119.187701}

\bibitem{Williams2012unconventional}
J.~Williams, A.~Bestwick, P.~Gallagher, S.~Hong, Y.~Cui, A.~Bleich,
  J.~Analytis, I.~Fisher, D.~Goldhaber-Gordon, Unconventional josephson effect
  in hybrid superconductor-topological insulator devices, Phys. Rev. Lett.
  109~(5) (2012) 56803.

\bibitem{Hart2014Induced}
S.~Hart, H.~Ren, T.~Wagner, P.~Leubner, M.~M{\"u}hlbauer, C.~Br{\"u}ne,
  H.~Buhmann, L.~W. Molenkamp, A.~Yacoby,
  \href{http://www.nature.com/nphys/journal/v10/n9/abs/nphys3036.html}{Induced
  superconductivity in the quantum spin hall edge}, Nat. Phys. 10~(9) (2014)
  638--643.
\newline\urlprefix\url{http://www.nature.com/nphys/journal/v10/n9/abs/nphys3036.html}

\bibitem{Sau2012realizing}
J.~Sau, S.~Sarma,
  \href{http://www.nature.com/ncomms/journal/v3/n7/abs/ncomms1966.html}{Realizing
  a robust practical majorana chain in a quantum-dot-superconductor linear
  array}, Nat. Commun. 3 (2012) 964.
\newline\urlprefix\url{http://www.nature.com/ncomms/journal/v3/n7/abs/ncomms1966.html}

\bibitem{Fulga2013adaptive}
I.~C. Fulga, A.~Haim, A.~R. Akhmerov, Y.~Oreg,
  \href{http://stacks.iop.org/1367-2630/15/i=4/a=045020}{Adaptive tuning of
  majorana fermions in a quantum dot chain}, New J. Phys. 15~(4) (2013) 045020.
\newline\urlprefix\url{http://stacks.iop.org/1367-2630/15/i=4/a=045020}

\bibitem{Zhang2016Majorana}
P.~Zhang, F.~Nori,
  \href{http://stacks.iop.org/1367-2630/18/i=4/a=043033}{Majorana bound states
  in a disordered quantum dot chain}, New Journal of Physics 18~(4) (2016)
  043033.
\newline\urlprefix\url{http://stacks.iop.org/1367-2630/18/i=4/a=043033}

\bibitem{Levine2017realizing}
Y.~Levine, A.~Haim, Y.~Oreg,
  \href{https://link.aps.org/doi/10.1103/PhysRevB.96.165147}{Realizing
  topological superconductivity with superlattices}, Phys. Rev. B 96 (2017)
  165147.
\newblock \href {http://dx.doi.org/10.1103/PhysRevB.96.165147}
  {\path{doi:10.1103/PhysRevB.96.165147}}.
\newline\urlprefix\url{https://link.aps.org/doi/10.1103/PhysRevB.96.165147}

\bibitem{Neupert2014wire}
T.~Neupert, C.~Chamon, C.~Mudry, R.~Thomale,
  \href{https://link.aps.org/doi/10.1103/PhysRevB.90.205101}{Wire
  deconstructionism of two-dimensional topological phases}, Phys. Rev. B 90
  (2014) 205101.
\newblock \href {http://dx.doi.org/10.1103/PhysRevB.90.205101}
  {\path{doi:10.1103/PhysRevB.90.205101}}.
\newline\urlprefix\url{https://link.aps.org/doi/10.1103/PhysRevB.90.205101}

\bibitem{Kane2002Fractional}
C.~L. Kane, R.~Mukhopadhyay, T.~C. Lubensky,
  \href{https://link.aps.org/doi/10.1103/PhysRevLett.88.036401}{Fractional
  quantum hall effect in an array of quantum wires}, Phys. Rev. Lett. 88 (2002)
  036401.
\newblock \href {http://dx.doi.org/10.1103/PhysRevLett.88.036401}
  {\path{doi:10.1103/PhysRevLett.88.036401}}.
\newline\urlprefix\url{https://link.aps.org/doi/10.1103/PhysRevLett.88.036401}

\bibitem{Klinovaja2013topologicalWireConstruc}
J.~Klinovaja, D.~Loss,
  \href{https://link.aps.org/doi/10.1103/PhysRevLett.111.196401}{Topological
  edge states and fractional quantum hall effect from umklapp scattering},
  Phys. Rev. Lett. 111 (2013) 196401.
\newblock \href {http://dx.doi.org/10.1103/PhysRevLett.111.196401}
  {\path{doi:10.1103/PhysRevLett.111.196401}}.
\newline\urlprefix\url{https://link.aps.org/doi/10.1103/PhysRevLett.111.196401}

\bibitem{Teo2014From}
J.~C.~Y. Teo, C.~L. Kane,
  \href{https://link.aps.org/doi/10.1103/PhysRevB.89.085101}{From luttinger
  liquid to non-abelian quantum hall states}, Phys. Rev. B 89 (2014) 085101.
\newblock \href {http://dx.doi.org/10.1103/PhysRevB.89.085101}
  {\path{doi:10.1103/PhysRevB.89.085101}}.
\newline\urlprefix\url{https://link.aps.org/doi/10.1103/PhysRevB.89.085101}

\bibitem{Seroussi2014topological}
I.~Seroussi, E.~Berg, Y.~Oreg,
  \href{https://link.aps.org/doi/10.1103/PhysRevB.89.104523}{Topological
  superconducting phases of weakly coupled quantum wires}, Phys. Rev. B 89
  (2014) 104523.
\newblock \href {http://dx.doi.org/10.1103/PhysRevB.89.104523}
  {\path{doi:10.1103/PhysRevB.89.104523}}.
\newline\urlprefix\url{https://link.aps.org/doi/10.1103/PhysRevB.89.104523}

\bibitem{Sagi2014nonabelian}
E.~Sagi, Y.~Oreg,
  \href{https://link.aps.org/doi/10.1103/PhysRevB.90.201102}{Non-abelian
  topological insulators from an array of quantum wires}, Phys. Rev. B 90
  (2014) 201102.
\newblock \href {http://dx.doi.org/10.1103/PhysRevB.90.201102}
  {\path{doi:10.1103/PhysRevB.90.201102}}.
\newline\urlprefix\url{https://link.aps.org/doi/10.1103/PhysRevB.90.201102}

\bibitem{Klinovaja2014Quantum}
J.~Klinovaja, Y.~Tserkovnyak,
  \href{https://link.aps.org/doi/10.1103/PhysRevB.90.115426}{Quantum spin hall
  effect in strip of stripes model}, Phys. Rev. B 90 (2014) 115426.
\newblock \href {http://dx.doi.org/10.1103/PhysRevB.90.115426}
  {\path{doi:10.1103/PhysRevB.90.115426}}.
\newline\urlprefix\url{https://link.aps.org/doi/10.1103/PhysRevB.90.115426}

\bibitem{Meng2014Time}
T.~Meng, E.~Sela,
  \href{https://link.aps.org/doi/10.1103/PhysRevB.90.235425}{Time reversal
  symmetry broken fractional topological phases at zero magnetic field}, Phys.
  Rev. B 90 (2014) 235425.
\newblock \href {http://dx.doi.org/10.1103/PhysRevB.90.235425}
  {\path{doi:10.1103/PhysRevB.90.235425}}.
\newline\urlprefix\url{https://link.aps.org/doi/10.1103/PhysRevB.90.235425}

\bibitem{Barkeshli2015physical}
M.~Barkeshli, J.~D. Sau, Physical architecture for a universal topological
  quantum computer based on a network of majorana nanowires, arXiv preprint
  arXiv:1509.07135.

\bibitem{Sagi2018spin}
E.~Sagi, H.~Ebisu, Y.~Tanaka, A.~Stern, Y.~Oreg,
  \href{https://arxiv.org/abs/1806.03304}{Spin liquids from majorana zero modes
  in a cooper box}, arXiv preprint arXiv:1806.03304.
\newline\urlprefix\url{https://arxiv.org/abs/1806.03304}

\bibitem{Chew2018fermionized}
A.~Chew, D.~F. Mross, J.~Alicea,
  \href{https://link.aps.org/doi/10.1103/PhysRevB.98.085143}{Fermionized
  parafermions and symmetry-enriched majorana modes}, Phys. Rev. B 98 (2018)
  085143.
\newblock \href {http://dx.doi.org/10.1103/PhysRevB.98.085143}
  {\path{doi:10.1103/PhysRevB.98.085143}}.
\newline\urlprefix\url{https://link.aps.org/doi/10.1103/PhysRevB.98.085143}

\bibitem{Reeg2017DIII}
C.~Reeg, C.~Schrade, J.~Klinovaja, D.~Loss,
  \href{https://link.aps.org/doi/10.1103/PhysRevB.96.161407}{Diii topological
  superconductivity with emergent time-reversal symmetry}, Phys. Rev. B 96
  (2017) 161407.
\newblock \href {http://dx.doi.org/10.1103/PhysRevB.96.161407}
  {\path{doi:10.1103/PhysRevB.96.161407}}.
\newline\urlprefix\url{https://link.aps.org/doi/10.1103/PhysRevB.96.161407}

\bibitem{manmana2012topological}
S.~R. Manmana, A.~M. Essin, R.~M. Noack, V.~Gurarie,
  \href{https://link.aps.org/doi/10.1103/PhysRevB.86.205119}{Topological
  invariants and interacting one-dimensional fermionic systems}, Phys. Rev. B
  86 (2012) 205119.
\newblock \href {http://dx.doi.org/10.1103/PhysRevB.86.205119}
  {\path{doi:10.1103/PhysRevB.86.205119}}.
\newline\urlprefix\url{https://link.aps.org/doi/10.1103/PhysRevB.86.205119}

\end{thebibliography}

\end{document}